\documentclass[12pt]{article}
\usepackage{chetnew}
\usepackage{amssymb,amsfonts}
\usepackage{textcomp,tikz,braket}
\usepackage{multirow}
\usepackage[all]{xy}
\usepackage{rotating}
\usepackage{stmaryrd}
\usepackage{bbold}
\usepackage[utf8]{inputenc}
\usepackage{slashed}

\usepackage{mathtools}

\def\one{{\,\hbox{1\kern-.8mm l}}}
\def\ceq{:=}

\newcommand{\su}{\mathfrak{su}}
\newcommand{\so}{\mathfrak{so}}

\newcommand{\normord}[1]{:\mathrel{#1}:}

\newcommand{\Tr}{\mathrm{Tr}}
\newcommand{\CB}{\mathcal{B}}
\newcommand{\CQ}{\mathcal{Q}}
\newcommand{\CD}{\mathcal{D}}
\newcommand{\CL}{\mathcal{L}}

\newcommand{\CC}{\mathcal{C}}
\newcommand{\CO}{\mathcal{O}}

\newcommand{\CN}{\mathcal{N}}
\newcommand{\CS}{\mathcal{S}}

\newcommand{\QQ}{\mathcal{Q}}
\newcommand{\QS}{\mathcal{S}}
\newcommand{\bA}{{\bf A}}
\newcommand{\bB}{{\bf B}}
\newcommand{\bC}{{\bf C}}

\newcommand{\PP}{\mathcal P}
\newcommand{\KK}{\mathcal K}

\newcommand{\HH}{\mathcal H}

\newcommand{\II}{\mathcal I}
\newcommand{\bOn}{{\bf 1}}
\newcommand{\bTw}{{\bf 2}}
\newcommand{\bTh}{{\bf 3}}
\newcommand{\bFo}{{\bf 4}}

\newcommand{\U}{\mathrm{U}}
\newcommand{\SU}{\mathrm{SU}}
\newcommand{\SO}{\mathrm{SO}}
\newcommand{\Sp}{\mathrm{Sp}}
\allowdisplaybreaks

\preprint{QMUL-PH-16-04 \\ EFI-16-08}

\title{Aspects of Superconformal Multiplets in ${\rm D}>4$}

\author{Matthew~Buican,$^{1,2\clubsuit}$ Joseph~Hayling$^{1\heartsuit}$ and Constantinos~Papageorgakis$^{1\spadesuit}$
}

\affiliation{$^1$ CRST and School of Physics and Astronomy\\ Queen Mary University of London, E1 4NS, UK\\
  $^2$ Enrico Fermi Institute and Department of Physics\\ The
  University of Chicago, IL 60637, USA\\
\emails{$^{\clubsuit}$m.buican@qmul.ac.uk,$^{\heartsuit}$j.a.hayling@qmul.ac.uk,
  $^{\spadesuit}$c.papageorgakis@qmul.ac.uk}}

\abstract{We explicitly construct and list all unitary superconformal multiplets, along with their index contributions, in five and six dimensions. From this data, we uncover various unifying themes in the representation theory of five- and six-dimensional superconformal field theories.  At the same time, we provide a detailed argument for the complete classification of unitary irreducible representations in five dimensions using a combination of physical and mathematical techniques.}

\date{\today}

\begin{document}

\maketitle

\setcounter{tocdepth}{2}
\tableofcontents

\newpage 

\section{Introduction and Summary}

By virtue of their high degree of symmetry, superconformal field theories (SCFTs) are somewhat simpler arenas in which to test and understand general ideas in quantum field theory (QFT) like duality \cite{Montonen:1977sn,Goddard:1976qe,Witten:1978mh,Seiberg:1994aj,Argyres:2007cn,Gaiotto:2009we} and emergent symmetry \cite{Green:2010da}. Moreover, since the endpoints of supersymmetric (SUSY) renormalisation group (RG) flows are often SCFTs, they constrain the asymptotics of QFT and give rise to striking manifestations of the idea of universality \cite{Seiberg:1994pq,Intriligator:1996ex}.

To construct an SCFT, we start with a superconformal algebra (SCA). In his pioneering work, Nahm showed that these algebras admit a simple classification \cite{Nahm:1977tg}. This list of allowed SCAs gives rise to important constraints even away from criticality.\foot{For example, Nahm's classification shows that six-dimensional $(1,1)$ QFTs do not flow to SCFTs at short distances.} At a superconformal point, we find the basic building blocks of the theory---the multiplets of local operators\foot{We should also supplement these degrees of freedom with the non-local operators of the theory.}---by studying the unitary irreducible representations (UIRs) of the SCA. 

These UIRs are of two general types: short representations and long
representations. Short UIRs have primaries that are annihilated by
certain non-trivial combinations of the Poincar\'e supercharges while
long representations do not. Moreover, short representations can
contribute to the superconformal index
\cite{Romelsberger:2005eg,Kinney:2005ej,Bhattacharya:2008zy}, can
realise non-trivial structures like chiral algebras
\cite{Beem:2014kka,Beem:2013sza} and chiral rings that enjoy various
non-renormalisation properties, can be used to study the structure of
anomalies \cite{Cordova:2015fha}, and can describe the SUSY-preserving
relevant and marginal deformations of SCFTs
\cite{Louis:2015mka,Cordova:2016xhm}. Furthermore, by understanding how short
representations recombine to form long representations one can hope,
when sufficient symmetry is present, to bootstrap non-trivial
correlation functions of local operators and perhaps even whole
theories (see \cite{Beem:2014kka,Beem:2015aoa} for important recent
progress on this front).

In this paper, we perform the conceptually straightforward, but
calculationally nontrivial, task of giving the level-by-level
construction of all UIRs for the five-dimensional $\mathcal N=1$ and
six-dimensional (1,0) and (2,0) SCAs (these are the only allowed SCAs
in five and six dimensions \cite{Nahm:1977tg}). We also calculate the
most general superconformal index associated with these
multiplets. Our approach throughout is based on the presentation and
conventions of \cite{Minwalla:1997ka,Bhattacharya:2008zy}.\foot{Note
  that a comprehensive classification of unitary irreducible
  representations (UIRs) for all SCAs was carried out in
  \cite{Dobrev:1985qv,Dobrev:1985vh,Dobrev:1985qz,Minwalla:1997ka,Dobrev:2002dt,
    Bhattacharya:2008zy} and further discussed in
  \cite{Cordova:2016xhm}. However, in this paper we add to these works
  by giving the level-by-level construction of the corresponding
  multiplets as well as the resulting superconformal index
  contributions. Part of this work was already done in
  \cite{Beem:2014kka} for the 6D $(2,0)$ SCA (but we will provide the
  full set of multiplets and index contributions for this algebra).}

We expect the results assembled
here to be useful for more detailed studies of the many
still-mysterious SCFTs in five and six dimensions (see, e.g., the
theories described in the classic works
\cite{Witten:1995zh,Seiberg:1996bd,Morrison:1996xf,Douglas:1996xp,Intriligator:1997pq}
and the more recent literature
\cite{Heckman:2015bfa,Bhardwaj:2015xxa}) as well as for more general
explorations of the space of SCFTs in these dimensions.

Before delving into technical details, we should note that although
the precise construction of the various multiplets depends on the
spacetime dimension and amount of supersymmetry, we find various
unifying themes in five and six dimensions. For example, we will see
that multiplets containing conserved currents or obeying equations of
motion cannot take part in recombination rules.\foot{Unlike in 4D,
  where one can sometimes tune an exactly marginal parameter and short
  multiplets may recombine into long ones, 5D and 6D superconformal
  theories are necessarily isolated
  \cite{Louis:2015mka,Cordova:2016xhm}. Therefore, recombination
  should be understood purely at the level of superconformal
  representation theory, i.e. how one can write a long multiplet in
  terms of short multiplets.} In particular, we will show that:

\begin{itemize}

\item[${\bf(i)}$] 

Multiplets containing higher-spin currents can never recombine into long multiplets. This statement is compatible with the fact that theories in 5D and 6D are isolated as SUSY theories \cite{Louis:2015mka,Cordova:2016xhm}, because it implies that there are no exactly marginal SUSY deformations of (almost) free theories.\foot{Here we use \cite{Maldacena:2011jn, Maldacena:2012sf, Alba:2015upa}.} This situation is unlike the one in four dimensions, where such pairing up is required in the decoupling limit of $\CN=2$ and $\CN=4$ superconformal gauge theories.

\item[${\bf(ii)}$] 

Multiplets containing conserved spin-two currents can never recombine
into long multiplets. This statement is also compatible with the
isolated nature of 5D and 6D theories \cite{Louis:2015mka,Cordova:2016xhm}, because it implies that there are no marginal SUSY couplings between general isolated SCFTs. In four dimensions, such recombination is required in coupling isolated interacting $\CN=2$ SCFTs (see, e.g., \cite{Gaiotto:2009we, Buican:2014hfa, DelZotto:2015rca} for examples of such couplings between theories and \cite{Buican:2016arp} for a more general discussion).

\item[${\bf(iii)}$] 

  Flavour-symmetry currents\foot{In this paper, we define these to be
    currents for symmetries that commute with the full SCA. Therefore, these currents do not sit in multiplets with higher-spin symmetries. Note that, as in four dimensions, there are spin-one currents that give rise to charges that commute with the $R$ symmetry and also sit in higher-spin multiplets, but the corresponding charges are necessarily part of a larger algebraic structure including supercharges and higher-spin charges. For example, we will see such a current in \eqref{hs5d}. This operator gives rise to a charge that acts on bosons but not on fermions.}
  are present in 5D and 6D (1,0) linear multiplets, which cannot
  recombine into long multiplets (for a discussion in the six-dimensional context, see \cite{CordovaTalk1}). This situation
  is analogous to the one in four-dimensional $\CN=2$ theories.
    
\item[${\bf(iv)}$] Certain classes of multiplets cannot appear in free
  SCFTs.\foot{Throughout this paper we only consider unitary
    theories.} These include certain 6D (1,0) $\CB$-type and
  $\CC$-type multiplets in Sec.~\ref{Sec:6D10} as well as some 6D
  (2,0) $\CB$-type, $\CC$-type and $\CD$-type multiplets in
  Sec.~\ref{Sec:6D20}.

\end{itemize}

The methodology we use to extract our results is rather general and
well established
\cite{Minwalla:1997ka,Dolan:2002zh,Dolan:2005wy,Bhattacharya:2008zy,Beem:2014kka,Cordova:2016xhm}. Indeed,
we use a simple Verma-module construction to obtain all irreducible
representations of the full SCA from irreducible representations of
its maximal compact subalgebra. The UIRs are labelled by highest
weights corresponding to superconformal primaries, from which all
descendants are recovered by the action of momentum operators and
supercharges. Hence, each UIR is uniquely identified by a string of
quantum numbers, which characterises the superconformal primary
state. As we described above, there are both long and short
multiplets. The short multiplets have null states, which can be
consistently deleted (hence the moniker, ``short''). A complete
classification of short UIRs can be obtained by imposing the condition
of unitarity. For special values of the quantum numbers characterising
short UIRs, additional null states can occur. The precise enumeration
and analysis of all such possibilities using unitarity is an intricate
task.

Once all null states have been identified, the
Racah--Speiser (RS) algorithm simplifies the multiplet construction
and clarifies the origin of equations of motion and conservation
equations, whenever these are present.\footnote{A concise summary of
  the Racah--Speiser algorithm can be found in App.~B of
  \cite{Dolan:2002zh}.} The RS algorithm provides a prescription for
the Clebsch--Gordan decomposition of states in representation space.
Since representations of the maximal compact subalgebra are labelled
by highest weights, these take values in the fundamental Weyl chamber
and the corresponding Dynkin labels are positive. After the
Clebsch--Gordan decomposition, a representation in the sum with
negative Dynkin labels lies outside the fundamental Weyl chamber and
can no longer label an irreducible representation. The RS prescription
involves applying successive Weyl reflections, which bounce the weight
vector off the boundaries of the Weyl chamber. Each time a Weyl
reflection is performed, the multiplicity of the representation flips
sign.  Therefore, if a representation is labelled by negative Dynkin
labels, it gets reflected back into the fundamental Weyl chamber up to
a sign. If it is labelled by a weight which lies exactly on the
boundary of the fundamental Weyl chamber, the state has zero
multiplicity and should be removed from the sum. A natural
interpretation for representations with negative multiplicities is in
terms of constraints imposed on operators inside the multiplet
\cite{Dolan:2002zh}.

Since we study the 5D $\CN=1$, 6D $(1,0)$, and 6D $(2,0)$ SCAs, our
presentation is split into three corresponding sections, all of which
are largely self-contained. The reader who is familiar with the
classification of UIRs and only interested in looking up the results
can proceed directly to the relevant tables. Each multiplet is
labelled by the quantum numbers designating its superconformal primary
and the shortening conditions the latter obeys. Some multiplets with
special values for their quantum numbers admit a distinct physical
interpretation; these are dealt with separately. For those interested
in the approach employed to obtain our diagrams, we provide a detailed
discussion for the case of the 5D $\mathcal N =1$ SCA in
Sec.~\ref{Sec:5DN1}, which extends naturally to 6D in
Sec.~\ref{Sec:6D10} and Sec.~\ref{Sec:6D20}. Throughout this analysis,
special emphasis is put on identifying operator constraints, whenever
present. Each section also contains expressions for recombination
rules and indices for the superconformal multiplets under study. A
short collection of simple applications arising from the results of
our analysis is presented in Sec.~\ref{Apps}. Finally,
Sec.~\ref{App:5Dproof} contains an argument for the complete
classification of 5D UIRs, which has been missing from the literature
(paying attention to some recent observations made in
\cite{Penedones:2015aga,Yamazaki:2016vqi,Oshima:2016gqy}).

We also include various appendices. App.~\ref{App:5DSCA},
\ref{superchar} and \ref{App:5DRS} contain conventions and results
which are necessary for our multiplet construction but would shift the
focus away from our aim in the main part of the text; SCA conventions
for five and six dimensions, the construction of supercharacters and
the superconformal index, as well as the relationship between the RS
algorithm and the identification of operator constraints.
App.~\ref{App:Index} collects the superconformal indices for all 5D
and 6D multiplets for quick reference. As the 6D (2,0) refined indices
are cumbersome, we only ever write down their Schur limit. However, we
also provide a complementary Mathematica file with all the refined
superconformal indices in five and six dimensions. Finally,
App.~\ref{(2,0) spectra} contains the explicit 6D (2,0) spectra, which
are too unwieldy to present in Sec.~\ref{Sec:6D20}.

\subsection*{Note added:}

{\it While finalising our construction of multiplets in ${\rm{\it D}}>4$,
  we became aware of an upcoming publication (\cite{Cordova:2016}
  cited in \cite{Cordova:2016xhm}). This upcoming work promises to be
  broader in scope than our own and have overlap with some of our
  constructions. Knowledge of the multiplet structure in five and six
  dimensional SCFTs, in \cite{Cordova:2016}, is essential background
  material for the results in \cite{Cordova:2015fha,Cordova:2016xhm};
  see also
  \cite{CordovaTalk1,IntriligatorTalk,DumitrescuTalk,CordovaTalk2,CordovaTalk3}.
 We would like to thank C. C\'ordova, T. Dumitrescu, and K. Intriligator for relevant correspondence, as well as for pointing out an error in the classification of 5D multiplets in a previous version of this paper.}

\section{Multiplets and Superconformal Indices for 5D
  $\mathbf{\mathcal{N}=1}$}\label{Sec:5DN1}

We begin by providing a systematic analysis of all short multiplets
admitted by the 5D $\mathcal{N}=1$ SCA, $F(4)$. This involves a
derivation of the superconformal unitarity bounds. By doing so we
reproduce the results of \cite{Bhattacharya:2008zy}. We then proceed
to write the complete multiplet spectra and compute their indices. Our
notation and conventions for the 5D SCA are provided in
App.~\ref{App:5DSCA}.

\subsection{UIR Building with Auxiliary Verma Modules}\label{suprep5D}

The superconformal primaries of the 5D SCA $F(4)$ are designated
$\ket{\Delta;l_1,l_2;k}$, where $\Delta $ is the conformal dimension,
$l_1\ge l_2>0$ are Lorentz symmetry quantum numbers in the orthogonal
basis and $k$ is an $R$-symmetry label. Each primary is in one-to-one
correspondence with a highest weight state of the maximal compact
subalgebra
$\mathfrak{so}(5)\oplus\mathfrak{so}(2)\oplus\mathfrak{su}(2)_R\subset
F(4)$.\footnote{The
  quantum numbers labelling the primary are eigenvalues for the
  Cartans of the maximal compact subalgebra in a particular basis.}
There are eight Poincar\'e and eight superconformal supercharges, denoted by
$\QQ_{ {\bf A} a }$ and $\QS_{ {\bf A} a}$ respectively---where
$a = 1,\cdots, 4$ is an $\mathfrak{so}(5)$ Lorentz spinor index and
$\bf A=1,2$ an index of $\mathfrak{su}(2)_R$. One also has five
momenta $\PP_\mu$ and special conformal generators $\mathcal K_\mu$,
where $\mu = 1,\cdots, 5$ is a Lorentz vector index. The
superconformal primary is annihilated by all $\QS_{ {\bf A}a}$ and
$\mathcal K_\mu$.  A basis for the representation space of $F(4)$ can
be constructed by considering the following Verma module
\begin{align}\label{supconfbasis1}
  \prod_{{\bf A}, a}(\QQ_{{\bf A} a})^{n_{{\bf A},a}}\prod_{\mu} \mathcal{P}_{\mu}^{\hphantom{\mu}n_\mu}
  \ket{\Delta;l_1,l_2;k}^{hw}
\end{align}
for some ordering of operators,\footnote{Any other ordering can be
  obtained using the superconformal algebra.} where
$n=\sum_{{\bf A} a} n_{{\bf A},a}$ and $\hat{n}=\sum_{\mu} n_{\mu}$
denote the ``level'' of a superconformal or conformal descendant
respectively. In order to obtain UIRs, the requirement of unitarity
needs to be imposed level-by-level on the Verma module. This leads to
bounds on the conformal dimension $\Delta$.

The highest $\mathfrak{su}(2)_R$-weight level-one superconformal-descendant states can be expressed in a particularly suitable
alternative basis as $\Lambda^a_\bOn \ket{\Delta;l_1,l_2;k}^{hw}$,
where we define
\begin{align}\label{biglambda}
\Lambda_{\bOn}^a := \sum_{b=1}^{a} \QQ_{\bOn b} \lambda^a_{b}\;.
\end{align}
The $\lambda^a_b$ are functions of the $\mathfrak{so}(5)$ quantum
numbers and Lorentz lowering operators. The combinations
\eqref{biglambda} have the property that, when acting upon a
conformal-primary highest-weight state of
$\mathfrak{so}(5)\oplus\mathfrak{so}(2)$, they produce another
conformal-primary highest-weight state. They can be uniquely
determined by imposing the requirement that all Lorentz raising
operators and $R$-symmetry raising operators annihilate
$\Lambda^a_{\bf A} \ket{\Delta;l_1,l_2;k}^{hw}$ and are given in
App.~\ref{App:Ltilde}. It turns out that the most stringent unitarity
bounds emerge by studying the norms of states constructed by acting
with the $\Lambda_\bOn^a$s on the superconformal primary.  We provide
a detailed argument in favour of this fact in Sec.~\ref{App:5Dproof}.

\subsection*{Ill-defined States}

The definition of the $\Lambda$ generators---as explicitly given in
App.~\ref{App:Ltilde}---is such that for certain values of the quantum
numbers the resulting state is not well defined. Consider e.g. the
state
\begin{align}\label{illdef}
  \Lambda_{\bOn}^2 |\Delta; l_1, l_2;k\rangle = \Big( \QQ_{\bOn 1
  }\lambda_1^1 + \QQ_{\bOn 2 }\lambda_1^2 \Big)  |\Delta; l_1,
  l_2;k\rangle= \Big( \QQ_{\bOn 1 } -\QQ_{\bOn 2 }\mathcal M_2^-
  \frac{1}{2 \mathcal H_2} \Big)  |\Delta; l_1, l_2;k\rangle \;,
\end{align}
where $\mathcal H_2 |\Delta; l_1, l_2;k\rangle = l_2 |\Delta; l_1, l_2;k\rangle $. The above is clearly ill-defined for $l_2 = 0$: Although $\mathcal M_2^-|\Delta; l_1,0;k\rangle = 0$ the factor of $1/l_2$ diverges and the norm of  \eqref{illdef} is indeterminate.  However, there exist cases where products of ill-defined $\Lambda$s can lead to well-defined states, through various cancellations. Hence, one has to perform a delicate analysis of such possibilities through explicit calculation.

This phenomenon will be very important in the classification of
unitarity bounds below, where one needs to evaluate the
norms of all well-defined, distinct (i.e. not related through
commutation relations) products of $\Lambda$s.\footnote{For example,
  in our upcoming discussion of unitarity bounds for $l_1 = l_2 =0$,
  $\Lambda_\bOn^3 \ket{\Delta ; 0,0;k}^{hw}$ and
  $\Lambda_\bOn^2 \ket{\Delta ; 0,0;k}^{hw}$ are individually ill
  defined, while
  $\Lambda_\bOn^2\Lambda_\bOn^3 \ket{\Delta ; 0,0;k}^{hw}$ is
  not. This can in turn lead to the wrong identification of shortening
  conditions, since
  $||\Lambda_\bOn^2\Lambda_\bOn^3 \ket{\Delta ; 0,0;k}^{hw}||^2=
  B^3(0,0,k)B^1(0,0,k)$, whereas
  $||\Lambda_\bOn^1\Lambda_\bOn^2 \ket{\Delta ; 0,0;k}^{hw}||^2=
  B^2(0,0,k)B^1(0,0,k)$, with the first one leading to more stringent
  unitarity bounds.}

\subsubsection*{Unitarity Bounds for $l_1\ge l_2 >0$}

We can calculate the norms of the superconformal descendant
states at level one to be
 \begin{align}
|| \Lambda^4_\bOn \ket{\Delta;l_1,l_2;k}^{hw}  ||^2 &= \left(\Delta -3 k-l_1-l_2-4\right)\frac{\left(2 l_1+3\right) \left(l_1+l_2+2\right) \left(2 l_2+1\right) }{4 \left(l_1+1\right) l_2 \left(l_1+l_2+1\right)}\;,\cr
|| \Lambda^3_\bOn \ket{\Delta;l_1,l_2;k}^{hw}  ||^2 &= \left(\Delta -3 k-l_1+l_2-3\right)\frac{\left(2 l_1+3\right) \left(l_1-l_2+1\right)}{2 \left(l_1+1\right) \left(l_1-l_2\right)}   \;,\cr
 || \Lambda^2_\bOn \ket{\Delta;l_1,l_2;k}^{hw}  ||^2 &= (\Delta -3 k+l_1-l_2-1)\frac{(2l_2+1)}{2l_2}\;,\cr
|| \Lambda^1_\bOn \ket{\Delta;l_1,l_2;k}^{hw}  ||^2 &=(\Delta -3 k+l_1+l_2)\;,
 \end{align}
 where we have normalised
 $\left\| \ket{\Delta;l_1,l_2;k}^{hw}\right\|^2=1$.  Observe that these
 norms are all of the form
\begin{align}
\left\| \Lambda^{a}_\bOn\ket{\Delta;l_1,l_2;k}^{hw}\right\|^2 = \Big(\Delta - f^a(l_1, l_2, k)\Big) g^a(l_1, l_2) =: B^a (l_1,l_2,k)g^a(l_1, l_2)\;,
\end{align}
where $g^a(l_1, l_2)$ is a positive-definite rational function in the
fundamental Weyl chamber, $l_1\ge l_2>0$.  Unitarity demands that the
norms are positive semidefinite and this imposes a bound on the
conformal dimension via the functions $B^a(l_1,l_2,k)$. The strongest
bound on the conformal dimension is provided by $B^4(l_1,l_2,k)\ge
0$. When $B^4(l_1, l_2, k) > 0$ the UIR can be obtained using
\eqref{supconfbasis1}. The resulting multiplet is called ``long'' and
labelled $\mathcal L$. 

When $B^4(l_1, l_2, k) = 0$ the state is null. This means that the
primary obeys the ``shortening condition''
$\Lambda_\bOn^4 \ket{\Delta;l_1,l_2;k}^{hw} = 0$. All such states can be
consistently removed from the superconformal representation. The
resulting multiplet is ``short'' and labelled as type $\mathcal A$.
Since it can be reached from a long multiplet by continuously dialling
$\Delta$ it is called a ``regular'' short multiplet. At higher levels,
$\prod_{a=1}^n \Lambda_\bOn^a 
  \ket{\Delta;l_1,l_2;k}^{hw}$
with $n>1$, the norms involve products of $B^a(l_1,l_2,k)$s and the
strongest bound still comes from $B^4(l_1,l_2,k)\ge 0$. Therefore,
there will be no change to the bounds obtained at level one.

\subsubsection*{Unitarity Bounds for $l_1> l_2 =0$}

We now turn to the special case with $l_1>0$, $l_2 = 0$, where the concept of ill-defined states becomes important. When $l_2 = 0$ the operator $\Lambda_\bOn^4$ is not well defined and we have to omit the level-one state $\Lambda_\bOn^4 \ket{\Delta ; l_1,0;k}^{hw}$ from our spectrum. Naively, the strongest bound then arises from the norm of the state $\Lambda_\bOn^3 \ket{\Delta ; l_1,0;k}^{hw}$. However, the level-two state $\Lambda_\bOn^3 \Lambda_\bOn^4 \ket{\Delta ; l_1,0;k}^{hw}$ is actually well defined, as can be explicitly checked. Its norm is proportional to
\begin{align}
\left\| \Lambda_\bOn^3 \Lambda_\bOn^4\ket{\Delta;l_1,0;k}^{hw}\right\|^2 \propto
  \left(\Delta -3 k-l_1-4\right) \left(\Delta -3 k-l_1-3\right) = B^3(l_1,0,k) B^4(l_1,0,k)
\end{align}
and the corresponding set of restrictions come from
$B^4 (l_1,0,k)\ge 0$ or $B^3 (l_1,0,k)=0$.

When $B^4(l_1,0,k) = 0$ one recovers a regular short representation of
type $\mathcal A$. Instead, one could also have $B^{3}(l_1, 0,k) = 0$;
this gives rise to the null state
$\Lambda_\bOn^3 \ket{\Delta ; l_1,0;k}^{hw}$.  Making that choice
leads to an ``isolated'' short multiplet of type
$\mathcal B$.\footnote{The name isolated is due to the fact that there
  are no states in the gap between $B^{3}(l_1, 0,k) = 0$ and
  $B^{4}(l_1, 0,k) \ge 0$.}

\subsubsection*{Unitarity Bounds for $l_1=l_2 =0$}

The same logic extends to $l_1 = l_2 = 0$: At level one the only
well-defined state is
$\Lambda_\bOn^1|\Delta;0,0;k\rangle^{hw}$. However, there exist
well-defined states at levels two and four, obtained by
$\Lambda_\bOn^2\Lambda_\bOn^3 \ket{\Delta ; 0,0;k}^{hw}$,
$\Lambda_\bOn^1 \Lambda_\bOn^2 \Lambda_\bOn^3
\Lambda_\bOn^4\ket{\Delta ; 0,0;k}^{hw}$.
These give rise to the conditions $B^1(0,0,k)= 0$,
$B^3(0,0,k)= 0$ or $B^4(0,0,k)\ge 0$ and lead to the new set of
isolated short multiplets $\mathcal D$.  We summarise their properties
and list all short multiplets for the 5D SCA in Table~\ref{Tab:5DN1}.

\subsubsection*{Additional Unitarity Bounds }

Finally, there are supplementary unitarity restrictions and associated
null states originating from conformal descendants. These have been
analysed in detail in \cite{Minwalla:1997ka,Dolan:2005wy}, the results
of which we use. Saturating a conformal bound results in a
``momentum-null'' state, where the corresponding shortening condition
is an operator constraint involving momentum analogues of the
superconformal $\Lambda$s \cite{Dolan:2005wy}. In that reference, a
prescription is given for removing the associated states,
$\PP_\mu \ket{\Delta;l_1,l_2;k}^{hw}$, from the auxiliary Verma-module
construction, again in analogy with the superconformal
procedure.\footnote{Note that this does not mean that all conformal
  descendants of a particular type should be removed from the set of
  local operators.}  However, we will choose not to exclude any
momenta from the basis of Verma-module generators
\eqref{supconfbasis1}. After using the RS algorithm this choice will
allow us to explicitly recover highest weight states corresponding to
the operator constraints from the general multiplet structure.

One can combine the conformal and superconformal bounds to predict
that operator constraints will appear in the following short
multiplets: 
\begin{align}\label{opcon5D}
\mathcal{B}[d_1,0;0]\;,\hspace{3mm}\mathcal{D}[0,0;\{1,2\}]\;.
\end{align}
The multiplet $\mathcal{D}[0,0;0]$ does not belong to this list as it
is the vacuum.

 \begin{table}[t]
\begin{center}
\begin{tabular}{|c|r|l|}
\hline
Multiplet & Shortening Condition & Conformal Dimension \\
\hline
\hline
$\mathcal{A}[l_1,l_2;k]$ & $\Lambda_\bOn^4\Psi=0$ & $\Delta = 3k +l_1 + l_2
                                               +4$ \\
$\mathcal{A}[l_1,0;k]$ & $\Lambda_\bOn^3 \Lambda_\bOn^4\Psi=0$ & $\Delta = 3k +l_1 +4$ \\
$\mathcal{A}[0,0;k]$ & $\Lambda_\bOn^1 \Lambda_\bOn^2 \Lambda_\bOn^3 \Lambda_\bOn^4\Psi=0$ & $\Delta = 3k  +4$\\
\hline
\hline
$\mathcal{B}[l_1,0;k]$ & $\Lambda_\bOn^3\Psi=0$ & $\Delta = 3k +l_1 +3$\\
$\mathcal{B}[0,0;k]$ & $ \Lambda_\bOn^2 \Lambda_\bOn^3\Psi=0$ & $\Delta
                                                               = 3k
                                                               +3$\\
\hline
\hline
$\mathcal{D}[0,0;k]$ & $\Lambda_\bOn^1\Psi=0$ & $\Delta =3k  $\\
\hline

\end{tabular}
\end{center}
\caption{\label{Tab:5DN1} A list of all short multiplets for the 5D
  $\mathcal N = 1$ SCA, along with the conformal dimension of the
  superconformal primary and the corresponding shortening condition. The $\Lambda_\bOn^a$ in the shortening
  conditions are defined in \eqref{biglambda} and
  \eqref{littlelambdaApp}. The first of these multiplets
  ($\mathcal A$) is a
  regular short representation, whereas  the rest ($\mathcal B,
  \mathcal D$) are isolated short representations. Here $\Psi$ denotes
  the superconformal primary state for each multiplet.}
\end{table}

\subsubsection*{Highest Weight Construction through the Auxiliary Verma Module}

The $\Lambda$ basis through which we obtained the unitarity bounds could in principle be used to construct the full multiplet. However, executing this procedure would require knowledge of the full Clebsh--Gordan decomposition for the resulting states. This is a very difficult task to carry out in practice. For that reason we will resort to constructing the highest weights of the superconformal representation using the auxiliary Verma module via the Racah--Speiser algorithm. This greatly simplifies the Glebsch--Gordan decomposition by implementing it at the level of highest weights.

Having motivated the use of the auxiliary Verma module basis, we will implement the conjectural recipe of \cite{Dolan:2002zh,Bianchi:2006ti,Bhattacharya:2008zy} to generate the spectrum; see also App.~C of \cite{Beem:2014kka}. According to these references, in addition to removing the supercharge associated with the shortening condition, one is instructed to also remove any other supercharge combination that annihilates the auxiliary primary. 

To make this point more transparent, let us consider the example of the $\mathcal B[l_1,0;k]$ multiplet. The shortening condition dictates that we remove $\Lambda^{\bOn}_3 \Psi \Rightarrow \QQ_{\bOn 3} \Psi_{\rm aux}$, where note that $\Psi$ and $\Psi_{\rm aux}$ have the same quantum numbers. Since $l_2 = 0$ we have that $\mathcal M_2^- \Psi_{\rm aux} =0 $ and therefore 
\begin{align}\label{Q13Q14}
0 =   \mathcal M_2^- \QQ_{\bOn 3}\Psi_{\rm aux} =\QQ_{\bOn 4}\Psi_{\rm aux}\;.
\end{align}
Hence we are required to also remove $\QQ_{\bOn 4}$ from the auxiliary Verma-module basis. Note that \eqref{Q13Q14} does {\it not} imply $\Lambda^{\bOn}_4\Psi=0$ but is merely a prescription for obtaining the correct set of highest weights. One could similarly use any lowering operator of the maximal compact subalgebra. E.g. for $k=0$ additional conditions can be generated by acting on the existing ones with $R$-symmetry lowering operators,\footnote{See e.g. the discussion in App.~6.2.1 of \cite{Beem:2014kka} in the context of the 6D (2,0) SCA.} resulting in the removal of more combinations of supercharges from the set of auxiliary Verma-module generators. We will mention explicitly the full set of such ``absent supercharges'' at the beginning of each case in our upcoming analysis.

We emphasise that this is an {\it auxiliary} Verma-module construction which leads to the same spectrum in terms of highest weights. If one is interested in the precise form of the operators, the much more involved $\Lambda$-basis should be used.

\subsection{The Procedure}

Based on the above ingredients, let us summarise our strategy for
constructing the superconformal UIRs:

\begin{enumerate}

\item For a given multiplet type, begin with a superconformal primary and consider the highest-weight component of the corresponding irreducible representation of the Lorentz and $R$-symmetry algebras.

\item Implement the conjecture of
  \cite{Dolan:2002zh,Bianchi:2006ti,Bhattacharya:2008zy,Beem:2014kka} to determine
  all combinations of supercharges which need to be removed from the
  auxiliary Verma-module basis \eqref{supconfbasis1}.

\item Use the remaining auxiliary Verma-module generators to determine the
  highest weights for all descendant states. This may result in some
  of the quantum numbers labelling the highest weight state becoming
  negative.

\item Apply the Racah--Speiser algorithm to recover a spectrum with only
  positive quantum numbers.\footnote{The details of the Racah--Speiser
    algorithm needed for this step can be found in
    App.~\ref{App:5DRS}.}  This could result in some states being
  projected out, while others acquiring a ``negative
  multiplicity''. The latter can cancel out against other states with
  the same quantum numbers but positive multiplicity.\footnote{There
    are instances when this general procedure leads to ambiguities,
    i.e. there is more than one choice for performing the
    cancellations; see also \cite{Cordova:2016xhm}. However, these can
    be resolved uniquely by the requirement that all highest weight
    states should be reached by the successive action of allowed
    supercharges on the superconformal primary. For the examples that
    we investigate in this article, this phenomenon only appears in
    the (2,0) SCA ($\mathcal B[c_1, c_2,0;0,0]$,
    $\mathcal C[c_1, 0,0;0,0]$), in which case the multipet spectra
    have also been compared to the construction by successive
    $\QQ$-actions.} Any remaining states with negative multiplicity
  can be interpreted as operator constraints. This conjectural identification
  follows \cite{Dolan:2002zh} and is based on a large number of
  examples, but can be additionally supported using supercharacters;
   c.f. App.~\ref{conserveq}.

 \item In some special cases, the supercharges that have been removed
   from the auxiliary Verma-module basis anticommute into momentum
   generators, which should also be removed. This has the effect of
   projecting out states corresponding to operator constraints.  The
   operator constraints can be restored using the discussion in
   App.~\ref{App:5DRS}.

\end{enumerate}

The spectrum of a given superconformal multiplet can always be
obtained following these steps and we believe the results (in all
$\mathrm{D}>4$ SCAs that we have considered) to be correct: They
satisfy the expected recombination rules, which have been checked
using supercharacters. Of the multiplets that do not appear in
recombination rules, for 5D $\mathcal N = 1$ and 6D $(1,0)$ SCAs, we
have also explicitly constructed the states using free fields. For 6D
$(2,0)$ the multiplets that do not appear have had the Schur limit of
their superconformal indices matched with the results of
\cite{Beem:2014kka}. Finally, an additional check on the computational
implementation of this procedure is via supercharacters, which can be
calculated in two ways: Either by evaluating them over the states of a
given short multiplet obtained using the RS algorithm or directly
using the Weyl character formula; c.f. App.~\ref{superchar}. Both of
these agree for all multiplets listed in our work.

\subsection{5D $\mathcal N=1 $ Multiplet Recombination Rules}

For the purposes of listing the recombination rules as well as for
explicitly constructing the multiplets, we will find it more
convenient to switch to the Dynkin basis for the various quantum
numbers. That is, we will use
\begin{align}\label{dynkin}
d_1 = l_1 -l_2\;,\hspace{3mm}d_2 = 2l_2\;,\hspace{3mm}K =2k\;.
\end{align}
Short multiplets can recombine to form long multiplets when the
conformal dimension for the latter approaches the unitarity bound,
that is when $\Delta + \epsilon \to \frac{3}{2}K +d_1 + d_2 +4$. It
can then be checked, using the results that we will present in the
following sections, that
\begin{align}\label{recomb rule}
\mathcal{L}[\Delta + \epsilon;d_1,d_2;K] &\xrightarrow{\epsilon\rightarrow 0} \mathcal{A}[d_1,d_2;K]\oplus \mathcal{A}[d_1,d_2-1;K+1]\;,\cr
\mathcal{L}[\Delta + \epsilon;d_1,0;K] &\xrightarrow{\epsilon\rightarrow 0} \mathcal{A}[d_1,0;K]\oplus \mathcal{B}[d_1-1,0;K+2]\;,\cr
\mathcal{L}[\Delta + \epsilon;0,0;K] &\xrightarrow{\epsilon\rightarrow 0} \mathcal{A}[0,0;K]\oplus \mathcal{D}[0,0;K+4]\;.
\end{align}
The following multiplets do not appear in a recombination rule:
\begin{align}
    &\mathcal{B}\left[d_1,0;\{0,1\}\right]\;,\cr
    &\mathcal{D}\left[0,0;\{0,1,2,3\}\right]\;.
\end{align}

\subsection{The 5D $\mathcal N =1$ Superconformal Index}\label{5DIndMain}

We define the superconformal index with respect to the supercharge
$\QQ_{\bOn 4}$, in accordance with \cite{Bhattacharya:2008zy,Kim:2012gu}. This is
given by
\begin{align}
  \label{eq:Main18}
  \II(x,y)=\Tr_{\HH}(-1)^F e^{-\beta \delta}  x^{\frac{2}{3}\Delta +\frac{1}{3}(d_1+d_2)}y^{d_1}\;,
\end{align}
where making use of the spin-statistics theorem the fermion number
is $F=d_2$ and the trace is over the Hilbert space of
operators of the theory. The states that are counted by this index
satisfy $\delta = 0$, where
\begin{align}
  \label{eq:Main17}
  \delta :=\{ \QQ_{\bOn 4},\QS_{\bTw 1}\}=\Delta - \frac{3}{2}K -d_1-d_2\;.
\end{align}
It is easy to see that as a result long multiplets can never
contribute to the index.  The charges $d_1$ and
$\frac{2}{3}\Delta +\frac{1}{3}(d_1+d_2)$ appearing in the exponents
of \eqref{eq:Main18} are eigenvalues for the generators commuting with
$\QQ_{\bOn 4}, \QS_{\bTw 1}$ and consequently with $\delta$.  In
practice, this index can be explicitly evaluated as a supercharacter
for each of the multiplets constructed below. A detailed construction
of characters for superconformal representations is reviewed in
App.~\ref{superchar}.

\subsection{Long Multiplets}\label{sec long mults 5d}

We can now go ahead with the explicit construction of multiplets.
Since long multiplets are not associated with any shortening
conditions, we can proceed as per \eqref{supconfbasis1} acting with
all supercharges and momenta on the superconformal primary to obtain
the unitary superconformal representation.

We will choose to group the supercharges together as
$Q=( \QQ_{\bA 1}, \QQ_{\bA 2})$ and
$\tilde{Q}=(\QQ_{\bA 3},\QQ_{\bA 4})$, purely for book-keeping
purposes. The explicit quantum numbers of these supercharges are given
by
\begin{align}\label{QQbars}
\QQ_{\bOn 1} &\sim (1)_{(0,1)}\;,\hspace{2mm}&&\QQ_{\bOn 2}\sim (1)_{(1,-1)}\;,\hspace{3mm}&&&\QQ_{\bOn 3}\sim (1)_{(-1,1)}\;,\hspace{3mm}&&&&\QQ_{\bOn 4}\sim (1)_{(0,-1)}\;,\nonumber\\
\QQ_{\bTw 1} &\sim (-1)_{(0,1)}\;,\hspace{2mm}&&\QQ_{\bTw 2}\sim (-1)_{(1,-1)}\;,\hspace{3mm}&&&\QQ_{\bTw 3}\sim (-1)_{(-1,1)}\;,\hspace{3mm}&&&&\QQ_{\bTw 4}\sim (-1)_{(0,-1)}\;.
\end{align}
With this information in hand, it is straightforward to map out their
action starting from a superconformal primary, labelled by
$(K)_{(d_1,d_2)}$:
\begin{align}\label{q actions long}
(K)_{(d_1,d_2)}&\xrightarrow{Q^{\phantom{2}}} (K+1)_{(d_1,d_2+1),(d_1+1,d_2-1)}\;,\;(K-1)_{(d_1,d_2+1),(d_1+1,d_2-1)}\;,\cr
&\xrightarrow{Q^2}(K+2)_{(d_1+1,d_2)}\;,\;(K)_{(d_1,d_2+2),(d_1+1,d_2)^2,(d_1+2,d_2-2)}\;,\;(K-2)_{(d_1+1,d_2)} \;,\cr
&\xrightarrow{Q^3}(K+1)_{(d_1+1,d_2+1),(d_1+2,d_2-1)}\;,\;(K-1)_{(d_1+1,d_2+1),(d_1+2,d_2-1)}\;,\cr
&\xrightarrow{Q^4}(K)_{(d_1+2,d_2)}\;,\cr
(K)_{(d_1,d_2)}&\xrightarrow{\tilde{Q}^{\phantom{2}}} (K+1)_{(d_1,d_2-1),(d_1-1,d_2+1)}\;,\;(K-1)_{(d_1,d_2-1),(d_1-1,d_2+1)}\;,\cr
&\xrightarrow{\tilde{Q}^2}(K+2)_{(d_1-1,d_2)}\;,\;(K)_{(d_1,d_2-2),(d_1-1,d_2)^2,(d_1-2,d_2+2)}\;,\;(K-2)_{(d_1-1,d_2)} \;,\cr
&\xrightarrow{\tilde{Q}^3}(K+1)_{(d_1-1,d_2-1),(d_1-2,d_2+1)}\;,\;(K-1)_{(d_1-1,d_2-1),(d_1-2,d_2+1)}\;,\cr
&\xrightarrow{\tilde{Q}^4}(K)_{(d_1-2,d_2)}\;,
\end{align}
where we have split the actions of $Q$ and $\tilde{Q}$ into two
``chains''. Since the Dynkin labels are generic, there is no need to
implement the RS algorithm. By definiton, these multiplets
do not contribute to the superconformal index.

\subsection{$\mathcal{A}$-type Multiplets}\label{A type multiplets 5d}

Recall from Table~\ref{Tab:5DN1} that $\mathcal{A}$-type multiplets
obey three types of shortening conditions depending on the quantum
numbers of the superconformal primary. These result in the removal of
the following combinations of supercharges from the basis of
auxiliary Verma-module generators \eqref{supconfbasis1}:\footnote{Once again, we
  are using a Dynkin basis for the quantum numbers.}
\begin{align}\label{Amults}
\mathcal{A}[d_1,d_2;K]:\hspace{3mm}&\QQ_{\bOn 4}\;,\cr
\mathcal{A}[d_1,0;K]:\hspace{3mm}&\QQ_{\bOn 3}\QQ_{\bOn 4}\;,\cr
\mathcal{A}[0,0;K]:\hspace{3mm}&\QQ_{\bOn 1}\QQ_{\bOn 2}\QQ_{\bOn 3}\QQ_{\bOn 4}\;.
\end{align} 

Let us consider the first case. On the one hand, acting with the allowed set of
supercharges yields the same result as found in \eqref{q actions long}
for the $Q$-supercharge set. On the other, for the $\tilde{Q}$-supercharge set we have
\begin{align}\label{tildQA}
(K)_{(d_1,d_2)}&\xrightarrow{\tilde{Q}^{\phantom{2}}} (K+1)_{(d_1-1,d_2+1)}\;,\;(K-1)_{(d_1,d_2-1),(d_1-1,d_2+1)}\;,\cr
&\xrightarrow{\tilde{Q}^2}(K)_{(d_1-2,d_2+2),(d_1-1,d_2)}\;,\;(K-2)_{(d_1-1,d_2)} \;,\cr
&\xrightarrow{\tilde{Q}^3}(K-1)_{(d_1-2,d_2+1)}\;.
\end{align}
The two remaining cases with $d_2=0$ and $d_1 = d_2 = 0$ can be
obtained by implementing the recipe at the end of Sec.~\ref{suprep5D},
by e.g. first constructing all the states using the combinations of
the $Q$-chain of Eq.~\eqref{q actions long} and $\tilde Q$-chain of
\eqref{tildQA} and then setting $d_2 = 0$. Applying the RS algorithm
will produce some negative-multiplicity states, all of which cancel
with positive-multiplicity states with the same quantum numbers. The
remaining states comprise the spectrum of the $\mathcal{A}[d_1,0;K]$
multiplet and one can proceed analogously for $\mathcal{A}[0,0;K]$.

For $K = 0$ and $d_1 , d_2$ generic the primary still lies above the unitarity bound for all conformal descendants and there are no momentum-null states. Moreover, one also needs to remove $\QQ_{\bTw 4}$, $\QQ_{\bTw 3}\QQ_{\bTw 4}$ and $\QQ_{\bTw 1}\QQ_{\bTw 2}\QQ_{\bTw 3}\QQ_{\bTw 4} $ from the construction of the respective multiplets for $d_1,d_2>0$, $d_2 =0$ and $d_1 = d_2 = 0$ using the auxiliary Verma module. The resulting spectrum is no different from setting $K=0$ and running the RS algorithm for $\mathfrak{su}(2)_R$.

The index over the spectrum of all $\mathcal A$-type multiplets is
given by
\begin{align}\label{5DAindex}
\mathcal{I}_{\mathcal{A}[d_1,d_2;K]}(x,y)=(-1)^{d_2+1}\frac{x^{d_1+d_2+K+4}}{\left(1-x y^{-1}\right) \left(1-x y\right)}\chi_{d_1}(y)\;,
\end{align}
by appropriately tuning $d_1,d_2$ and $K$, including $d_2 = 0$ and
$d_1 = d_2 = 0$, where we have used the $\mathfrak{su}(2)$ character for the
spin-$\frac{l}{2}$ representation
\begin{align}
\chi_{l}(y)=\frac{y^{l+1}-y^{-l-1}}{y-y^{-1}}\;.
\end{align}
One readily sees that \eqref{5DAindex} is compatible with the recombination rule \eqref{recomb rule}:
\begin{align}
\lim_{\epsilon\rightarrow 0}\mathcal{I}_{\mathcal{L}[\Delta+\epsilon;d_1,d_2;K]}(x,y)&=\mathcal{I}_{\mathcal{A}[d_1,d_2;K]}(x,y)+\mathcal{I}_{\mathcal{A}[d_1,d_2-1;K+1]}(x,y)=0\;.
\end{align}

\subsection{$\mathcal{B}$-type Multiplets}\label{5DBtype}

For $\mathcal{B}$-type multiplets, the supercharges that need to be
removed from the auxiliary Verma-module basis \eqref{supconfbasis1} due to null states are
\begin{align}\label{Bmults}\begin{split}
\mathcal{B}[d_1,0;K]:\hspace{3mm}&\QQ_{\bOn 3}\;,\\
\mathcal{B}[0,0;K]:\hspace{3mm}&\QQ_{\bOn 2}\QQ_{\bOn 3}\;.
\end{split}
\end{align} 

For $\mathcal B[d_1,0;K]$, following the argument in \eqref{Q13Q14}, one finds that $\QQ_{\bOn 4}$ should also be removed from the basis of auxiliary Verma-module generators. Acting on the primary---while keeping the quantum numbers generic---one has the same action as \eqref{q actions long} for $Q$, while the $\tilde{Q}$-chain is
\begin{align}
(K)_{(d_1,d_2)}&\xrightarrow{\tilde{Q}^{\phantom{2}}} (K-1)_{(d_1,d_2-1),(d_1-1,d_2+1)}\;,\cr
&\xrightarrow{\tilde{Q}^2}(K-2)_{(d_1-1,d_2)}\;.
\end{align}
We may then combine the action of these supercharges to construct the
following grid; acting with $Q$ is captured by
\emph{southwest} motion on the diagram, while $\tilde{Q}$ by
\emph{southeast} motion. This module is well defined for all
$d_1\geq 1$ values:
\begin{equation}
{\scriptsize\begin{tikzpicture}
\draw [dashed, ultra thin] (0,0) -- (-6,-6);
\draw [dashed, ultra thin] (0,0) -- (3,-3);
\draw [dashed, ultra thin] (-1.5,-1.5) -- (1.5,-4.5);
\draw [dashed, ultra thin] (-3,-3) -- (0,-6);
\draw [dashed, ultra thin] (1.5,-1.5) -- (-4.5,-7.5);
\draw [dashed, ultra thin] (3,-3) -- (-3,-9);
\draw [dashed, ultra thin] (-4.5,-4.5) -- (-1.5,-7.5);
\draw [dashed, ultra thin] (-6,-6) -- (-3,-9);
\node at (-9.5,1) {$\Delta$};
\node at (-9.5,0) {$3+d_1+\frac{3 K}{2}$};
\node at (-9.5,-1.5) {$\frac{7}{2}+d_1+\frac{3 K}{2}$};
\node at (-9.5,-3) {$4+d_1+\frac{3 K}{2}$};
\node at (-9.5,-4.5) {$\frac{9}{2}+d_1+\frac{3 K}{2}$};
\node at (-9.5,-6) {$5+d_1+\frac{3 K}{2}$};
\node at (-9.5,-7.5) {$\frac{11}{2}+d_1+\frac{3 K}{2}$};
\node at (-9.5,-9) {$6+d_1+\frac{3 K}{2}$};
\draw [fill=white] (0.8,0.3) rectangle (-0.8,-0.3);
\node at (0,0) {$(K)_{(d_1,0)}$};
\draw [fill=white] (-2.5,-1.8) rectangle (-0.5,-1.2);
\node at (-1.5,-1.5) {$(K\pm 1)_{(d_1,1)}$};
\draw [fill=white] (2.7,-1.8) rectangle (0.3,-1.2);
\node at (1.5,-1.5) {$(K- 1)_{(d_1-1,1)}$};
\draw [fill=white] (-4.25,-2.45) rectangle (-1.85,-3.55);
\node at (-3,-2.75) {$(K\pm 2)_{(d_1+1,0)}$};
\node at (-3,-3.25) {$(K)_{(d_1,2),(d_1+1,0)}$};
\draw [fill=white] (-1.6,-2.45) rectangle (1.6,-3.55);
\node at (0,-2.75) {$(K)_{(d_1-1,2),(d_1,0)}$};
\node at (0,-3.25) {$(K- 2)_{(d_1-1,2),(d_1,0)}$};
\draw [fill=white] (4.1,-2.7) rectangle (1.9,-3.3);
\node at (3,-3) {$(K-2)_{(d_1-1,0)}$};
\draw [fill=white] (-5.6,-4.2) rectangle (-3.4,-4.8);
\node at (-4.5,-4.5) {$(K\pm 1)_{(d_1+1,1)}$};
\draw [fill=white] (-3,-3.7) rectangle (0,-5.3);
\node at (-1.5,-4) {$(K\pm 1)_{(d_1,1)}$};
\node at (-1.5,-4.5) {$(K- 1)_{(d_1,1),(d_1-1,3)}$};
\node at (-1.5,-5) {$(K- 3)_{(d_1,1)}$};
\draw [fill=white] (0.3,-3.95) rectangle (2.7,-5.05);
\node at (1.5,-4.25) {$(K-1)_{(d_1-1,1)}$};
\node at (1.5,-4.75) {$(K-3)_{(d_1-1,1)}$};
\draw [fill=white] (-6.9,-5.7) rectangle (-5.1,-6.3);
\node at (-6,-6) {$(K)_{(d_1 +2,0)}$};
\draw [fill=white] (-4.7,-5.45) rectangle (-1.7,-6.55);
\node at (-3.2,-5.75) {$(K-2)_{(d_1,2),(d_1+1,0)}$};
\node at (-3.2,-6.25) {$(K)_{(d_1,2),(d_1+1,0)}$};
\draw [fill=white] (-1.4,-5.45) rectangle (1.6,-6.55);
\node at (0.1,-5.75) {$(K-2)_{(d_1 -1,2),(d_1,0)}$};
\node at (0.1,-6.25) {$(K-4),(K)_{(d_1,0)}$};
\draw [fill=white] (-5.5,-7.2) rectangle (-3.5,-7.8);
\node at (-4.5,-7.5) {$(K-1)_{(d_1+1)}$};
\draw [fill=white] (-2.8,-7.2) rectangle (0.3,-7.8);
\node at (-1.25,-7.5) {$(K-1),(K-3)_{(d_1,1)}$};
\draw [fill=white] (-4.2,-8.7) rectangle (-1.8,-9.3);
\node at (-3,-9) {$(K-2)_{(d_1+1,0)}$};
\end{tikzpicture}}
\end{equation}

Let us next look at some special values of the $R$-symmetry quantum
number. For $K=1$ the states with values $(K-4)$ and $(K-3)$ are
reflected to $-(1)$ and $-(0)$ respectively via the RS algorithm,
where they subsequently cancel with other descendants with identical
quantum numbers but non-negative multiplicity. The $(K-2)$ states are
simply deleted as they lie on the boundary of the Weyl chamber for
$K=1$. Likewise if $K=2$, the $(K-3)$ states are deleted and the
$(K-4)$ state is reflected to $-(0)$ where it cancels against another
state with non-negative multiplicity.

Following this reasoning, it quickly becomes obvious that only through
setting $K=0$ does one end up with negative multiplicities being
present after cancellations, which we can observe from the last two
levels of the module. In that case, the $(K-1)$ highest weights are
deleted to leave the states $-[11/2+d_1; d_1,1;1]$ and
$-[6+d_1;d_1+1,0;0]$. We will study this $\mathcal{B}[d_1,0;0]$
multiplet separately below.

The second type of $\mathcal{B}$-multiplet to consider is
$\mathcal{B}[0,0;K]$, for which one is instructed to remove from the
auxiliary Verma-module basis \eqref{supconfbasis1} the supercharge
combinations
$\QQ_{\bOn 2}\QQ_{\bOn 3}$, $\QQ_{\bOn 2}\QQ_{\bOn 4}$ and $\QQ_{\bOn
  3}\QQ_{\bOn 4}$.  The superconformal representation constructed
using the remaining supercharges is

\begin{equation}
{\scriptsize\begin{tikzpicture}
\draw [dashed, ultra thin] (0,0) -- (-6,-6);
\draw [dashed, ultra thin] (-1.5,-1.5) -- (0,-3);
\draw [dashed, ultra thin] (-3,-3) -- (0,-6);
\draw [dashed, ultra thin] (0,-3) -- (-4.5,-7.5);
\draw [dashed, ultra thin] (0,-6) -- (-3,-9);
\draw [dashed, ultra thin] (-4.5,-4.5) -- (-1.5,-7.5);
\draw [dashed, ultra thin] (-6,-6) -- (-3,-9);
\node at (-8.5,1) {$\Delta$};
\node at (-8.5,0) {$3+\frac{3 K}{2}$};
\node at (-8.5,-1.5) {$\frac{7}{2}+\frac{3 K}{2}$};
\node at (-8.5,-3) {$4+\frac{3 K}{2}$};
\node at (-8.5,-4.5) {$\frac{9}{2}+\frac{3 K}{2}$};
\node at (-8.5,-6) {$5+\frac{3 K}{2}$};
\node at (-8.5,-7.5) {$\frac{11}{2}+\frac{3 K}{2}$};
\node at (-8.5,-9) {$6+\frac{3 K}{2}$};
\draw [fill=white] (-0.8,0.3) rectangle (0.8,-0.3);
\node at (0,0) {$(K)_{(0,0)}$};
\draw [fill=white] (-2.4,-1.2) rectangle (-0.6,-1.8);
\node at (-1.5,-1.5) {$(K\pm 1)_{(0,1)}$};
\draw [fill=white] (-4,-2.45) rectangle (-2,-3.55);
\node at (-3,-2.75) {$(K)_{(0,2),(1,0)}$};
\node at (-3,-3.25) {$(K\pm 2)_{(1,0)}$};
\draw [fill=white] (-1.2,-2.7) rectangle (1.2,-3.3);
\node at (0,-3) {$(K- 2),(K)_{(0,0)}$};
\draw [fill=white] (-5.4,-4.2) rectangle (-3.6,-4.8);
\node at (-4.5,-4.5) {$(K\pm 1)_{(1,1)}$};
\draw [fill=white] (-3.1,-3.95) rectangle (0,-5.05);
\node at (-1.5,-4.25) {$2(K-1),(K+1)_{(0,1)}$};
\node at (-1.5,-4.75) {$(K-3)_{(0,1)}$};
\draw [fill=white] (-1.5,-5.45) rectangle (1.5,-6.55);
\node at (0,-5.75) {$(K-4),(K-2)_{(0,0)}$};
\node at (0,-6.25) {$(K)_{(0,0)}$};
\draw [fill=white] (-4.3,-5.45) rectangle (-1.7,-6.55);
\node at (-3,-5.75) {$(K-2)_{(1,0),(0,2)}$};
\node at (-3,-6.25) {$(K)_{(1,0),(0,2)}$};
\draw [fill=white] (-6.7,-5.7) rectangle (-5.3,-6.3);
\node at (-6,-6) {$(K)_{(2,0)}$};
\draw [fill=white] (-3,-7.2) rectangle (-0,-7.8);
\node at (-1.5,-7.5) {$(K-1),(K-3)_{(0,1)}$};
\draw [fill=white] (-5.4,-7.2) rectangle (-3.6,-7.8);
\node at (-4.5,-7.5) {$(K-1)_{(1,1)}$};
\draw [fill=white] (-3.9,-8.7) rectangle (-2.1,-9.3);
\node at (-3,-9) {$(K-2)_{(1,0)}$};
\end{tikzpicture}}
\end{equation}

This UIR can also be obtained from $\mathcal{B}[d_1,0;K]$ by setting
$d_1=0$.  The same arguments as in the $\mathcal{B}[d_1,0;K]$ case can
be applied to study the behaviour of the multiplet for concrete $K$
values. Again, we find that non-cancelling negative-multiplicity
states only appear at $K=0$.

The superconformal index for all values of $d_1$ and $K>0$ is calculated to be
\begin{align}
\mathcal{I}_{\mathcal{B}[d_1,0;K]}(x,y)=\frac{x^{ d_1+K+3}}{\left(1-x y^{-1}\right) \left(1-x y\right)}\chi_{d_1+1}(y)\;.
\end{align}
One observes that this satisfies the recombination rules \eqref{recomb
  rule}:
\begin{align}
\lim_{\epsilon\rightarrow 0}\mathcal{I}_{\mathcal{L}[\Delta+\epsilon;d_1,0;K]}(x,y)&=\mathcal{I}_{\mathcal{A}[d_1,0;K]}(x,y)+\mathcal{I}_{\mathcal{B}[d_1-1,0;K+2]}(x,y)=0\;.
\end{align}

\subsubsection{Higher-Spin-Current Multiplets: $\mathcal{B}[d_1,0;0]$}\label{HSB5D}

Let us now address the special case with $K=0$. For $d_1\neq 0$ this
family of $\mathcal{B}$-type multiplets contains higher-spin currents
and has a primary corresponding to the symmetric traceless
representation of $\so(5)$. One finds that $\QQ_{\bA 3}$ and
$\QQ_{\bA 4}$ need to be removed from the auxiliary Verma-module basis
\eqref{supconfbasis1}. Therefore the entire representation is built by
just acting on the superconformal primary with the set of $Q$s from
\eqref{QQbars} and is

\begin{equation}\label{Bphys}
\begin{tikzpicture}
\draw [dashed, ultra thin] (0,0) -- (-6,-6);
\node at (-8.5,1) {$\Delta$};
\node at (-8.5,0) {$3+d_1$};
\node at (-8.5,-1.5) {$\frac{7}{2}+d_1$};
\node at (-8.5,-3) {$4+d_1$};
\node at (-8.5,-4.5) {$\frac{9}{2}+d_1$};
\node at (-8.5,-6) {$5+d_1$};
\draw [fill=white] (-0.9,0.4) rectangle (0.9,-0.4);
\node at (0,0) {$(0)_{(d_1,0)}$};
\draw [fill=white] (-4,-2.35) rectangle (-2,-3.65);
\node at (-3,-2.75) {$(2)_{(d_1+1,0)}$};
\node at (-3,-3.25) {$(0)_{(d_1,2)}$};
\draw [fill=white] (-2.4,-1.1) rectangle (-0.6,-1.9);
\node at (-1.5,-1.5) {$(1)_{(d_1,1)}$};
\draw [fill=white] (-5.5,-4.1) rectangle (-3.5,-4.9);
\node at (-4.5,-4.5) {$(1)_{(d_1+1,1)}$};
\draw [fill=white] (-7,-5.6) rectangle (-5,-6.4);
\node at (-6,-6) {$(0)_{(d_1+2,0)}$};
\end{tikzpicture}
\end{equation} 

This result seems to contradict \eqref{opcon5D}, which predicted the
presence of operator constraints. However, note that the absent $\QQ$s
anticommute into $\PP_5$, which has therefore been implicitly removed
from the auxiliary Verma-module generators. This has the effect of projecting
out states corresponding to operator constraints; we will henceforth
refer to the remaining states as ``reduced states''.

The operator constraints can actually be
restored---c.f. App.~\ref{App:5DRS}---by utilising the following
relationship between characters
\begin{align}
\hat{\chi}[\Delta;d_1,d_2;K]=\chi[\Delta;d_1,d_2;K]-\chi [\Delta+1;d_1-1,d_2;K]\;.
\end{align}
Here the $\hat{\chi}$ and $\chi$ correspond to taking a character of
the superconformal representation without/with the $\PP_5$
respectively. We can appropriately account for the negative
contributions to the RHS via negative-multiplicity states. The full
multiplet can then be expressed as:

\begin{equation}\label{Bneg}
\begin{tikzpicture}
\draw [dashed, ultra thin] (0,0) -- (-6,-6);
\node at (-8.5,1) {$\Delta$};
\node at (-8.5,0) {$3+d_1$};
\node at (-8.5,-1.5) {$\frac{7}{2}+d_1$};
\node at (-8.5,-3) {$4+d_1$};
\node at (-8.5,-4.5) {$\frac{9}{2}+d_1$};
\node at (-8.5,-6) {$5+d_1$};
\node at (-8.5,-7.5) {$\frac{11}{2}+d_1$};
\node at (-8.5,-9) {$6+d_1$};
\draw [fill=white] (-0.9,0.4) rectangle (0.9,-0.4);
\node at (0,0) {$(0)_{(d_1,0)}$};
\draw [fill=white] (-4,-2.35) rectangle (-2,-3.65);
\node at (-3,-2.75) {$(2)_{(d_1+1,0)}$};
\node at (-3,-3.25) {$(0)_{(d_1,2)}$};
\draw [fill=white] (-2.4,-1.1) rectangle (-0.6,-1.9);
\node at (-1.5,-1.5) {$(1)_{(d_1,1)}$};
\draw [fill=white] (-5.5,-4.1) rectangle (-3.5,-4.9);
\node at (-4.5,-4.5) {$(1)_{(d_1+1,1)}$};
\draw [fill=white] (-7,-5.6) rectangle (-5,-6.4);
\node at (-6,-6) {$(0)_{(d_1+2,0)}$};
\draw [dashed, ultra thin] (3,-3) -- (-3,-9);
\draw [fill=white] (1.9,-3.4) rectangle (4.1,-2.6);
\node at (3,-3) {$-(0)_{(d_1-1,0)}$};
\draw [fill=white] (-1.1,-5.35) rectangle (1.1,-6.65);
\node at (0,-5.75) {$-(2)_{(d_1,0)}$};
\node at (0,-6.25) {$-(0)_{(d_1-1,2)}$};
\draw [fill=white] (0.4,-4.1) rectangle (2.6,-4.9);
\node at (1.5,-4.5) {$-(1)_{(d_1-1,1)}$};
\draw [fill=white] (-2.5,-7.1) rectangle (-0.5,-7.9);
\node at (-1.5,-7.5) {$-(1)_{(d_1,1)}$};
\draw [fill=white] (-4.1,-8.6) rectangle (-1.9,-9.4);
\node at (-3,-9) {$-(0)_{(d_1+1,0)}$};
\end{tikzpicture}
\end{equation} 

An example of the above $[3+d_1;d_1,0;0]$ primary is the
operator
\begin{align}\label{hs5d}
\mathcal{O}_{\mu_1 \cdots \mu_{d_1}}= \epsilon_{\bA \bB} \phi^{\bA} \overset\leftrightarrow{\partial}_{\mu_1} \cdots \overset\leftrightarrow{\partial}_{\mu_{d_1}} \phi^{\bB}\;~,
\end{align}
where $\phi^A$ is a free hypermultiplet scalar. This object satisfies the generalised conservation equation
\begin{align}
\partial^\mu \mathcal{O}_{\mu \mu_2 \dots \mu_{d_1 -1}}=0\;,
\end{align}
which is identified with the state $-[4+d_1;d_1-1,0;0]$.

It is interesting to point out that for $d_1=1$ the superconformal
primary is an $R$-neutral $\Delta=4$ conserved current, which is not
itself a higher-spin current. The higher-spin currents are instead
found as its descendants. This logic also shows that the commutator of
the conserved charge, $\mathcal{T}$, associated with the primary has a non-vanishing commutator with the supercharges
\begin{equation}\label{nosymm}
[\CQ_{{\bf1}1},\mathcal T]=\mathcal T_{{\bf1}1}'~,
\end{equation}
where $\mathcal T_{{\bf1}1}'$ is the charge associated with the
level-one descendant current. This commutator explains why we do not
refer to $\mathcal T$ as a flavour symmetry.

The index is of course insensitive to the above discussion; it gives
the same answer when evaluated either on \eqref{Bphys} or \eqref{Bneg}
and that is
\begin{align}
\mathcal{I}_{\mathcal{B}[d_1,0;0]}(x,y)=\frac{x^{d_1+3}}{\left(1-xy^{-1}\right) \left(1-x y\right)}\chi_{d_1+1}(y)\;.
\end{align}

\subsubsection{The Stress-Tensor Multiplet: $\mathcal{B}[0,0;0]$}

Let us finally turn to the case where one also sets $d_1=0$. The
supercharges removed from the auxiliary Verma-module basis \eqref{supconfbasis1}
should be those for $\mathcal{B}[0,0;K]$, but since $K=0$ one can also
act on its shortening conditions with the $R$-symmetry lowering
operator. This results in needing to remove
$\QQ_{{\bf 1}a}\QQ_{{\bf 1} b}$, $\QQ_{{\bf 2}a}\QQ_{{\bf 1} b}$ and
$\QQ_{{\bf 2}a}\QQ_{{\bf 2} b}$ with $b>a>1$. Consequently the module is:
\begin{equation}
\begin{tikzpicture}
\draw [dashed, ultra thin] (0,0) -- (-6,-6);
\draw [dashed, ultra thin] (0,-6) -- (-3,-9);
\node at (-7.5,1) {$\Delta$};
\node at (-7.5,0) {$3$};
\node at (-7.5,-1.5) {$\frac{7}{2}$};
\node at (-7.5,-3) {$4$};
\node at (-7.5,-4.5) {$\frac{9}{2}$};
\node at (-7.5,-6) {$5$};
\node at (-7.5,-7.5) {$\frac{11}{2}$};
\node at (-7.5,-9) {$6$};
\draw [fill=white] (-0.8,0.4) rectangle (0.8,-0.4);
\node at (0,0) {$(0)_{(0,0)}$};
\draw [fill=white] (-3.8,-2.35) rectangle (-2.2,-3.65);
\node at (-3,-2.75) {$(2)_{(1,0)}$};
\node at (-3,-3.25) {$(0)_{(0,2)}$};
\draw [fill=white] (-2.2,-1.1) rectangle (-0.8,-1.9);
\node at (-1.5,-1.5) {$(1)_{(0,1)}$};
\draw [fill=white] (-5.3,-4.1) rectangle (-3.7,-4.9);
\node at (-4.5,-4.5) {$(1)_{(1,1)}$};
\draw [fill=white] (-6.8,-5.6) rectangle (-5.2,-6.4);
\node at (-6,-6) {$(0)_{(2,0)}$};
\draw [dashed, ultra thin] (0,-6) -- (-3,-9);
\draw [fill=white] (-0.9,-5.6) rectangle (0.9,-6.4);
\node at (0,-6) {$-(2)_{(0,0)}$};
\draw [fill=white] (-2.4,-7.1) rectangle (-0.6,-7.9);
\node at (-1.5,-7.5) {$-(1)_{(0,1)}$};
\draw [fill=white] (-3.9,-8.6) rectangle (-2.1,-9.4);
\node at (-3,-9) {$-(0)_{(1,0)}$};
\end{tikzpicture}
\end{equation} 

We recognise that this multiplet contains the $R$-symmetry current,
the supersymmetry current and the stress tensor, as well as their corresponding
equations of motion. In particular, we identify:
\begin{align}
[4;1,0;2]&:\hspace{3mm}J^{(\bA \bB)}_\mu\;,&-[5;0,0;2]&:\hspace{3mm}\partial^\mu J^{(\bA \bB)}_\mu=0\;,\nonumber\\
[9/2;1,1;1]&:\hspace{3mm}S^{\bA}_{\mu a}\;,&-[11/2;0,1;1]&:\hspace{3mm}\partial^\mu S^{\bA}_{\mu a}=0\;,\\
[5;2,0;0]&:\hspace{3mm}\Theta_{\mu\nu}\;,&-[6;1,0;0]&:\hspace{3mm}\partial^\mu \Theta_{\mu\nu}=0\;.\nonumber
\end{align}
This multiplet also contains three states that do not obey
conservation equations.

The index over this multiplet counts just two components of the
$R$-symmetry current, $J^{\bOn\bOn}_{1}$ and $J^{\bOn \bOn}_{4}$. It
is given by
\begin{align}
\mathcal{I}_{\mathcal{B}[0,0;0]}(x,y)=\frac{x^3}{\left(1-xy^{-1}\right) \left(1-x y\right)}\chi_{1}(y)\;.
\end{align}

\subsection{$\mathcal D$-type Multiplets}

For the $\mathcal{D}[0,0;K]$ multiplet unitarity requires that the
conformal dimension of the primary is $\Delta = \frac{3K}{2}$. The
null state associated with this condition instructs us to remove
$\QQ_{\bOn 1}$ from the basis of auxiliary Verma-module generators, but due to
having $d_1=d_2=0$ one actually needs to remove the larger set of
supercharges $\QQ_{\bOn a}$, $a = 1,\cdots,4$. In fact, in this case
$\Lambda_\bOn^1 \Psi = \QQ_{\bOn 1}\Psi = 0$ and one has the shortening
condition
\begin{align}
  \QQ_{\bOn a}\Psi = 0\;,
\end{align}
which renders the multiplet $\frac{1}{2}$ BPS.

The action of the remaining supercharges on a primary state where the
quantum numbers are kept generic is then
\begin{align}
(K)_{(d_1,d_2)}&\xrightarrow{{Q}^{\phantom{2}}} (K-1)_{(d_1,d_2+1),(d_1+1,d_2-1)}\;,\cr
&\xrightarrow{{Q}^2}(K-2)_{(d_1+1,d_2)}\;,\cr
(K)_{(d_1,d_2)}&\xrightarrow{\tilde{Q}^{\phantom{2}}} (K-1)_{(d_1,d_2-1),(d_1-1,d_2+1)}\;,\cr
&\xrightarrow{\tilde{Q}^2}(K-2)_{(d_1-1,d_2)}\;.
\end{align}
After setting $d_1=d_2=0$ and employing RS the full
multiplet is
\begin{equation}\label{D00K}
\begin{tikzpicture}
\draw [dashed, ultra thin] (0,0) -- (-3,-3);
\draw [dashed, ultra thin] (-1.5,-1.5) -- (0,-3);
\draw [dashed, ultra thin] (-3,-3) -- (0,-6);
\node at (-5.5,1) {$\Delta$};
\node at (-5.5,0) {$\frac{3 K}{2}$};
\node at (-5.5,-1.5) {$\frac{3 K}{2}+\frac{1}{2}$};
\node at (-5.5,-3) {$\frac{3 K}{2}+1$};
\node at (-5.5,-4.5) {$\frac{3 K}{2}+\frac{3}{2}$};
\node at (-5.5,-6) {$\frac{3 K}{2}+2$};
\draw [fill=white] (0.8,0.4) rectangle (-0.8,-0.4);
\node at (0,0) {$(K)_{(0,0)}$};
\draw [fill=white] (1.1,-2.6) rectangle (-1.1,-3.4);
\node at (0,-3) {$(K-2)_{(0,0)}$};
\draw [fill=white] (-4.1,-2.6) rectangle (-1.9,-3.4);
\node at (-3,-3) {$(K-2)_{(1,0)}$};
\draw [fill=white] (-2.6,-1.1) rectangle (-0.4,-1.9);
\node at (-1.5,-1.5) {$(K-1)_{(0,1)}$};
\draw [fill=white] (-2.6,-4.1) rectangle (-0.4,-4.9);
\node at (-1.5,-4.5) {$(K-3)_{(0,1)}$};
\draw [fill=white] (-1.1,-5.6) rectangle (1.1,-6.4);
\node at (0,-6) {$(K-4)_{(0,0)}$};
\end{tikzpicture}
\end{equation} 
For low enough values of $K$ there can be additional reflections. If
$K=1$ or $K=2$ these will correspond to operator constraints, to be
discussed below. If $K=3$ then the only problematic state will be the
level four $[0,0;K-4]$, which becomes $[0,0;-1]$; this is on the
boundary of the Weyl chamber, and hence will also be deleted.

The index for this multiplet is 
\begin{align}
\mathcal{I}_{\mathcal{D}[0,0;K]}(x,y)=\frac{x^{K}}{\left(1-xy^{-1}\right) \left(1- x y\right)}\;.
\end{align}
This satisfies a recombination rule
from \eqref{recomb rule}:
\begin{align}
\lim_{\epsilon\rightarrow 0}\mathcal{I}_{\mathcal{L}[\Delta+\epsilon;0,0;K]}(x,y)&=\mathcal{I}_{\mathcal{A}[0,0;K]}(x,y)+\mathcal{I}_{\mathcal{D}[0,0;K+4]}(x,y)=0\;.
\end{align}

\subsubsection{The Hypermultiplet: $\mathcal{D}[0,0;1]$}

When $K=1$ no additional shortening conditions arise, therefore we may
proceed directly using \eqref{D00K}. One recovers:
\begin{equation}
\begin{tikzpicture}
\draw [dashed, ultra thin] (0,0) -- (-1.5,-1.5);
\draw [dashed, ultra thin] (-1.5,-4.5) -- (0,-6);
\node at (-5.5,1) {$\Delta$};
\node at (-5.5,0) {$\frac{3}{2}$};
\node at (-5.5,-1.5) {$2$};
\node at (-5.5,-3) {$\frac{5}{2}$};
\node at (-5.5,-4.5) {$3$};
\node at (-5.5,-6) {$\frac{7}{2}$};
\draw [fill=white] (-0.8,-0.4) rectangle (0.8,0.4);
\node at (0,0) {$(1)_{(0,0)}$};
\draw [fill=white] (-2.3,-1.1) rectangle (-0.7,-1.9);
\node at (-1.5,-1.5) {$(0)_{(0,1)}$};
\draw [fill=white] (-2.4,-4.1) rectangle (-0.6,-4.9);
\node at (-1.5,-4.5) {$-(0)_{(0,1)}$};
\draw [fill=white] (-0.9,-6.4) rectangle (0.9,-5.6);
\node at (0,-6) {$-(1)_{(0,0)}$};
\end{tikzpicture}
\end{equation} 
We can recognise these highest weight states as
\begin{align}\begin{split}
[3/2;0,0;1]&:\hspace{3mm}\phi^{\bA}\;,\\
[2;0,1;0]&:\hspace{3mm}\lambda_a\;,\\
-[3;0,1;0]&:\hspace{3mm}\slashed \partial^{ab}\overline\lambda_{b}=0\;,\\
-[7/2;0,0;1]&:\hspace{3mm}\partial^2\phi^{\bA}=0\;.\\
\end{split}
\end{align}

The index for this multiplet counts the first operator in the primary,
$\phi^\bOn$ along with the $\PP_1$ and $\PP_2$ conformal
descendants. Therefore we have
\begin{align}
\mathcal{I}_{\mathcal{D}[0,0;1]}(x,y)=\frac{x}{\left(1-xy^{-1}\right) \left(1- x y\right)}\;.
\end{align}
This is the single-letter index obtained in
\cite{Kim:2012gu,Rodriguez-Gomez:2013dpa}.

\subsubsection{The Flavour-Current Multiplet: $\mathcal{D}[0,0;2]$}

One of the $\mathcal{D}$-type multiplets predicted to contain
operator constraints is $\mathcal{D}[0,0;2]$. As above, we may jump
straight into the multiplet structure by setting $K=2$ in the
$\mathcal{D}[0,0;K]$ module. This produces
\begin{equation}
\begin{tikzpicture}
\draw [dashed, ultra thin] (0,0) -- (-3,-3);
\draw [dashed, ultra thin] (-1.5,-1.5) -- (0,-3);
\node at (-4.5,1) {$\Delta$};
\node at (-4.5,0) {$3$};
\node at (-4.5,-1.5) {$\frac{7}{2}$};
\node at (-4.5,-3) {$4$};
\node at (-4.5,-4.5) {$\frac{9}{2}$};
\node at (-4.5,-6) {$5$};
\draw [fill=white] (-0.8,-0.4) rectangle (0.8,0.4);
\node at (0,0) {$(2)_{(0,0)}$};
\draw [fill=white] (-0.8,-2.6) rectangle (0.8,-3.4);
\node at (0,-3) {$(0)_{(0,0)}$};
\draw [fill=white] (-3.8,-2.6) rectangle (-2.2,-3.4);
\node at (-3,-3) {$(0)_{(1,0)}$};
\draw [fill=white] (-2.3,-1.1) rectangle (-0.7,-1.9);
\node at (-1.5,-1.5) {$(1)_{(0,1)}$};
\draw [fill=white] (-0.9,-6.4) rectangle (0.9,-5.6);
\node at (0,-6) {$-(0)_{(0,0)}$};
\end{tikzpicture}
\end{equation} 
We recognise these fields to be a scalar  $\mu^{({\bf AB})}$ in the $\mathbf{3}$ of
$\su(2)_R$, a symplectic Majorana fermion
$\psi^{\bf A}_{a}$ in the $\mathbf{2}$ of $\su(2)_R$, a vector $J_\mu$
and an R-neutral scalar $M$. Furthermore the negative-multiplicity
state is the equation of motion for the vector current
$\partial^\mu J_\mu=0$. This multiplet is also known as the linear
multiplet and appeared in the UV symmetry-enhancement discussion of
\cite{Tachikawa:2015mha}.

The corresponding index is:
\begin{align}
\mathcal{I}_{\mathcal{D}[0,0;2]}(x,y)=\frac{x^{2}}{\left(1-xy^{-1}\right) \left(1- x y\right)}\;.
\end{align}

This concludes our listing of superconformal multiplets for the 5D SCA.

\section{On the Complete Classification of UIRs for the 5D SCA}\label{App:5Dproof}

We will next give an argument that the conditions imposed in
Sec.~\ref{suprep5D} on the dimension of the superconformal
primary---giving rise to the $\mathcal{A}[d_1,d_2;K]$,
$\mathcal{B}[d_1,0;K]$ and $\mathcal{D}[0,0;K]$ short-multiplet
types---are necessary and sufficient for unitarity. To the best of
our knowledge, there is no such argument in the literature for 5D; for
proofs in 4D and 6D see
\cite{Dobrev:1985vh,Dobrev:1985qv,Dobrev:1985qz,Dobrev:2002dt}.

Our argument will proceed in three steps. We first use the results of
\cite{Oshima:2016gqy} to establish when the Verma module can admit
null states (i.e., when the Verma module is reducible). We then show
that only the 5D multiplet types
$\mathcal A, \mathcal B, \mathcal D, \mathcal L $ can be unitary
(determining necessity). Finally, we argue that all multiplets of the
above type are indeed unitary (determining sufficiency).

\subsection{Reducibility Conditions for $F(4)$}\label{reducVerm}

We begin by determining the necessary and sufficient conditions for
when a representation of the 5D SCA can contain null states (i.e., when the $F(4)$ parabolic Verma module is reducible). This question has recently been revisited in \cite{Yamazaki:2016vqi,Oshima:2016gqy}, the results of which we use. Towards that end, let us first establish some
notation following, e.g., \cite{Kac:1977qb,Frappat:1996pb}.

The 5D SCA, $F(4)$, is a Lie superalgebra for which we choose the
following simple-root decomposition
\begin{align}
\Pi = \left\{ \alpha_1 = \delta\;,\;\alpha_2 = \frac{1}{2}\left(-\delta + \epsilon_1 -\epsilon_2-\epsilon_3\right)\;,\;\alpha_3 = \epsilon_3\;,\;\alpha_4 =\epsilon_2 - \epsilon_3\;\right\}\;,
\end{align}
with corresponding Dynkin diagram and Cartan matrix:
\begin{center}
\begin{picture}(140,20)
\thicklines
\put(0,0){\circle{14}}
\put(42,0){\circle{14}}
\put(84,0){\circle{14}}
\put(126,0){\circle{14}}
\put(0,15){\makebox(0.4,0.6){$\alpha_1$}}
\put(42,15){\makebox(0.4,0.6){$\alpha_2$}}
\put(84,15){\makebox(0.4,0.6){$\alpha_3$}}
\put(126,15){\makebox(0.4,0.6){$\alpha_4$}}
\put(27,0){\line(-1,1){10}}\put(27,0){\line(-1,-1){10}}
\put(37,-5){\line(1,1){10}}\put(37,5){\line(1,-1){10}}
\put(6,-4){\line(1,0){30}}
\put(7,0){\line(1,0){28}}
\put(6,4){\line(1,0){30}}
\put(49,0){\line(1,0){28}}
\put(90,-3){\line(1,0){30}}
\put(90,3){\line(1,0){30}}
\put(101,0){\line(1,1){10}}\put(101,0){\line(1,-1){10}}
\end{picture}
\qquad \qquad  
$ [a_{ij}] = \left(\begin{array}{rrrr} 
2 & -1 & 0 & 0 \cr -3 & 0 & 1 & 0 \cr 
0 & -1 & 2 & -2 \cr 0 & 0 & -1 & 2 \cr
\end{array}\right)$
\end{center}
The $\alpha_1, \alpha_3,\alpha_4\in \Pi_{\overline 0}$ are even simple
roots, while $\alpha_2\in \Pi_{\overline 1}$ is an odd simple
root. These have been expressed in an orthogonal basis in terms of
$\epsilon_i$, for $i=1,2,3$, and $\delta$ such that
$(\epsilon_i, \epsilon_j) = \delta_{ij}$, $(\delta,\delta) = -3$ and
$(\epsilon_i, \delta) = 0$.

The simple roots can be used to obtain the even and odd positive roots
of $F(4)$
\begin{align}\label{+veevenodd}
\Phi^+_{\bar{0}}=\left\{\delta\;,\epsilon_i \pm \epsilon_j\text{ (for $i<j$)}\;,\epsilon_{i}\right\}\;,\hspace{10mm}\Phi^+_{\bar{1}}=\left\{ 
 \frac{1}{2}\Big( \pm\delta +\epsilon_1 \pm \epsilon_2\pm \epsilon_3\Big)\right\}\;,
\end{align}
where the signs are not correlated. One also typically defines the set
$\overline \Phi_{\overline 0}$ and the set of isotropic roots
$\overline \Phi_{\overline 1}$ through
\begin{align}
  \overline  \Phi_{\overline 0} := \{ \alpha\in \Phi_{\overline 0}~ |~
  \alpha/2 \notin \Phi_{\overline 1}\}\subseteq\Delta_{\overline 0}\;,\qquad   \overline  \Phi_{\overline 1} := \{ \alpha\in \Phi_{\overline 1}~ |~
  2\alpha \notin \Phi_{\overline 0}\}\subseteq\Delta_{\overline 1}~,
\end{align}
although it is easy to see that in our case
$\overline \Phi_{\overline 0} = \Phi_{\overline 0}$ and
$\overline \Phi_{\overline 1} = \Phi_{\overline 1}$.

The difference between the half-sum of the positive even roots
$\rho_{\overline 0}$ and half-sum of the positive odd roots
$\rho_{\overline 1}$ denotes the Weyl vector, which can be evaluated
to be
\begin{align}
\rho := \rho_{\bar{0}}-\rho_{\bar{1}}=\frac{1}{2}(\delta + \epsilon_1 + 3\epsilon_2 + \epsilon_3)\;.
\end{align}

A highest weight representation of $F(4)$, $\lambda$, may be expanded
in terms of the fundamental weights of the bosonic subalgebra
$\mathfrak g_{\overline 0} = \mathfrak{su}(2)_R\oplus \mathfrak{so}(5,2)$ as
\begin{align}
\lambda = \sum_{a}\omega_a H_a\;,
\end{align}
where the $\omega_a$ for $a=1$ and $a = 2,3,4$ are the fundamental
weights associated with the simple roots of $\mathfrak{su}(2)_R$ and
$\mathfrak{so}(5,2)$ respectively and the $H_a$ are the Cartans. For
each of these bosonic subalgebras, the fundamental weights are related
to the simple roots via the inverse of the Cartan matrix. For
$\mathfrak{su} (2)_R$ and $\mathfrak{so}(5,2)$  this allows us to
express
\begin{align}
\omega_1 = \frac{1}{2}\delta\qquad \textrm{and} \qquad \omega_2 =\epsilon_1\;,\quad \omega_3 = \epsilon_1 + \epsilon_2\;,
  \quad \omega_4 =\frac{1}{2}(\epsilon_1 + \epsilon_2 + \epsilon_3)\;.
\end{align}
Since in our discussion thus far we have been labelling
representations in terms of their
$\mathfrak l = \mathfrak{su}(2)_R\oplus
\mathfrak{so}(5)\oplus\mathfrak{so}(2)\subset \mathfrak g$
Dynkin labels, we will also need the $\mathfrak{so}(5)$ fundamental
weights
\begin{align}
  \hat{\omega}_1 = \epsilon_2\;\qquad   \hat{\omega}_2 =\frac{1}{2}(
  \epsilon_2 + \epsilon_3)\;,
\end{align}
which we use to write
\begin{align}
  \lambda = \frac{1}{2}\delta H_1 + \omega_2 (H_2 + H_3 +\frac{1}{2} H_4) + \hat{\omega}_1 H_3 + \hat{\omega}_2 H_4\;.
\end{align}
Now, by definition we have that 
\begin{align}
  H_1 = K\;,\qquad H_3 = d_1\;,\qquad H_4 = d_2\;,
\end{align}
since $H_{1,3,4}$ multiply the fundamental weights of $\mathfrak{su}(2)_R$
and $\mathfrak{so}(5)$ respectively. We can naturally assign the
remaining label to the conformal dimension
$H_2 + H_3 +\frac{1}{2} H_4 = -\Delta $, where the minus sign is there
to account for the signature of $\mathfrak{so}(5,2)$. This means that
\begin{align}
 H_2 = -\Delta - d_1 - \frac{1}{2}d_2\;. 
\end{align}
Expressing the highest weight using the above in the orthogonal basis 
\begin{align}
\lambda =\delta \frac{K}{2} -\epsilon_1 \Delta + \epsilon_2 \left(d_1 +\frac{d_2}{2}\right)+\epsilon_3 \frac{d_2}{2} \;.
\end{align}

Finally, the following two sets have to be defined before proceeding,
where we have adapted the definitions of \cite{Oshima:2016gqy} to the case of $F(4)$:
\begin{align}
  \Psi_{\lambda,{\rm iso}} := \{ \alpha \in \Phi_{\overline
  1}^+ ~|~ (\lambda + \rho , \alpha ) = 0\} 
\end{align}
and

\begin{align}
  \Psi_{\lambda,{\rm non-iso}} :=   \Big\{ \alpha \in  \Phi_{n,\overline
  0}^+ ~\Big|~ n_\alpha := \frac{2(\lambda + \rho , \alpha )}{(\alpha,\alpha)} \in
  \mathbb Z_{>0} \Big\} \;,
\end{align}
where $\Phi_{n,\bar{0}}:=\Phi_n \cap 
\Phi_{\bar{0}}$,
with $\Phi_n = \Phi^+\backslash\Phi_l$. Hence we have 
\begin{align}
\Phi_{\mathfrak l} &= \left\{\pm \delta\;,\pm\epsilon_2 \pm \epsilon_3,\pm
   \epsilon_{2},\pm \epsilon_3\right\}\;,\cr
\Phi^+_{n,\bar{0}} & =\{\epsilon_1\pm \epsilon_2,\epsilon_1\pm\epsilon_3,\epsilon_1\}\;\;,
\end{align}
where the signs are not correlated.

With this information at hand, we can now apply the criteria of
\cite{Oshima:2016gqy} regarding the necessary and sufficient
conditions for irreducibility:

\begin{itemize}
\item {\it If an $F(4)$ Verma module is irreducible, then
  $\Psi_{\lambda,\rm{iso}} = \varnothing$ and for all $\alpha \in
  \Psi_{\lambda,\rm{non-iso}}$ there exists $\beta\in
  \Phi_{\overline 0}$ such that $(\lambda + \rho,\beta) = 0$.} (\cite{Oshima:2016gqy}, Proposition 4)

\item {\it An $F(4)$ Verma module is irreducible if 
  $\Psi_{\lambda,\rm{iso}} = \varnothing$, and, for all $\alpha \in
  \Psi_{\lambda,\rm{non-iso}}$, there exists $\beta\in
  \Phi_{\overline 0}$ such that $(\lambda + \rho,\beta) = 0$ and
  $s_\alpha(\beta):=\beta - \frac{2(\beta, \alpha)}{(\alpha,\alpha)}\alpha \in \Phi_{\mathfrak l}$.} (\cite{Oshima:2016gqy}, Proposition~5)

\end{itemize}

From the first bullet point above we immediately determine that the module is reducible when
$\Psi_{\lambda,\rm{iso}} \neq \varnothing$. This phenomenon occurs when eight
conditions are satisfied, one for each of the positive odd roots,
\begin{align}\label{allfs}
  \Delta = \frac{3K}{2}-d_1-d_2:=
  f^1\;, &&\qquad  \Delta = -\frac{3K}{2}-d_1-d_2-3:=\tilde{f}^1\;,\cr
  \Delta= \frac{3K}{2}-d_1+1:= f^2 \;, &&\qquad \Delta = -\frac{3K}{2}-d_1-2:= \tilde{f}^2\;,\cr
                                                 \Delta =
                                                 \frac{3K}{2}+d_1+3:= f^3\;,&&\qquad \Delta =
                                                                               -\frac{3K}{2}+d_1:= \tilde{f}^3\;,\cr
                                                                               \Delta = \frac{3K}{2}+d_1+d_2 + 4:= f^4\;,&&\qquad\Delta = -\frac{3K}{2}+d_1+d_2 + 1:= \tilde{f}^4\;. 
\end{align}
These equations correspond exactly to the conditions for which the eight
level-one norms become null, Eq.~\eqref{norms app}.

Note that, according to the second bullet point above, there can exist additional reducible representations only if
$\Psi_{\lambda,\rm{non-iso}} \neq \varnothing$. This latter condition is equivalent to the following equations with $n_\alpha\in \mathbb Z_{>0}$: 
\begin{align}\label{nonisononempty}
n_{\epsilon_1} &= 1 - 2\Delta\;,\cr
n_{\epsilon_1 + \epsilon_2} &= d_1 +\frac{d_2}{2}+2 - \Delta\;,\cr
n_{\epsilon_1 - \epsilon_2} &= - d_1 -\frac{d_2}{2} -1 - \Delta\;,\cr
n_{\epsilon_1 + \epsilon_3} &= \frac{d_2}{2} + 1 - \Delta \;,\cr
n_{\epsilon_1 - \epsilon_3} &=-\frac{d_2}{2}- \Delta \;.
\end{align}

However, the conformal dimensions of these modules are below the unitarity bounds associated with conformal descendants:
\begin{align}
\text{Scalar}:&\hspace{3mm}\Delta\geq \frac{3}{2}\;,\cr
\text{Operator with spin $\frac{d_2}{2}$}:&\hspace{3mm}\Delta\geq 2+\frac{d_2}{2}\;,\cr
\text{Composite operator with spin $d_1+\frac{d_2}{2}$}:&\hspace{3mm}\Delta\geq 3+d_1+\frac{d_2}{2}\;.
\end{align}
Hence, these additional modules are not unitary and will
not be useful for the discussion of UIRs.

We conclude that $F(4)$ admits reducible modules precisely when the eight
level-one norms become null, as predicted by \cite{Minwalla:1997ka}.

\subsection{An Argument for Necessity}

For the necessity part of the argument we will begin from the various
unitarity bounds for the $\mathcal{A},\mathcal{B}$ and
$\mathcal{D}$-type multiplets. As shown in Sec.~\ref{suprep5D}, these
can be expressed as
\begin{align}\label{unitbnd}
  \Delta =  f^a(d_1,d_2,K) \qquad \mathrm{with}\qquad a=1,3,4\;,
\end{align}
where the $f^a(d_1,d_2,K)$ are given in \eqref{allfs} and we observe
that $f_4>f_3>f_2>f_1$. We will next show that there can be no UIRs of
the 5D SCA except for the ones satisfying \eqref{unitbnd} and
$\Delta>f^4$, i.e. that there is at least one negative-norm state in
the intervals $(f^3,f^4)$, $(f^1,f^3)$ and $(-\infty ,f^1)$.

Towards that end, consider the norms of the
following well-defined states:
\begin{align}
\mathbb{F}_4&=\|\Lambda_{\bOn}^4 \Lambda_{\bOn}^3\Lambda_{\bOn}^2
              \Lambda_{\bOn}^1 |\Delta;d_1, d_2;K\rangle^{hw}\|^2 =
              (\Delta-f^4)(\Delta-f^3)(\Delta-f^2)(\Delta-f^1)\;,\cr
\mathbb{F}_3&=\|\Lambda_{\bOn}^3
              \Lambda_{\bOn}^1 |\Delta;d_1, d_2;K\rangle^{hw}\|^2 = (\Delta-f^3)(\Delta-f^1)\;,\cr
\mathbb{F}_2&=\|\Lambda_{\bOn}^2\Lambda_{\bOn}^3 |\Delta;0, 0;K\rangle^{hw}\|^2
              =(\Delta-f^3)(\Delta-f^1)\;,\cr
\mathbb{F}_1&=\|\Lambda_{\bOn}^1 |\Delta;d_1, d_2;K\rangle^{hw}\|^2 =(\Delta-f^1)\;.
\end{align}

\begin{enumerate}

\item ${\bf d_1,d_2\neq 0}$ : $\mathbb{F}_a$ is positive for
  $\Delta > f^4$, this describes a long multiplet
  $\mathcal{L}[\Delta;d_1,d_2;K]$. When $\Delta = f^4$ this describes
  an $\mathcal{A}[d_1,d_2;K]$ multiplet. If $f^3<\Delta<f^4$ then
  $\mathbb{F}_4$ is negative. If $f^1<\Delta<f^3$ then $\mathbb{F}_3$
  is negative. The condition $0<\Delta <f^1$ would result in the
  positivity of $\mathbb{F}_4$ and $\mathbb{F}_3$, but leads to
  $\mathbb{F}_1$ being negative. $\Delta<0$ is forbidden by unitarity
  since it would result in a negative conformal dimension for the
  superconformal primary. If $\Delta = f^a$ for $a=1,2,3$ then,
  despite the fact that $\mathbb{F}_4 =0$, the level-one norm of the
  state $\Lambda_{\bOn}^4\Psi$ will be negative, and is therefore not
  allowed.

\item ${\bf d_1 \neq 0, d_2=0 }$ : The same logic can be applied to
  this case when taking $d_2= 0$. However, alongside $\Delta=f^4$, it
  is possible to have $\Delta=f^3$, describing
  $\mathcal{B}[d_1,0;K]$. This is because the level-one state
  $\Lambda_{\bOn}^4 |\Delta;d_1, 0;K\rangle^{hw}$ does not exist as it
  is ill-defined for $d_2=0$, hence there would be no negative-norm
  state associated with it.

\item ${\bf d_1=d_2=0}$ : There is now one additional way to saturate
  a unitarity bound. The states
  $\Lambda_{\bOn}^{a=2,3,4}|\Delta;0, 0;K\rangle^{hw}$ are all
  ill-defined for these $d_i$ values. Therefore one can achieve
  $\mathbb{F}_2=0$ with $\Delta = f^1$, which describes a
  $\mathcal{D}[0,0;K]$ multiplet, in addition to  $\Delta =  f^3$,
  which describes the $\mathcal{B}[0,0;K]$ multiplet.

\end{enumerate}

We conclude that the necessary conditions for unitarity are
$\Delta \geq f^4$ for $d_1,d_2\neq 0$; $\Delta\geq f^4$ or
$\Delta = f^3$ for $d_1\neq 0$, $d_2=0$; $\Delta\geq f^4$ or
$\Delta = f^{a=1,3}$ for $d_1=d_2=0$, in line with the classification
of Sec.~\ref{suprep5D}.\footnote{In a previous version of this paper,
  an additional necessary condition for unitarity had been incorrectly
  identified for $d_1 = d_2 = 0$ at $\Delta = f^2$. We thank the
  authors of \cite{Cordova:2016} for bringing this to our attention.}

\subsection{An Argument for Sufficiency}

For this part one needs to show that all states contained in the
$\mathcal A$, $\mathcal B$, $\mathcal D$ short multiplets and the long
multiplets $\mathcal L$ are indeed unitary.

\begin{enumerate}

\item ${\bf \mathcal{D}[0,0;K]}$ : All superconformal-primary norms
  for this multiplet can be explicitly calculated using the results of
  App.~\ref{App:5DSCA} and determined to be positive.\footnote{The
    relevant Mathematica notebooks can be obtained from the authors
    upon request.} As an additional check, one recognises that in a
  free theory one can construct the $\mathcal{D}[0,0;K]$ multiplets
  from $K$ free hypermultiplets, $\mathcal{D}[0,0;1]$, and therefore
  the former are necessarily unitary.

\item ${\bf \mathcal{B}[d_1,0;K]}$ : These multiplets can admit a
  realisation in terms of free hypermultiplets. Similarly to
  \eqref{hs5d}, for a free theory we may write a superconformal
  primary with quantum numbers $[3+d_1+\frac{3K}{2};d_1,0;K]$ as
\begin{align}
\mathcal{O}^{(\bA_1\cdots\bA_{K})}_{\mu_1 \cdots \mu_{d_1}}= \epsilon_{\bB \bC} \phi^{\bB} \overset\leftrightarrow{\partial}_{\mu_1} \cdots \overset\leftrightarrow{\partial}_{\mu_{d_1}} \phi^{\bC}\phi^{(\bA_1}\cdots\phi^{\bA_K)}\;.
\end{align}
Therefore, this multiplet must
necessarily be unitary. 

\item ${\bf \mathcal{A}[d_1,d_2;K]}$ : From the set of the
  $\mathcal{A}[d_1,d_2;K]$ multiplets, only the $\mathcal{A}[d_1,1;K]$
  can admit a free-field realisation. For those cases and in a free theory, the primary can be
  expressed as
\begin{align}
\mathcal{O}^{(\bA_1\cdots\bA_{K})}_{a,\mu_1 \cdots \mu_{d_1}}=  \epsilon_{\bB \bC} \phi^{\bB} \overset\leftrightarrow{\partial}_{\mu_1} \cdots \overset\leftrightarrow{\partial}_{\mu_{d_1}} \phi^{\bC}\lambda_a\phi^{(\bA_1}\cdots\phi^{\bA_K)}\;.
\end{align}
However, we can argue for the unitarity of other choices of quantum
numbers using the recombination rules
\begin{align}\label{app recomb}
  \mathcal{L}[\Delta + \epsilon;d_1,d_2;K] &\xrightarrow{\epsilon\rightarrow 0} \mathcal{A}[d_1,d_2;K]\oplus \mathcal{A}[d_1,d_2-1;K+1]~,
\end{align}
for $\Delta = 4+d_1+d_2+\frac{3K}{2}$ as follows.

Let us formally denote the subsets of long-multiplet states related to
the $\mathcal A[d_1,d_2;K]$ and $\mathcal A[d_1,d_2-1;K+1]$ ones above
as
$\Psi_{\mathcal L_{A_1},\;\mathcal
  L_{A_2}}\xrightarrow{\epsilon\rightarrow 0} \Psi_{A_1,\; A_2}$;
that is, the states in $\Psi_{\mathcal{L}_{A_1}}$ can be completed into
a long multiplet by including the states
$\Psi_{\mathcal{L}_{A_2}}$. We may therefore write
\begin{align}
\Psi_{\mathcal L_{A_2}} =\Lambda_{\bOn}^4\Psi_{\mathcal L_{A_1}}~,
\end{align}
for all states in $\Psi_{\mathcal{L}_{A_2}}$. Their norms can be explicitly calculated and are related by
\begin{align}
\|\Psi_{\mathcal L_{A_2}}\|^2 = \epsilon\frac{\left(2 d_1+d_2+3\right) \left(d_1+2\right) \left(d_2+1\right) }{2 \left(d_1+d_2/2+1\right) d_2 \left(d_1+d_2+1\right)}\|\Psi_{\mathcal L_{A_1}}\|^2\;,
\end{align}
where note that the coefficient on the RHS is always positive. 

Since the multiplet $\mathcal{A}[d_1,1;K]$ can appear in the recombination rule \eqref{app recomb} as 
\begin{align}
\mathcal{L}[\Delta + \epsilon;d_1,1;K] &\xrightarrow{\epsilon\rightarrow 0} \mathcal{A}[d_1,1;K]\oplus \mathcal{A}[d_1,0;K+1]\;,
\end{align}
we can use the above argument to conclude that in the $\epsilon\to 0$
limit $\mathcal{A}[d_1,0;K+1]$ is also unitary. Moreover, one can also
write
\begin{align}
\mathcal{L}[\Delta + \epsilon;d_1,2;K-1] &\xrightarrow{\epsilon\rightarrow 0} \mathcal{A}[d_1,2;K-1]\oplus \mathcal{A}[d_1,1;K]~,
\end{align} 
and use the same logic to conclude that the multiplet
$\mathcal{A}[d_1,2;K]$ is unitary. This method can be used recursively
to demonstrate that all $\mathcal{A}[d_1,d_2;K]$ are unitary with the
exception of $\mathcal{A}[0,0;0]$. However, the unitarity of $\mathcal{A}[0,0;0]$ can be argued for by appealing to the index.

To that end, recall the index of a unitary SCFT is a sum over
contributions from individual local operators of the theory
transforming in UIRs. Now, the index for a 5D SCFT with gauge group
$\SU (2)$ and global symmetry group $E_2 = \SU(2) \times \U(1)$ was
evaluated in \cite{Kim:2012gu}; see also
\cite{Hwang:2014uwa}. Moreover, the arguments in \cite{Kim:2012gu}
suggest that this result can be understood as an index coming from one
of Seiberg's unitary 5D SCFTs \cite{Seiberg:1996bd} that appear in the
low-energy limit of string theory. Assuming this identification is
correct (or, least restrictively, that the index computed in
\cite{Kim:2012gu} corresponds to an index of some unitary theory with
$E_2$ global symmetry), then it is straightforward to show that
$\mathcal{A}[0,0;0]$ is also a unitary representation.

To understand this last statement, note that the authors of \cite{Kim:2012gu} found the index in question admits the following expansion
\begin{align}\label{Kor index}
\mathcal I(x,y) &=1+[1+\chi_2 (q)]x^2+\chi_1 (y)[2+\chi_2 (q)]x^3\cr
&\qquad \qquad +\bigg(\chi_2(y)\left[2+\chi_2 (q)\right]+1-\chi_{\bFo}(f)\bigg)x^4+\mathcal{O}(x^5)\;,
\end{align}
where $q$ is a fugacity associated with the instanton contributions and
\begin{align}
\chi_{\bFo}(f)=(e^{\frac{i\rho}{2}}+e^{\frac{-i\rho}{2}})\chi_1(q)~,
\end{align}
for some $\U(1)$ flavour charge $\rho$. It is then possible to rewrite this expression in terms of contributions from local operators as follows
\begin{align}
\mathcal I(x,y) & =1+[1+\chi_2
                  (q)]\mathcal{I}_{\mathcal{D}[0,0;2]}(x,y)+\mathcal{I}_{\mathcal{B}[0,0;0]}(x,y)+\chi_3(q)\mathcal{I}_{\mathcal{D}[0,0;4]}\cr
& \qquad\qquad+\chi_{\bFo}(f)\mathcal{I}_{\mathcal{A}[0,0;0]}(x,y)+\mathcal{O}(x^5)\;.
\end{align}
Note that the $-\chi_{\bFo}(f)x^4$ term in \eqref{Kor index} can only
be accounted for by using the index for $\mathcal{A}[0,0;0]$ since all
other indices at $\mathcal{O}(x^4)$ either have positive sign or
include $y$-dependent factors.  We therefore conclude that
$\mathcal{A}[0,0;0]$ is also unitary. 

\item ${\bf \mathcal{L}[f_4+\epsilon;d_1,d_2;K]}$ : For the long
  multiplets approaching the unitarity bound
  $\Delta = 4+d_1+d_2+\frac{3K}{2}$, one can conclude that, since the
  $\mathcal{L}[f_4 + \epsilon;d_1,d_2\neq 0;K]$ can be written as a
  direct sum of two unitary multiplets as in \eqref{app recomb}, it
  too must be unitary. The remaining two recombination rules of
  \eqref{recomb rule}
\begin{align}\begin{split}
\mathcal{L}[\Delta + \epsilon;d_1,0;K] &\xrightarrow{\epsilon\rightarrow 0} \mathcal{A}[d_1,0;K]\oplus \mathcal{B}[d_1-1,0;K+2]\;,\\
\mathcal{L}[\Delta + \epsilon;0,0;K] &\xrightarrow{\epsilon\rightarrow 0} \mathcal{A}[0,0;K]\oplus \mathcal{D}[0,0;K+4]\;,\end{split}
\end{align}
can be similarly used to determine that all
$\mathcal{L}[\Delta+\epsilon;d_1,d_2;K]$ are unitary.

\item ${\bf \mathcal{L}[\Delta;d_1,d_2;K]}$ : In the neighbourhood
  $\Delta = 4+d_1+d_2+\frac{3K}{2}+\epsilon=f^4+\epsilon$, every state
  has positive norm. Since the norm of a state is a smooth function of
  $\Delta$, in order for it to become negative as we move arbitrarily
  far from $f^4$, it must first become null. By definition this scenario would
  lead to a reducible Verma module. However, in Sec.~\ref{reducVerm}, we
  determined all the values for which such a situation can arise. It is easy to
  see that $f^4$ is the largest of these and therefore the norm can
  never be negative.
\end{enumerate}

We therefore conclude that the multiplets
$\mathcal{L}[\Delta;d_1,d_2;K]$, $\mathcal{A}[d_1,d_2;K]$,
$\mathcal{B}[d_1,0;K]$ and $\mathcal{D}[0,0;K]$ are unitary hence
concluding our argument.

\section{Multiplets and Superconformal Indices for 6D
  (1,0)}\label{Sec:6D10}

We now switch to six dimensions. In this section we provide a
systematic analysis of all superconformal multiplets admitted by the
6D $\mathcal{N}=(1,0)$ algebra, $\mathfrak{osp}(8^*|2)$.\footnote{Some
  of the multiplets up to spin 2 discussed in this section have been previously
  constructed using the dual gravity description in \cite{Passias:2016fkm}.} Since the
method for building UIRs is completely analogous to
Sec.~\ref{suprep5D} we will be brief and merely sketch the derivation
of the unitarity bounds obtained in detail by
\cite{Bhattacharya:2008zy}. We then proceed to write out the complete
set of spectra for superconformal multiplets along with their
corresponding superconformal indices.

\subsection{UIR Building with Auxiliary Verma Modules}\label{super conformal rep building 6D}

The superconformal primaries of the algebra $\mathfrak{osp}(8^*|2)$
are designated $\ket{\Delta ; c_1, c_2, c_3; K}$ and labelled by the
conformal dimension $\Delta$, the Lorentz quantum numbers for $\su(4)$
in the Dynkin basis $c_i$ and an $R$-symmetry label $K$. Each primary
is in one-to-one correspondence with a highest weight labeling
irreducible representations of the maximal compact subalgebra
$\mathfrak{so}(6)\oplus\mathfrak{so}(2)\oplus\mathfrak{su}(2)_R\subset\mathfrak{osp}(8^*|2)$. There
are eight Poincar\'e and superconformal supercharges, denoted by
$\QQ_{\bA a}$ and $\QS_{\bA \dot a}$---where $a,\dot a=1,\cdots,4$ are
$\su(4)$ (anti)fundamental indices and $\bA=1,2$ an index of
$\su(2)_R$. One also has six momenta $\PP_\mu$ and special conformal
generators $\mathcal{K}_\mu$, where $\mu =1,\cdots,6$ is a Lorentz
vector index.  The superconformal primary is annihilated by all
$\QS_{\bA \dot a}$ and $\mathcal{K}_\mu$. A basis for the
representation space of $\mathfrak{osp}(8^*|2)$ can be constructed by
considering the following Verma
module \begin{align}\label{supconfbasis 6D (1,0)} \prod_{{\bf A},
    a}(\QQ_{{\bf A} a})^{n_{{\bf A},a}}\prod_{\mu}
  \mathcal{P}_{\mu}^{\hphantom{\mu}n_\mu}
  \ket{\Delta;c_1,c_2,c_3;K}^{hw}\;,
\end{align}
where $n=\sum_{{\bf A} a} n_{{\bf A},a}$ and
$\hat{n}=\sum_{\mu} n_{\mu}$ denote the level of a superconformal or
conformal descendant respectively. UIRs can be obtained from the above
after also imposing the requirement of unitarity level-by-level.

We will now review the conditions imposed by unitarity, starting with
the superconformal descendants \cite{Bhattacharya:2008zy}.  For
descendants of level $n>0$ it suffices to calculate the norms of
states in the highest weight of $\su(2)_R$, since these provide the
most stringent set of unitarity bounds
\cite{Minwalla:1997ka,Dobrev:2002dt,Bhattacharya:2008zy}.\footnote{In
  the language of \cite{Oshima:2016gqy}, the proof for the 6D SCAs in
  \cite{Dobrev:2002dt} only deals with the conditions associated with
  $\Psi_{\lambda,{\rm iso}}\neq \varnothing$. We have investigated the
  possible reducible modules coming from
  $\Psi_{\lambda,{\rm non-iso}}\neq \varnothing$ and, similar to our
  findings in Sec.~\ref{App:5Dproof}, all predicted cases are below
  the unitarity bounds associated with conformal descendants.} Hence
the unitarity bounds stem from the study of superconformal descendants
arising from the action of supercharges of the form $\QQ_{\bOn a}$.
Before proceeding, it is convenient to define the basis
$A^a \ket{\Delta;c_1,c_2,c_3;K}^{hw}$, where
\begin{align}\label{Aas}
A^a =\sum^{a}_{b=1} \QQ_{\bOn b} \Upsilon^a_b\;.
\end{align}
The $\Upsilon^a_b$ are functions of $\mathfrak u(4)$ Cartans and
Lorentz lowering operators, the explicit form of which can be found in
\cite{Bhattacharya:2008zy}. They map highest weights to highest
weights.

One then conducts a level-by-level analysis of the norms for the
superconformal descendants, while also keeping track of whether the
states $A^a \ket{\Delta;c_1,c_2,c_3;K}^{hw}$ are well defined when
setting $c_i=0$. This gives rise to a set of regular and isolated
short multiplets, obeying certain shortening conditions
\cite{Bhattacharya:2008zy}.  The precise form of the $A^a$ is
unimportant; they have the same quantum numbers as the supercharge
$\QQ_{\bOn a}$.  The shortening conditions can then be translated into
absent generators in the auxiliary Verma module \eqref{supconfbasis 6D
  (1,0)} and we are instructed to remove the corresponding
combinations of supercharges in a straightforward way. These results
are summarised in Table \ref{Tab:6D(1,0)}.

 \begin{table}[t]
\begin{center}
\begin{tabular}{|c|r|l|}
\hline
Multiplet & Shortening Condition & Conformal Dimension\\
\hline
\hline
$\mathcal{A}[c_1,c_2,c_3;K]$ & $A^4\Psi=0$ & $\Delta = 2K+ \frac{c_1}{2}+c_2+\frac{3c_3}{2}+6$ \\
$\mathcal{A}[c_1,c_2,0;K]$ & $A^3 A^4\Psi=0$ & $\Delta =2K+\frac{c_1}{2}+c_2+6$ \\
$\mathcal{A}[c_1,0,0;K]$ & $A^2 A^3 A^4\Psi=0$ & $\Delta =  2K+\frac{c_1}{2}+6$\\
$\mathcal{A}[0,0,0;K]$ & $A^1 A^2 A^3 A^4\Psi=0$ & $\Delta = 2K+6$\\
\hline
\hline
$\mathcal{B}[c_1,c_2,0;K]$ & $A^3\Psi=0$ & $\Delta = 2K+\frac{c_1}{2}+c_2+4$\\
$\mathcal{B}[c_1,0,0;K]$ & $A^2 A^3\Psi=0$ & $\Delta =2K+\frac{c_1}{2}+4$\\
$\mathcal{B}[0,0,0;K]$ & $A^1 A^2 A^3\Psi=0$ & $\Delta =2K+4$\\
\hline
\hline
$\mathcal{C}[c_1,0,0;K]$ & $A^2\Psi=0$ & $\Delta =2K+\frac{c_1}{2}+2  $\\
$\mathcal{C}[0,0,0;K]$ & $A^1 A^2\Psi=0$ & $\Delta =2K+2  $\\
\hline
\hline
$\mathcal{D}[0,0,0;K]$ & $A^1\Psi=0$ & $\Delta =2K  $\\
\hline

\end{tabular}
\end{center}
\caption{\label{Tab:6D(1,0)} A list of all short multiplets for the 6D
  $(1,0)$ SCA, along with the conformal dimension of the
  superconformal primary and the corresponding shortening
  condition. The $A^a$ in the shortening conditions are defined in
  \eqref{Aas} and \cite{Bhattacharya:2008zy}. The first of these multiplets
  ($\mathcal A$) is a
  regular short representation, whereas  the rest ($\mathcal B, \mathcal C,
  \mathcal D$) are isolated short representations. Here $\Psi$ denotes
  the superconformal primary state for each multiplet.}
\end{table}

Additional absent supercharges can be obtained when $K=0$ by acting on the
existing null states with Lorentz and $R$-symmetry lowering operators. As a
result additional supercharges need to be removed from
\eqref{supconfbasis 6D (1,0)}. These occurrences will be dealt with on
a case-by-case basis.

Finally, there can be supplementary unitarity restrictions and
associated null states arising from considering conformal
descendants. These have been studied in detail in
\cite{Minwalla:1997ka,Dolan:2005wy}. In analogy with the 5D approach,
we will choose not to exclude any momenta from the basis of auxiliary
Verma-module generators \eqref{supconfbasis 6D (1,0)}; with the help
of the RS algorithm this will account for states corresponding to
equations of motion and conservation equations.

We can combine the information from the conformal and superconformal
unitarity bounds to predict when a multiplet will contain operator
constraints. These are found to be
\begin{align}\label{6D10coneq}
\mathcal{B}[c_1,c_2,0;0]\;,\hspace{3mm}\mathcal{C}[c_1,0,0;\{0,1\}]\;,\hspace{3mm}\mathcal{D}[0,0,0;\{1,2\}]
\end{align}
and will be treated separately in the following sections. The
multiplet $\mathcal{D}[0,0,0;0]$ does not belong to this list as it is
the vacuum, which is annihilated by all supercharges and momenta.

\subsection{6D (1,0) Recombination Rules}

Short multiplets can recombine into a long multiplet $\mathcal{L}$. This occurs when the conformal dimension of $\mathcal{L}$ approaches the unitarity bound, that is when $\Delta +\epsilon \to 2K +\frac{1}{2}(c_1 +2c_2+3c_3)+6$. We find that 
\begin{align}\label{recomb rule 6d n=1}
\mathcal{L}[\Delta + \epsilon;c_1,c_2,c_3;K] &\xrightarrow{\epsilon\rightarrow 0} \mathcal{A}[c_1,c_2,c_3;K]\oplus \mathcal{A}[c_1,c_2,c_3-1;K+1]\;,\cr
\mathcal{L}[\Delta + \epsilon;c_1,c_2,0;K] &\xrightarrow{\epsilon\rightarrow 0} \mathcal{A}[c_1,c_2,0;K]\oplus \mathcal{B}[c_1,c_2-1,0;K+2]\;,\cr
\mathcal{L}[\Delta + \epsilon;c_1,0,0;K] &\xrightarrow{\epsilon\rightarrow 0} \mathcal{A}[c_1,0,0;K]\oplus \mathcal{C}[c_1-1,0,0;K+3]\;,\cr
\mathcal{L}[\Delta + \epsilon;0,0,0;K] &\xrightarrow{\epsilon\rightarrow 0} \mathcal{A}[0,0,0;K]\oplus \mathcal{D}[0,0,0;K+4]\;.
\end{align}
These identities can be explicitly checked, e.g. by using
supercharacters for the multiplets that we discuss below.

A small number of short multiplets do not appear in any recombination rule. These are
\begin{align}\label{6D10const}
&\mathcal{B}\left[c_1,c_2,0;\{0,1\}\right]\;,\cr
&\mathcal{C}\left[c_1,0,0;\{0,1,2\}\right]\;,\cr
&\mathcal{D}\left[0,0,0;\{0,1,2,3\}\right]\;.
\end{align}

\subsection{The 6D (1,0) Superconformal Index}\label{6D10IndMain}

We define the 6D (1,0) superconformal index with respect to the supercharge $\QQ_{\bOn 4}$, in accordance with \cite{Bhattacharya:2008zy}. This is given by 
\begin{align}
  \label{(1,0) index}
  \II(p,q,s)=\Tr_{\HH}(-1)^F e^{-\beta \delta}  q^{\Delta-\frac{1}{2}K}p^{c_2}s^{c_1}\;,
\end{align}
where the fermion number is $F=c_1+c_3$. The states that are counted satisfy $\delta = 0$, where
\begin{align}
  \label{(1,0) delta}
  \delta :=\{ \QQ_{\bOn 4}, \QS_{\bTw \dot 4}\}=\Delta- 2 K -\frac{1}{2}(c_1+2c_2+3c_3)\;.
\end{align}
The Cartan combinations $\Delta-\frac{1}{2}K$, $c_2$ and $c_1$ are
generators of the subalgebra that commutes with
$\QQ_{\bOn 4},\QS_{\bTw \dot 4}$ and $\delta$. This index can be evaluated as
a 6D supercharacter, as discussed in App.~\ref{superchar}.

\subsection{Long Multiplets}

Long multiplets are generated by the action of all supercharges.  We
will choose to group them into $Q=( \QQ_{ \bA 1},\QQ_{\bA 2})$
and $\tilde{Q}=(\QQ_{ \bA 3},\QQ_{\bA 4})$, with their individual
quantum numbers given by
\begin{align}\label{6DQQtilde}
\QQ_{\bOn 1} &\sim (1)_{(1,0,0)}\;,\hspace{2mm}&&\QQ_{\bOn 2}\sim (1)_{(-1,1,0)}\;,\hspace{3mm}&&&\QQ_{\bOn 3}\sim (1)_{(0,-1,1)}\;,\hspace{3mm}&&&&\QQ_{\bOn 4}\sim (1)_{(0,0,-1)}\;,\nonumber\\
\QQ_{\bTw 1} &\sim (-1)_{(1,0,0)}\;,\hspace{2mm}&&\QQ_{\bTw 2}\sim (-1)_{(-1,1,0)}\;,\hspace{3mm}&&&\QQ_{\bTw 3}\sim (-1)_{(0,-1,1)}\;,\hspace{3mm}&&&&\QQ_{\bTw 4}\sim (-1)_{(0,0,-1)}\;.
\end{align}
The action of these supercharges on a superconformal primary $(K)_{(c_1,c_2,c_3)}$ is given by 
\begin{align}\label{q actions long 6d N=1}
(K)_{(c_1,c_2,c_3)}&\xrightarrow{Q^{\phantom{2}}} (K\pm 1)_{(c_1+1,c_2,c_3),(c_1-1,c_2+1,c_3)}\;,\cr
&\xrightarrow{Q^2}(K\pm 2)_{(c_1,c_2+1,c_3)}\;,\;(K)_{(c_1+2,c_2,c_3),(c_1,c_2+1,c_3)^2,(c_1-2,c_2+2,c_3)}\;,\cr
&\xrightarrow{Q^3}(K\pm 1)_{(c_1+1,c_2+1,c_3),(c_1-1,c_2+2,c_3)}\;,\cr
&\xrightarrow{Q^4}(K)_{(c_1,c_2+2,c_3)}\;,\cr
(K)_{(c_1,c_2,c_3)}&\xrightarrow{\tilde{Q}^{\phantom{2}}} (K\pm 1)_{(c_1,c_2-1,c_3+1),(c_1,c_2,c_3-1)}\;,\cr
&\xrightarrow{\tilde{Q}^2}(K\pm 2)_{(c_1,c_2-1,c_3)}\;,\;(K)_{(c_1,c_2-2,c_3+2),(c_1,c_2-1,c_3)^2,(c_1,c_2,c_3-2)}\;,\cr
&\xrightarrow{\tilde{Q}^3}(K\pm 1)_{(c_1,c_2-1,c_3-1),(c_1,c_2-2,c_3+1)}\;,\cr
&\xrightarrow{\tilde{Q}^4}(K)_{(c_1,c_2-2,c_3)}\;.
\end{align}

\subsection{$\mathcal{A}$-type Multiplets}

Recall from Table~\ref{Tab:6D(1,0)} that the $\mathcal{A}$--multiplets
obey four kinds of shortening conditions. These result in the removal
of the following combinations of supercharges from the basis of Verma
module generators \eqref{supconfbasis 6D (1,0)}:
\begin{align}\label{410}
\mathcal{A}[c_1,c_2,c_3;K]&: \QQ_{\bOn 4}\;, \cr
\mathcal{A}[c_1,c_2,0;K]&: \QQ_{\bOn 3}\QQ_{\bOn 4}\;,\cr
\mathcal{A}[c_1,0,0;K]&: \QQ_{\bOn 2}\QQ_{\bOn 3}\QQ_{\bOn 4}\;,\cr
\mathcal{A}[0,0,0;K]&: \QQ_{\bOn 1}\QQ_{\bOn 2}\QQ_{\bOn 3}\QQ_{\bOn 4}\;.
\end{align}
With regards to the spectrum, let us first consider the
$\mathcal{A}[c_1,c_2,c_3;K]$ multiplet. The action of the $Q$-set of
supercharges is the same as for the long multiplet \eqref{q actions
  long 6d N=1}, however since we are also instructed to remove
$\QQ_{\bOn 4}$ the $\tilde{Q}$-chain becomes
\begin{align}\label{A6D10}
(K)_{(c_1,c_2,c_3)}&\xrightarrow{\tilde{Q}^{\phantom{2}}} (K\pm 1)_{(c_1,c_2-1,c_3+1)}\;,\;(K-1)_{(c_1,c_2,c_3-1)}\;,\cr
&\xrightarrow{\tilde{Q}^2}(K)_{(c_1,c_2-2,c_3+2),(c_1,c_2-1,c_3)}\;,\;(K-2)_{(c_1,c_2-1,c_3)} \;,\cr
&\xrightarrow{\tilde{Q}^3}(K-1)_{(c_1,c_2-2,c_3+1)}\;.
\end{align}
The remaining four cases can be obtained straightforwardly in a
similar way. The resulting multiplet spectra turn out to be the same
as starting with \eqref{A6D10}, substituting for specific $c_i$ values
and running the RS algorithm.

For $K=0$ one should also remove the additional supercharge
combinations obtained by acting with the $R$-symmetry lowering operators
on the supercharges mentioned in \eqref{410} from the basis of
auxiliary Verma-module generators. Once again, the resulting spectra
are equivalent to starting with the $K\neq 0$ multiplets, explicitly
setting $K=0$ and running the RS algorithm for $\su(2)_R$.

The superconformal index for all values of $c_i$ and $K$ is given by
\begin{align}
\mathcal{I}_{\mathcal{A}[c_1,c_2,c_3;K]}(p,q,s)=\frac{(-1)^{c_1+c_3}q^{6+\frac{3K}{2}+\frac{1}{2}(c_1+2c_2+3c_3)}}{(p q-1) \left(p^2-s\right) (p s-1) \left(p-s^2\right) (q-s) (p-q s)}\nonumber\\
\times\Bigg\{p^{-c_2+1} s^{c_2+4} \left(p^{-c_1}-p s^{c_1+1}\right)
+ p^{c_2+4} s^{c_1+4}\nonumber\\ -p^{-c_1+2} s^{2-c_2}+p^{c_1+4} s^{-c_1+1} \left(s^{-c_2}-s p^{c_2+1}\right)\Bigg\}\;.
\end{align}
This index satisfies the following  recombination rules
\begin{align}
\lim_{\epsilon\rightarrow 0}\mathcal{I}_{\mathcal{L}[\Delta+\epsilon;c_1,c_2,c_3;K]}(p,q,s)=\mathcal{I}_{\mathcal{A}[c_1,c_2,c_3;K]}(p,q,s)+\mathcal{I}_{\mathcal{A}[c_1,c_2,c_3-1;K+1]}(p,q,s)=0\;.
\end{align}

\subsection{$\mathcal{B}$-type Multiplets}\label{6D10B}

For the $\mathcal{B}$-type multiplets, the supercharges that need to
be removed from the basis \eqref{supconfbasis 6D (1,0)} are 
\begin{align}\begin{split}
\mathcal{B}[c_1,c_2,0;K]&: \QQ_{\bOn 3}\;,\\
\mathcal{B}[c_1,0,0;K]&: \QQ_{\bOn 2}\QQ_{\bOn 3}\;,\\
\mathcal{B}[0,0,0;K]&: \QQ_{\bOn 1}\QQ_{\bOn 2}\QQ_{\bOn 3}\;.\end{split}
\end{align}

For the first type of multiplet, $\mathcal{B}[c_1,c_2,0;K]$, one
should also remove $\QQ_{\bOn 4}$. Acting on a primary with generic quantum
numbers, $(K)_{(c_1,c_2,c_3)}$, one has the same action as \eqref{q
  actions long 6d N=1} for $Q$. The $\tilde{Q}$-chain is modified to
\begin{align}\begin{split}
(K)_{(c_1,c_2,c_3)}&\xrightarrow{\tilde{Q}^{\phantom{2}}} (K-1)_{(c_1,c_2-1,c_3+1),(c_1,c_2,c_3-1)}\;,\\
&\xrightarrow{\tilde{Q}^2}(K-2)_{(c_1,c_2-1,c_3)}\;.
\end{split}
\end{align}
We can represent this multiplet on a grid, where acting with $Q$s
corresponds to {\it southwest} motion on the diagram, while acting
with $\tilde{Q}$ to {\it southeast} motion

\begin{equation}\label{B[c1,c2,0;K spec]}
{\scriptsize
\begin{tikzpicture}
\draw [dashed, ultra thin] (0,0) -- (-7.2,-7.2);
\draw [dashed, ultra thin] (0,0) -- (3.6,-3.6);
\draw [dashed, ultra thin] (-1.8,-1.8) -- (1.8,-5.4);
\draw [dashed, ultra thin] (-3.6,-3.6) -- (0,-7.2);
\draw [dashed, ultra thin] (-5.4,-5.4) -- (-1.8,-9);
\draw [dashed, ultra thin] (-7.2,-7.2) -- (-3.6,-10.8);
\draw [dashed, ultra thin] (1.8,-1.8) -- (-5.4,-9);
\draw [dashed, ultra thin] (3.6,-3.6) -- (-3.6,-10.8);
\node at (-10.5,1) {$\Delta$};
\node at (-10.5,0) {$4+2K+\frac{c1}{2}+c_2$};
\node at (-10.5,-1.8) {$\frac{9}{2}+2K+\frac{c_1}{2}+c_2$};
\node at (-10.5,-3.6) {$5+2K+\frac{c_1}{2}+c_2$};
\node at (-10.5,-5.4) {$\frac{11}{2}+2K+\frac{c_1}{2}+c_2$};
\node at (-10.5,-7.2) {$6+2K+\frac{c_1}{2}+c_2$};
\node at (-10.5,-9) {$\frac{13}{2}+2K+\frac{c_1}{2}+c_2$};
\node at (-10.5,-10.8) {$7+2K+\frac{c_1}{2}+c_2$};
\draw [fill=white] (0.9,0.3) rectangle (-0.9,-0.3);
\node at (0,0) {$(K)_{(c_1,c_2,0)}$};
\draw [fill=white] (-0.3,-2.35) rectangle (-3.3,-1.25);
\node at (-1.8,-1.55) {$(K\pm 1)_{(c_1+1,c_2,0)}$};
\node at (-1.8,-2.05) {$(K\pm 1)_{(c_1-1,c_2+1,0)}$};
\draw [fill=white] (0.4,-2.15) rectangle (3.1,-1.45);
\node at (1.8,-1.8) {$(K-1)_{(c_1,c_2-1,1)}$};
\draw [fill=white] (-1.9,-2.8) rectangle (-5.2,-4.35);
\node at (-3.6,-3.1) {$(K)_{(c_1+2,c_2,0)}$};
\node at (-3.6,-3.6) {$(K\pm2),(K)^2_{(c_1,c_2+1,0)}$};
\node at (-3.6,-4.1) {$(K)_{(c_1-2,c_2+2,0)}$};
\draw [fill=white] (1.8,-2.8) rectangle (-1.8,-4.35);
\node at (0,-3.1) {$(K),(K-2)_{(c_1+1,c_2-1,1)}$};
\node at (0,-3.6) {$(K)_{(c_1-1,c_2,1)}$};
\node at (0,-4.1) {$(K-2)_{(c_1-1,c_2,1)}$};
\draw [fill=white] (4.6,-3.25) rectangle (2.1,-3.85);
\node at (3.4,-3.6) {$(K-2)_{(c_1,c_2-1,0)}$};
\draw [fill=white] (-6.9,-4.8) rectangle (-3.9,-5.9);
\node at (-5.4,-5.15) {$(K\pm 1)_{(c_1+1,c_2+1,0)}$};
\node at (-5.4,-5.65) {$(K\pm 1)_{(c_1-1,c_2+2,0)}$};
\draw [fill=white] (-3.75,-4.6) rectangle (0.1,-6.15);
\node at (-1.8,-4.9) {$(K-1)_{(c_1+2,c_2-1,1)}$};
\node at (-1.8,-5.4) {$(K-3),(K\pm1)_{(c_1,c_2,1)}$};
\node at (-1.8,-5.9) {$(K-1)_{(c_1-2,c_2+1,1),(c_1,c_2,1)}$};
\draw [fill=white] (0.3,-4.8) rectangle (4.4,-5.9);
\node at (2.3,-5.15) {$(K-1),(K-3)_{(c_1+1,c_2-1,0)}$};
\node at (2.3,-5.65) {$(K-1),(K-3)_{(c_1-1,c_2,0)}$};
\draw [fill=white] (-8.2,-6.9) rectangle (-6.1,-7.45);
\node at (-7.2,-7.2) {$(K)_{(c_1,c_2+2,0)}$};
\draw [fill=white] (-5.9,-6.65) rectangle (-2,-7.7);
\node at (-3.9,-6.95) {$(K),(K-2)_{(c_1+1,c_2,1)}$};
\node at (-3.9,-7.45) {$(K),(K-2)_{(c_1-1,c_2+1,1)}$};
\draw [fill=white] (-1.8,-6.4) rectangle (2.2,-7.95);
\node at (0.2,-6.7) {$(K-2)_{(c_1+2,c_2-1,0)}$};
\node at (0.2,-7.2) {$(K-4),(K-2),(K)_{(c_1,c_2,0)}$};
\node at (0.2,-7.7) {$(K-2)_{(c_1-2,c_2+1,0),(c_1,c_2,0)}$};
\draw [fill=white] (-6.7,-9.25) rectangle (-4.1,-8.75);
\node at (-5.4,-9) {$(K-1)_{(c_1,c_2+1,1)}$};
\draw [fill=white] (-3.8,-8.45) rectangle (0.2,-9.5);
\node at (-1.8,-8.75) {$(K-1),(K-3)_{(c_1+1,c_2,0)}$};
\node at (-1.8,-9.25) {$(K-1),(K-3)_{(c_1-1,c_2+1,0)}$};
\draw [fill=white] (-4.9,-10.5) rectangle (-2.2,-11.05);
\node at (-3.6,-10.8) {$(K-2)_{(c_1,c_2+1,0)}$};
\end{tikzpicture}}
\end{equation}

The next multiplet type is $\mathcal{B}[c_1,0,0;K]$, where one is instructed to remove $\QQ_{\bOn 2}\QQ_{\bOn 3}$ due to the shortening condition. Since $c_2=c_3=0$, one can deduce that $\QQ_{\bOn 2}\QQ_{\bOn 4}$ and $\QQ_{\bOn 3}\QQ_{\bOn 4}$ should also be removed. The multiplet spectrum is given by

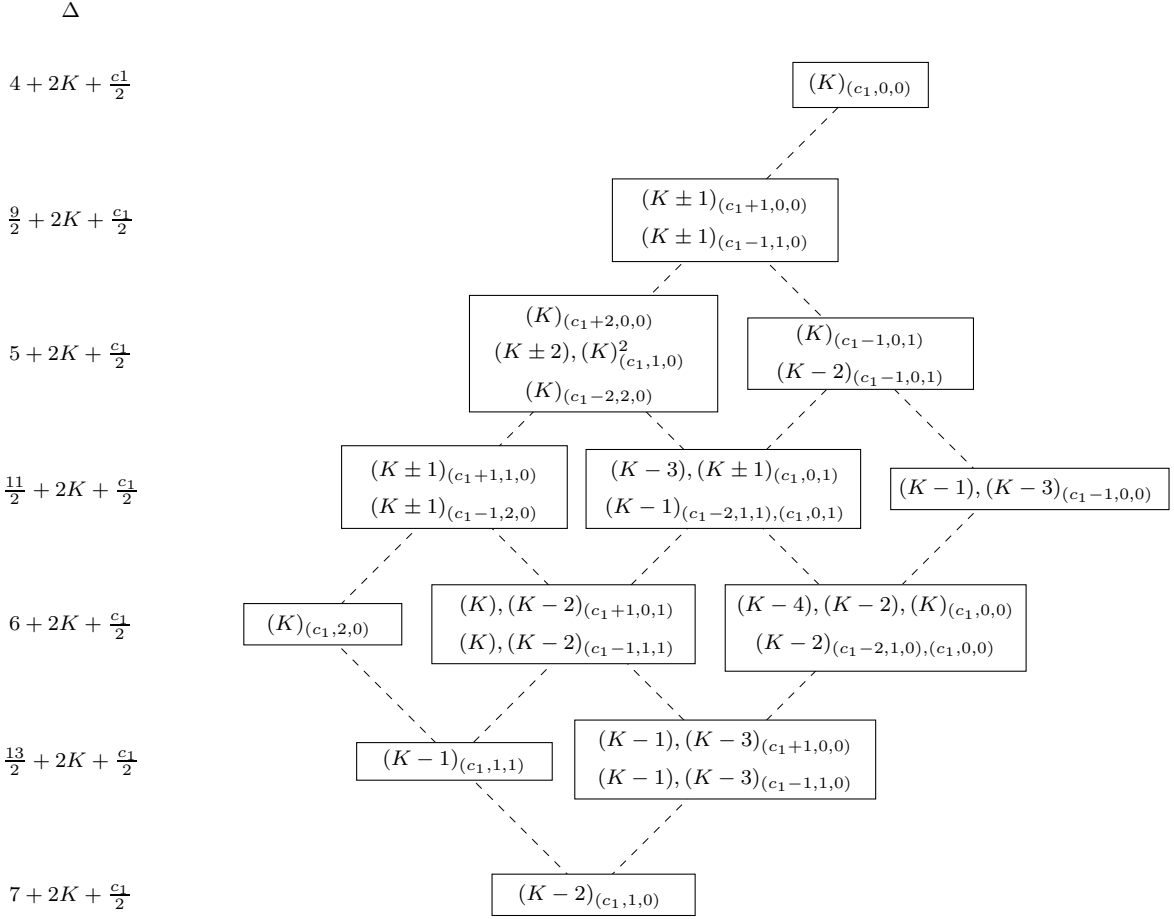
\begin{equation}\label{B[c1,0,0;K spec]}
{\scriptsize
\begin{tikzpicture}
\draw [dashed, ultra thin] (0,0) -- (-7.2,-7.2);
\draw [dashed, ultra thin] (-1.8,-1.8) -- (1.8,-5.4);
\draw [dashed, ultra thin] (-3.6,-3.6) -- (0,-7.2);
\draw [dashed, ultra thin] (-5.4,-5.4) -- (-1.8,-9);
\draw [dashed, ultra thin] (-7.2,-7.2) -- (-3.6,-10.8);
\draw [dashed, ultra thin] (0,-3.6) -- (-5.4,-9);
\draw [dashed, ultra thin] (1.8,-5.4) -- (-3.6,-10.8);
\node at (-10.5,1) {$\Delta$};
\node at (-10.5,0) {$4+2K+\frac{c1}{2}$};
\node at (-10.5,-1.8) {$\frac{9}{2}+2K+\frac{c_1}{2}$};
\node at (-10.5,-3.6) {$5+2K+\frac{c_1}{2}$};
\node at (-10.5,-5.4) {$\frac{11}{2}+2K+\frac{c_1}{2}$};
\node at (-10.5,-7.2) {$6+2K+\frac{c_1}{2}$};
\node at (-10.5,-9) {$\frac{13}{2}+2K+\frac{c_1}{2}$};
\node at (-10.5,-10.8) {$7+2K+\frac{c_1}{2}$};
\draw [fill=white] (0.9,0.3) rectangle (-0.9,-0.3);
\node at (0,0) {$(K)_{(c_1,0,0)}$};
\draw [fill=white] (-0.3,-2.35) rectangle (-3.3,-1.25);
\node at (-1.8,-1.55) {$(K\pm 1)_{(c_1+1,0,0)}$};
\node at (-1.8,-2.05) {$(K\pm 1)_{(c_1-1,1,0)}$};
\draw [fill=white] (-1.9,-2.8) rectangle (-5.2,-4.35);
\node at (-3.6,-3.1) {$(K)_{(c_1+2,0,0)}$};
\node at (-3.6,-3.6) {$(K\pm2),(K)^2_{(c_1,1,0)}$};
\node at (-3.6,-4.1) {$(K)_{(c_1-2,2,0)}$};
\draw [fill=white] (1.5,-3.1) rectangle (-1.5,-4.05);
\node at (0,-3.35) {$(K)_{(c_1-1,0,1)}$};
\node at (0,-3.85) {$(K-2)_{(c_1-1,0,1)}$};
\draw [fill=white] (-6.9,-4.8) rectangle (-3.9,-5.9);
\node at (-5.4,-5.15) {$(K\pm 1)_{(c_1+1,1,0)}$};
\node at (-5.4,-5.65) {$(K\pm 1)_{(c_1-1,2,0)}$};
\draw [fill=white] (-3.65,-4.85) rectangle (0,-5.9);
\node at (-1.8,-5.15) {$(K-3),(K\pm1)_{(c_1,0,1)}$};
\node at (-1.8,-5.65) {$(K-1)_{(c_1-2,1,1),(c_1,0,1)}$};
\draw [fill=white] (0.4,-5.1) rectangle (4.1,-5.65);
\node at (2.2,-5.4) {$(K-1),(K-3)_{(c_1-1,0,0)}$};
\draw [fill=white] (-8.2,-6.9) rectangle (-6.1,-7.45);
\node at (-7.2,-7.2) {$(K)_{(c_1,2,0)}$};
\draw [fill=white] (-5.7,-6.65) rectangle (-2.2,-7.7);
\node at (-3.9,-6.95) {$(K),(K-2)_{(c_1+1,0,1)}$};
\node at (-3.9,-7.45) {$(K),(K-2)_{(c_1-1,1,1)}$};
\draw [fill=white] (-1.8,-6.65) rectangle (2.2,-7.8);
\node at (0.2,-6.95) {$(K-4),(K-2),(K)_{(c_1,0,0)}$};
\node at (0.2,-7.45) {$(K-2)_{(c_1-2,1,0),(c_1,0,0)}$};
\draw [fill=white] (-6.7,-9.25) rectangle (-4.1,-8.75);
\node at (-5.4,-9) {$(K-1)_{(c_1,1,1)}$};
\draw [fill=white] (-3.8,-8.45) rectangle (0.2,-9.5);
\node at (-1.8,-8.75) {$(K-1),(K-3)_{(c_1+1,0,0)}$};
\node at (-1.8,-9.25) {$(K-1),(K-3)_{(c_1-1,1,0)}$};
\draw [fill=white] (-4.9,-10.5) rectangle (-2.2,-11.05);
\node at (-3.6,-10.8) {$(K-2)_{(c_1,1,0)}$};
\end{tikzpicture}}
\end{equation}

The last $\mathcal{B}$-type multiplet to consider is
$\mathcal{B}[0,0,0;K]$. Its shortening condition is
$A^1A^2A^3\Psi =0$ implying that $\QQ_{\bOn 1}\QQ_{\bOn 2}\QQ_{\bOn 3}$ shoud be removed from the auxiliary Verma-module basis. One further deduces that 
$\QQ_{\bOn a}\QQ_{\bOn b}\QQ_{\bOn c}$ should also be removed from the set of generators since $c_i=0$. The multiplet spectrum is determined to be
  
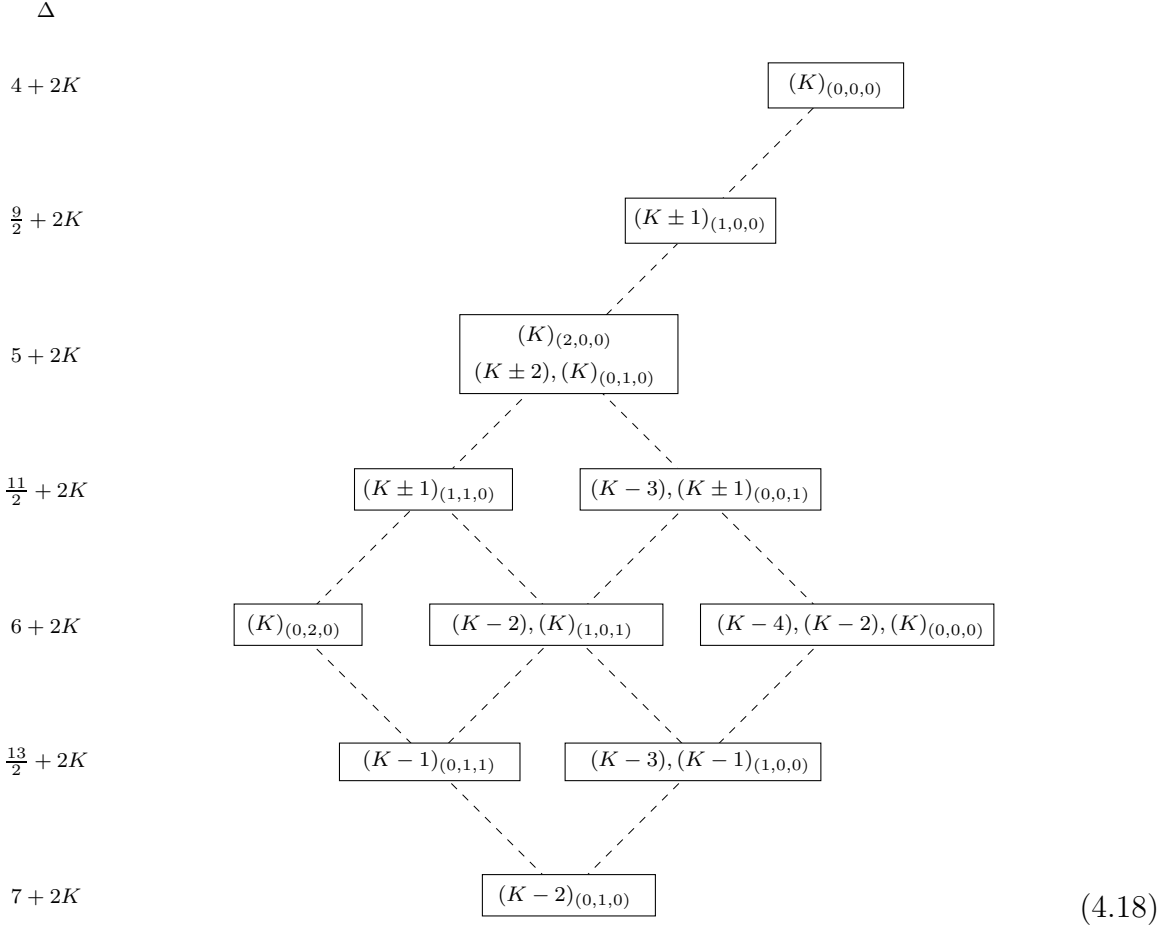
\begin{equation}\label{B[0,0,0;K spec]}
{\scriptsize
\begin{tikzpicture}
\draw [dashed, ultra thin] (0,0) -- (-7.2,-7.2);
\draw [dashed, ultra thin] (-3.6,-3.6) -- (0,-7.2);
\draw [dashed, ultra thin] (-5.4,-5.4) -- (-1.8,-9);
\draw [dashed, ultra thin] (-7.2,-7.2) -- (-3.6,-10.8);
\draw [dashed, ultra thin] (-1.8,-5.4) -- (-5.4,-9);
\draw [dashed, ultra thin] (0,-7.2) -- (-3.6,-10.8);
\node at (-10.5,1) {$\Delta$};
\node at (-10.5,0) {$4+2K$};
\node at (-10.5,-1.8) {$\frac{9}{2}+2K$};
\node at (-10.5,-3.6) {$5+2K$};
\node at (-10.5,-5.4) {$\frac{11}{2}+2K$};
\node at (-10.5,-7.2) {$6+2K$};
\node at (-10.5,-9) {$\frac{13}{2}+2K$};
\node at (-10.5,-10.8) {$7+2K$};
\draw [fill=white] (0.9,0.3) rectangle (-0.9,-0.3);
\node at (0,0) {$(K)_{(0,0,0)}$};
\draw [fill=white] (-0.8,-1.5) rectangle (-2.8,-2.1);
\node at (-1.8,-1.8) {$(K\pm 1)_{(1,0,0)}$};
\draw [fill=white] (-2.1,-3.05) rectangle (-5,-4.1);
\node at (-3.6,-3.35) {$(K)_{(2,0,0)}$};
\node at (-3.6,-3.85) {$(K\pm 2),(K)_{(0,1,0)}$};
\draw [fill=white] (-6.4,-5.1) rectangle (-4.3,-5.65);
\node at (-5.4,-5.4) {$(K\pm 1)_{(1,1,0)}$};
\draw [fill=white] (-3.4,-5.1) rectangle (-0.2,-5.65);
\node at (-1.8,-5.4) {$(K-3),(K\pm 1)_{(0,0,1)}$};
\draw [fill=white] (-8,-6.9) rectangle (-6.3,-7.45);
\node at (-7.2,-7.2) {$(K)_{(0,2,0)}$};
\draw [fill=white] (-5.4,-6.9) rectangle (-2.3,-7.45);
\node at (-3.9,-7.2) {$(K-2),(K)_{(1,0,1)}$};
\draw [fill=white] (-1.8,-6.9) rectangle (2.1,-7.45);
\node at (0.2,-7.2) {$(K-4),(K-2),(K)_{(0,0,0)}$};
\draw [fill=white] (-6.6,-9.25) rectangle (-4.2,-8.75);
\node at (-5.4,-9) {$(K-1)_{(0,1,1)}$};
\draw [fill=white] (-3.6,-8.75) rectangle (-0.2,-9.25);
\node at (-1.8,-9) {$(K-3),(K-1)_{(1,0,0)}$};
\draw [fill=white] (-4.7,-10.5) rectangle (-2.4,-11.05);
\node at (-3.6,-10.8) {$(K-2)_{(0,1,0)}$};
\end{tikzpicture}}
\end{equation}

The index over $\mathcal B$-type multiplets for all values of
$c_i$ and $K\neq 0$  is given by 
\begin{align}
\mathcal{I}_{\mathcal{B}[c_1,c_2,0;K]}(p,q,s)=(-1)^{c_1+1}q^{4+\frac{3K}{2}+\frac{c_1}{2}+c_2}\Bigg\{\frac{p^{-c_2} s^{c_2+5} \left(p^{-c_1}-p s^{c_1+1}\right)-p^{-c_1+2} s^{1-c_2}}{(p q-1) \left(p^2-s\right) (p s-1) \left(p-s^2\right) (q-s) (p-q s)}\nonumber\\
+ \frac{p^{c_2+5} s^{c_1+4}+p^{c_1+4} s^{-c_1} \left(s^{-c_2}-s^2 p^{c_2+2}\right)}{(p q-1) \left(p^2-s\right) (p s-1) \left(p-s^2\right) (q-s) (p-q s)}\Bigg\}\;.
\end{align}

\subsubsection{Higher-Spin-Current Multiplets: $\mathcal{B}[c_1,c_2,0;0]$}\label{higher spin 6d (1,0)}

In Eq.~\eqref{6D10coneq} we claimed that the special subset of
$\mathcal{B}$-type multiplets with $K=0$ should contain operator
constraints. Recall that for $\mathcal{B}[c_1,c_2,0;K]$ one should
remove $\QQ_{\bOn 4}$ and $\QQ_{\bOn 3}$ from the basis of
auxiliary Verma-module generators. Since $K=0$ we also remove two more supercharges,
$\QQ_{\bTw 4}$ and $\QQ_{\bTw 3}$. That is, we should completely
remove the set of $\tilde{Q}$ supercharges of \eqref{6DQQtilde} and
the spectrum is generated solely by acting with the set of all $Q$s:

\begin{equation}\label{B[c1,c2,0;0 spec]}
{\scriptsize
\begin{tikzpicture}
\draw [dashed, ultra thin] (0,0) -- (-7.2,-7.2);
\node at (-10.5,1) {$\Delta$};
\node at (-10.5,0) {$4+\frac{c1}{2}+c_2$};
\node at (-10.5,-1.8) {$\frac{9}{2}+\frac{c_1}{2}+c_2$};
\node at (-10.5,-3.6) {$5+\frac{c_1}{2}+c_2$};
\node at (-10.5,-5.4) {$\frac{11}{2}+\frac{c_1}{2}+c_2$};
\node at (-10.5,-7.2) {$6+\frac{c_1}{2}+c_2$};
\draw [fill=white] (0.9,0.3) rectangle (-0.9,-0.3);
\node at (0,0) {$(0)_{(c_1,c_2,0)}$};
\draw [fill=white] (-0.5,-2.35) rectangle (-3,-1.25);
\node at (-1.8,-1.55) {$(1)_{(c_1+1,c_2,0)}$};
\node at (-1.8,-2.05) {$(1)_{(c_1-1,c_2+1,0)}$};
\draw [fill=white] (-2.1,-2.8) rectangle (-5,-4.35);
\node at (-3.6,-3.1) {$(0)_{(c_1+2,c_2,0)}$};
\node at (-3.6,-3.6) {$(2),(0)_{(c_1,c_2+1,0)}$};
\node at (-3.6,-4.1) {$(0)_{(c_1-2,c_2+2,0)}$};
\draw [fill=white] (-6.6,-4.8) rectangle (-4.1,-5.9);
\node at (-5.4,-5.15) {$(1)_{(c_1+1,c_2+1,0)}$};
\node at (-5.4,-5.65) {$(1)_{(c_1-1,c_2+2,0)}$};
\draw [fill=white] (-8.2,-6.9) rectangle (-6.1,-7.45);
\node at (-7.2,-7.2) {$(0)_{(c_1,c_2+2,0)}$};
\end{tikzpicture}}
\end{equation}

This result seems to contradict \eqref{6D10coneq}, which predicted the
presence of operator constraints. However, note that the absent $\QQ$s
anticommute into $\PP_6$, which has therefore been implicitly removed
from the auxiliary Verma-module generators. This has the effect of projecting
out states corresponding to operator constraints and hence the
spectrum only contains reduced states. The operator constraints can
be restored using the dictionary developed in
App.~\ref{App:5DRS}. This can be done using the character expression
\begin{align}
\hat{\chi}[\Delta;c_1,c_2,c_2;K]=\chi[\Delta;c_1,c_2,c_3;K]-\chi[\Delta+1;c_1,c_2-1,c_3; K]\;,
\end{align}
where the hat indicates a character of the reduced
(i.e. $\PP_6$-removed) Verma module. The result is

\begin{equation}\label{B[c1,c2,0;0 spec eom]}
{\scriptsize
\begin{tikzpicture}
\draw [dashed, ultra thin] (0,0) -- (-7.2,-7.2);
\draw [dashed, ultra thin] (3.6,-3.6) -- (-3.6,-10.8);
\node at (-10.5,1) {$\Delta$};
\node at (-10.5,0) {$4+\frac{c1}{2}+c_2$};
\node at (-10.5,-1.8) {$\frac{9}{2}+\frac{c_1}{2}+c_2$};
\node at (-10.5,-3.6) {$5+\frac{c_1}{2}+c_2$};
\node at (-10.5,-5.4) {$\frac{11}{2}+\frac{c_1}{2}+c_2$};
\node at (-10.5,-7.2) {$6+\frac{c_1}{2}+c_2$};
\node at (-10.5,-9) {$\frac{13}{2}+\frac{c_1}{2}+c_2$};
\node at (-10.5,-10.8) {$7+\frac{c_1}{2}+c_2$};
\draw [fill=white] (0.9,0.3) rectangle (-0.9,-0.3);
\node at (0,0) {$(0)_{(c_1,c_2,0)}$};
\draw [fill=white] (-0.5,-2.35) rectangle (-3,-1.25);
\node at (-1.8,-1.55) {$(1)_{(c_1+1,c_2,0)}$};
\node at (-1.8,-2.05) {$(1)_{(c_1-1,c_2+1,0)}$};
\draw [fill=white] (-2.1,-2.8) rectangle (-5,-4.35);
\node at (-3.6,-3.1) {$(0)_{(c_1+2,c_2,0)}$};
\node at (-3.6,-3.6) {$(2),(0)_{(c_1,c_2+1,0)}$};
\node at (-3.6,-4.1) {$(0)_{(c_1-2,c_2+2,0)}$};
\draw [fill=white] (-6.6,-4.8) rectangle (-4.1,-5.9);
\node at (-5.4,-5.15) {$(1)_{(c_1+1,c_2+1,0)}$};
\node at (-5.4,-5.65) {$(1)_{(c_1-1,c_2+2,0)}$};
\draw [fill=white] (-8.2,-6.9) rectangle (-6.1,-7.45);
\node at (-7.2,-7.2) {$(0)_{(c_1,c_2+2,0)}$};
\draw [fill=white] (2.5,-3.3) rectangle (4.7,-3.9);
\node at (3.6,-3.6) {$-(0)_{(c_1,c_2-1,0)}$};
\draw [fill=white] (3.1,-5.95) rectangle (0.6,-4.85);
\node at (1.8,-5.15) {$-(1)_{(c_1+1,c_2-1,0)}$};
\node at (1.8,-5.65) {$-(1)_{(c_1-1,c_2,0)}$};
\draw [fill=white] (1.5,-6.4) rectangle (-1.5,-7.95);
\node at (0,-6.7) {$-(0)_{(c_1+2,c_2-1,0)}$};
\node at (0,-7.2) {$-(2),-(0)_{(c_1,c_2,0)}$};
\node at (0,-7.7) {$-(0)_{(c_1-2,c_2+1,0)}$};
\draw [fill=white] (-3,-8.4) rectangle (-0.5,-9.5);
\node at (-1.8,-8.75) {$-(1)_{(c_1+1,c_2,0)}$};
\node at (-1.8,-9.25) {$-(1)_{(c_1-1,c_2+1,0)}$};
\draw [fill=white] (-4.6,-10.5) rectangle (-2.5,-11.05);
\node at (-3.6,-10.8) {$-(0)_{(c_1,c_2+1,0)}$};
\end{tikzpicture}}
\end{equation}

An example of a superconformal primary for a $\mathcal{B}[0,c_2,0;0]$
multiplet with $c_2>0$ is
\begin{align}
\mathcal{O}_{\mu_1 \cdots \mu_{c_2}}= \epsilon_{\bf A B} \phi^{\bA} \overset\leftrightarrow{\partial}_{\mu_1} \cdots \overset\leftrightarrow{\partial}_{\mu_{c_2}} \phi^{\bB}\;,
\end{align}
where $\phi^A$ is a free hypermultiplet scalar. It is interesting to
point out that for $c_1=1,c_2=0$ or $c_1=0,c_2=1$ the superconformal
primary is not higher spin. The higher-spin currents are instead found
as their descendants. Moreover, the superconformal primary in the
multiplet of $\mathcal B[0,1,0;0]$ is a $\Delta= 5$, R-neutral
conserved current. However, as explained around \eqref{nosymm} in the
case of 5D, this is not a flavour current.

The corresponding superconformal index for any $c_1, c_2\geq0$ reads
\begin{align}
\mathcal{I}_{\mathcal{B}[c_1,c_2,0;0]}(p,q,s)=(-1)^{c_1+1}q^{4+\frac{1}{2}(c_1+2c_2)}\Bigg\{\frac{p^{-c_2} s^{c_2+5} \left(p^{-c_1}-p s^{c_1+1}\right)-p^{-c_1+2} s^{1-c_2}}{(p q-1) \left(p^2-s\right) (p s-1) \left(p-s^2\right) (q-s) (p-q s)}\nonumber\\
+ \frac{p^{c_2+5} s^{c_1+4}+p^{c_1+4} s^{-c_1} \left(s^{-c_2}-s^2 p^{c_2+2}\right)}{(p q-1) \left(p^2-s\right) (p s-1) \left(p-s^2\right) (q-s) (p-q s)}\Bigg\}\;.
\end{align}

\subsubsection{The Stress-Tensor Multiplet: $\mathcal{B}[0,0,0;0]$}

The last $\mathcal{B}$-type multiplet of note is the stress
tensor. Ordinarily, the shortening conditions require that we remove
$\QQ_{\bOn a}\QQ_{\bOn b}\QQ_{\bOn c}$ from the basis of generators,
but since $K=0$ we also need to remove contributions obtained by
acting repeatedly with $R$-symmetry lowering operators. The resulting
multiplet is therefore

\begin{equation}\label{stress tensor spec]}
\begin{tikzpicture}
\draw [dashed, ultra thin] (0,0) -- (-7.2,-7.2);
\draw [dashed, ultra thin] (0,-7.2) -- (-3.6,-10.8);
\node at (-10.5,1) {$\Delta$};
\node at (-10.5,0) {$4$};
\node at (-10.5,-1.8) {$\frac{9}{2}$};
\node at (-10.5,-3.6) {$5$};
\node at (-10.5,-5.4) {$\frac{11}{2}$};
\node at (-10.5,-7.2) {$6$};
\node at (-10.5,-9) {$\frac{13}{2}$};
\node at (-10.5,-10.8) {$7$};
\draw [fill=white] (0.9,0.4) rectangle (-0.9,-0.3);
\node at (0,0) {$(0)_{(0,0,0)}$};
\draw [fill=white] (-0.9,-2.1) rectangle (-2.7,-1.4);
\node at (-1.8,-1.8) {$(1)_{(1,0,0)}$};
\draw [fill=white] (-2.7,-2.95) rectangle (-4.5,-4.15);
\node at (-3.6,-3.35) {$(0)_{(2,0,0)}$};
\node at (-3.6,-3.85) {$(2)_{(0,1,0)}$};
\draw [fill=white] (-1,-6.8) rectangle (1,-7.55);
\node at (-0,-7.2) {$-(2)_{(0,0,0)}$};
\draw [fill=white] (-6.2,-5) rectangle (-4.5,-5.75);
\node at (-5.4,-5.4) {$(1)_{(1,1,0)}$};
\draw [fill=white] (-2.8,-8.6) rectangle (-0.7,-9.35);
\node at (-1.8,-9) {$-(1)_{(1,0,0)}$};
\draw [fill=white] (-8.1,-6.8) rectangle (-6.2,-7.55);
\node at (-7.2,-7.2) {$(0)_{(0,2,0)}$};
\draw [fill=white] (-4.6,-10.4) rectangle (-2.5,-11.15);
\node at (-3.6,-10.8) {$-(0)_{(0,1,0)}$};
\end{tikzpicture}
\end{equation}

We recognise these states as being associated with the fields
\begin{align}
[5;0,1,0;2]&:\hspace{3mm}J^{(\bA\bB)}_\mu\;,&-[6;0,0,0;2]&:\hspace{3mm}\partial^\mu J^{(\bA\bB)}_\mu=0\;,\nonumber\\
[11/2;1,1,0;1]&:\hspace{3mm}S^{\bA}_{\mu a}\;,&-[13/2;1,0,0;1]&:\hspace{3mm}\partial^\mu S^{\bA}_{\mu a}=0\;,\\
[6;0,2,0;0]&:\hspace{3mm}\Theta_{\mu\nu}\;,&-[7;0,1,0;0]&:\hspace{3mm}\partial^\mu \Theta_{\mu\nu}=0\;,\nonumber
\end{align}
namely the 6D $R$-symmetry current, supersymmetry current and stress
tensor. We also have three states that do not obey equations of
motion. These are
\begin{align}
[4;0,0,0;0]&:\hspace{3mm}\Sigma\;,\cr
[9/2;1,0,0;1]&:\hspace{3mm}\zeta^{\bA}_{a}\;,\cr
[5;2,0,0;0]&:\hspace{3mm}Z^+_{(ab)}\;,
\end{align}
where $+$ denotes the selfdual part of the operator.

The index over this stress-tensor multiplet is
\begin{align}\label{index for stress}
\mathcal{I}_{\mathcal{B}[0,0,0;0]}(p,q,s)=q^{4}\frac{s^{-1}+p+p^{-1}s}{(1-p q) (1-q s^{-1}) (1-p^{-1}q s)}\;,
\end{align}
counting three components of the $R$-symmetry current plus conformal descendants.

\subsection{$\mathcal{C}$-type Multiplets}

The two distinct $\mathcal{C}$-type multiplets are
$\mathcal{C}[c_1,0,0;K]$ and $\mathcal{C}[0,0,0;K]$, with the
requirement that we remove $\QQ_{\bOn a}$ for $a\neq 1$ and
$\QQ_{\bOn a}\QQ_{\bOn b}$ respectively from the basis of auxiliary Verma-module
generators.  On the one hand, since the generators $\QQ_{\bOn 3}$ and
$\QQ_{\bOn 4}$ are absent, the set of available $\tilde{Q}$s is the
same as for the $\mathcal{B}[c_1,c_2,0;K]$ case, that is
$\QQ_{\bTw 3}$, $\QQ_{\bTw 4}$. On the other, only $\QQ_{\bOn 2}$ is
removed from the set of $Q$s. Therefore the two resulting chains of
supercharge actions on a generic state $(K)_{(c_1,c_2,c_3)}$ are
\begin{align}\begin{split}
(K)_{(c_1,c_2,c_3)}&\xrightarrow{Q^{\phantom{2}}} (K\pm 1)_{(c_1+1,c_2,c_3)}\;,(K-1)_{(c_1-1,c_2+1,c_3)}\;,\\
&\xrightarrow{Q^2}(K- 2)_{(c_1,c_2+1,c_3)}\;,\;(K)_{(c_1+2,c_2,c_3),(c_1,c_2+1,c_3)}\;,\\
&\xrightarrow{Q^3}(K- 1)_{(c_1+1,c_2+1,c_3)}\;,\\
(K)_{(c_1,c_2,c_3)}&\xrightarrow{\tilde{Q}^{\phantom{2}}} (K-1)_{(c_1,c_2-1,c_3+1),(c_1,c_2,c_3-1)}\;,\\
&\xrightarrow{\tilde{Q}^2}(K-2)_{(c_1,c_3+1,c_3)}\;.
\end{split}
\end{align}
After substituting in the relevant $c_i$ values for the primary and
implementing the RS algorithm we obtain for $\mathcal{C}[c_1,0,0;K]$

\begin{equation}\label{C[c1,0,0;K spec]}
{\scriptsize
\begin{tikzpicture}
\draw [dashed, ultra thin] (0,0) -- (-4.5,-4.5);
\draw [dashed, ultra thin] (-1.5,-1.5) -- (1.5,-4.5);
\draw [dashed, ultra thin] (-3,-3) -- (0,-6);
\draw [dashed, ultra thin] (0,-3) -- (-3,-6);
\draw [dashed, ultra thin] (1.5,-4.5) -- (-1.5,-7.5);
\draw [dashed, ultra thin] (-4.5,-4.5) -- (-1.5,-7.5);
\node at (-7.5,1) {$\Delta$};
\node at (-7.5,0) {$2+2K+\frac{c1}{2}$};
\node at (-7.5,-1.5) {$\frac{5}{2}+2K+\frac{c_1}{2}$};
\node at (-7.5,-3) {$3+2K+\frac{c_1}{2}$};
\node at (-7.5,-4.5) {$\frac{7}{2}+2K+\frac{c_1}{2}$};
\node at (-7.5,-6) {$4+2K+\frac{c_1}{2}$};
\node at (-7.5,-7.5) {$\frac{9}{2}+2K+\frac{c_1}{2}$};
\draw [fill=white] (-0.9,-0.3) rectangle (0.9,0.3);
\node at (0,0) {$(K)_{(c_1,0,0)}$};
\draw [fill=white] (-2.7,-0.95) rectangle (-0.3,-2.05);
\node at (-1.5,-1.25) {$(K\pm 1)_{(c_1+1,0,0)}$};
\node at (-1.5,-1.75) {$(K- 1)_{(c_1-1,1,0)}$};
\draw [fill=white] (-4.35,-2.45) rectangle (-1.65,-3.55);
\node at (-3,-2.75) {$(K-2),(K)_{(c_1,1,0)}$};
\node at (-3,-3.25) {$(K)_{(c_1+2,0,0)}$};
\draw [fill=white] (-1.2,-2.7) rectangle (1.2,-3.3);
\node at (-0,-3) {$(K-2)_{(c_1-1,0,1)}$};
\draw [fill=white] (-5.7,-4.2) rectangle (-3.3,-4.8);
\node at (-4.5,-4.5) {$(K-1)_{(c_1+1,1,0)}$};
\draw [fill=white] (-3.1,-4.2) rectangle (0.1,-4.8);
\node at (-1.5,-4.5) {$(K-1),(K-3)_{(c_1,0,1)}$};
\draw [fill=white] (0.3,-4.2) rectangle (2.7,-4.8);
\node at (1.5,-4.5) {$(K-3)_{(c_1-1,0,0)}$};
\draw [fill=white] (-4.2,-5.7) rectangle (-1.8,-6.3);
\node at (-3,-6) {$(K-2)_{(c_1+1,0,1)}$};
\draw [fill=white] (1.6,-5.7) rectangle (-1.6,-6.3);
\node at (0,-6) {$(K-2),(K-4)_{(c_1,0,0)}$};
\draw [fill=white] (-2.7,-7.2) rectangle (-0.3,-7.8);
\node at (-1.5,-7.5) {$(K-3)_{(c_1+1,0,0)}$};
\end{tikzpicture}}
\end{equation}

The corresponding superconformal index is
\begin{align}\label{C[c_1,0,0;K] index}
\mathcal{I}_{\mathcal{C}[c_1,0,0;K]}(p,q,s)=(-1)^{c_1}q^{2+\frac{3K}{2}+\frac{c_1}{2}} \frac{  p^{ c_1+5} s^{-c_1} (p s-1)+p^{2} s^{c_1+5} \left(s-p^2\right)+p^{-c_1} s^{2} \left(p-s^2\right)}{(p q-1) \left(p^2-s\right) (p s-1) \left(p-s^2\right) (q-s) (q s-p)}\;,
\end{align}
which can be used in conjunction with $\mathcal{A}[c_1,0,0;K]$ to
verify the recombination rules
\begin{align}
\lim_{\epsilon\rightarrow 0}\mathcal{I}_{\mathcal{L}[\Delta+\epsilon;c_1,0,0;K]}(p,q,s)=\mathcal{I}_{\mathcal{A}[c_1,0,0;K]}(p,q,s)+\mathcal{I}_{\mathcal{C}[c_1-1,0,0;K+3]}(p,q,s)=0\;.
\end{align}

Turning our attention to $\mathcal{C}[0,0,0;K]$, we recall that the
shortening conditions require the combinations
$\QQ_{\bOn a}\QQ_{\bOn b}$ to be absent from the basis of
auxiliary Verma-module generators. The resulting spectrum is alternatively obtained by
setting $c_1=0$ in $\mathcal{C}[c_1,0,0;K]$ and running the RS
algorithm. This is a simple task: the lowest value of $c_1$ that
appears in $\mathcal{C}[c_1,0,0;K]$ is $c_1-1$ and only these states
are deleted when setting $c_1=0$---they will be on the boundary of the
Weyl chamber. There is no need to perform any of the more elaborate
Weyl reflections on any other state.

Therefore the spectrum for $\mathcal{C}[0,0,0;K]$ is given by 

\begin{equation}
{\scriptsize
\begin{tikzpicture}
\draw [dashed, ultra thin] (0,0) -- (-4.5,-4.5);
\draw [dashed, ultra thin] (-3,-3) -- (0,-6);
\draw [dashed, ultra thin] (-1.5,-4.5) -- (-3,-6);
\draw [dashed, ultra thin] (0,-6) -- (-1.5,-7.5);
\draw [dashed, ultra thin] (-4.5,-4.5) -- (-1.5,-7.5);
\node at (-7.5,1) {$\Delta$};
\node at (-7.5,0) {$2+2K$};
\node at (-7.5,-1.5) {$\frac{5}{2}+2K$};
\node at (-7.5,-3) {$3+2K$};
\node at (-7.5,-4.5) {$\frac{7}{2}+2K$};
\node at (-7.5,-6) {$4+2K$};
\node at (-7.5,-7.5) {$\frac{9}{2}+2K$};
\draw [fill=white] (-0.8,-0.3) rectangle (0.8,0.3);
\node at (0,0) {$(K)_{(0,0,0)}$};
\draw [fill=white] (-2.5,-1.2) rectangle (-0.5,-1.8);
\node at (-1.5,-1.5) {$(K\pm 1)_{(1,0,0)}$};
\draw [fill=white] (-4.3,-2.45) rectangle (-1.7,-3.55);
\node at (-3,-2.75) {$(K-2),(K)_{(0,1,0)}$};
\node at (-3,-3.25) {$(K)_{(2,0,0)}$};
\draw [fill=white] (-5.5,-4.2) rectangle (-3.5,-4.8);
\node at (-4.5,-4.5) {$(K-1)_{(1,1,0)}$};
\draw [fill=white] (-3,-4.2) rectangle (-0,-4.8);
\node at (-1.5,-4.5) {$(K-1),(K-3)_{(0,0,1)}$};
\draw [fill=white] (-4,-5.7) rectangle (-2,-6.3);
\node at (-3,-6) {$(K-2)_{(1,0,1)}$};
\draw [fill=white] (1.6,-5.7) rectangle (-1.6,-6.3);
\node at (0,-6) {$(K-2),(K-4)_{(0,0,0)}$};
\draw [fill=white] (-0.5,-7.2) rectangle (-2.5,-7.8);
\node at (-1.5,-7.5) {$(K-3)_{(1,0,0)}$};
\end{tikzpicture}}
\end{equation}
The associated index for $K>1$ evaluates to
\begin{align}\label{index for C[0,0,0;K]}
\mathcal{I}_{\mathcal{C}[0,0,0;K]}(p,q,s)=-q^{2+\frac{3 K}{2}}\frac{s^{-1} p+s+p^{-1}}{(1-p q) (1-q s^{-1}) (1-p^{-1}q s)}\;.
\end{align}

The cases with $K =0,1$ are predicted to contain operator constraints
from \eqref{6D10coneq} and will be dealt with separately below.

\subsubsection{$\mathcal{C}[c_1,0,0;0]$}

For this class of multiplets, $K = 0$ and we obtain additional
shortening conditions. These can be translated into the requirement
that $\QQ_{\bA a}$ for $a\neq 1$ be removed from the basis of
auxiliary Verma-module generators. The latter then in turn imply that
$\PP_{23}\sim \{\QQ_{\bOn 2},\QQ_{\bTw 3} \}$,
$ \PP_{24}\sim \{\QQ_{\bOn 2},\QQ_{\bTw 4}\}$ and
$ \PP_{34}\sim \{\QQ_{\bOn 3},\QQ_{\bTw 4}\}$ have also been removed
from the auxiliary Verma-module basis. In vector notation, these
momentum operators correspond respectively to $\PP_3$, $\PP_5$ and
$\PP_6$.  The implication of this fact is that the multiplet
construction will only reproduce the reduced states, with the
operator constraints having been projected out. Thus the
$\mathcal{C}[c_1,0,0;0]$ multiplet is very simply:

\begin{eqnarray}\label{c100}
\begin{tikzpicture}
\draw [dashed, ultra thin] (0,0) -- (-3,-3);
\node at (-6,1) {$\Delta$};
\node at (-6,0) {$2+\frac{c_1}{2}$};
\node at (-6,-1.5) {$\frac{5}{2}+\frac{c_1}{2}$};
\node at (-6,-3) {$3+\frac{c_1}{2}$};
\draw [fill=white] (-0.9,-0.4) rectangle (0.9,0.4);
\node at (0,0) {$(0)_{(c_1,0,0)}$};
\draw [fill=white] (-2.6,-1.1) rectangle (-0.4,-1.9);
\node at (-1.5,-1.5) {$(1)_{(c_1+1,0,0)}$};
\draw [fill=white] (-4.1,-2.6) rectangle (-1.9,-3.4);
\node at (-3,-3) {$(0)_{(c_1+2,0,0)}$};
\end{tikzpicture}
\end{eqnarray}

The operator constraints can be restored by means of
App.~\ref{App:5DRS}. Implementing this would result in introducing---for
each of the three states in \eqref{c100}---the following combinations
\begin{align}
(K)^\Delta_{(c_1,0,0)}:\qquad -(K)^{\Delta +1}_{(c_1-1,0,1)}+(K)^{\Delta +2}_{(c_1-2,1,0)}-(K)^{\Delta +3}_{(c_1-2,0,0)}\;,
\end{align}
for a total of nine additional states corresponding to 
equations of motion. As a side comment, note that this spectrum is not
the one we would have obtained had we just set $K=0$ in
\eqref{C[c1,0,0;K spec]} and implemented the RS algorithm. As such,
the index that one obtains is different to just setting $K=0$ in
\eqref{C[c_1,0,0;K] index}, and is given by
\begin{align}
\mathcal{I}_{\mathcal{C}[c_1,0,0;0]}=(-1)^{c_1}q^{2+\frac{c_1}{2}}\Bigg\{\frac{p^{c_1+6} s^{1-c_1}-p^{c_1+5} s^{-c_1}+s^2 p^{-c_1} \left(p-s^2\right)-p^4 s^{c_1+5}+p^2 s^{c_1+6}}{(p q-1) \left(p^2-s\right) (p s-1) \left(p-s^2\right) (q-s) (q s-p)}\nonumber\\
+ \frac{  p^{ c_1+5} s^{-c_1} (p s-1)+p^{2} s^{c_1+5} \left(s-p^2\right)+p^{-c_1} s^{2} \left(p-s^2\right)}{(p q-1) \left(p^2-s\right) (p s-1) \left(p-s^2\right) (q-s) (q s-p)}\Bigg\}\;.
\end{align}

\subsubsection{The Free-Tensor Multiplet: $\mathcal{C}[0,0,0;0]$}

When $c_1=0$ we recover the free-tensor multiplet. In this case the
lowering operator of $\su(2)_R$ acts on the null-state condition
$\QQ_{\bOn 1}\QQ_{\bOn 2}\Psi_{\rm aux} =0$ to create additional shortening conditions. This
translates into the requirement that $\QQ_{\bOn a}\QQ_{\bOn b}$,
$\QQ_{\bTw a}\QQ_{\bOn b}$ and $\QQ_{\bTw a}\QQ_{\bTw b}$ for
$a\neq b$ should be removed from the basis of auxiliary Verma-module
generators. The multiplet can be constructed using the remaining
supercharges and is given by

\begin{equation}\label{free tensor mult diagram N=1}
\begin{tikzpicture}
\draw [dashed, ultra thin] (0,0) -- (-3,-3);
\node at (-4.5,1) {$\Delta$};
\node at (-4.5,0) {$2$};
\node at (-4.5,-1.5) {$\frac{5}{2}$};
\node at (-4.5,-3) {$3$};
\draw [fill=white] (-0.9,-0.4) rectangle (0.9,0.4);
\node at (0,0) {$(0)_{(0,0,0)}$};
\draw [fill=white] (-2.4,-1.1) rectangle (-0.6,-1.9);
\node at (-1.5,-1.5) {$(1)_{(1,0,0)}$};
\draw [fill=white] (-3.9,-2.6) rectangle (-2.1,-3.4);
\node at (-3,-3) {$(0)_{(2,0,0)}$};
\end{tikzpicture}
\end{equation}
	
We identify the states with the following fields 
\begin{align}
[2;0,0;0;0]&:\hspace{3mm}\varphi\;,\cr
[5/2;1,0,0;1]&:\hspace{3mm}\lambda^{\bf A}_{a}\;,\cr
[3;2,0,0;0]&: \hspace{3mm}H^+_{[\mu\nu\rho]}\;.
\end{align}
Similar to the previous subsection, \eqref{free tensor mult diagram
  N=1} only contains reduced states and no equations of motion.
When the latter are restored by means of App.~\ref{App:5DRS} we
recover the expected
\begin{align}\label{field EoM free tensor}\begin{split}
\partial^2 \varphi =0\;:& \hspace{3mm}-[4;0,0,0;0]\;,\\
\slashed{\partial}^{\dot a\dot b}\tilde\lambda_{\dot b}^{\bA} =0\;:& \hspace{3mm}-[7/2;0,0,1;1]\;,\\
\partial_{[\sigma}H^+_{\mu\nu\rho]}=0\;:& \hspace{3mm}-[4;1,0,1;0]+[5;0,1,0;0]-[6;0,0,0;0]\;.
\end{split}
\end{align}

The index for this configuration evaluates to
\begin{align}
\mathcal{I}_{\mathcal{C}[0,0,0;0]}(p,q,s)=-q^{2}\frac{s^{-1} p+s+p^{-1}-q}{(1-p q) (1-q s^{-1}) (1-p^{-1}q s)}\;,
\end{align}
counting three components of the fermion $\lambda^{\bf A}_a$ alongside
its equation of motion.

\subsubsection{Higher-Spin-Current Multiplets: $\mathcal{C}[c_1,0,0;1]$}

These multiplets are simple to construct since, contrary to the above
case, we need not act with the $R$-symmetry lowering operator. As such,
we can take the spectrum of \eqref{C[c1,0,0;K spec]} and set $K=1$
without incurring new shortening conditions. The resulting $\mathcal{C}[c_1,0,0;1]$ multiplet is

\begin{equation}\label{C[c1,0,0;1 spec]}
\begin{tikzpicture}
\draw [dashed, ultra thin] (0,0) -- (-4.5,-4.5);
\draw [dashed, ultra thin] (1.5,-4.5) -- (-1.5,-7.5);
\node at (-7.5,1) {$\Delta$};
\node at (-7.5,0) {$4+\frac{c1}{2}$};
\node at (-7.5,-1.5) {$\frac{9}{2}+\frac{c_1}{2}$};
\node at (-7.5,-3) {$5+\frac{c_1}{2}$};
\node at (-7.5,-4.5) {$\frac{11}{2}+\frac{c_1}{2}$};
\node at (-7.5,-6) {$6+\frac{c_1}{2}$};
\node at (-7.5,-7.5) {$\frac{13}{2}+\frac{c_1}{2}$};
\draw [fill=white] (-0.9,-0.4) rectangle (0.9,0.4);
\node at (0,0) {$(1)_{(c_1,0,0)}$};
\draw [fill=white] (-3,-0.85) rectangle (0,-2.15);
\node at (-1.5,-1.25) {$(2),(0)_{(c_1+1,0,0)}$};
\node at (-1.5,-1.75) {$(0)_{(c_1-1,1,0)}$};
\draw [fill=white] (-4.1,-2.35) rectangle (-1.9,-3.65);
\node at (-3,-2.75) {$(1)_{(c_1,1,0)}$};
\node at (-3,-3.25) {$(1)_{(c_1+2,0,0)}$};
\draw [fill=white] (-5.6,-4.1) rectangle (-3.4,-4.9);
\node at (-4.5,-4.5) {$(0)_{(c_1+1,1,0)}$};
\draw [fill=white] (0.3,-4.1) rectangle (2.7,-4.9);
\node at (1.5,-4.5) {$-(0)_{(c_1-1,0,0)}$};
\draw [fill=white] (1,-6.4) rectangle (-1,-5.6);
\node at (0,-6) {$-(1)_{(c_1,0,0)}$};
\draw [fill=white] (-2.7,-7.1) rectangle (-0.3,-7.9);
\node at (-1.5,-7.5) {$-(0)_{(c_1+1,0,0)}$};
\end{tikzpicture}
\end{equation}

which includes conservation equations. It is worth pointing out that
the multiplet with $c_1 = 1$ contains a $\Delta = 5 $, R-neutral
conserved current as a level one descendant. However, we see that
there is also a spin-2 current with $\Delta=6$ at level three.

The corresponding index evaluates to 
\begin{align}\label{C[c_1,0,0;1] index}
\mathcal{I}_{\mathcal{C}[c_1,0,0;1]}(p,q,s)=q^{\frac{7}{2}+\frac{c_1}{2}} \frac{  p^{ c_1+5} s^{-c_1} (p s-1)+p^{2} s^{c_1+5} \left(s-p^2\right)+p^{-c_1} s^{2} \left(p-s^2\right)}{(p q-1) \left(p^2-s\right) (p s-1) \left(p-s^2\right) (q-s) (q s-p)}\;.
\end{align}
Note that this index is valid for $c_1=0$. This is because
setting $K=1$ does not introduce any additional shortening conditions. However, when $c_1=0$ this multiplet does not contain currents with spin $j\geq 2$, and will be discussed below.

\subsubsection{``Extra''-Supercurrent Multiplet: $\mathcal{C}[0,0,0;1]$}

These multiplets can be constructed by substituting $K=1$
into the spectrum of $\mathcal{C}[0,0,0;K]$. The result is

\begin{equation}\label{C[0,0,0;1 spec]}
\begin{tikzpicture}
\draw [dashed, ultra thin] (0,0) -- (-4.5,-4.5);
\draw [dashed, ultra thin] (0,-6) -- (-1.5,-7.5);
\node at (-7.5,1) {$\Delta$};
\node at (-7.5,0) {$4$};
\node at (-7.5,-1.5) {$\frac{9}{2}$};
\node at (-7.5,-3) {$5$};
\node at (-7.5,-4.5) {$\frac{11}{2}$};
\node at (-7.5,-6) {$6$};
\node at (-7.5,-7.5) {$\frac{13}{2}$};
\draw [fill=white] (-0.9,-0.4) rectangle (0.9,0.4);
\node at (0,0) {$(1)_{(0,0,0)}$};
\draw [fill=white] (-2.8,-1.1) rectangle (-0.2,-1.85);
\node at (-1.5,-1.5) {$(2),(0)_{(1,0,0)}$};
\draw [fill=white] (-3.9,-2.35) rectangle (-2.1,-3.65);
\node at (-3,-2.75) {$(1)_{(0,1,0)}$};
\node at (-3,-3.25) {$(1)_{(2,0,0)}$};
\draw [fill=white] (-5.4,-4.1) rectangle (-3.6,-4.9);
\node at (-4.5,-4.5) {$(0)_{(1,1,0)}$};
\draw [fill=white] (1,-6.4) rectangle (-1,-5.6);
\node at (0,-6) {$-(1)_{(0,0,0)}$};
\draw [fill=white] (-2.5,-7.1) rectangle (-0.5,-7.9);
\node at (-1.5,-7.5) {$-(0)_{(1,0,0)}$};
\end{tikzpicture}
\end{equation}

This multiplet contains a conserved $K=1$ spin-1 current and a
conserved R-neutral spin-$\frac{3}{2}$ current. These generate
additional global and supersymmetry transformations, which---as we
will discuss in Sec.~\ref{susy1020}---are necessary for the
description of the (2,0) stress-tensor multiplet in the language of
the (1,0) SCA.

Its index is then given by
\begin{align}\label{index for C[0,0,0;1]}
\mathcal{I}_{\mathcal{C}[0,0,0;1]}(p,q,s)=-q^{\frac{7}{2}}\frac{s^{-1} p+s+p^{-1}}{(1-p q) (1-q s^{-1}) (1-p^{-1}q s)}\;.
\end{align}

\subsection{$\mathcal{D}$-type Multiplets}

We finally turn to the $\mathcal D[0,0,0;K]$ multiplets. These are the
simplest of the 6D $(1,0)$ SCA for generic values of $K$. Since the
shortening condition is $A^1 \Psi = \QQ_{\bOn 1}\Psi=0$, acting with
all Lorentz lowering operators leads to
\begin{align}
\QQ_{\bOn a}\Psi=0
\end{align}
and hence the multiplet is $\frac{1}{2}$--BPS. 

Starting with a generic superconformal primary state
$(K)_{(c_1,c_2,c_3)}$, the two chains obtained from the action of
$Q$s and $\tilde{Q}$s are
\begin{align}\begin{split}
(K)_{(c_1,c_2,c_3)}&\xrightarrow{{Q}^{\phantom{2}}} (K-1)_{(c_1+1,c_2,c_3),(c_1-1,c_2+1,c_3)}\;,\\
&\xrightarrow{{Q}^2}(K-2)_{(c_1,c_2+1,c_3)}\;,\\
(K)_{(c_1,c_2,c_3)}&\xrightarrow{\tilde{Q}^{\phantom{2}}} (K-1)_{(c_1,c_2-1,c_3+1),(c_1,c_2,c_3-1)}\;,\\
&\xrightarrow{\tilde{Q}^2}(K-2)_{(c_1,c_3+1,c_3)}\;.
\end{split}
\end{align}
One then needs to set $c_i=0$ and run the RS algorithm. Upon doing so,
the result is

\begin{equation}
\begin{tikzpicture}
\draw [dashed, ultra thin] (0,0) -- (-3,-3);
\draw [dashed, ultra thin] (-3,-3) -- (0,-6);
\node at (-5.5,1) {$\Delta$};
\node at (-5.5,0) {$2K$};
\node at (-5.5,-1.5) {$2K+\frac{1}{2}$};
\node at (-5.5,-3) {$2K+1$};
\node at (-5.5,-4.5) {$2K+\frac{3}{2}$};
\node at (-5.5,-6) {$2K+2$};
\draw [fill=white] (-0.9,-0.4) rectangle (0.9,0.4);
\node at (0,0) {$(K)_{(0,0,0)}$};
\draw [fill=white] (-4.2,-3.4) rectangle (-1.8,-2.6);
\node at (-3,-3) {$(K-2)_{(0,1,0)}$};
\draw [fill=white] (-2.7,-1.1) rectangle (-0.3,-1.9);
\node at (-1.5,-1.5) {$(K-1)_{(1,0,0)}$};
\draw [fill=white] (-2.7,-4.1) rectangle (-0.3,-4.9);
\node at (-1.5,-4.5) {$(K-3)_{(0,0,1)}$};
\draw [fill=white] (-1.2,-5.6) rectangle (1.2,-6.4);
\node at (0,-6) {$(K-4)_{(0,0,0)}$};
\end{tikzpicture}
\end{equation} 

We notice that for $K\leq 2$ the above will lead to
negative-multiplicity representations via the RS algorithm and hence
operator constraints, to be discussed in the next section.

The superconformal index for $K>2$ is given by
\begin{align}
\mathcal{I}_{\mathcal{D}[0,0,0;K]}(p,q,s)=\frac{q^{\frac{3 K}{2}}}{(1-p q) (1-q s^{-1}) (1-p^{-1} q s)}\;.
\end{align}

\subsubsection{The Hypermultiplet: $\mathcal{D}[0,0,0;1]$}

Setting $K=1$ in the $\mathcal{D}[0,0,0;K]$ multiplet will not incur
any nontrivial changes to the spectrum, as it does not lead to
additional shortening conditions. As a result, we may simply write the
multiplet out as

\begin{equation}
\begin{tikzpicture}
\draw [dashed, ultra thin] (0,0) -- (-1.5,-1.5);
\draw [dashed, ultra thin] (-1.5,-4.5) -- (0,-6);
\node at (-5.5,1) {$\Delta$};
\node at (-5.5,0) {$2$};
\node at (-5.5,-1.5) {$\frac{5}{2}$};
\node at (-5.5,-3) {$3$};
\node at (-5.5,-4.5) {$\frac{7}{2}$};
\node at (-5.5,-6) {$4$};
\draw [fill=white] (-0.8,-0.4) rectangle (0.8,0.4);
\node at (0,0) {$(1)_{(0,0,0)}$};
\draw [fill=white] (-2.3,-1.1) rectangle (-0.7,-1.9);
\node at (-1.5,-1.5) {$(0)_{(1,0,0)}$};
\draw [fill=white] (-2.4,-4.1) rectangle (-0.6,-4.9);
\node at (-1.5,-4.5) {$-(0)_{(0,0,1)}$};
\draw [fill=white] (-0.9,-6.4) rectangle (0.9,-5.6);
\node at (0,-6) {$-(1)_{(0,0,0)}$};
\end{tikzpicture}
\end{equation} 

These states can be interpreted as a scalar $\phi^{\bA}$ with its
associated Klein--Gordon equation $\partial^2 \phi^{\bA}=0$ and a
fermion $\psi_a$ alongside the Dirac equation
$\slashed{\partial}^{\dot a\dot b}\tilde\psi_{\dot b} =0$.

The corresponding index is given by
\begin{align}
\mathcal{I}_{\mathcal{D}[0,0,0;1]}(p,q,s)=\frac{q^{\frac{3 }{2}}}{(1-p q) (1-q s^{-1}) (1-p^{-1} q s)}\;.
\end{align}

\subsubsection{The Flavour-Current Multiplet: $\mathcal{D}[0,0,0;2]$}

As above, we can simply substitute $K=2$ into $\mathcal{D}[0,0,0;K]$
to generate this spectrum, as there are no additional shortening
conditions. The result is

\begin{equation}
\begin{tikzpicture}
\draw [dashed, ultra thin] (0,0) -- (-3,-3);
\node at (-5.5,1) {$\Delta$};
\node at (-5.5,0) {$4$};
\node at (-5.5,-1.5) {$\frac{9}{2}$};
\node at (-5.5,-3) {$5$};
\node at (-5.5,-4.5) {$\frac{11}{2}$};
\node at (-5.5,-6) {$6$};
\draw [fill=white] (-0.8,-0.4) rectangle (0.8,0.4);
\node at (0,0) {$(2)_{(0,0,0)}$};
\draw [fill=white] (-3.8,-2.6) rectangle (-2.2,-3.4);
\node at (-3,-3) {$(0)_{(0,1,0)}$};
\draw [fill=white] (-2.3,-1.1) rectangle (-0.7,-1.9);
\node at (-1.5,-1.5) {$(1)_{(1,0,0)}$};
\draw [fill=white] (-0.9,-6.4) rectangle (0.9,-5.6);
\node at (0,-6) {$-(0)_{(0,0,0)}$};
\end{tikzpicture}
\end{equation} 

These states can be identified with the operators $\mu^{(\bA\bB)}$,
$\psi^{\bA}_a$ and the $\mathfrak{su}(2)_R$-singlet conserved current
$J_\mu$, with $\partial^\mu J_\mu =0$. This multiplet is also known as
a linear multiplet and appears in
\cite{Howe:1987ik,Kuzenko:2015xiz,Tachikawa:2015mha}.

The index for this multiplet is given by
\begin{align}
\mathcal{I}_{\mathcal{D}[0,0,0;2]}(p,q,s)=\frac{q^3}{(1-p q) (1-q s^{-1}) (1-p^{-1} q s)}\;.
\end{align}

This concludes our discussion of the superconformal multiplets for the
6D (1,0) SCA.

\section{Multiplets and Superconformal Indices for $\mathbf{6D}$
  (2,0)}\label{Sec:6D20}

We lastly turn to the construction of superconformal multiplets for
the 6D (2,0) SCA, $\mathfrak{osp}(8^*|4)$. Since we now have sixteen
Poincar\'e supercharges, the UIRs will be much larger compared to the
ones obtained in Sec.~\ref{Sec:5DN1} and Sec.~\ref{Sec:6D10} and
representing them diagrammatically would not be particularly
instructive. Similarly, the full expressions for the most general
(``refined'') superconformal indices are unwieldy. Instead we will
choose to detail the multiplet types, their shortening conditions and
the ``Schur'' limit of their indices, although we do include the
refined versions of the index in the accompanying Mathematica
notebook.\footnote{A discussion of the null states for the 6D (2,0)
  SCA, along with the calculation of Schur indices for the various
  short multiplets, can be found in App.~C of
  \cite{Beem:2014kka}. Here we additionally construct the full
  multiplets with an emphasis on the equations of motion and
  conservation equations. We also provide the refined indices for all
  multiplets.}

\subsection{UIR Building with Auxiliary Verma Modules}

The superconformal primaries of the algebra $\mathfrak{osp}(8^*|4)$
are designated $\ket{\Delta;c_1,c_2,c_3;d_1,d_2}$ and labelled by the
conformal dimension $\Delta$, the Lorentz quantum numbers for
$\mathfrak{su(4)}$ in the Dynkin basis $c_i$ and the $R$-symmetry
quantum numbers in the Dynkin basis $d_i$. Each primary is in
one-to-one correspondence with a highest weight labeling irreducible
representations of the maximal compact subalgebra
$\mathfrak{so}(6)\oplus\mathfrak{so}(2)\oplus\mathfrak{so}(5)_R\subset\mathfrak{osp}(8^*|4)$.
There are sixteen Poincar\'e and superconformal supercharges, denoted
by $\QQ_{\bA a}$ and $\QS_{\bA \dot a}$, where $\dot a ,a=1,\cdots,4$
are (anti)fundamental indices of $\su(4)$ and $\bA =1,\cdots,4$ a
spinor index of $\so(5)_R$. One also has six momenta $\PP_\mu$ and
special conformal generators $\mathcal{K}_\mu$, where $\mu$ is a
vector index of the Lorentz group, $\mu =1,\cdots,6$. The
superconformal primary is annihilated by all $\QS_{\bA \dot a}$ and
$\mathcal{K}_\mu$. A basis for the representation space of
$\mathfrak{osp}(8^*|4)$ can be constructed by considering the Verma
module
\begin{align}\label{6d (2,0) conf basis}
  \prod_{{\bf A}, a}(\QQ_{{\bf A} a})^{n_{{\bf A},a}}\prod_{\mu} \mathcal{P}_{\mu}^{\hphantom{\mu}n_\mu}
  \ket{\Delta;c_1,c_2,c_3;d_1,d_2}^{hw}\;,
\end{align}
where $n=\sum_{{\bf A} a} n_{{\bf A},a}$ and
$\hat{n}=\sum_{\mu} n_{\mu}$ denote the level of a superconformal or
conformal descendant respectively.  In order to obtain UIRs, the
requirement of unitarity needs to be imposed level-by-level on the
Verma module. This leads to bounds on the conformal dimension
$\Delta$.

Starting with the superconformal descendants, the conditions imposed
by unitarity can be deduced as follows. In principle, one needs to
calculate the norms of superconformal descendants for $n>0$.  However,
since it is sufficient to perform this analysis in the highest weight
of the $R$-symmetry group
\cite{Minwalla:1997ka,Dobrev:2002dt,Bhattacharya:2008zy}, the results
of Sec.~\ref{super conformal rep building 6D} can be easily imported
to the (2,0) case and we may still use the basis
$A^{a_i}\cdots A^{a_j}
\ket{\Delta;c_1,c_2,c_3;d_1,d_2}^{hw}$.\footnote{The
  analysis in \cite{Minwalla:1997ka,Dobrev:2002dt,Bhattacharya:2008zy}
  was performed for generic $R$-symmetry
  $\mathfrak{sp}(\mathcal{N})$. The construction is largely focussed
  around the Lorentz Cartans and raising operators.}  We need only
convert the Cartan for the highest weight of $\su(2)_R$ to $\so(5)_R$,
which is done by simply replacing $K\rightarrow d_1+d_2$. This gives
rise to a group of similar short multiplets, which we collect in Table
\ref{Tab:6D(2,0)}. We will provide more details regarding the
generators that are absent from the auxiliary Verma module when
discussing individual multiplets.  Furthermore, it will be important
to clarify which additional absent generators can occur from tuning
the $R$-symmetry quantum numbers, $d_1$ and $d_2$. These turn out to
be far more intricate than for (1,0).

 \begin{table}[t]
\begin{center}
\begin{tabular}{|c|r|l|}
  \hline
  Multiplet & Shortening Condition &Conformal Dimension \\
  \hline
  \hline
  $\mathcal{A}[c_1,c_2,c_3;d_1,d_2]$ & $A^4 \Psi = 0$&$\Delta = 2d_1+2d_2+ \frac{c_1}{2}+c_2+\frac{3c_3}{2}+6$ \\
  $\mathcal{A}[c_1,c_2,0;d_1,d_2]$& $A^3 A^4 \Psi = 0$& $\Delta
                                                        =2d_1+2d_2+\frac{c_1}{2}+c_2+6$
                                                       \\
  $\mathcal{A}[c_1,0,0;d_1,d_2]$ & $A^2 A^3 A^4 \Psi = 0$& $\Delta =  2d_1+2d_2+\frac{c_1}{2}+6$\\
  $\mathcal{A}[0,0,0;d_1,d_2]$  & $A^1A^2A^3A^4 \Psi = 0$& $\Delta = 2d_1+2d_2+6$\\
  \hline
  \hline
  $\mathcal{B}[c_1,c_2,0;d_1,d_2]$   & $A^3  \Psi = 0$& $\Delta = 2d_1+2d_2+\frac{c_1}{2}+c_2+4$\\
  $\mathcal{B}[c_1,0,0;d_1,d_2]$ & $A^2 A^3 \Psi = 0$& $\Delta =2d_1+2d_2+\frac{c_1}{2}+4$\\
  $\mathcal{B}[0,0,0;d_1,d_2]$ & $A^1 A^2 A^3 \Psi = 0$& $\Delta =2d_1+2d_2+4$\\
  \hline
  \hline
  $\mathcal{C}[c_1,0,0;d_1,d_2]$ & $A^2  \Psi = 0$  & $\Delta =2d_1+2d_2+\frac{c_1}{2}+2  $\\
  $\mathcal{C}[0,0,0;d_1,d_2]$ & $A^1 A^2  \Psi = 0$ & $\Delta =2d_1+2d_2+2  $\\
  \hline
  \hline
  $\mathcal{D}[0,0,0;d_1,d_2]$ & $A^1\Psi = 0$ & $\Delta =2d_1+2d_2  $\\
  \hline
\end{tabular}
\end{center}
\caption{\label{Tab:6D(2,0)} A list of all short multiplets for the 6D
  $(2,0)$ SCA, along with the shortening condition and  conformal dimension of the
  superconformal primary. The $A^a$ in the shortening conditions are
  defined in \eqref{Aas} and \cite{Bhattacharya:2008zy}. The first of these multiplets
  ($\mathcal A$) is a
  regular short representation, whereas  the rest ($\mathcal B, \mathcal C,
  \mathcal D$) are isolated short representations. Here $\Psi$ denotes
  the superconformal primary state for each multiplet.}
\end{table}

The null states arising from conformal descendants are identical to
Sec.~\ref{super conformal rep building 6D}.  One can combine the
information from the conformal and superconformal unitarity bounds to
predict when a multiplet will contain operator
constraints\cite{Minwalla:1997ka,Dolan:2005wy}.  These special cases
are
\begin{align}\label{6D20cons}
\mathcal{B}[c_1,c_2,0;0,0]\;,\hspace{3mm}\mathcal{C}[c_1,0,0;d_1,d_2]\hspace{3mm}\text{for $d_1+d_2\leq 1$}\;,\hspace{3mm}\mathcal{D}[0,0,0;d_1,d_2]\hspace{3mm}\text{for $d_1+d_2 \leq 2$}\;.
\end{align}
The multiplet $\mathcal{D}[0,0,0;0,0]$ does not belong to this list as
it is the vacuum, which is annihilated by all supercharges and
momenta.

\subsection{6D (2,0) Recombination Rules}

Short multiplets can recombine into a long multiplet $\mathcal{L}$.
This occurs when the conformal dimension of $\mathcal{L}$ approaches
the unitarity bound, that is when
$\Delta +\epsilon \to 2d_1 +2d_2 +\frac{1}{2}(c_1 +2c_2+3c_3)+6$. As
in \cite{Beem:2014kka}, we find that
\begin{align}\label{recomb rule 6d n=2}
\mathcal{L}[\Delta + \epsilon;c_1,c_2,c_3;d_1,d_2] &\xrightarrow{\epsilon\rightarrow 0} \mathcal{A}[c_1,c_2,c_3;d_1,d_2]\oplus \mathcal{A}[c_1,c_2,c_3-1;d_1,d_2+1]\;,\cr
\mathcal{L}[\Delta + \epsilon;c_1,c_2,0;d_1,d_2] &\xrightarrow{\epsilon\rightarrow 0} \mathcal{A}[c_1,c_2,0;d_1,d_2]\oplus \mathcal{B}[c_1,c_2-1,0;d_1,d_2+2]\;,\cr
\mathcal{L}[\Delta + \epsilon;c_1,0,0;d_1,d_2] &\xrightarrow{\epsilon\rightarrow 0} \mathcal{A}[c_1,0,0;d_1,d_2]\oplus \mathcal{C}[c_1-1,0,0;d_1,d_2+3]\;,\cr
\mathcal{L}[\Delta + \epsilon;0,0,0;d_1,d_2] &\xrightarrow{\epsilon\rightarrow 0} \mathcal{A}[0,0,0;d_1,d_2]\oplus \mathcal{D}[0,0,0;d_1,d_2+4]\;.
\end{align}

A small number of short multiplets do not appear in a recombination rule. These are
\begin{align}
  &\mathcal{B}\left[c_1,c_2,0;d_1,\{0,1\}\right]\;,\cr
&\mathcal{C}\left[c_1,0,0;d_1,\{0,1,2\}\right]\;,\cr
&\mathcal{D}\left[0,0,0;d_1,\{0,1,2,3\}\right]\;.
\end{align}

\subsection{The 6D (2,0) Superconformal Index}\label{6D20IndMain}

We define the 6D (2,0) superconformal index with respect to the
supercharge $\QQ_{\bTw 4}$. This is given by 
\begin{align}
  \label{(2,0) index}
  \II(p,q,s,t)=\Tr_{\HH}(-1)^F e^{-\beta \delta}q^{\Delta-d_1 - \frac{1}{2}d_2}p^{c_2+ c_3 -
  d_2}t^{-d_1 -d_2}s^{c_1+c_2}\;,  
\end{align}
where the fermion number is $F=c_1+c_3$. The states that are
counted satisfy $\delta =0$, with
\begin{align}
  \label{2,0 delta}
  \delta :=\{ \QQ_{\bTw 4},\QS_{\bTh \dot 4}\}= \Delta - 2d_1-\frac{1}{2}(
  c_1 +2 c_2 +3 c_3)\;.
\end{align}
The exponents appearing in \eqref{(2,0) index} are the eigenvalues for
the generators commuting with $\QQ_{\bTw 4}$, $\QS_{\bTh \dot 4}$ and
$\delta$. This index can be evaluated as a 6D supercharacter; this is
discussed in our App.~\ref{superchar} and App.~C of
\cite{Beem:2014kka}.

\subsubsection{The 6D Schur limit} 

The authors of \cite{Beem:2014kka} define the 6D Schur limit by taking
$t\to 1$ in \eqref{(2,0) index}.  Under the trace the resulting index
can be rewritten as
\begin{align}
  \label{eq:112}
  \II^{Schur}(p,q,s)=\Tr_\HH(-1)^Fe^{-\beta\;\delta}q^{\Delta-d_1 - \frac{1}{2}d_2}p^{\delta'}s^{c_1+c_2}~,
\end{align}
where
\begin{align}
  \label{eq:113}
  \delta'\ceq   \{\QQ_{\bOn 2},\QS_{\bFo \dot 2}\}= \Delta - 2 d_1 - 2d_2 -\frac{1}{2}(c_1 - 2c_2 - c_3) \;.
\end{align}
It can be seen that all exponents in \eqref{eq:112} commute with the
supercharge $\QQ_{\bOn 2}$ and the 6D Schur index consequently counts
operators annihilated by two supercharges. As a result the above index
is independent of both $e^{-\beta}$ and $p$ and the trace can be taken
over operators satisfying $\delta = 0 = \delta'$:
\begin{align}
  \label{eq:115}
  \II^{Schur}(q,s)=\Tr_{\HH_{\delta =0=
  \delta'}}(-1)^Fq^{\Delta-d_1-\frac{1}{2}d_2}s^{c_1+c_2}~.
\end{align}
The operators contributing to the index in this limit satisfy
\begin{align}\label{schurcond}
  \Delta &= 2 d_1 + d_2 + \frac{1}{2} (c_1 +c_3)\;,\cr
  d_2 & = c_2 + c_3\;. 
\end{align}

\subsection{Long Multiplets}

Long multiplets are constructed by acting with all supercharges on a
superconformal primary $(d_1, d_2)_{(c_1, c_2, c_3)}$. This leads to
lengthy expressions which, although not presented here, are available
from the authors upon request. For book-keeping purposes we will group
the supercharges into $Q = (\QQ_{\bTw a},\QQ_{\bTh a})$ and
$\tilde{Q} = (\QQ_{\bOn a},\QQ_{\bFo a})$ for the remaining of the 6D
(2,0) discussion.

\subsection{$\mathcal{A}$-type Multiplets}\label{6D20A}

Recall from Table~\ref{Tab:6D(2,0)} that the $\mathcal A$-type
multiplets obey four kinds of shortening conditions. These result in
the removal of the following supercharges from the basis of Verma
module generators in \eqref{6d (2,0) conf basis}
\begin{align}
\mathcal{A}[c_1,c_2,c_3;d_1,d_2]&: \QQ_{\bOn 4}\;, \cr
\mathcal{A}[c_1,c_2,0;d_1,d_2]&: \QQ_{\bOn 3}\QQ_{\bOn 4}\;,\cr
\mathcal{A}[c_1,0,0;d_1,d_2]&: \QQ_{\bOn 2}\QQ_{\bOn 3}\QQ_{\bOn 4}\;,\cr
\mathcal{A}[0,0,0;d_1,d_2]&: \QQ_{\bOn 1}\QQ_{\bOn 2}\QQ_{\bOn 3}\QQ_{\bOn 4}\;.
\end{align}
The $\mathcal A$-type multiplets cannot contain operator constraints
for any $d_1$ or $d_2$.

One can arrive at new shortening conditions---and as a result a
reduced number of $\QQ$-generators---for $d_1\neq 0$, $d_2 =0$ and
$d_1=d_2=0$. Consider for instance the case
$\mathcal{A}[c_1,c_2,c_3;d_1,0]$. From Table~\ref{Tab:6D(2,0)} the
null state reads $A^4 \Psi =0$. In the auxiliary Verma-module basis,
this corresponds to $\QQ_{\bOn 4}\Psi_{\rm aux}=0$. However, since
$d_2 =0$ one also finds that
$ \mathcal R_2^- \QQ_{\bOn 4} \Psi_{\rm aux} =0$. This corresponds to
having to additionally remove $\QQ_{\bTw 4}$ from our auxiliary
Verma-module basis. Following on from this, when $d_1=0$ the operator
$\mathcal R^-_1$ also annihilates the superconformal primary and two
new conditions are obtained:
$\mathcal R_1^- \mathcal R_2^- \QQ_{\bOn 4}\Psi_{\rm aux}=0$ and
$ \mathcal R_2^- \mathcal R_1^- \mathcal R_2^- \QQ_{\bOn 4}\Psi_{\rm
  aux}=0$.  These correspond to also removing $\QQ_{\bTh 4}$ and
$\QQ_{\bFo 4}$ respectively from the basis of Verma-module generators.

All $\mathcal A$-type multiplets have zero contribution to the Schur
limit of the 6D (2,0) index.  We provide their spectra in App.~\ref{A
  (2,0) spectra}.

\subsection{$\mathcal{B}$-type Multiplets}\label{(2,0) B mult}

For the $\mathcal B$-type multiplets, the supercharges that need to
be removed from the auxiliary Verma-module basis due to null states \eqref{6d (2,0) conf basis} are
 \begin{align}\begin{split}
\mathcal{B}[c_1,c_2,0;d_1, d_2]&: \QQ_{\bOn 3}\;,\\
\mathcal{B}[c_1,0,0;d_1, d_2]&: \QQ_{\bOn 2}\QQ_{\bOn 3}\;,\\
\mathcal{B}[0,0,0;d_1, d_2]&: \QQ_{\bOn 1}\QQ_{\bOn 2}\QQ_{\bOn 3}\;.\end{split}
\end{align}

For the first type of multiplet, $\mathcal B[c_1,c_2,0;d_1,d_2]$, one
should also remove $\QQ_{\bOn 4}$.  A similar argument can be made regarding
$\mathcal{B}[c_1,0,0;d_1,d_2]$ to reach the conclusion that
$\QQ_{\bOn 2}\QQ_{\bOn 3}$, $\QQ_{\bOn 2}\QQ_{\bOn 4}$ and
$\QQ_{\bOn 3}\QQ_{\bOn 4}$ should be removed from the auxiliary Verma-module
basis. For $\mathcal{B}[0,0,0;d_1,d_2]$ these additional supercharges
are $\QQ_{\bOn a}\QQ_{\bOn b}\QQ_{\bOn c}$ with $a\neq b \neq c$,
exactly as in Sec.~\ref{6D10B}.

There are three distinct sub-cases that need to be considered when
dialling $d_1$, $d_2$.  These are:

\begin{enumerate}
\item $\mathcal{B}[c_1,c_2,0;d_1,1]$:

  When $d_2=1$ we find that $(\mathcal R_2^-)^2 \Psi_{\rm aux} =0$. This leads
  to the combination $\QQ_{\bTw 3}\QQ_{\bTw 4}$ being additionally
  removed from the basis of auxiliary Verma-module generators
  \cite{Beem:2014kka}.

\item $\mathcal{B}[c_1,c_2,0;d_1,0]$: 

  Having $d_2 =0$ implies that $\mathcal R_2^- \Psi_{\rm aux} =0$. This means
  that $\mathcal R_2^- \QQ_{\bOn 3} \Psi_{\rm aux} =0 = \mathcal R_2^- \QQ_{\bOn 4} \Psi_{\rm aux}$ and we consequently we remove $\QQ_{\bTw 3}$ and $\QQ_{\bTw 4}$ from the basis of auxiliary Verma-module
  generators.

\item $\mathcal{B}[c_1,c_2,0;0,0]$: 

  When $d_1=d_2=0$ all $R$-symmetry lowering operators annihilate the
  primary. Hence we can apply the above logic while including the
  lowering operator $\mathcal R_1^-$. One finds that the supercharges
  $\QQ_{\bTh 3}$, $\QQ_{\bFo 3}$, $\QQ_{\bTh 4}$, $\QQ_{\bFo 4}$ should also be removed from the basis.

\end{enumerate}

According to \eqref{6D20cons} the multiplets
$\mathcal{B}[c_1,c_2,0;0,0]$ should contain operator
constraints. However, the removal of $\QQ_{\bA 3}$ and $\QQ_{\bA 4}$
from the basis of generators implicitly also removes
\begin{align}\label{Pprediction}
\PP_{34}&=\{ \QQ_{\bOn 3},\QQ_{\bFo 4}\}=\{ \QQ_{\bTw 3},\QQ_{\bTh 4}\} \propto \PP_6\;.
\end{align}
This corresponds to projecting out all states associated with a
conservation equation from the UIR and the resulting module will not
include negative-multiplicity representations. The operator
constraints can be restored using the dictionary of
App.~\ref{App:5DRS}.

The spectra of the $Q$- and $\tilde{Q}$-actions for $\mathcal B$-type
multiplets can be found in App.~\ref{B (2,0) spectra}. Their
contribution to the Schur limit of the 6D (2,0) superconformal index
is vanishing, with the exception of $\mathcal{B}[c_1,c_2,0;d_1,0]$,
for which
\begin{align}\label{schur b d1 gen}
\mathcal{I}^{Schur}_{\mathcal{B}[c_1,c_2,0;d_1,0]}(q,s)=(-1)^{c_1}\frac{q^{4+d_1+\frac{c_1}{2}+c_2} }{1-q}\chi_{c_1 + 1}(s)\;.
\end{align}

\subsubsection{Higher-Spin-Current Multiplets: $\mathcal{B}[c_1,c_2,0;0,0]$}\label{HS6D20}

Recall that we are only constructing the auxiliary Verma module with the
supercharges $\QQ_{\bA 1}$ and $\QQ_{\bA 2}$. Their action on a
superconformal primary with generic Dynkin labels
$(d_1, d_2)_{(c_1, c_2, c_3)}$ are given by
\begin{align}\label{QQ B actions}
(d_1, d_2)_{(c_1, c_2, c_3)}\xrightarrow{\QQ}& (d_1-1, d_2+1),(d_1+1,d_2-1)_{(c_1+1, c_2, c_3),(c_1-1, c_2+1, c_3)}\;,\cr
\xrightarrow{\QQ^2} &(d_1,d_2)_{(c_1+2, c_2, c_3),(c_1-2, c_2+2, c_3),(c_1,c_2+1,c_3)^2},(d_1-2,d_2+2)_{(c_1,c_2+1,c_3)},\cr
\hphantom{\xrightarrow{\QQ^2}}& (d_1+2,d_2-2)_{(c_1,c_2+1,c_3)}\;,\cr
\xrightarrow{\QQ^3} &(d_1+1,d_2-1),(d_1-1,d_2+1)_{(c_1+1, c_2+1, c_3),(c_1-1, c_2+2, c_3)}\;,\cr
\xrightarrow{\QQ^4}&(d_1,d_2)_{(c_1,c_2+2,c_3)}\;,\cr
(d_1, d_2)_{(c_1, c_2, c_3)}\xrightarrow{\tilde{Q}}& (d_1, d_2+1),(d_1,d_2-1)_{(c_1+1, c_2, c_3),(c_1-1, c_2+1, c_3)}\;,\cr
\xrightarrow{\tilde{Q}^2} &(d_1,d_2)_{(c_1+2, c_2, c_3),(c_1-2, c_2+2, c_3),(c_1,c_2+1,c_3)^2},(d_1,d_2+2)_{(c_1,c_2+1,c_3)}\;,\cr
\hphantom{\xrightarrow{\tilde{Q}^2}}& (d_1,d_2-2)_{(c_1,c_2+1,c_3)}\;,\cr
\xrightarrow{\tilde{Q}^3} &(d_1,d_2-1),(d_1,d_2+1)_{(c_1+1, c_2+1, c_3),(c_1-1, c_2+2, c_3)}\;,\cr
\xrightarrow{\tilde{Q}^4}&(d_1,d_2)_{(c_1,c_2+2,c_3)}\;.
\end{align}
The full representation is then built from these chains of
supercharges. Clearly since $d_1=d_2 = 0$ we will only be using the
$\mathfrak{so}(5)$ Weyl reflections in the implementation of the RS
algorithm. Denoting the action of $Q = (\QQ_{\bTw a},\QQ_{\bTh a})$ as
moving {\it southwest} on the diagram and
$\tilde{Q} = (\QQ_{\bOn a},\QQ_{\bFo a})$ as moving {\it southeast},
we can represent this multiplet as:
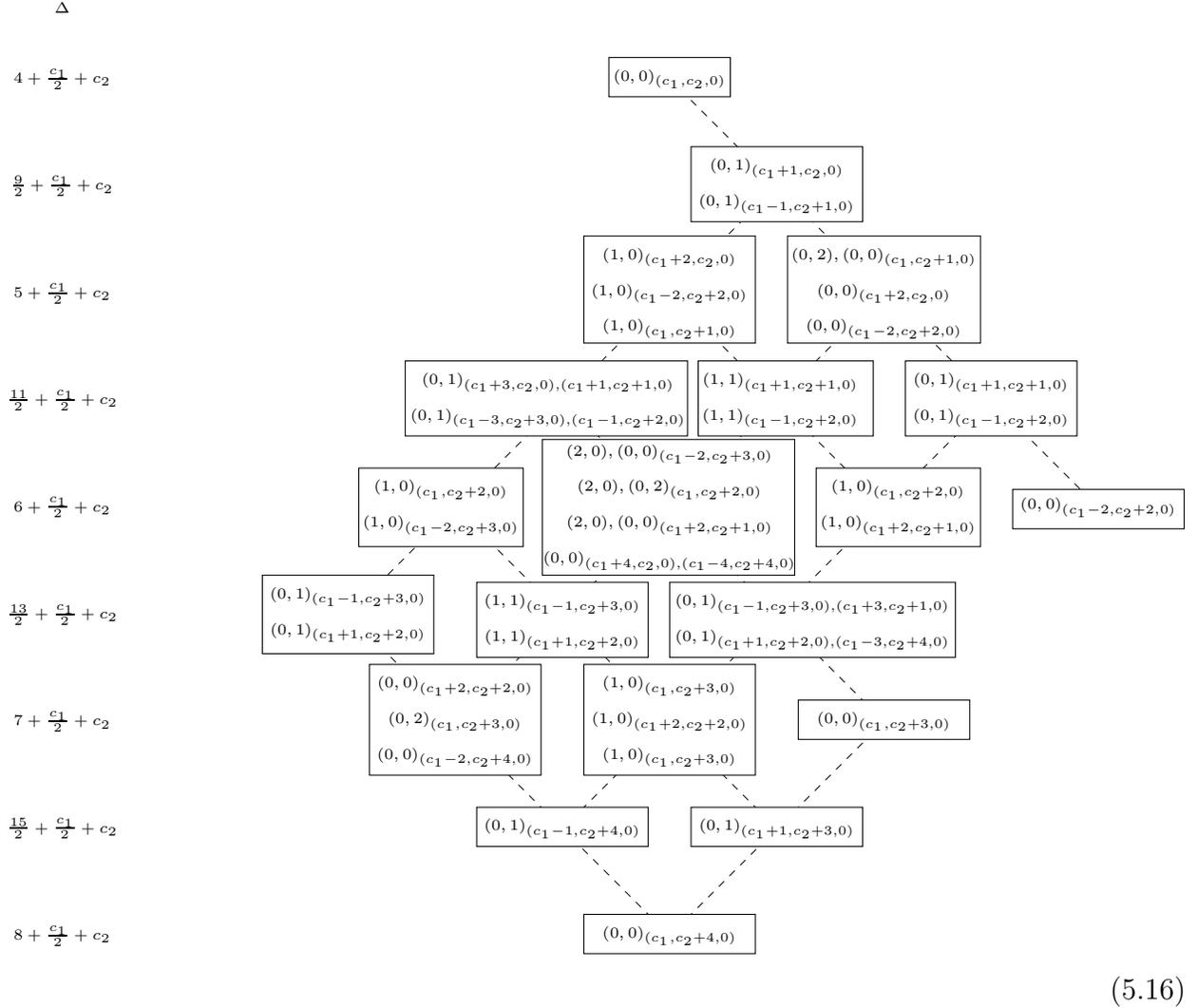
\begin{equation}
{\tiny\begin{tikzpicture}
\draw [dashed, ultra thin] (0,0) -- (6,-6);
\draw [dashed, ultra thin] (0,-3) -- (3,-6);
\draw [dashed, ultra thin] (-1.5,-4.5) -- (3,-9);
\draw [dashed, ultra thin] (-3,-6) -- (1.5,-10.5);
\draw [dashed, ultra thin] (-4.5,-7.5) -- (0,-12);
\draw [dashed, ultra thin] (1.5,-1.5) -- (-4.5,-7.5);
\draw [dashed, ultra thin] (3,-3) -- (-3,-9);
\draw [dashed, ultra thin] (4.5,-4.5) -- (-1.5,-10.5);
\draw [dashed, ultra thin] (3,-9) -- (-0,-12);
\node at (-8.5,1) {$\Delta$};
\node at (-8.5,0) {$4+\frac{c_1}{2}+c_2$};
\node at (-8.5,-1.5) {$\frac{9}{2}+\frac{c_1}{2}+c_2$};
\node at (-8.5,-3) {$5+\frac{c_1}{2}+c_2$};
\node at (-8.5,-4.5) {$\frac{11}{2}+\frac{c_1}{2}+c_2$};
\node at (-8.5,-6) {$6+\frac{c_1}{2}+c_2$};
\node at (-8.5,-7.5) {$\frac{13}{2}+\frac{c_1}{2}+c_2$};
\node at (-8.5,-9) {$7+\frac{c_1}{2}+c_2$};
\node at (-8.5,-10.5) {$\frac{15}{2}+\frac{c_1}{2}+c_2$};
\node at (-8.5,-12) {$8+\frac{c_1}{2}+c_2$};
\draw [fill=white] (0.85,0.3) rectangle (-0.85,-0.25);
\node at (0,0) {$(0,0)_{(c_1,c_2,0)}$};
\draw [fill=white] (2.7,-0.95) rectangle (0.3,-2);
\node at (1.5,-1.25) {$(0,1)_{(c_1+1,c_2,0)}$};
\node at (1.5,-1.75) {$(0,1)_{(c_1-1,c_2+1,0)}$};
\draw [fill=white] (4.35,-2.2) rectangle (1.65,-3.7);
\node at (3,-2.5) {$(0,2),(0,0)_{(c_1,c_2+1,0)}$};
\node at (3,-3) {$(0,0)_{(c_1+2,c_2,0)}$};
\node at (3,-3.5) {$(0,0)_{(c_1-2,c_2+2,0)}$};
\draw [fill=white] (-1.2,-2.2) rectangle (1.2,-3.7);
\node at (0,-2.5) {$(1,0)_{(c_1 + 2, c_2, 0)}$};
\node at (0,-3) {$(1,0)_{(c_1-2 , c_2+2, 0)}$};
\node at (0,-3.5) {$(1,0)_{(c_1 , c_2+1, 0)}$};
\draw [fill=white] (-3.7,-3.95) rectangle (0.25,-5);
\node at (-1.7,-4.25) {$(0,1)_{(c_1+3,c_2,0),(c_1+1,c_2+1,0)}$};
\node at (-1.7,-4.75) {$(0,1)_{(c_1-3,c_2+3,0),(c_1-1,c_2+2,0)}$};
\draw [fill=white] (3.3,-3.95) rectangle (5.7,-5);
\node at (4.5,-4.75) {$(0,1)_{(c_1-1,c_2+2,0)}$};
\node at (4.5,-4.25) {$(0,1)_{(c_1+1,c_2+1,0)}$};
\draw [fill=white] (0.4,-3.95) rectangle (2.85,-5);
\node at (1.55,-4.25) {$(1,1)_{(c_1+1,c_2+1,0)}$};
\node at (1.55,-4.75) {$(1,1)_{(c_1-1,c_2+2,0)}$};
\draw [fill=white] (2.05,-5.45) rectangle (4.35,-6.55);
 \node at (3.2,-5.75) {$(1,0)_{(c_1,c_2+2,0)}$};
 \node at (3.2,-6.25) {$(1,0)_{(c_1+2,c_2+1,0)}$};
\draw [fill=white] (-1.78,-5.05) rectangle (1.75,-6.95);
 \node at (0,-5.25) {$(2,0),(0,0)_{(c_1-2,c_2+3,0)}$};
 \node at (0,-5.75) {$(2,0),(0,2)_{(c_1,c_2+2,0)}$};
 \node at (0,-6.25) {$(2,0),(0,0)_{(c_1+2,c_2+1,0)}$};
 \node at (0,-6.75) {$(0,0)_{(c_1+4,c_2,0),(c_1-4,c_2+4,0)}$};
\draw [fill=white] (-2.05,-5.45) rectangle (-4.35,-6.55);
 \node at (-3.2,-5.75) {$(1,0)_{(c_1,c_2+2,0)}$};
 \node at (-3.2,-6.25) {$(1,0)_{(c_1-2,c_2+3,0)}$};
\draw [fill=white] (4.8,-5.75) rectangle (7.2,-6.3);
 \node at (6,-6) {$(0,0)_{(c_1-2,c_2+2,0)}$};
\draw [fill=white] (-3.3,-6.95) rectangle (-5.7,-8.05);
\node at (-4.5,-7.25) {$(0,1)_{(c_1-1,c_2+3,0)}$};
\node at (-4.5,-7.75) {$(0,1)_{(c_1+1,c_2+2,0)}$};
\draw [fill=white] (0,-7.05) rectangle (4,-8.1);
\node at (2,-7.35) {$(0,1)_{(c_1-1,c_2+3,0),(c_1+3,c_2+1,0)}$};
\node at (2,-7.85) {$(0,1)_{(c_1+1,c_2+2,0),(c_1-3,c_2+4,0)}$};
\draw [fill=white] (-2.7,-7.05) rectangle (-0.3,-8.1);
\node at (-1.5,-7.35) {$(1,1)_{(c_1-1,c_2+3,0)}$};
\node at (-1.5,-7.85) {$(1,1)_{(c_1+1,c_2+2,0)}$};
\draw [fill=white] (-4.2,-8.2) rectangle (-1.8,-9.75);
\node at (-3,-8.5) {$(0,0)_{(c_1+2,c_2+2,0)}$};
\node at (-3,-9.00) {$(0,2)_{(c_1,c_2+3,0)}$};
\node at (-3,-9.5) {$(0,0)_{(c_1-2,c_2+4,0)}$};
\draw [fill=white] (-1.2,-8.2) rectangle (1.2,-9.75);
\node at (0,-9) {$(1,0)_{(c_1+2,c_2+2,0)}$};
\node at (0,-9.5) {$(1,0)_{(c_1,c_2+3,0)}$};
\node at (0,-8.5) {$(1,0)_{(c_1,c_2+3,0)}$};
\draw [fill=white] (1.8,-8.7) rectangle (4.2,-9.25);
\node at (3,-9) {$(0,0)_{(c_1,c_2+3,0)}$};
\draw [fill=white] (2.7,-10.2) rectangle (0.3,-10.75);
\node at (1.5,-10.5) {$(0,1)_{(c_1+1,c_2+3,0)}$};
\draw [fill=white] (-2.7,-10.2) rectangle (-0.3,-10.75);
\node at (-1.5,-10.5) {$(0,1)_{(c_1-1,c_2+4,0)}$};
\draw [fill=white] (-1.2,-11.7) rectangle (1.2,-12.25);
\node at (0,-12) {$(0,0)_{(c_1,c_2+4,0)}$};
\end{tikzpicture}}
\end{equation}

This is the reduced spectrum because, as predicted from the discussion around \eqref{Pprediction}, there are
no negative-multiplicity states.  In
order to restore them we use the character relation
\begin{align}
\hat{\chi}[\Delta;c_1,c_2,c_3;d_1,d_2]=\chi[\Delta;c_1,c_2,c_3;d_1,d_2]- \chi[\Delta+1;c_1,c_2-1,c_3; d_1,d_2]\;,
\end{align}
where the hat denotes a character over the reduced (i.e.
$\PP_6$--removed) Verma module.

Since there are many states in the reduced spectrum, reconstructing
the full multiplet with the operator constraints included would be
rather unwieldy. We will however note that states with $c_2 \neq 0$
will pair up with their conservation equation according to

\begin{align}
\begin{tikzpicture}
\draw [dashed, ultra thin] (0,0) -- (-1.5,-1.5);
\draw [dashed, ultra thin] (0,0) -- (1.5,-1.5);
\draw [dashed, ultra thin] (-1.5,-1.5) -- (0,-3);
\draw [dashed, ultra thin] (1.5,-1.5) -- (0,-3);
\node at (-3,0) {$\Delta$};
\node at (-3,-1.5) {$\Delta+\frac{1}{2}$};
\node at (-3,-3) {$\Delta+1$};
\draw [fill=white] (-1.3,-0.4) rectangle (1.3,0.4);
\node at (0,0) {$(d_1,d_2)_{(c_1,c_2,0)}$};
\draw [fill=white] (-1.7,-2.6) rectangle (1.7,-3.4);
\node at (0,-3) {$-(d_1,d_2)_{(c_1,c_2-1,0)}$};
\end{tikzpicture}
\end{align}

This leads to the observation that, for arbitrary values of $c_1$ and
$c_2$, we have an infinite family of conserved currents which have
higher spin.

A subset of these conserved higher-spin currents are the ones
belonging to the multiplet $\mathcal{B}[0,c_2,0;0,0]$. This multiplet
has a superconformal primary in the rank-$c_2$ symmetric traceless
representation of $\mathfrak{su}(4)$, which corresponds to the
higher-spin currents that one expects to find in the free 6D $(2,0)$
theory.

For example, we may take
\begin{align}
\mathcal{O}_{\mu_1 \cdots \mu_{c_2}}= \sum_{{\bf I}} \Phi^{\bf I} \overset\leftrightarrow{\partial}_{\mu_1} \cdots \overset\leftrightarrow{\partial}_{\mu_{c_2}} \Phi^{\bf I} ,
\end{align}
where ${\bf I} = 1,\cdots, 5$ is an $\so (5)_R$ vector index, and $\Phi^I$ is a free-tensor primary. Therefore, this object satisfies the conservation
equation
\begin{align}
\partial^\mu \mathcal{O}_{\mu \mu_2 \dots \mu_{c_2 -1}}=0\;.
\end{align}

For generic $c_1, c_2$, the Schur index for this type of multiplet is given by
\begin{align}
\mathcal{I}^{Schur}_{\mathcal{B}[c_1,c_2,0;0,0]}(q,s)=(-1)^{c_1}\frac{q^{\frac{c_1}{2}+c_2+4} }{1-q}\chi_{c_1 + 1}(s)\;.
\end{align}

\subsection{$\mathcal{C}$-type Multiplets}

From Table~\ref{Tab:6D(2,0)} the two distinct $\mathcal{C}$-type
multiplets are $\mathcal{C}[c_1,0,0;d_1,d_2]$ and
$\mathcal{C}[0,0,0;d_1,d_2]$. Upon repeating the null-state analysis,
one finds that for generic values of $d_1, d_2$ one is required to
remove $\QQ_{\bOn a}$ for $a\neq 1$ and $\QQ_{\bOn a}\QQ_{\bOn b}$
respectively from the basis of auxiliary Verma-module generators.

One also obtains additional absent auxiliary Verma-module generators for certain values of $d_1$
and $d_2$. The procedure for identifying these is the same as the one
presented in Sec.~\ref{(2,0) B mult}, so we simply summarise the
additional set of $\QQ$s that are to be removed:
\begin{align}\label{CadditionalQ}
\mathcal{C}[c_1,0,0;d_1,2]:&\hspace{5mm}\QQ_{\bTw 2}\QQ_{\bTw 3}\QQ_{\bTw 4}\;,\cr
\mathcal{C}[c_1,0,0;d_1,1]:&\hspace{5mm}\QQ_{\bTw a}\QQ_{\bTw b}\hspace{3mm}\text{for  $a\neq b\neq 1$}\;,\cr
\mathcal{C}[c_1,0,0;d_1,0]:&\hspace{5mm}\QQ_{\bTw a}\hspace{3mm}\text{for $a\neq 1$}\;, \cr
\mathcal{C}[c_1,0,0;0,0]:&\hspace{5mm}\QQ_{\bA a}\hspace{3mm}\text{for $a\neq 1$}\;.
\end{align}
We provide the spectra for the above in App.~\ref{C (2,0)
  spectra}. 

There are three combinations of $d_1, d_2$ for which the multiplet
contains operator constraints. The first two
$\mathcal{C}[c_1,0,0;1,0]$, $\mathcal{C}[c_1,0,0;0,1]$ appear in
App.~\ref{C (2,0) spectra}. The case of $\mathcal{C}[c_1,0,0;0,0]$
will be discussed separately below.

The only non-vanishing Schur indices for the $\mathcal{C}$--multiplets are 
\begin{align}\begin{split}
\mathcal{I}^{Schur}_{\mathcal{C}[c_1,0,0;d_1,1]}(q,s)&=(-1)^{c_1+1}\frac{q^{\frac{7}{2}+d_1+\frac{c_1}{2}}}{1-q}\chi_{c_1 + 1}(s)\;,\\
\mathcal{I}^{Schur}_{\mathcal{C}[c_1,0,0;d_1,0]}(q,s)&=(-1)^{c_1}\frac{q^{2+d_1+\frac{c_1}{2}}}{1-q}\chi_{c_1 + 2}(s)\;.\end{split}
\end{align}

\subsubsection{$\mathcal{C}[c_1,0,0;0,0]$}\label{C mult EoM detail}

Recall from \eqref{CadditionalQ} that the associated Verma module is constructed by
the action of $\QQ_{\bA 1}$.  Since we are removing all other
supercharges, we also have to remove several momenta from the basis of
generators of the Verma
module. These are
\begin{align}
\PP_{23}&=\{ \QQ_{\bOn 2},\QQ_{\bFo 3}\}=\{ \QQ_{\bTw 2},\QQ_{\bTh
          3}\}\propto \PP_3,\cr
\PP_{24}&=\{ \QQ_{\bOn 2},\QQ_{\bFo 4}\}=\{ \QQ_{\bTw 2},\QQ_{\bTh
          4}\}\propto \PP_5,\cr
\PP_{34}&=\{ \QQ_{\bOn 3},\QQ_{\bFo 4}\}=\{ \QQ_{\bTw 3},\QQ_{\bTh
          4}\}\propto \PP_6\;,
\end{align}
where in the last column we have converted to Lorentz vector
indices. Thus for generic Dynkin labels the actions of the
supercharges lead to
\begin{align}\label{QQ C actions}
(d_1, d_2)_{(c_1, c_2, c_3)}&\xrightarrow{Q} (d_1-1, d_2+1),(d_1+1,d_2-1)_{(c_1+1, c_2, c_3)},\cr
&\xrightarrow{Q^2} (d_1,d_2)_{(c_1+2, c_2, c_3)},\cr
(d_1, d_2)_{(c_1, c_2, c_3)}&\xrightarrow{\tilde{Q}} (d_1, d_2+1),(d_1,d_2-1)_{(c_1+1, c_2, c_3)},\cr
&\xrightarrow{\tilde{Q}^2} (d_1,d_2)_{(c_1+2, c_2, c_3)}.
\end{align}
Upon replacing the actual values of the Dynkin labels and running
the RS algorithm, the physical spectrum is given by

\begin{equation}
\begin{tikzpicture}
\draw [dashed, ultra thin] (0,0) -- (3,-3);
\draw [dashed, ultra thin] (-1.5,-4.5) -- (0,-6);
\draw [dashed, ultra thin] (1.5,-1.5) -- (-1.5,-4.5);
\node at (-5.5,1) {$\Delta$};
\node at (-5.5,0) {$2+\frac{c_1}{2}$};
\node at (-5.5,-1.5) {$\frac{5}{2}+\frac{c_1}{2}$};
\node at (-5.5,-3) {$3+\frac{c_1}{2}$};
\node at (-5.5,-4.5) {$\frac{7}{2}+\frac{c_1}{2}$};
\node at (-5.5,-6) {$4+\frac{c_1}{2}$};
\draw [fill=white] (-1.3,-0.4) rectangle (1.3,0.4);
\node at (0,0) {$(0,0)_{(c_1,0,0)}$};
\draw [fill=white] (0.2,-1.1) rectangle (2.8,-1.9);
\node at (1.5,-1.5) {$(0,1)_{(c_1+1,0,0)}$};
\draw [fill=white] (-1.3,-2.6) rectangle (1.3,-3.4);
\node at (0,-3) {$(1,0)_{(c_1+2,0,0)}$};
\draw [fill=white] (4.3,-2.6) rectangle (1.7,-3.4);
\node at (3,-3) {$(0,0)_{(c_1+2,0,0)}$};
\draw [fill=white] (-2.8,-4.1) rectangle (-0.2,-4.9);
\node at (-1.5,-4.5) {$(0,1)_{(c_1+3,0,0)}$};
\draw [fill=white] (-1.3,-5.6) rectangle (1.3,-6.4);
\node at (0,-6) {$(0,0)_{(c_1+4,0,0)}$};
\end{tikzpicture}
\end{equation}

As in the $\mathcal{B}[c_1,c_2,0;0,0]$ case, we can restore the
negative-multiplicity states by making use of the character relation
\begin{align}\label{bianchi rep}
\hat{\chi}[\Delta;c_1,0,0;d_1,d_2]=&\chi [\Delta;c_1,0,0;d_1,d_2] - \chi [\Delta+1;c_1-1,0,1;d_1,d_2]\cr &+ \chi [\Delta+2;c_1-2,1,0;d_1,d_2] - \chi [\Delta+3;c_1-2,0,0;d_1,d_2]\;, 
\end{align}
where a hat denotes the $\PP$-reduced character. This will lead to a
set of equations of motion.

The index over this multiplet in the Schur limit is
\begin{align}
\mathcal{I}^{Schur}_{\mathcal{C}[c_1,0,0;0,0]}(q,s)=(-1)^{c_1}\frac{q^{\frac{c_1}{2}+2}}{1-q}\chi_{c_1 + 2}(s)\;.
\end{align}

\subsection{$\mathcal D$-type multiplets}\label{D20}

These multiplets, summarised in Table~\ref{Tab:6D(2,0)}, are the
smallest of the $\mathfrak{osp}(8^*|4)$ algebra. The associated null
state is $A^1 \Psi =0$, which implies that $\QQ_{\bOn a}\Psi_{\rm aux}=0$ and
the multiplet is thus $\frac{1}{4}$-BPS.

For $d_2 \leq 3$, one also needs to remove the following combinations of supercharges
from the basis of auxiliary Verma-module generators:
\begin{align}
\mathcal{D}[0,0,0;d_1,3]  &:\hspace{5mm} \QQ_{\bTw 1}\QQ_{\bTw 2}\QQ_{\bTw 3} \QQ_{\bTw 4}\;,\cr
\mathcal{D}[0,0,0;d_1,2] &:\hspace{5mm} \QQ_{\bTw a}\QQ_{\bTw b}\QQ_{\bTw c}\;\cr
\mathcal{D}[0,0,0;d_1,1] &:\hspace{5mm} \QQ_{\bTw a}\QQ_{\bTw b}\;,\cr
\mathcal{D}[0,0,0;d_1,0] &:\hspace{5mm}\QQ_{\bTw a}\;.
\end{align}
When $d_1=d_2=0$ this corresponds to the vacuum, because all
supercharges annihilate the superconformal primary.

For generic $d_1, d_2$ we have the $Q$-chain 
\begin{align}\label{Qs on D}
(d_1, d_2)_{(c_1, c_2, c_3)}&\xrightarrow{Q} (d_1, d_2-1)_{(c_1+1, c_2, c_3),(c_1-1, c_2+1, c_3),(c_1, c_2-1, c_3+1),(c_1, c_2, c_3-1)}\;,\cr
&\xrightarrow{Q^2} (d_1, d_2-2)_{(c_1, c_2+1, c_3),(c_1+1, c_2-1, c_3+1),(c_1+1, c_2, c_3-1),(c_1-1, c_2, c_3+1)}\;,\cr
&\phantom{\xrightarrow{Q^2}\hspace{2.5mm}}(d_1, d_2-2)_{(c_1-1,c_2 + 1, c_3 -1),(c_1, c_2-1, c_3)}\;,\cr
&\xrightarrow{Q^3} (d_1, d_2-3)_{(c_1, c_2, c_3+1),(c_1, c_2+1, c_3-1),(c_1+1, c_2-1, c_3),(c_1-1, c_2, c_3)}\;,\cr
&\xrightarrow{Q^4} (d_1, d_2-4)_{(c_1, c_2, c_3)}\;. 
\end{align}
The action of the $\tilde{Q}$ supercharges will be dealt with on a
case-by-case basis as we dial $d_2$, and provided in App.~\ref{D (2,0)
  Spectra}. The only exception is the case $d_2=0$, which will be
detailed below.

The non-zero contributions to the Schur limit of the index 
are given by
\begin{align}
\mathcal{I}^{Schur}_{\mathcal{D}[0,0,0;d_1,2]}(q,s)&=\frac{q^{3+d_1}}{1-q}
                                                     \qquad
                                                     \mathrm{for}\quad
                                                     d_1\ge 0\;,\cr
\mathcal{I}^{Schur}_{\mathcal{D}[0,0,0;d_1,1]}(q,s)&=-\frac{q^{\frac{3}{2}+d_1}}{1-q}\chi_{1}(s)
                                                     \qquad
                                                     \mathrm{for}\quad
                                                     d_1\ge 0\;,\cr
\mathcal{I}^{Schur}_{\mathcal{D}[0,0,0;d_1,0]}(q,s)&=\frac{q^{d_1}}{1-q}\qquad\mathrm{for}
                                                     \quad d_1>0 \;.
\end{align}

The $\mathcal{D}$-type multiplets contain negative-multiplicity
states for $d_1 + d_2 \leq 2$. These include the free-tensor and
stress-tensor multiplet.

\subsubsection{Half--BPS Multiplets: $\mathcal{D}[0,0,0;d_1,0]$}

Recall that in this case we are prescribed to remove $\QQ_{\bTw a}$
from the basis of auxiliary Verma-module generators. This is because
$A^1\Psi = \QQ_{\bOn 1} \Psi= 0 $ and using $R$-symmetry lowering operators
we find that all $\QQ_{\bOn a}$, $\QQ_{\bTw a}$ annihilate the
primary. As a result, the multiplet is $\frac{1}{2}$--BPS. The set of
$\tilde{Q}$s consist entirely of $\QQ_{\bTh a}$ supercharges. Acting on a generic superconformal primary
state $(d_1, d_2)_{(c_1, c_2, c_3)}$ with the $\tilde{Q}$s yields
\begin{align}\label{Q3 actions}
(d_1, d_2)_{(c_1, c_2, c_3)}&\xrightarrow{\tilde{Q}} (d_1-1, d_2+1)_{(c_1+1, c_2, c_3),(c_1-1, c_2+1, c_3),(c_1, c_2-1, c_3+1),(c_1, c_2, c_3-1)}\;,\cr
&\xrightarrow{\tilde{Q}^2} (d_1-2, d_2+2)_{(c_1, c_2+1, c_3),(c_1+1, c_2-1, c_3+1),(c_1+1, c_2, c_3-1),(c_1, c_2-1, c_3)}\;,\cr
&\phantom{\xrightarrow{\tilde{Q}^2}\hspace{2.5mm}}(d_1-2, d_2+2)_{(c_1-1,c_2 + 1, c_3 -1),(c_1-1, c_2, c_3+1)}\;,\cr
&\xrightarrow{\tilde{Q}^3} (d_1-3, d_2+3)_{(c_1, c_2, c_3+1),(c_1, c_2+1, c_3-1),(c_1+1, c_2-1, c_3),(c_1-1, c_2, c_3)}\;,\cr
&\xrightarrow{\tilde{Q}^4} (d_1-4, d_2+4)_{(c_1, c_2, c_3)}\;. 
\end{align}
The Verma module is then built out of $\QQ_{\bTh a}$, $\QQ_{\bFo a}$
and we obtain

\begin{equation} \begin{tikzpicture}
 \draw [dashed, ultra thin] (0,0) -- (-6,-6); 
 \draw [dashed, ultra thin] (-1.5,-1.5) -- (0,-3);
 \draw [dashed, ultra thin] (-3,-3) -- (0,-6); 
 \draw [dashed, ultra thin] (-4.5,-4.5) -- (0,-9);
 \draw [dashed, ultra thin] (-6,-6) -- (0,-12);
 \draw [dashed, ultra thin] (0,-3) -- (-4.5,-7.5);
 \draw [dashed, ultra thin] (0,-6) -- (-3,-9); 
 \draw [dashed, ultra thin] (0,-9) -- (-1.5,-10.5);
 \node at (-9.5,1) {$\Delta$}; 
 \node at (-9.5,0) {$2d_1$};
 \node at (-9.5,-1.5) {$2d_1 +\frac{1}{2}$};
 \node at (-9.5,-3) {$2d_1+1$};
 \node at (-9.5,-4.5) {$2d_1 +\frac{3}{2}$};
 \node at (-9.5,-6) {$2d_1+2$};
 \node at (-9.5,-7.5) {$2d_1 +\frac{5}{2}$}; 
 \node at (-9.5,-9) {$2d_1+3$};
 \node at (-9.5,-10.5) {$2d_1 +\frac{7}{2}$}; 
 \node at (-9.5,-12) {$2d_1+4$};
\draw [fill=white] (-1.2,-0.3) rectangle (1.2,0.4);
 \node at (0,0) {$(d_1,0)_{(0,0,0)}$}; 
\draw [fill=white] (-2.9,-1.1) rectangle (-0.1,-1.9);
 \node at (-1.5,-1.5) {$(d_1-1,1)_{(1,0,0)}$};
\draw [fill=white] (-1.4,-2.6) rectangle (1.4,-3.4);
 \node at (0,-3) {$(d_1-1,0)_{(2,0,0)}$};
\draw [fill=white] (-4.4,-2.6) rectangle (-1.6,-3.4);
 \node at (-3,-3) {$(d_1-2,2)_{(0,1,0)}$};
\draw [fill=white] (-5.9,-4.1) rectangle (-3.1,-4.9);
 \node at (-4.5,-4.5) {$(d_1-3,3)_{(0,0,1)}$};
\draw [fill=white] (-2.9,-4.1) rectangle (-0.1,-4.9);
 \node at (-1.5,-4.5) {$(d_1-2,1)_{(1,1,0)}$};
 \draw [fill=white] (-7.4,-5.6) rectangle (-4.6,-6.4);
 \node at (-6,-6) {$(d_1-4,4)_{(0,0,0)}$};
 \draw [fill=white] (-4.4,-5.6) rectangle (-1.6,-6.4);
 \node at (-3,-6) {$(d_1-3,2)_{(1,0,1)}$};
 \draw [fill=white] (-1.4,-5.6) rectangle (1.4,-6.4);
 \node at (0,-6) {$(d_1-2,0)_{(0,2,0)}$};
 \draw [fill=white] (-5.9,-7.1) rectangle (-3,-7.9);
 \node at (-4.5,-7.5) {$(d_1-4,3)_{(1,0,0)}$};
 \draw [fill=white] (-2.9,-7.1) rectangle (-0.1,-7.9);
 \node at (-1.5,-7.5) {$(d_1-3,1)_{(0,1,1)}$};
 \draw [fill=white] (-4.4,-8.6) rectangle (-1.6,-9.4);
 \node at (-3,-9) {$(d_1-4,2)_{(0,1,0)}$};
 \draw [fill=white] (-1.4,-8.6) rectangle (1.4,-9.4);
 \node at (0,-9) {$(d_1-3,0)_{(0,0,2)}$};
 \draw [fill=white] (-2.9,-10.1) rectangle (0.1,-10.9);
 \node at (-1.5,-10.5) {$(d_1-4,1)_{(0,0,1)}$};
 \draw [fill=white] (-1.4,-11.6) rectangle (1.4,-12.4);
 \node at (0,-12) {$(d_1-4,0)_{(0,0,0)}$};
\end{tikzpicture} \end{equation}

\subsubsection{The Free-Tensor Multiplet:
  $\mathcal{D}[0,0,0;1,0]$}

This spectrum can be obtained by substituting $d_1=1$ into that of
$\mathcal{D}[0,0,0;d_1,0]$ and running the RS algorithm. The resulting
states match a scalar $\Phi^{\bf I}$, ${\bf I} = 1,\cdots, 5$, a
fermion $\lambda_a^\bA$ and a selfdual tensor $H_{[\mu\nu\rho]}^+$.
The negative-multiplicity representations are naturally matched with
the equations of motion $\partial^2 \Phi^{\bf I}=0$,
$\slashed{\partial}^{\dot a\dot b}\tilde\lambda_{\dot b}^\bA=0$ and the Bianchi identity for the
selfdual tensor $\partial_{[\mu}H^+_{\nu \rho \sigma]}=0$.

The full multiplet is given by

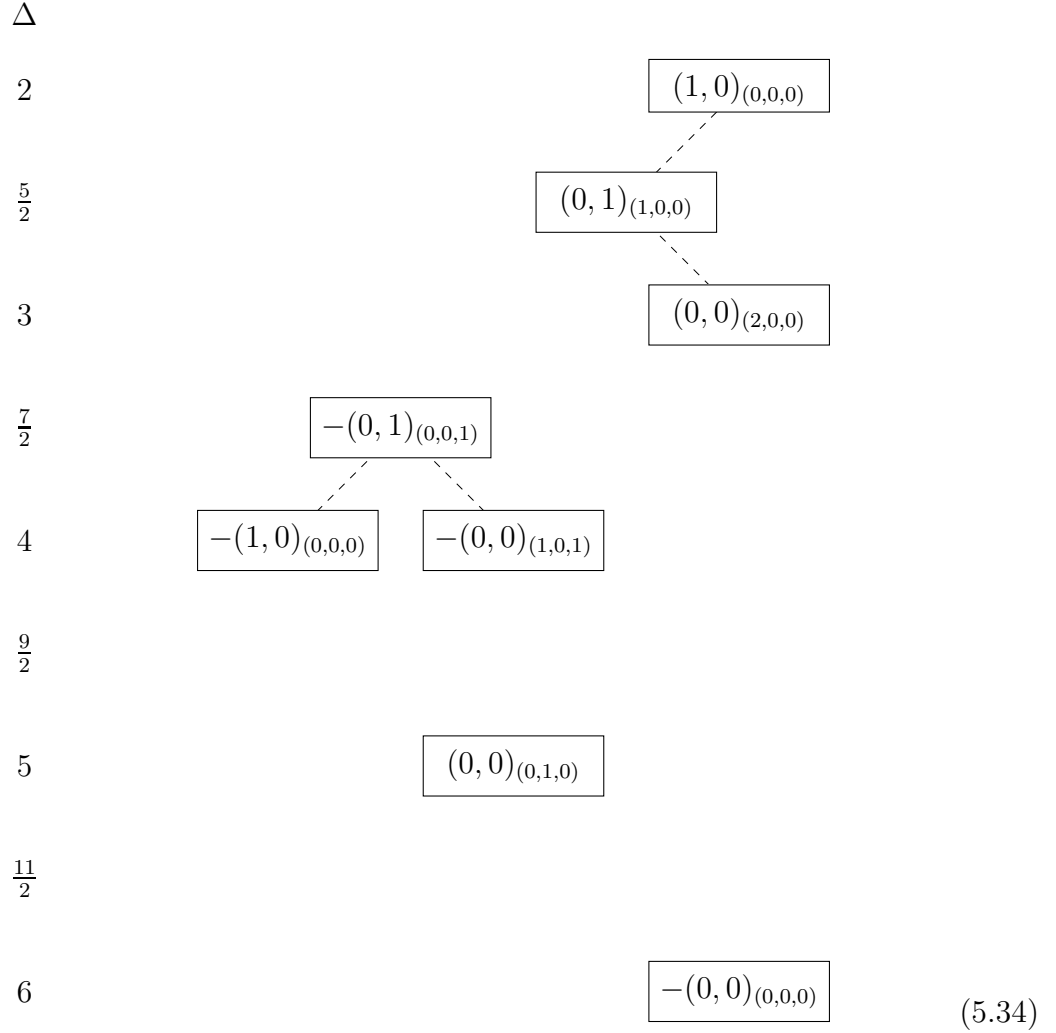
\begin{equation} \begin{tikzpicture}
 \draw [dashed, ultra thin] (0,0) -- (-1.5,-1.5); 
 \draw [dashed, ultra thin] (-1.5,-1.5) -- (0,-3);
 \draw [dashed, ultra thin] (-4.5,-4.5) -- (-3,-6);
 \draw [dashed, ultra thin] (-4.5,-4.5) -- (-6,-6);
 \node at (-9.5,1) {$\Delta$}; 
 \node at (-9.5,0) {$2$};
 \node at (-9.5,-1.5) {$\frac{5}{2}$};
 \node at (-9.5,-3) {$3$};
 \node at (-9.5,-4.5) {$\frac{7}{2}$};
 \node at (-9.5,-6) {$4$};
 \node at (-9.5,-7.5) {$\frac{9}{2}$}; 
 \node at (-9.5,-9) {$5$};
 \node at (-9.5,-10.5) {$\frac{11}{2}$}; 
 \node at (-9.5,-12) {$6$};
\draw [fill=white] (-1.2,-0.3) rectangle (1.2,0.4);
 \node at (0,0) {$(1,0)_{(0,0,0)}$}; 
\draw [fill=white] (-2.7,-1.1) rectangle (-0.3,-1.9);
 \node at (-1.5,-1.5) {$(0,1)_{(1,0,0)}$};
\draw [fill=white] (-1.2,-2.6) rectangle (1.2,-3.4);
 \node at (0,-3) {$(0,0)_{(2,0,0)}$};
\draw [fill=white] (-5.7,-4.1) rectangle (-3.3,-4.9);
 \node at (-4.5,-4.5) {$-(0,1)_{(0,0,1)}$};
 \draw [fill=white] (-7.2,-5.6) rectangle (-4.8,-6.4);
 \node at (-6,-6) {$-(1,0)_{(0,0,0)}$};
 \draw [fill=white] (-4.2,-5.6) rectangle (-1.8,-6.4);
 \node at (-3,-6) {$-(0,0)_{(1,0,1)}$};
 \draw [fill=white] (-4.2,-8.6) rectangle (-1.8,-9.4);
 \node at (-3,-9) {$(0,0)_{(0,1,0)}$};
 \draw [fill=white] (-1.2,-11.6) rectangle (1.2,-12.4);
 \node at (0,-12) {$-(0,0)_{(0,0,0)}$};
\end{tikzpicture} \end{equation}

The refined index over this multiplet is compact enough to be
presented in full
\begin{align}
  \label{eq:19}
  \II_{\mathcal{D}[0,0,0;1,0]}(p,q,s,t) =
  \frac{qt^{-1} + q^2p^2t -
  q^2p(s + p + s^{-1})+ q^3p^2}{(1-q)(1-qps)(1-qp
  s^{-1})}\;.
\end{align}

\subsubsection{The Stress-Tensor Multiplet: $\mathcal{D}[0,0,0;2,0]$}

This spectrum can be obtained by substituting $d_1=2$ into that of
$\mathcal{D}[0,0,0;d_1,0]$ and running the RS algorithm. 
The resulting representation is

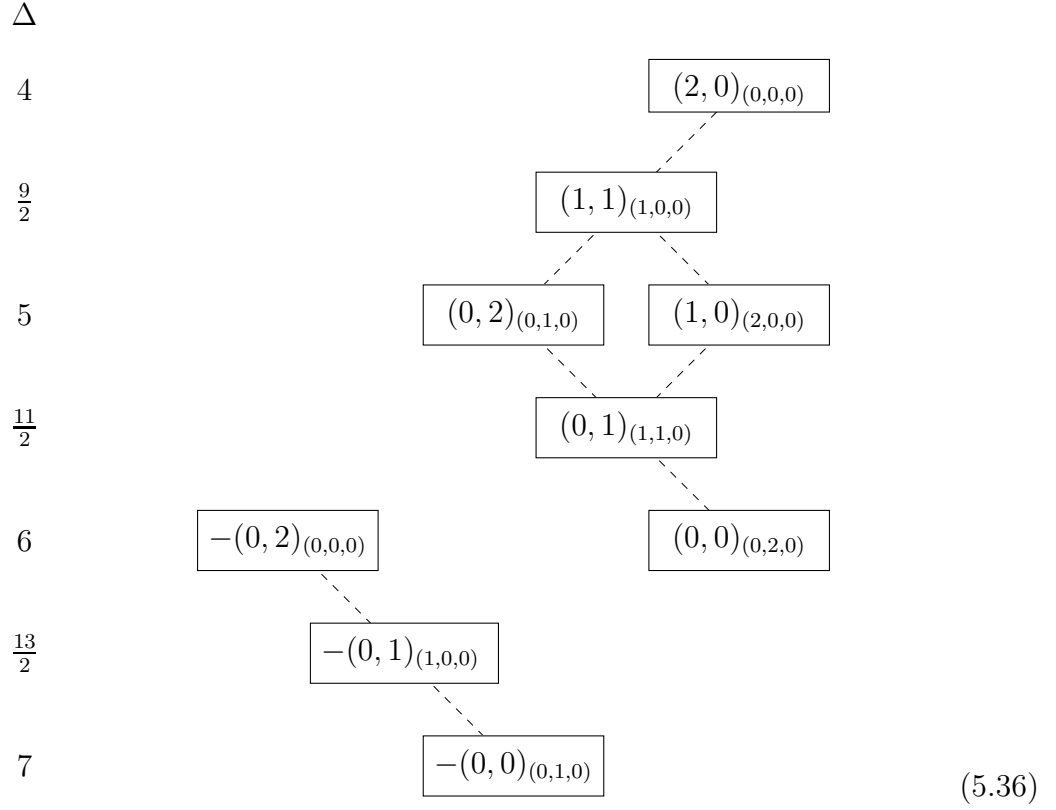
\begin{equation} \begin{tikzpicture}
 \draw [dashed, ultra thin] (0,0) -- (-3,-3); 
 \draw [dashed, ultra thin] (-1.5,-1.5) -- (0,-3);
 \draw [dashed, ultra thin] (-3,-3) -- (0,-6); 
 \draw [dashed, ultra thin] (-6,-6) -- (-3,-9);
 \draw [dashed, ultra thin] (0,-3) -- (-1.5,-4.5); 
 \node at (-9.5,1) {$\Delta$}; 
 \node at (-9.5,0) {$4$};
 \node at (-9.5,-1.5) {$\frac{9}{2}$};
 \node at (-9.5,-3) {$5$};
 \node at (-9.5,-4.5) {$\frac{11}{2}$};
 \node at (-9.5,-6) {$6$};
 \node at (-9.5,-7.5) {$\frac{13}{2}$}; 
 \node at (-9.5,-9) {$7$};
\draw [fill=white] (-1.2,-0.3) rectangle (1.2,0.4);
 \node at (0,0) {$(2,0)_{(0,0,0)}$}; 
\draw [fill=white] (-2.7,-1.1) rectangle (-0.3,-1.9);
 \node at (-1.5,-1.5) {$(1,1)_{(1,0,0)}$};
\draw [fill=white] (-1.2,-2.6) rectangle (1.2,-3.4);
 \node at (0,-3) {$(1,0)_{(2,0,0)}$};
\draw [fill=white] (-4.2,-2.6) rectangle (-1.8,-3.4);
 \node at (-3,-3) {$(0,2)_{(0,1,0)}$};
\draw [fill=white] (-2.7,-4.1) rectangle (-0.3,-4.9);
 \node at (-1.5,-4.5) {$(0,1)_{(1,1,0)}$};
 \draw [fill=white] (-7.2,-5.6) rectangle (-4.8,-6.4);
 \node at (-6,-6) {$-(0,2)_{(0,0,0)}$};
 \draw [fill=white] (-1.2,-5.6) rectangle (1.2,-6.4);
 \node at (0,-6) {$(0,0)_{(0,2,0)}$};
 \draw [fill=white] (-5.7,-7.1) rectangle (-3.2,-7.9);
 \node at (-4.5,-7.5) {$-(0,1)_{(1,0,0)}$};
 \draw [fill=white] (-4.2,-8.6) rectangle (-1.8,-9.4);
 \node at (-3,-9) {$-(0,0)_{(0,1,0)}$};
\end{tikzpicture} \end{equation}

The
superconformal primary is now the diboson
\begin{align}
\mathcal{O}^{\bf IJ}:=\hspace{1.5mm}\normord{\Phi^{({\bf I}}\Phi^{{\bf
  J})}}\;.
\end{align} 
We can match the remaining states with the following currents and
their conservation equations
\begin{align}
[5;0,1,0;0,2]&:\hspace{3mm}J^{(\bA\bB)}_\mu\;,&-[6;0,0,0;0,2]&:\hspace{3mm}\partial^\mu J^{(\bA\bB)}_\mu=0\;,\nonumber\\
[11/2;1,1,0;0,1]&:\hspace{3mm}S^{\bA}_{\mu a}\;,&-[13/2;1,0,0;0,1]&:\hspace{3mm}\partial^\mu S^{\bA}_{\mu a}=0\;,\\
[6;0,2,0;0,0]&:\hspace{3mm}\Theta_{\mu\nu}\;,&-[7;0,1,0;0,0]&:\hspace{3mm}\partial^\mu \Theta_{\mu\nu}=0\;,\nonumber
\end{align}
namely the $R$-symmetry current, supersymmetry current and stress tensor
respectively. We also have three states with no associated
conservation equations. These are
\begin{align}\begin{split}
[4;0,0,0;2,0]&:\hspace{3mm}\Phi^{({\bf I}}\Phi^{{\bf J})}\;,\\
[9/2;1,0,0;1,1]&:\hspace{3mm}\Phi^{{\bf I}} \lambda^{\bA}_{a}\;,\\
[5;2,0,0;1,0]&:\hspace{3mm}H^{+}_{[\mu\nu\rho]}\Phi^{\bf I}\;.\\
\end{split}
\end{align}
The refined index for this multiplet is calculated to be 
\begin{align}
  \mathcal{I}_{\mathcal{D}[0,0,0;2,0]}(p,q,s,t)= -\frac{pq^3(st^{-1}+s^{-1}t^{-1}+p t^{-1}+p^3qt+p^2qst+p^2 qs^{-1}t)}{(1-q)(1-qps)(1-qp
  s^{-1})}\nonumber\\
  +\frac{q^2t^{-2}+p^2q^3(1+p^2q t^2)+p^2q^4(1+ps+p s^{-1})}{(1-q)(1-qps)(1-qp
  s^{-1})}\;.
\end{align}

\subsubsection{$\mathcal{D}[0,0,0;1,1]$}

For this multiplet we are prescribed to remove the combinations
$\QQ_{\bTw a}\QQ_{\bTw b}$ from the basis of $\QQ$-generators of the
Verma module. Acting on the superconformal primary with the remaining
set of $\tilde{Q}$ supercharges gives rise to \begin{align}\label{Q3Q2
    actions}\begin{split}
    (d_1, d_2)_{(c_1, c_2, c_3)}&\xrightarrow{\tilde{Q}} (d_1-1, d_2+1),(d_1+1,d_2-1)_{(c_1+1, c_2, c_3),(c_1-1, c_2+1, c_3),(c_1, c_2-1, c_3+1),(c_1, c_2, c_3-1)},\\
    &\xrightarrow{\tilde{Q}^2} (d_1-2, d_2+2),(d_1,d_2)_{(c_1, c_2+1, c_3),(c_1+1, c_2-1, c_3+1),(c_1+1, c_2, c_3-1),(c_1-1, c_2, c_3+1)},\\
    &\phantom{\xrightarrow{\tilde{Q}^2}\hspace{2.5mm}}(d_1-2, d_2+2),(d_1,d_2)_{(c_1-1,c_2 + 1, c_3 -1),(c_1, c_2-1, c_3)},\\
    &\phantom{\xrightarrow{\tilde{Q}^2}\hspace{2.5mm}}(d_1,d_2)_{(c_1+2, c_2, c_3),(c_1-2, c_2+2, c_3),(c_1, c_2-2, c_3+2),(c_1, c_2, c_3-2)},\\
    &\xrightarrow{\tilde{Q}^3} (d_1-3, d_2+3),(d_1-1, d_2+1)_{(c_1, c_2, c_3+1),(c_1, c_2+1, c_3-1),(c_1+1, c_2-1, c_3),(c_1-1, c_2, c_3)},\\
    &\phantom{\xrightarrow{\tilde{Q}^2}\hspace{2.5mm}}(d_1-1,d_2+1)_{(c_1+1,c_2+1, c_3),(c_1-1,c_2+2,c_3),(c_1+2,c_2-1,c_3+1),(c_1-2,c_2+1,c_3+1)},\\
    &\phantom{\xrightarrow{\tilde{Q}^2}\hspace{2.5mm}}(d_1-1,d_2+1)_{(c_1+2,c_2, c_3-1),(c_1+1,c_2,c_3-2),(c_1-2,c_2+2,c_3-1),(c_1-1,c_2+1,c_3-2)},\\
    &\phantom{\xrightarrow{\tilde{Q}^2}\hspace{2.5mm}}(d_1-1,d_2+1)_{(c_1,c_2-1, c_3-1),(c_1-1,c_2-1,c_3+2),(c_1,c_2-2,c_3+1)},\\
    &\xrightarrow{\tilde{Q}^4} (d_1-4, d_2+4),(d_1-2,d_2+2)_{(c_1, c_2, c_3)},\\
    &\phantom{\xrightarrow{\tilde{Q}^4}\hspace{2.5mm}}(d_1-2,d_2+2)_{(c_1+1,c_2, c_3+1),(c_1-1,c_2+1,c_3+1),(c_1,c_2-1,c_3+2),(c_1+1,c_2+1,c_3-1),},\\
    &\phantom{\xrightarrow{\QQ^4}\hspace{2.5mm}}(d_1-2,d_2+2)_{(c_1,c_2+1, c_3-2),(c_1+2,c_2-1,c_3),(c_1,c_2-1,c_3+2),(c_1+1,c_2-1,c_3-1)},\\
    &\phantom{\xrightarrow{\QQ^4}\hspace{2.5mm}}(d_1-2,d_2+2)_{(c_1-1,c_2-1, c_3+1),(c_1-1,c_2,c_3-1),(c_1-1,c_2+2,c_3-1),(c_1-2,c_2+1,c_3)},\\
    &\xrightarrow{\tilde{Q}^5} (d_1-3, d_2+3)_{(c_1+1, c_2, c_3),(c_1-1, c_2+1, c_3),(c_1, c_2-1, c_3+1),(c_1, c_2, c_3-1)}\;.\\
  \end{split} \end{align} The action of the $Q$ supercharges is still
the one presented in \eqref{Qs on D}. Combining the two, the full
module is given by 

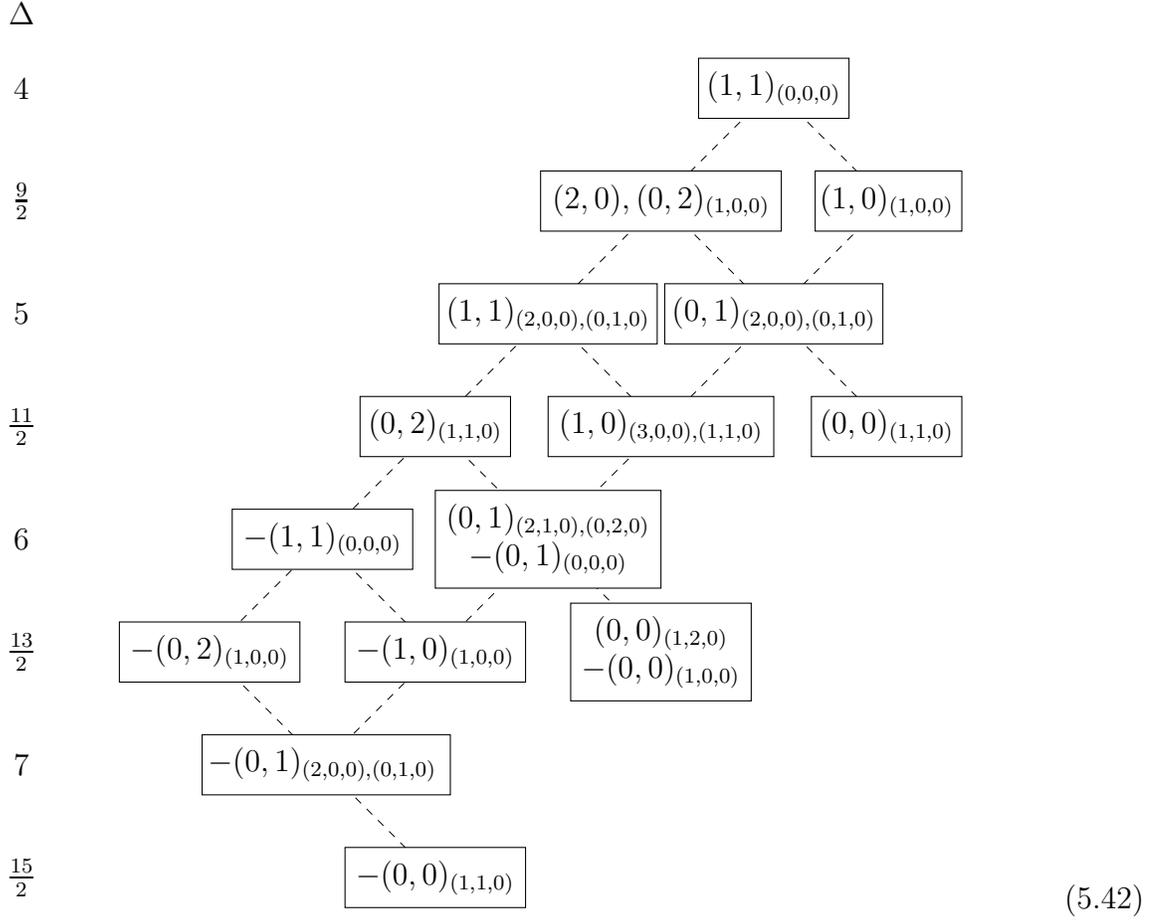
\begin{align} \begin{tikzpicture} \draw [dashed,
    ultra thin] (0,0) -- (-7.5,-7.5); \draw [dashed, ultra thin] (0,0)
    -- (1.5,-1.5); \draw [dashed, ultra thin] (-1.5,-1.5) --
    (1.5,-4.5); \draw [dashed, ultra thin] (-3,-3) -- (-1.5,-4.5);
    \draw [dashed, ultra thin] (-4.5,-4.5) -- (-1.5,-7.5); \draw
    [dashed, ultra thin] (-6,-6) -- (-4.5,-7.5); \draw [dashed, ultra
    thin] (1.5,-1.5) -- (-6,-9); \draw [dashed, ultra thin]
    (-7.5,-7.5) -- (-4.5,-10.5); \node at (-10,1)
    {$\Delta$}; \node at (-10,0)
    {$4$}; \node at (-10,-1.5)
    {$\frac{9}{2}$}; \node at (-10,-3)
    {$5$}; \node at (-10,-4.5)
    {$\frac{11}{2}$}; \node at (-10,-6)
    {$6$}; \node at (-10,-7.5)
    {$\frac{13}{2}$}; \node at (-10,-9)
    {$7$}; \node at (-10,-10.5)
    {$\frac{15}{2}$}; \draw [fill=white] (-1,-0.4) rectangle (1,0.4);
    \node at (0,0)
    {$(1,1)_{(0,0,0)}$}; \draw [fill=white] (-3.1,-1.1) rectangle
    (0.1,-1.9); \node at (-1.5,-1.5)
    {$(2,0),(0,2)_{(1,0,0)}$}; \draw [fill=white] (0.55,-1.1)
    rectangle (2.5,-1.9); \node at (1.5,-1.5)
    {$(1,0)_{(1,0,0)}$}; \draw [fill=white] (-4.45,-2.6) rectangle
    (-1.55,-3.4); \node at (-3,-3)
    {$(1,1)_{(2,0,0),(0,1,0)}$}; \draw [fill=white] (-1.45,-2.6)
    rectangle (1.45,-3.4); \node at (0,-3)
    {$(0,1)_{(2,0,0),(0,1,0)}$}; \draw [fill=white] (-5.5,-4.1)
    rectangle (-3.5,-4.9); \node at (-4.5,-4.5)
    {$(0,2)_{(1,1,0)}$}; \draw [fill=white] (-3,-4.1) rectangle
    (0,-4.9); \node at (-1.5,-4.5)
    {$(1,0)_{(3,0,0),(1,1,0)}$}; \draw [fill=white] (0.5,-4.1)
    rectangle (2.5,-4.9); \node at (1.5,-4.5)
    {$(0,0)_{(1,1,0)}$}; \draw [fill=white] (-7.2,-5.6) rectangle
    (-4.8,-6.4); \node at (-6,-6)
    {$-(1,1)_{(0,0,0)}$}; \draw [fill=white] (-4.5,-5.35) rectangle
    (-1.5,-6.65); \node at (-3,-5.75)
    {$(0,1)_{(2,1,0),(0,2,0)}$}; \node at (-3,-6.25)
    {$-(0,1)_{(0,0,0)}$}; \draw [fill=white] (-8.7,-7.1) rectangle
    (-6.3,-7.9); \node at (-7.5,-7.5)
    {$-(0,2)_{(1,0,0)}$}; \draw [fill=white] (-5.7,-7.1) rectangle
    (-3.3,-7.9); \node at (-4.5,-7.5)
    {$-(1,0)_{(1,0,0)}$}; \draw [fill=white] (-2.7,-6.85) rectangle
    (-0.3,-8.15); \node at (-1.5,-7.25)
    {$(0,0)_{(1,2,0)}$}; \node at (-1.5,-7.75)
    {$-(0,0)_{(1,0,0)}$}; \draw [fill=white] (-7.6,-8.6) rectangle
    (-4.3,-9.4); \node at (-6,-9)
    {$-(0,1)_{(2,0,0),(0,1,0)}$}; \draw [fill=white] (-5.7,-10.1)
    rectangle (-3.3,-10.9); \node at (-4.5,-10.5)
    {$-(0,0)_{(1,1,0)}$}; \end{tikzpicture} \end{align}

\subsubsection{$\mathcal{D}[0,0,0;0,1]$}

The shortening conditions for this multiplet are similar to the
$\mathcal{D}[0,0,0;1,1]$ case; they follow from \eqref{Q3Q2
  actions} by setting $d_1=0$. The full spectrum of states, including
those corresponding to operator constraints, is given by the diagram:

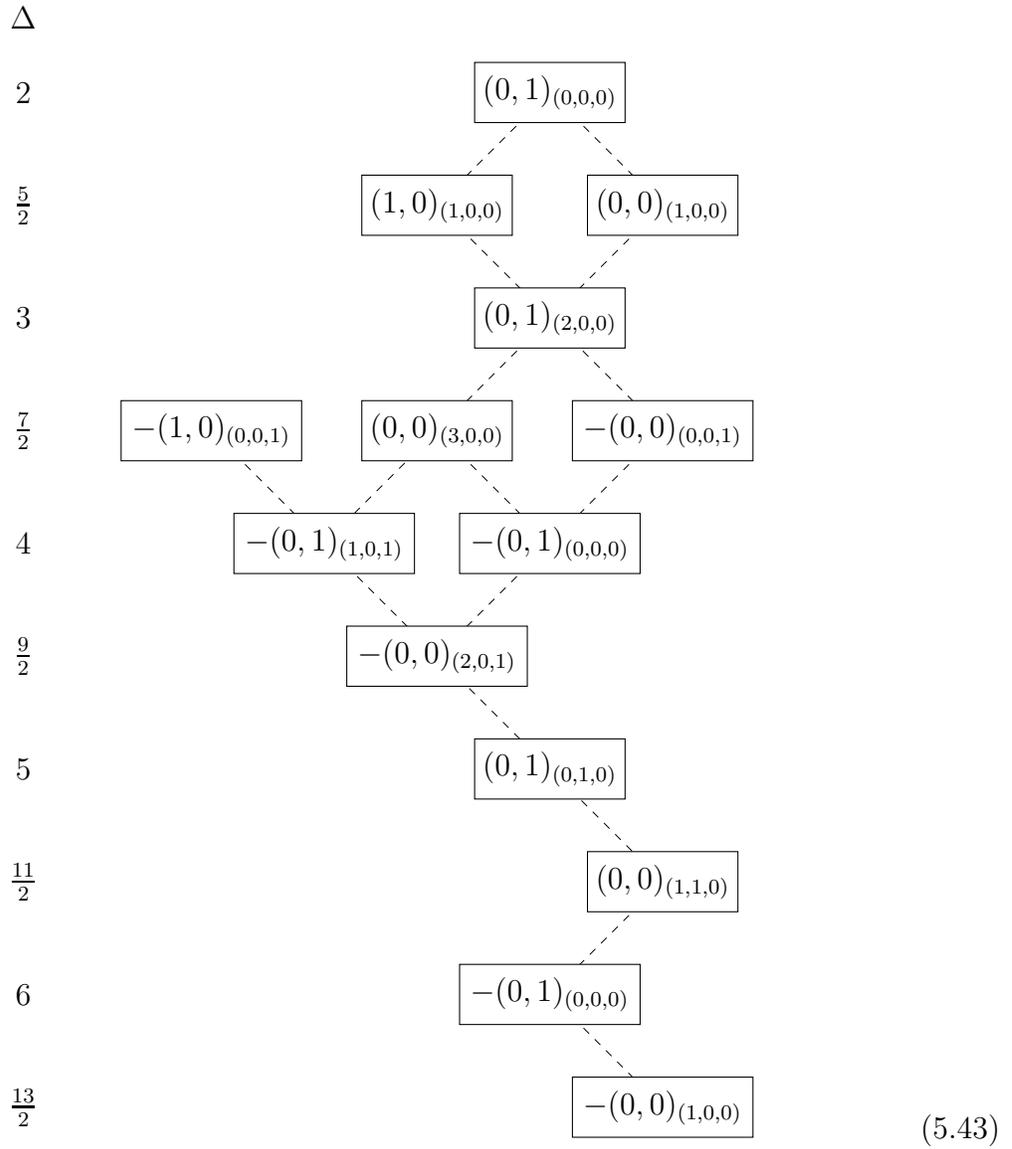
\begin{equation}
\begin{tikzpicture}
\draw [dashed, ultra thin] (0,0) -- (-1.5,-1.5);
\draw [dashed, ultra thin] (0,0) -- (1.5,-1.5);
\draw [dashed, ultra thin] (-1.5,-1.5) -- (1.5,-4.5);
\draw [dashed, ultra thin] (-1.5,-4.5) -- (0,-6);
\draw [dashed, ultra thin] (-4.5,-4.5) -- (1.5,-10.5);
\draw [dashed, ultra thin] (1.5,-10.5) -- (0,-12);
\draw [dashed, ultra thin] (1.5,-1.5) -- (-3,-6);
\draw [dashed, ultra thin] (1.5,-4.5) -- (-1.5,-7.5);
\draw [dashed, ultra thin] (0,-12) -- (1.5,-13.5);
\node at (-7,1) {$\Delta$};
\node at (-7,0) {$2$};
\node at (-7,-1.5) {$\frac{5}{2}$};
\node at (-7,-3) {$3$};
\node at (-7,-4.5) {$\frac{7}{2}$};
\node at (-7,-6) {$4$};
\node at (-7,-7.5) {$\frac{9}{2}$};
\node at (-7,-9) {$5$};
\node at (-7,-10.5) {$\frac{11}{2}$};
\node at (-7,-12) {$6$};
\node at (-7,-13.5) {$\frac{13}{2}$};
\draw [fill=white] (-1,-0.4) rectangle (1,0.4);
\node at (0,0) {$(0,1)_{(0,0,0)}$};
\draw [fill=white] (-2.5,-1.1) rectangle (-0.5,-1.9);
\node at (-1.5,-1.5) {$(1,0)_{(1,0,0)}$};
\draw [fill=white] (2.5,-1.1) rectangle (0.5,-1.9);
\node at (1.5,-1.5) {$(0,0)_{(1,0,0)}$};
\draw [fill=white] (1,-2.6) rectangle (-1,-3.4);
\node at (0,-3) {$(0,1)_{(2,0,0)}$};
\draw [fill=white] (0.3,-4.1) rectangle (2.7,-4.9);
\node at (1.5,-4.5) {$-(0,0)_{(0,0,1)}$};
\draw [fill=white] (-5.7,-4.1) rectangle (-3.3,-4.9);
\node at (-4.5,-4.5) {$-(1,0)_{(0,0,1)}$};
\draw [fill=white] (-2.5,-4.1) rectangle (-0.5,-4.9);
\node at (-1.5,-4.5) {$(0,0)_{(3,0,0)}$};
\draw [fill=white] (-1.2,-5.6) rectangle (1.2,-6.4);
\node at (0,-6) {$-(0,1)_{(0,0,0)}$};
\draw [fill=white] (-4.2,-5.6) rectangle (-1.8,-6.4);
\node at (-3,-6) {$-(0,1)_{(1,0,1)}$};
\draw [fill=white] (-2.7,-7.1) rectangle (-0.3,-7.9);
\node at (-1.5,-7.5) {$-(0,0)_{(2,0,1)}$};
\draw [fill=white] (-1,-8.6) rectangle (1,-9.4);
\node at (-0,-9) {$(0,1)_{(0,1,0)}$};
\draw [fill=white] (0.5,-10.1) rectangle (2.5,-10.9);
\node at (1.5,-10.5) {$(0,0)_{(1,1,0)}$};
\draw [fill=white] (-1.2,-11.6) rectangle (1.2,-12.4);
\node at (-0,-12) {$-(0,1)_{(0,0,0)}$};
\draw [fill=white] (0.3,-13.1) rectangle (2.7,-13.9);
\node at (1.5,-13.5) {$-(0,0)_{(1,0,0)}$};
\end{tikzpicture}
\end{equation}

\subsubsection{$\mathcal{D}[0,0,0;0,2]$}

The Verma module is obtained through the set of
$Q$-actions of \eqref{Qs on D} along with the set of
$\tilde{Q}$-actions of Table~\ref{D spec (2,0) d2=2 QT}; the latter can be
found in App.~\ref{D (2,0) Spectra}. Using these one has

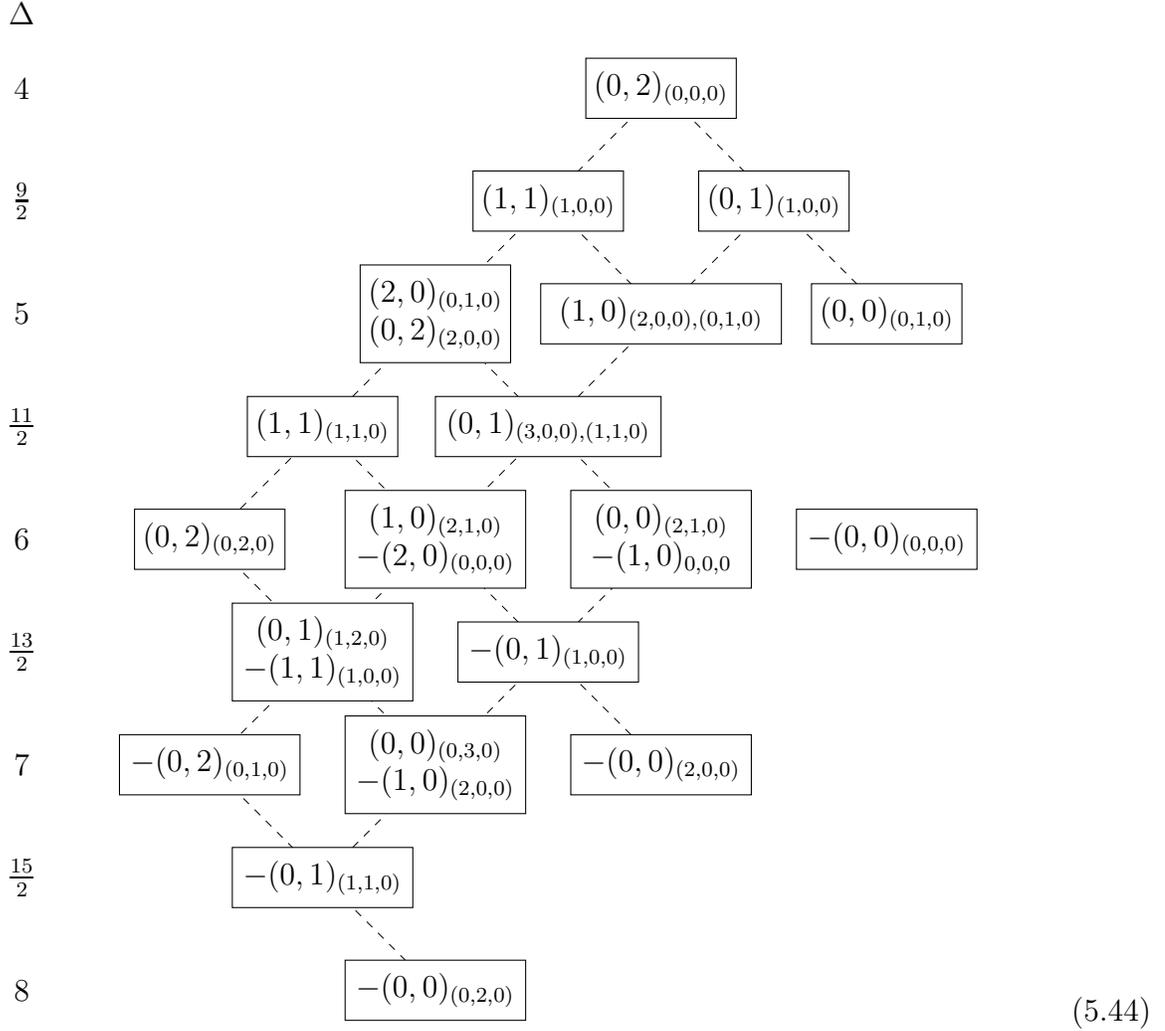
\begin{equation}
\begin{tikzpicture}
\draw [dashed, ultra thin] (0,0) -- (-6,-6);
\draw [dashed, ultra thin] (0,0) -- (3,-3);
\draw [dashed, ultra thin] (-1.5,-1.5) -- (0,-3);
\draw [dashed, ultra thin] (-3,-3) -- (0,-6);
\draw [dashed, ultra thin] (-4.5,-4.5) -- (0,-9);
\draw [dashed, ultra thin] (-6,-6) -- (-3,-9);
\draw [dashed, ultra thin] (1.5,-1.5) -- (-6,-9);
\draw [dashed, ultra thin] (0,-6) -- (-4.5,-10.5);
\draw [dashed, ultra thin] (-6,-9) -- (-3,-12);
\node at (-8.5,1) {$\Delta$};
\node at (-8.5,0) {$4$};
\node at (-8.5,-1.5) {$\frac{9}{2}$};
\node at (-8.5,-3) {$5$};
\node at (-8.5,-4.5) {$\frac{11}{2}$};
\node at (-8.5,-6) {$6$};
\node at (-8.5,-7.5) {$\frac{13}{2}$};
\node at (-8.5,-9) {$7$};
\node at (-8.5,-10.5) {$\frac{15}{2}$};
\node at (-8.5,-12) {$8$};
\draw [fill=white] (-1,-0.4) rectangle (1,0.4);
\node at (0,0) {$(0,2)_{(0,0,0)}$};
\draw [fill=white] (2.5,-1.1) rectangle (0.5,-1.9);
\node at (1.5,-1.5) {$(0,1)_{(1,0,0)}$};
\draw [fill=white] (-2.5,-1.1) rectangle (-0.5,-1.9);
\node at (-1.5,-1.5) {$(1,1)_{(1,0,0)}$};
\draw [fill=white] (2,-2.6) rectangle (4,-3.4);
\node at (3,-3) {$(0,0)_{(0,1,0)}$};
\draw [fill=white] (1.8,-5.6) rectangle (4.2,-6.4);
\node at (3,-6) {$-(0,0)_{(0,0,0)}$};
\draw [fill=white] (-4,-2.35) rectangle (-2,-3.65);
\node at (-3,-2.75) {$(2,0)_{(0,1,0)}$};
\node at (-3,-3.25) {$(0,2)_{(2,0,0)}$};
\draw [fill=white] (-4.2,-5.35) rectangle (-1.8,-6.65);
\node at (-3,-5.75) {$(1,0)_{(2,1,0)}$};
\node at (-3,-6.25) {$-(2,0)_{(0,0,0)}$};
\draw [fill=white] (-1.6,-2.6) rectangle (1.6,-3.4);
\node at (0,-3) {$(1,0)_{(2,0,0),(0,1,0)}$};
\draw [fill=white] (-1.2,-5.35) rectangle (1.2,-6.65);
\node at (0,-5.75) {$(0,0)_{(2,1,0)}$};
\node at (0,-6.25) {$-(1,0)_{0,0,0}$};
\draw [fill=white] (-3,-4.1) rectangle (0,-4.9);
\node at (-1.5,-4.5) {$(0,1)_{(3,0,0),(1,1,0)}$};
\draw [fill=white] (-2.7,-7.1) rectangle (-0.3,-7.9);
\node at (-1.5,-7.5) {$-(0,1)_{(1,0,0)}$};
\draw [fill=white] (-5.5,-4.1) rectangle (-3.5,-4.9);
\node at (-4.5,-4.5) {$(1,1)_{(1,1,0)}$};
\draw [fill=white] (-5.7,-6.85) rectangle (-3.3,-8.15);
\node at (-4.5,-7.25) {$(0,1)_{(1,2,0)}$};
\node at (-4.5,-7.75) {$-(1,1)_{(1,0,0)}$};
\draw [fill=white] (-1.2,-8.6) rectangle (1.2,-9.4);
\node at (0,-9) {$-(0,0)_{(2,0,0)}$};
\draw [fill=white] (-7,-5.6) rectangle (-5,-6.4);
\node at (-6,-6) {$(0,2)_{(0,2,0)}$};
\draw [fill=white] (-7.2,-8.6) rectangle (-4.8,-9.4);
\node at (-6,-9) {$-(0,2)_{(0,1,0)}$};
\draw [fill=white] (-4.2,-8.35) rectangle (-1.8,-9.65);
\node at (-3, -8.75) {$(0,0)_{(0,3,0)}$};
\node at (-3,-9.25) {$-(1,0)_{(2,0,0)}$};
\draw [fill=white] (-5.7,-10.1) rectangle (-3.3,-10.9);
\node at (-4.5,-10.5) {$-(0,1)_{(1,1,0)}$};
\draw [fill=white] (-4.2,-11.6) rectangle (-1.8,-12.4);
\node at (-3,-12) {$-(0,0)_{(0,2,0)}$};
\end{tikzpicture}
\end{equation}

This concludes our discussion of 6D (2,0) superconformal multiplets. 

\section{Some Initial Applications}\label{Apps}

We finally mention some brief applications of the results that we have
presented thus far, while leaving a more in-depth exploration for
future work. We study aspects of flavour symmetries and flavour
anomalies. We also rule out the presence of certain multiplets in free
theories and provide evidence that certain 6D (1,0) multiplets can
consistently combine to form 6D (2,0) multiplets.

\subsection{Flavour symmetries}

Both the 5D $\CN=1$ and 6D $(1,0)$ algebras admit flavour symmetries.\foot{Here we define such symmetries to be those that commute with the superconformal algebra. See the discussion around \eqref{nosymm} for an explanation of why flavour symmetries do not sit in multiplets with higher-spin symmetries.}  Therefore it is interesting to ask which representations are allowed to transform nontrivially under these symmetries.

For example, in 4D $\CN=2$ theories, it is known that $\CN=2$ chiral
operators do not transform under flavour symmetries
\cite{Buican:2013ica,Buican:2014qla}. Since scalar $\CN=2$ chiral
operators parametrise the Coulomb branch (when it exists), this
statement is consistent with the fact that these theories do not have any massless matter (besides free vector multiplets) at generic points on the Coulomb branch. Below, we will explore similar physical constraints in 5D and 6D.

Let us begin by considering 6D $(1,0)$ theories. Many of these
theories have tensor branches. Along these moduli spaces, the
$\mathfrak{sp}(1)_R$ and flavour symmetries are unbroken, and free-tensor multiplets play an important role. It is therefore reasonable to assume that there are flavour and $\mathfrak{sp}(1)_R$-neutral superconformal primaries in short multiplets that (partially) describe the physics on this branch of the moduli space. 

To understand the last point in more detail, let us study short multiplets with an $\mathfrak{sp}(1)_R$-neutral primary, $\CO_I$. Let us further suppose that $\CO_I$ transforms linearly under a representation, $R_{\CO}$, of the theory's flavour symmetry group, $G$. We will study the conditions under which superconformal representation theory forces this representation to be trivial, i.e. $R_{\CO}={\bf 1}$. One crucial observation for us is that flavour Ward identities require the following leading OPE between the flavour symmetry current, $j^\alpha_{ab}$, associated with $G$ and $\CO_I$:
\eqn{
j^\alpha_{ab}(x)\CO_I(0)\sim-i{t^{\alpha\ J}_{\CO I}}{x_{ab}\over x^6}\CO_J+\cdots~.
}[jOOPE]
In this equation, the ellipsis contains less singular terms, and $t^\alpha_{\CO}$ is the matrix for the representation, $R_{\CO}$, that $\CO_{I}$ transforms under. For now we will assume $\CO_I$ is a Lorentz scalar, but we will relax this condition later. One important point that will come into play below is that $j^\alpha_{ab}$ in \eqref{jOOPE} is a level-two superconformal descendant of the $\Sp(1)_R$ spin-1 moment map primary
\eqn{
j_{ab}^\alpha(x)=\CQ_{{\bf A}a}\CQ_{{\bf B}b} \left(J^{\alpha,{\bf AB}}(x)\right)~.
}[Jjrelat]

To understand why certain operators are forbidden by the
superconformal algebra from transforming linearly under flavour
symmetry, it will be important to relate \eqref{jOOPE} to an OPE of
superconformal primaries. In particular, we will need the fact that
\eqn{\begin{split} \CS^a_{\bf A}\left(J^{\alpha{\bf
          AB}}(x)\right)\equiv\left[\CS^{a}_{\bf A}, J^{\alpha{\bf
          AB}}(x)\right]&=-ix^{bc}\left(\delta^a_b\CQ_{{\bf
          A}c}-\delta^a_c\CQ_{{\bf A}b}\right)\left(J^{\alpha{\bf
          AB}}(x)\right)\cr & =-2ix^{ab}\CQ_{{\bf A}
      b}\left(J^{\alpha{\bf AB}}(x)\right)~,
\end{split}}[Saction]
where we have used the superconformal algebra and the form of the translated operator
\eqn{
J^{\alpha{\bf AB}}(x)=e^{-iP\cdot x}J^{\alpha{\bf AB}}(0)e^{iP\cdot x}~.
}[translate]
Using \eqref{Saction}, we can also deduce that
\eqn{
\CS^c_{\bf B}\CS^a_{\bf A}\left(J^{\alpha{\bf AB}}(x)\right)=-2ix^{ab}\CQ_{{\bf A}b}\CS^c_{\bf B}\left(J^{\alpha{\bf AB}}(x)\right)=-4x^{ab}x^{cd}j_{bd}^{\alpha}(x)~.
}[Saction2]

From this discussion, it follows that we can extract the OPE in \eqref{jOOPE} by acting with the special supercharges on the superconformal primary OPE
\eqn{
x^{ab}x^{cd}j_{bd}^{\alpha}(x)\CO_I(0)=\CS^c_{\bf B}\CS^a_{\bf A}\left(J^{\alpha{\bf AB}}(x)\CO_I(0)\right)\sim-it^{\alpha\ J}_{\CO I}{x^{ab}x^{cd}x_{bd}\over x^6}\CO_J(0)+\cdots~.
}[relprimOPE]
In order to produce the correct leading singularity in \eqref{relprimOPE}, the superconformal primary OPE must then contain terms of the form
\eqn{
J^{\alpha{\bf AB}}(x)\CO_I(0)\supset-it^{\alpha\ J}_{\CO I}{x^{ab}\over x^4}\CQ_a^{\bf A}\CQ_b^{\bf B}\CO_J(0)+\cdots~.
}[primOPE]
This discussion implies the following results:

\begin{itemize}
\item[$\bullet$] It is straightforward to check that in the
  $\CC[0,0,0;0]$ multiplet there are no operators of the type written
  on the RHS of \eqref{primOPE} since there are no $\mathfrak{sp} (1)_R$ spin-1 descendants at level two. Therefore, these multiplets cannot transform linearly under any flavour symmetries in an SCFT. Since adding spin does not allow for such a descendant, we conclude that all the $\CC[c_1,0,0;0]$ multiplets cannot transform under nontrivial linear representations of the flavour symmetry.
\item[$\bullet$] Note that the absence of a conformal primary with the
  quantum numbers of the operator on the RHS of \eqref{primOPE} is a
  sufficient but not necessary condition for the multiplet to be
  flavour neutral. For example, the stress-tensor multiplet contains
  an $\mathfrak{sp}(1)_R$ current which has the correct quantum
  numbers. However, in this case, since the stress tensor generates
  translations, it must be flavour neutral. Still, in free theories,
  with a collection of hypermultiplets, $\phi^{i\bf A}\in\CD[0,0,0;1]$ (here $i=1,\cdots, N$ indexes the hypermultiplets) we can construct $\CB[0,0,0;0]$ multiplets that are not flavour neutral as follows
\eqn{
\CO^{ij}=\epsilon_{\bf AB}\phi^{i{\bf A}}\phi^{j{\bf B}}\in\CB[0,0,0;0]~.
}[flavT]
\item[$\bullet$] We can also play the same game in 5D, since the R
  symmetry is the same, and the flavour moment maps are again
  superconformal primaries of $\mathfrak{su} (2)_R$ spin-1 multiplets
  containing the flavour currents. It is straightforward to check that
  the above argument does not rule out flavour transformations for any
  unitary representations.
\end{itemize}

Note that the above discussion on the triviality of certain linear
representations does not preclude the possibility that certain
multiplets transform non-linearly when we turn on background gauge
fields for global symmetries. Indeed, the free-tensor fields,
$\varphi^{I}\in\CC[0,0,0;0]$, often must transform non-linearly in order
to match nontrivial 't Hooft anomalies for RG flows onto the tensor
branch of certain interacting $(1,0)$ theories
\cite{Ohmori:2014kda,Intriligator:2014eaa,Cordova:2015fha} (our
discussion below mostly follows \cite{Intriligator:2014eaa,
  Ohmori:2014kda}). For example, the $\varphi^I$ multiplets can be
used to match a discrepancy in the UV and IR anomalies of the form
\eqn{
 \Delta I_8={1\over2\pi}\Omega_{IJ} X_4^I\wedge X_4^J~.  
}[dffanom] In
this equation, $\Delta I_8$ is the naive change in the anomaly
polynomial eight-form between the UV and the IR, $\Omega_{IJ}$ a
positive-definite symmetric matrix, and the four-form, $X_4^I$, has a
contribution from the flavour symmetry background fields, $F_i$, 
\eqn{
  X^I_4\supset \sum_i n^{Ii} c_2(F_i)~, 
} 
where $c_2$ is a Chern class
for the background flavour symmetry. In order to make up for the
discrepancy in \eqref{dffanom} there must be a coupling of the tensor
field two-form, $B_2^I$, to $X_4^J$ in the IR effective action 
\eqn{
  \delta\CL\sim\Omega_{IJ} B_2^I\wedge X_4^J~, 
}[GSmech] 
where the necessary anomalous variation is generated by requiring that
$\delta B_2^I=-\pi X_2^{I}$, where
\eqn{
 \delta X_3^I=\mathrm d
  X_2^I~, \ \ \ X_4^I=\mathrm dX_3^I~.  
}[defnX2] 
In particular, we see that the $\CC[0,0,0;0]$ multiplet transforms
since 
\eqn{ \delta
  H_3=-\pi\delta X_3^I\ne0~.  
}[trans] 
This discussion does not contradict our previous statements because
the $\CC[0,0,0;0]\ni H_3^I$ representations do not transform linearly
under the flavour symmetry (in fact, they do not transform at all if
we turn off the background flavour gauge fields).

\subsection{Supersymmetry Enhancement of 6D $(1,0)$ Multiplets}\label{susy1020}

An important consistency condition for our 6D $\CN = 1$ multiplets is
that they can combine to form $\CN = 2$ multiplets. For instance, one
should be able to construct the additional supersymmetry and
$R$-symmetry currents of the $\CN = 2$ SCA from the multiplets of
Sec.~\ref{Sec:6D10}.

It will be instructive to first show that two hypermultiplets and a
tensor multiplet of the 6D $\CN = 1$ SCA combine to form a single
tensor multiplet in $\CN = 2$. We need only consider the $R$ symmetry,
since the Lorentz quantum numbers are the same for both cases. We
identify the $\su(2)_R\subset \so(5)_R$ by choosing $K=d_1$; note
that as a result distinct states for $\CN = 2$ can give rise to the
same state for $\CN = 1$. The free tensor in $\CN = 2$ has the $R$-symmetry
highest weight $(1,0)$, which corresponds to a representation with
Dynkin values $(1,0)$, $(-1,2)$, $(0,0)$, $(1,-2)$ and
$(-1,0)$. Reducing this to $\su(2)_R$ results in the states
$2\times (1)$, $(0)$ and $2\times(-1)$. Another way of writing this is
in terms of three modules, two with highest weights labelled by $(1)$ and one
labelled by $(0)$. Thus the superconformal primary of
$\mathcal{D}[0,0,0;1,0]$ can be identified with two primaries from the
$\mathcal{N}=1$ hypermultiplet and one from the $\mathcal{N}=1$ tensor
multiplet. Indeed one can repeat this for every state in
$\mathcal{D}[0,0,0;1,0]$ and the result is that the entire multiplet
is written as
\begin{align}
\mathcal{D}[0,0,0;1,0]\simeq 2\mathcal{D}[0,0,0;1]\oplus \mathcal{C}[0,0,0;0]\;,
\end{align}
where the equivalence is up to the identification $d_1=K$.

One can similarly rewrite the stress-tensor multiplet
of $\mathcal{N}=2$, as a collection of $\mathcal{N}=1$ multiplets. The
result is that 
\begin{align}\label{decomp}
\mathcal{D}[0,0,0;2,0]\simeq 3\mathcal{D}[0,0,0;2]\oplus 2\mathcal{C}[0,0,0;1]\oplus\mathcal{B}[0,0,0;0]\;,
\end{align}
where the equivalence is again up to the identification $d_1=K$,
confirming the original expectation.\foot{For example, note that we find precisely the expected decomposition of the $\mathfrak{so}(5)_R$ currents (recall that these are level two superconformal descendants of the $\CD[0,0,0;2,0]$ multiplet) in terms of representations of $\mathfrak{su}(2)_R$: ${\bf10}=3\times{\bf1}+2\times{\bf 2}+1\times{\bf3}$.}
\clearpage

\ack{ \bigskip We would like to thank D.~Berman, E.~Hughes, B.~Penante
  and B.~van~Rees for helpful discussions and comments. J.~H. would
  like to thank Humboldt University of Berlin for their
  hospitality. The work of M.~B. is partially
  supported by the Royal Society under the grant \lq\lq New
  Constraints and Phenomena in Quantum Field Theory," and by the
  U.S. Department of Energy under grant DE-SC0009924. The work of
  J.~H. is supported by an STFC research studentship. C.~P. is supported by the Royal Society through a
  University Research Fellowship.}

\begin{appendix}

\section{The Superconformal Algebra in 5D and 6D}\label{App:5DSCA}

In this appendix we collect our notation and conventions for the 5D
and 6D SCAs. For a fully detailed account of conformal and
superconformal algebras in various dimensions we refer the interested
reader to \cite{Minwalla:1997ka, Dolan:2005wy}.

\subsection{The Conformal Algebra in $D$ Dimensions}

The conformal group in $D$ dimensions is locally isomorphic to
$\SO(D,2)$ and generated by $\frac{1}{2}D(D-1)$ Lorentz generators
$M_{\mu\nu}$, $D$ momenta $\PP_\mu$ and special conformal generators
$\KK_\mu$, and the conformal Hamiltonian $H$.  The associated Lie
algebra is defined by the commutation relations
\begin{align}\label{conformal com relations}
[M_{\mu\nu},M_{\kappa\lambda}]&=i(\delta_{\mu\kappa}M_{\nu\lambda}-\delta_{\mu\lambda}M_{\nu\kappa} - \delta_{\nu\kappa}M_{\mu\lambda}+\delta_{\nu\lambda}M_{\mu\kappa})\;,\cr
[M_{\mu\nu},\PP_\kappa]&=i(\delta_{\mu\kappa}\PP_\nu-\delta_{\nu\kappa}\PP_\mu)\;,\cr
[M_{\mu\nu},\KK_\kappa]&=i(\delta_{\mu\kappa}\KK_\nu-\delta_{\nu\kappa}\KK_\mu)\;,\cr
[\KK_\mu,\PP_\nu]&=-2i M_{\mu\nu} + 2\delta_{\mu\nu} H\;,\cr
[H,\PP_\mu]&=\PP_\mu\;,\cr
[H,\KK_\mu]&=-\KK_\mu\;,
\end{align}
with all other commutators vanishing.

We note here that the subsequent analysis is performed in the
orthogonal basis of quantum numbers, since the Lorentz
raising/lowering operators, as well as $\PP_\mu$ and $\KK_\mu$ have very natural
representations in orthogonal root space.

\subsection{The 5D Superconformal Algebra}

Extending the 5D conformal algebra to include supersymmetry is
achieved by adding to our existing set (now with $D=5$) the generators
of supertranslations $\QQ_{{\bf A}a}$ and superconformal translations
$\QS_{{\bf A } a}$. These are equipped with a Lorentz spinor index
$a=1,\cdots,4$ and an $\mathfrak{su}(2)_R$ index
${\bf A}=1,2$. Their associated Clifford algebras are generated by
$\Gamma_\mu$ and $\tilde{\Gamma}_\mu$, respectively.\footnote{See
  App.~\ref{conventions} for our gamma-matrix conventions.}  The
collection of bosonic and fermionic generators build the $F(4)$
superconformal algebra, the bosonic part of which is
$\mathfrak{so}(5,2)\oplus \mathfrak{su}(2)_R$.

First, it will be useful to relate $M_{\mu\nu}$ to the Lorentz raising
(lowering) operators $\mathcal{M}_i^{\pm}$, associated with the
positive (negative) simple roots, and the Cartans $\mathcal H_i$ (the eigenvalues of which are $l_i$), where $i=1,2$. We do so using the relations 
\begin{align}
\mathcal{M}_1^{\pm}&=M_{13}\pm iM_{23} \mp i M_{14} + M_{24}\;,\cr
\mathcal{M}_2^{\pm} &= M_{35} \pm iM_{45}\;,\cr
\mathcal H_i &= M_{2i-1 \; 2i}\;,\hspace{5mm}\text{such that $[\mathcal H_i,\mathcal H_j]=0$}\;.
\end{align}
The algebra is now extended to include the following commutation
relations---see e.g. \cite{Minwalla:1997ka}:
\begin{align}\label{SCFT Coms 5d}
&[M_{\mu\nu},\QQ_{{\bf A} a}]=\frac{i}{4}[\Gamma_\mu
,\Gamma_\nu]_a^{\phantom{a}b} \QQ_{{\bf
    A}b}\;, &&[M_{\mu\nu},\QS_{{\bf
    A}a}]=\frac{i}{4}[\tilde{\Gamma}_\mu
,\tilde{\Gamma}_\nu]_{a}^{\phantom{a}{b}} \QS_{{\bf A}{b}}\;,\cr
&[H,\QQ_{{\bf A}a}]=\frac{1}{2}\QQ_{{\bf
    A}a}\;,&&[H,\QS_{{\bf A}{a}}]=-\frac{1}{2}\QS_{{\bf A}{a}}\;,\cr
&[\PP_\mu , \QQ_{{\bf A}a}]=0\;,&&[\PP_\mu , \QS_{\bA
{a}}]=i(\tilde{\Gamma}_\mu\tilde \Gamma_5)_{{a}}^{\hphantom{a}b} \QQ_{{\bf A}b}\;,\cr
&[\KK_\mu , \QQ_{{\bf A}a}]=i (\Gamma_\mu \Gamma_5)_a^{\hphantom{a}{b}}\QS_{{\bf A}{b}}\;,&&[\KK_\mu , \QS_{{\bf A}{a}}]=0\;.
\end{align}
The generators of the $\mathfrak{su}(2)_R$ algebra are denoted $T_m$ for $m=1,2,3$. They act on the supercharges according to 
\begin{align}\label{r sym 5d data}
[T_m, \QQ_{{\bf A}a}]&=\left(\frac{\sigma_m}{2}\right)_{\bf
  A}^{\hphantom{\bf A}{\bf B}}\QQ_{{\bf B}a}\;,\cr
[T_m, \QS_{{\bf A} a}]&=\left(\frac{\sigma_m}{2}\right)_{\bf
  A}^{\hphantom{\bf A}{\bf B}}\QS_{{\bf B} a}\;,
\end{align}
where $\sigma_m$ are the Pauli matrices. In fact we may compactly write these $R$-symmetry generators as 
\begin{align}
[R_{\bf A}^{\phantom{\bf A}\bf B}]=\left[(T^m \sigma_m)_{\bf
  A}^{\phantom{\bf A}\bf B}\right]=\begin{pmatrix}\hat{\mathcal R}&\mathcal R^+\\ \mathcal
R^-&-\hat{\mathcal R}\end{pmatrix}\;,
\end{align}
with the algebra 
\begin{align}
[R_{\bf A}^{\phantom{\bf A}\bf B},R_{\bf C}^{\phantom{\bf C}\bf D}]=\delta_{\bf C}^{\bf B} R_{\bf A }^{\phantom{\bf A}\bf D}-\delta_{\bf A}^{\bf D} R_{\bf C}^{\phantom{\bf C}\bf B}\;.
\end{align}
From the above, we may infer that
\begin{align}
[\mathcal R^+,\QQ_{\bOn a}]=0\;,\hspace{3mm}[\mathcal R^-,\QQ_{\bOn
  a}]& =\QQ_{\bTw
  a}\;,\hspace{3mm}[\hat{\mathcal R},\QQ_{\bOn
       a}]=\frac{1}{2}\QQ_{\bOn a}\;,\cr
[\mathcal R^+,\QS_{\bOn  a}]=0\;,\hspace{3mm}[\mathcal R^-,\QS_{\bOn
  a}]& =\QS_{\bTw
 a}\;,\hspace{3mm}[\hat{\mathcal R},\QS_{\bOn
            a}]=\frac{1}{2}\QS_{\bOn a}\;.
\end{align}
We denote the eigenvalue of $\hat{\mathcal R}$  in the orthogonal basis
to be $k$. The odd elements of the 5D
superconformal algebra satisfy
\begin{align}
  \left\{ \QQ_{{\bf A}a},\QQ_{{\bf B}b}\right\}&=(\Gamma^\mu \PP_\mu
                                                 K)_{a b} \epsilon_{\bf AB}\;,\cr
                                                 \left\{ \QS_{{\bf A} a},\QS_{{\bf B} b}\right\}&=(\tilde{\Gamma}^\mu
                                                                                                          \KK_\mu K)_{ a  b} \epsilon_{\bf AB}\;,\cr
                                                                                                          \left\{ \QQ_{{\bf A}a},\QS_{{\bf C} c}\right\}&=\left[\delta_{\bf
    A}^{\bf B}M_{a}^{\hphantom{a}b} +\delta_{\bf
    A}^{\bf B}\delta_{a}^{b}H-3R_{\bf A}^{\phantom{\bf A}\bf B}
  \delta_{a}^{b} \right](i \epsilon_{\bf BC}\Gamma^5 K)_{bc}\;,
\end{align}
where $K$ is the charge-conjugation matrix and $\epsilon_{\bA\bB}$ is the antisymmetric $2\times 2$ matrix such that $\epsilon_{\bOn\bTw}=1$. It is straightforward to check that these matrices are given by 
\begin{align}\label{Q anti coms}
\big[(\Gamma^\mu \PP_\mu K)_{a b}\big]&=\begin{pmatrix}0&\PP_1&i \PP_{2}&-\PP_3\\
-\PP_{1}&0&-\PP_3&i\PP_{4}\\
-i\PP_{2}& \PP_3&0&\PP_{5}\\
\PP_3&-i\PP_{4}&-\PP_{5}&0\end{pmatrix}\;,\cr
\big[(\tilde{\Gamma}^\mu \KK_\mu K)_{ a b}\big]&=
\begin{pmatrix}
 0 & \KK_{1} & i \KK_{2} & \KK_3 \\
 -\KK_{1} & 0 & \KK_3 & i \KK_{4} \\
 -i\KK_{2} & -\KK_3 & 0 &\KK_{5}\\
-\KK_3 & -i \KK_{4} &-\KK_{5}& 0 \\
\end{pmatrix}\;.
\end{align}
We have also used the matrix $M_a^{\hphantom{a}b}$, which is defined by
\begin{align}
&[M_{a}^{\hphantom{a}b}]=\frac{i}{4}[\Gamma^\mu,\Gamma^\nu]_{a}^{\hphantom{a}b}
  M_{\mu\nu}\cr
&=\begin{pmatrix}
\mathcal H_1+\mathcal H_2& \mathcal{M}_2^+&-\frac{1}{2}[\mathcal{M}_1^+,\mathcal{M}_2^+]&-\frac{1}{2}[[\mathcal{M}_1^+,\mathcal{M}_2^+],\mathcal{M}_2^+]\\
\mathcal{M}_2^-&\mathcal H_1 - \mathcal H_2 &\mathcal{M}_1^+ & \frac{1}{2}[\mathcal{M}_1^+,\mathcal{M}_2^+] \\
\frac{1}{2}[\mathcal{M}_1^-,\mathcal{M}_2^-] & \mathcal{M}_1^-&-\mathcal H_1+\mathcal H_2 &\mathcal{M}_2^+\\
-\frac{1}{2}[[\mathcal{M}_1^-,\mathcal{M}_2^-],\mathcal{M}_2^-]&-\frac{1}{2}[\mathcal{M}_1^-,\mathcal{M}_2^-]& \mathcal{M}_2^-&-\mathcal H_1 -\mathcal H_2 \end{pmatrix}\;.\cr
\end{align}
The supercharges have the following conjugation relations \cite{Minwalla:1997ka}
\begin{align}
\QQ_{{\bf A} a}&=i \epsilon_{\bf
                 AB}(K\Gamma_5^T)_{ab}{\QS^\dagger}^{{\bf B} b}
                 \;,\hspace{5mm}\QS_{{\bf A} a}=-i
                 \epsilon_{\bf
                 AB}(K\tilde{\Gamma}_5^T)_{a b}{\QQ^\dagger}^{{\bf B}b}\;,
\end{align}
which allows us to rewrite the $\{ \QQ,\QS\}$ anticommutator as
\begin{align}
\{ \QQ_{{\bf A} a},{\QQ^{\dagger}}^{ {\bf
  B} b}\}&=\delta_{\bf A}^{\bf B}M_{a}^{\hphantom{a}b} +\delta_{\bf
         A}^{\bf B}\delta_{a}^{\hphantom{a}b}H-3R_{\bf
           A}^{\phantom{\bf A} \bf B} \delta_{a}^{\hphantom{a}b}\;.
\end{align}

\subsection{Gamma-Matrix Conventions in 5D}\label{conventions}

Here we collect our gamma-matrix conventions. For Euclidean $\mathfrak{so}(5)$ spinors we have 
\begin{align}
  \Gamma^1 = \begin{pmatrix} 0 & \mathbb{1}_{2\times 2}\cr \mathbb{1}_{2\times 2} &0\end{pmatrix}\;,\hspace{3mm} \Gamma^2=\begin{pmatrix} 0& i\hspace{0.5mm} \mathbb{1}_{2\times 2}\\-i \hspace{0.5mm}\mathbb{1}_{2\times 2}&0\end{pmatrix}\;,\hspace{3mm}\Gamma^3 = \begin{pmatrix} \sigma_2 &0\\ 0 &-\sigma_2\end{pmatrix}\;,\cr
\Gamma^4 = \begin{pmatrix}\sigma_1 & 0\\0&-\sigma_1\end{pmatrix}\;,\hspace{3mm}\Gamma^5 =\begin{pmatrix}-\sigma_3 &0\\ 0&\sigma_3\end{pmatrix}\;.
\end{align}
These are supplemented by the charge conjugation matrix
\begin{align}
K=\begin{pmatrix}0&i \sigma_2\\ i\sigma_2&0\end{pmatrix}\;.
\end{align}
Note that $\tilde\Gamma^{1,2,3,4} =\Gamma^{1,2,3,4}$, while
$\tilde{\Gamma}^5 = -\Gamma^5$. Each set generates the Euclidean
Clifford algebra in five dimensions, that is they satisfy the
relations
\begin{align}
\{ \Gamma^\mu,\Gamma^\nu\}= 2 \delta^{\mu\nu}\;,\hspace{2mm}(\Gamma^\mu)^\dagger = \Gamma^\mu\;,\hspace{2mm} \Gamma^1 \Gamma^2 \Gamma^3 \Gamma^4 \Gamma^5 =\mathbb{1}_{4\times 4}\;,\hspace{2mm}K^T=-K\;,\hspace{2mm}(K \Gamma^\mu)^T=-K\Gamma^\mu
\end{align}
and similarly for the $\tilde{\Gamma}$ matrices.

\subsection{The $\Lambda$-basis of Generators}\label{App:Ltilde}

There exists a natural basis for constructing the UIRs of the 5D SCA,
in which particular combinations of supercharges map highest weights
of the maximal compact subalgebra,
$\mathfrak{so}(5)\oplus \mathfrak{so}(2)\oplus\mathfrak{su}(2)_R$, to other highest weights. We
define
\begin{align}\label{biglambdaApp}
\Lambda^a_\bOn := \sum_{b=1}^{a} \QQ_{\bOn b}
  \lambda^a_{b}\qquad\textrm{and}\qquad \Lambda^a_\bTw := \sum_{b=1}^{a} \QQ_{\bTw b}
  \lambda^a_{b} - \sum_{b=1}^{a} \QQ_{\bOn b}\mathcal R^-
  \lambda^a_{b} \frac{1}{2 \hat{\mathcal R}}\;.
\end{align}
The $\lambda^a_b$ are given by
\begin{align}\label{littlelambdaApp}
\lambda^{a}_{a}&= 1\;,\hspace{3mm}\text{where $a$ is not summed over}\;,\cr
\lambda^{2}_{1}&= -\mathcal M_2^-\frac{1}{2\mathcal H_2}\;,\cr
\lambda^{3}_{1}&= \left(-\mathcal M_1^-
                 \mathcal M_2^- + \mathcal M_2^- \mathcal M_1^- \frac{(\mathcal H_1 -\mathcal H_2 +1)}{(\mathcal H_1 -\mathcal H_2)}\right)\frac{1}{4(\mathcal H_1+1)}\;,\cr
\lambda^{3}_{2}&=-\mathcal M_1^- \frac{1}{2(\mathcal H_1-\mathcal H_2)} \;,\cr
\lambda^{4}_{1}&=\Big(- \mathcal M_2^-
                 \mathcal M_1^- \mathcal M_2^-\frac{(2+\mathcal H_1 + 5 \mathcal H_2
                 +4 \mathcal H_1\hspace{0,5mm}\mathcal H_2)}{(\mathcal
                 H_1+1) \mathcal H_2}+\mathcal M_1^- (\mathcal
                 M_2^-)^2 \frac{(2 \mathcal H_2+1)}{\mathcal H_2 }\cr
&\qquad \qquad \qquad \qquad \qquad \qquad \qquad \qquad
                 +(\mathcal M_2^-)^2 \mathcal M_1^- \frac{(2 \mathcal H_1+3)}{(\mathcal H_1+1)}\Big) \frac{1}{8(\mathcal H_1+\mathcal H_2+1)}\;,\cr
\lambda^{4}_{2}&=\left(-\mathcal M_2^- \mathcal M_1^-+ \mathcal M_1^- \mathcal M_2^-\frac{\left(\mathcal H_2+1\right)}{\mathcal H_2} \right) \frac{1}{4(\mathcal H_1+1)}\;,\cr
\lambda^{4}_{3}&=-\mathcal M_2^- \frac{1}{2 \mathcal H_2}\;.
\end{align}
With the help of the above it is straightforward to calculate all
level-one norms:
 \begin{align}\label{norms app}
|| \Lambda^4_\bOn \ket{\Delta;l_1,l_2;k}^{hw}  ||^2 &= \left(\Delta -3 k-l_1-l_2-4\right)\frac{\left(2 l_1+3\right) \left(l_1+l_2+2\right) \left(2 l_2+1\right) }{4 \left(l_1+1\right) l_2 \left(l_1+l_2+1\right)}\;,\cr
|| \Lambda^3_\bOn \ket{\Delta;l_1,l_2;k}^{hw}  ||^2 &= \left(\Delta -3 k-l_1+l_2-3\right)\frac{\left(2 l_1+3\right) \left(l_1-l_2+1\right)}{2 \left(l_1+1\right) \left(l_1-l_2\right)}   \;,\cr
 || \Lambda^2_\bOn \ket{\Delta;l_1,l_2;k}^{hw}  ||^2 &= (\Delta -3 k+l_1-l_2-1)\frac{(2l_2+1)}{2l_2}\;,\cr
|| \Lambda^1_\bOn \ket{\Delta;l_1,l_2;k}^{hw}  ||^2 &=(\Delta -3
                                                 k+l_1+l_2)\cr
|| \Lambda^4_\bTw \ket{\Delta;l_1,l_2;k}^{hw}  ||^2 &=(\Delta + 3 k  - l_1 - l_2 - 1 ) \frac{\left(2 l_1+3\right) \left(l_1+l_2+2\right) \left(2 l_2+1\right) }{4 \left(l_1+1\right) l_2 \left(l_1+l_2+1\right)}\Big(1+\frac{1}{2k}\Big) \;,\cr
|| \Lambda^3_\bTw \ket{\Delta;l_1,l_2;k}^{hw}  ||^2 &=(\Delta + 3 k  - l_1 + l_2  ) \frac{\left(2 l_1+3\right) \left(l_1-l_2+1\right)}{2 \left(l_1+1\right) \left(l_1-l_2\right)} \Big(1+\frac{1}{2k}\Big)  \;,\cr
 || \Lambda^2_\bTw \ket{\Delta;l_1,l_2;k}^{hw}  ||^2 &=(\Delta + 3 k  + l_1 -l_2 +2 ) \frac{(2l_2+1)}{2l_2} \Big(1+\frac{1}{2k}\Big)\;,\cr
|| \Lambda^1_\bTw \ket{\Delta;l_1,l_2;k}^{hw}  ||^2 &=(\Delta + 3 k  + l_1 +l_2 +3)  \Big(1+\frac{1}{2k}\Big)
\;,
 \end{align}
 where we have normalised
 $\left\| \ket{\Delta;l_1,l_2;k}^{hw}\right\|^2=1$.

\subsection{The 6D Superconformal Algebras}

The superconformal algebra in 6D is
$\mathfrak{osp}(8^*|2\mathcal{N})$, the bosonic part of which is
$\mathfrak{so}(6,2)\oplus \mathfrak{sp}(\mathcal{N})_R$.\footnote{Note
  that we will carry out this discussion using the $\so(6)$ algebra in
  the orthogonal basis, instead of $\su(4)$ algebra in the Dynkin
  basis. The dictionary between the two will be provided at the end.}
The set of conformal generators is extended to include the generators
of supersymmetry $\QQ_{\bA a}$ and superconformal translations
$\QS_{\bA \dot{a}}$. The Lorentz spinor index ranges
from $a,\dot{a}=1,\cdots,4$ (the dotted spinor index refers to the
fact that it is an antifundamental index), while the
$\mathfrak{sp}(\mathcal{N})_R$ index ranges from
$\bA = 1,\cdots, 2\mathcal{N}$. Their associated Clifford algebras are
generated by $\Gamma_\mu$ and $\tilde{\Gamma}_\mu$, respectively, the
conventions of which are detailed in App.~\ref{conventions 6D}.

Since their algebras are very similar, we first list the features that
are shared by both $\mathcal{N}=1$ and $\mathcal N = 2$ cases.  Due to the
fact that spinors in 6D are pseudo-real, they satisfy the reality
conditions
\begin{align}
  \QQ_{{\bf A} a}&=i \Omega_{\bf
                 AB}(K\Gamma_6^T)_{a\dot{b}}{\QS^\dagger}^{{\bf B}\dot{b}
                 }\;,\hspace{5mm}\QS_{{\bf A}\dot{a}}=-i
                 \Omega_{\bf
                 AB}(K\tilde{\Gamma}_6^T)_{\dot{a} b}{\QQ^\dagger}^{{\bf B}b}\;,
\end{align}
where $\Omega_{\bA \bB}$ is the appropriate antisymmetric matrix in
$2\mathcal{N}$ dimensions.\footnote{For $\mathcal N = 1$ one has
  $\Omega_{\bA \bB} = \epsilon_{\bA \bB}$, the 2D antisymmetric
  matrix. For $\mathcal{N}=2$, $\Omega_{\bA \bB}$ is the 4D
  symplectic matrix with
  $\Omega_{\bOn \bFo} = -\Omega_{ \bFo \bOn}=\Omega_{\bTw
    \bTh}=-\Omega_{\bTh\bTw }=1$
  and all other components vanishing.} The commutation relations of the
$\QQ$ and $\QS$ with the 6D conformal algebra are given by
\begin{align}\label{SCFT Coms 6d}
&[M_{\mu\nu},\QQ_{{\bf A} a}]=\frac{i}{4}[\Gamma_\mu
,\Gamma_\nu]_a^{\phantom{a}b} \QQ_{{\bf
    A}b}\;, &&[M_{\mu\nu},\QS_{{\bf
    A}\dot{a}}]=\frac{i}{4}[\tilde{\Gamma}_\mu
,\tilde{\Gamma}_\nu]_{\dot{a}}^{\phantom{a}\dot{b}} \QS_{{\bf A}\dot{b}}\;,\cr
&[H,\QQ_{{\bf A}a}]=\frac{1}{2}\QQ_{{\bf
    A}a}\;,&&[H,\QS_{{\bf A}\dot{a}}]=-\frac{1}{2}\QS_{{\bf A}\dot{a}}\;,\cr
&[\PP_\mu , \QQ_{{\bf A}a}]=0\;,&&[\PP_\mu , \QS_{\bA
\dot{a}}]=i(\tilde{\Gamma}_\mu\tilde \Gamma_6)_{\dot{a}}^{\hphantom{a}b} \QQ_{{\bf A}b}\;,\cr
&[\KK_\mu , \QQ_{{\bf A}a}]=i (\Gamma_\mu \Gamma_6)_a^{\hphantom{a}\dot{b}}\QS_{{\bf A}\dot{b}}\;,&&[\KK_\mu , \QS_{{\bf A}\dot{a}}]=0\;.
\end{align}

It is helpful to define the Lorentz raising/lowering operators
$\mathcal{M}_i^{\pm}$ and Cartans $\mathcal H_i$---the eigenvalues of
which are $h_i$ in the orthogonal basis of $\so(6)$---in terms of
$M_{\mu\nu}$. The relations are provided below:
\begin{align}\begin{split}
    \mathcal{M}_1^{\pm}&=M_{13}\pm iM_{23} \mp i M_{14} + M_{24}\;,\\
    \mathcal{M}_2^{\pm}&=M_{35}\pm iM_{45} \mp i M_{36} + M_{46}\;,\\
    \mathcal{M}_3^{\pm}&=M_{35}\pm iM_{45} \pm i M_{36} - M_{46}\;,\\
    \mathcal H_i &= M_{2i-1 \; 2i}\;,\hspace{5mm}\text{such that
      $[\mathcal H_i,\mathcal H_j]=0$}\;.
\end{split}
\end{align}
For a superconformal algebra, all supercharges must have the same
chirality \cite{Minwalla:1997ka}, which we choose to be
positive. Therefore we define the projector
$P_+ =\frac{1}{2} (1+\Gamma_7)$ and have that
\begin{align}
\left\{ \QQ_{{\bf A}a},\QQ_{{\bf B}b}\right\}&=(P_+\Gamma^\mu \PP_\mu
K)_{a b} \Omega_{\bf AB}\;,\cr
\left\{ \QS_{{\bf A}\dot a},\QS_{{\bf B}\dot b}\right\}&=(P_+\tilde{\Gamma}^\mu
\KK_\mu K)_{\dot a \dot b}\Omega_{\bf AB}\;.
\end{align}
We may also use the projector $P_+$ to define $M_a^{\hphantom{a}b}$ as
\begin{align}
&[M_{a}^{\hphantom{a}b}]=-\frac{i}{4}(P_+)_a^{\hphantom{a}c}[\Gamma^\mu,\Gamma^\nu]_{c}^{\hphantom{c}b}
  M_{\mu\nu}\cr
&=\begin{pmatrix}
\mathcal H_1+\mathcal H_2+\mathcal H_3&i \mathcal{M}_3^-&-\frac{1}{2}[\mathcal{M}_1^-,\mathcal{M}_3^-]&\frac{1}{4}[\mathcal{M}_1^-,[\mathcal{M}_2^-,\mathcal{M}_3^-]]\\
-i\mathcal{M}_3^+&\mathcal H_1-\mathcal H_2-\mathcal H_3&i \mathcal{M}_1^- & \frac{1}{2}[\mathcal{M}_1^-,\mathcal{M}_2^-] \\
\frac{1}{2}[\mathcal{M}_1^+,\mathcal{M}_3^+] & -i \mathcal{M}_1^+& -\mathcal H_1+\mathcal H_2-\mathcal H_3&i\mathcal{M}_2^-\\
\frac{1}{4}[\mathcal{M}_1^+,[\mathcal{M}_2^+,\mathcal{M}_3^+]]&-\frac{1}{2}[\mathcal{M}_1^+,\mathcal{M}_2^+]& -i\mathcal{M}_2^+&-\mathcal H_1-\mathcal H_2+\mathcal H_3 \end{pmatrix}\;.\nonumber\\
\end{align}

\subsubsection{$\mathcal{N}=1$}

For this case, the $R$-symmetry algebra is $\mathfrak{sp}(1)_R \simeq
\su(2)_R$.
Conveniently, all the information we require about $\su(2)_R$ has
already been provided in the 5D $\mathcal{N}=1$ discussion,
specifically \eqref{r sym 5d data} onwards. We may therefore use the
previously-defined $R_{\bA \bB}$ and write the $\{\QQ, \QQ^\dagger\}$
anticommutator as
\begin{align}
\{ \QQ_{{\bf A} a},{\QQ^{\dagger}}^{ {\bf
  B}b}\}&=\delta_{\bf A}^{\bf B}M_{a}^{\hphantom{a}b} +\delta_{\bf
         A}^{\bf B}(P_+)_{a}^{\hphantom{a}b}H-4R_{\bf
          A}^{\hphantom{\bf A}\bf B} (P_+)_{a}^{\hphantom{a}b}\;.
\end{align}
The expression \eqref{(1,0) index} we used for the 6D $(1,0)$ index is then, in
the orthogonal basis,
\begin{align}
\delta :=\{ \QQ_{\bOn 4},\QS_{\bTw \dot 4}\}=\{ \QQ_{\bOn 4},{\QQ^{\dagger}}^{\bOn 4}\}=\Delta- 4k -(h_1+h_2-h_3)\;.
\end{align}

\subsubsection{$\mathcal{N}=2$}

For this case, the $R$-symmetry algebra is $\mathfrak{sp}(2)_R \simeq \so(5)_R$. The
matrix $R_{\bA}^{\hphantom{\bA}{\bB}}$ for $\mathfrak{so}(5)_R$ is
\begin{align}
[R_{\bA}^{\hphantom{\bA}{\bB}}]=\begin{pmatrix}
\mathcal J_1+\mathcal J_2& \mathcal R_2^+&-\frac{1}{2}[\mathcal R_1^+,\mathcal R_2^+]&-\frac{1}{2}[[\mathcal R_1^+,\mathcal R_2^+],\mathcal R_2^+]\\
\mathcal R_2^-&\mathcal J_1 - \mathcal J_2 &\mathcal R_1^+ & \frac{1}{2}[\mathcal R_1^+,\mathcal R_2^+] \\
\frac{1}{2}[\mathcal R_1^-,\mathcal R_2^-] & \mathcal R_1^-&-\mathcal J_1+\mathcal J_2 &\mathcal R_2^+\\
-\frac{1}{2}[[\mathcal R_1^-,\mathcal R_2^-],\mathcal R_2^-]&-\frac{1}{2}[\mathcal R_1^-,\mathcal R_2^-]& \mathcal R_2^-&-\mathcal J_1 -\mathcal J_2 \end{pmatrix}\;,
\end{align}
where the $\mathcal J_i$ are the Cartans of $\so(5)_R$---the eigenvalues of
which are $j_i$---and $\mathcal R_i^{\pm}$ are the raising/lowering operators. This allows the $\{\QQ, \QQ^\dagger\}$ anticommutator to be written as \cite{Minwalla:1997ka}
\begin{align}
\{ \QQ_{{\bf A} a},{\QQ^{\dagger}}^{{\bf
  B}b }\}&=\delta_{\bf A}^{{\bB}} M_{a}^{\hphantom{a}b} +\delta_{\bf
         A}^{{\bB}}(P_+)_{a}^{\hphantom{a}b}H-2 R_{\bf A}^{\hphantom{\bA}{\bB}} (P_+)_{a}^{\hphantom{a}b}\;.
\end{align}
Hence the $\delta$ we used for the 6D (2,0) index is in the orthogonal basis
\begin{align}
  \delta :=\{\QQ_{\bTw 4}, \QS_{\bTh \dot 4},\}=\{
  \QQ_{\bTw 4},{\QQ^{\dagger}}^{\bTw 4}\}= \Delta - 2j_1+2j_2 -(h_1+h_2-h_3)\;.
\end{align}

To make contact with the main part of this paper, we convert from the
orthogonal bases to the Dynkin bases using the following expressions
for the $\so(6)$---to $\su(4)$ Dynkin---, $\so(5)$ and $\su(2)$
orthogonal Cartans respectively:
\begin{align}\begin{split}
&h_1 = \frac{1}{2}(c_1+2c_2+c_3)\;,\hspace{3mm}h_2=\frac{1}{2}(c_1+c_3)\;,\hspace{3mm}h_3=\frac{1}{2}(c_1-c_3)\;,\\
&j_1 = d_1 + \frac{1}{2}d_2\;,\hspace{3mm}j_2 = \frac{1}{2}d_2\;,\\
&k = \frac{1}{2}K\;.\end{split}
\end{align}

\subsection{Gamma-Matrix Conventions in 6D}\label{conventions 6D}

Here we collect our gamma-matrix conventions. For Euclidean $\mathfrak{so}(6)$ spinors we have 
\begin{align}\begin{split}
\Gamma^1 =\sigma_1 \otimes \mathbb{1}_{2\times 2}\otimes
\mathbb{1}_{2\times 2}\;,\hspace{3mm} \Gamma^2=\sigma_2 \otimes
\mathbb{1}_{2\times 2}\otimes \mathbb{1}_{2\times
  2}\;,\hspace{3mm}\Gamma^3 =\sigma_3 \otimes \sigma_1 \otimes \mathbb{1}_{2\times 2}\;,\\
\Gamma^4 = \sigma_3 \otimes\sigma_2\otimes \mathbb{1}_{2\times 2}\;,\hspace{3mm}\Gamma^5 = \sigma_3 \otimes\sigma_3\otimes\sigma_1\;,\hspace{3mm}\Gamma^6 = \sigma_3 \otimes\sigma_3\otimes\sigma_2\;.
\end{split}
\end{align}
We also have $\Gamma^7=i\Gamma^1\Gamma^2\Gamma^3\Gamma^4\Gamma^5\Gamma^6$, which can be equivalently defined as  
\begin{align}
\Gamma^7 = \sigma_3 \otimes\sigma_3\otimes\sigma_3\;.
\end{align}
Since we are in even dimensions, we have a choice between which charge conjugation matrices to use, $K_{(\pm)}$. These have the properties $K_{(\pm)}^T = \mp K_{(\pm)}$ and $K_{(\pm)}^2 = \mp 1$. We choose $K_{(-)}$, but to lighten the notation, we will omit the subscript. We define our $K_{(-)}=K$ as 
\begin{align}
K=(-i\sigma_2) \otimes \sigma_1 \otimes (-i\sigma_2)\;,
\end{align}
Note that $\Gamma^{1,2,3,4,5} =\tilde{\Gamma}^{1,2,3,4,5}$, while
$\tilde{\Gamma}^6 = -\Gamma^6$. They each generate the Euclidean
Clifford algebra in five dimensions. The $\Gamma$s satisfy the relations
\begin{align}
\{ \Gamma^\mu,\Gamma^\nu\}= 2 \delta^{\mu\nu}\;,\hspace{2mm}(\Gamma^\mu)^\dagger = \Gamma^\mu\;,\hspace{2mm}K (\Gamma^\mu)^T K^{-1}=-\Gamma^\mu\;.
\end{align}
and similarly for the $\tilde{\Gamma}$ matrices.

\section{Multiplet Supercharacters}\label{superchar}

Consider a representation of a Lie algebra $\mathfrak g$ with highest weight $\lambda$. The Weyl character formula is given by \cite{fuchs2003symmetries}
\begin{align}\label{Weyl Kac Character}
\chi_{\lambda}=\frac{\sum_{w \in
  W}\text{sgn}(w)e^{w(\lambda + \rho)}}{e^{ \rho} \prod_{\alpha \in \Phi_{-}}(1-e^{\alpha})},
\end{align}
where $W$ is the Weyl group of the Lie algebra root system and $\rho$
is the half sum of the positive roots $\Phi_{+}$.  Note that
$\text{sgn}(w)=(-1)^{l(w)}$ where $l(w)$ is the length of the Weyl
group element, i.e. how many simple reflections it is comprised of.

One can alternatively obtain this formula using a Verma-module
construction: Decompose the algebra $\frak{g}$ as
\begin{align}\label{vermadecomp}
\mathfrak{g}=\Phi_+ \oplus \mathfrak{h}\oplus\Phi_-\;,
\end{align}
where $\mathfrak{h}$ corresponds to the Cartan subalgebra and $\Phi_-$
($\Phi_+$ ) are the negative (positive) roots. We construct the Verma
module $\mathcal{V}$ corresponding to some highest (lowest) weight
$\ket{\lambda}$ by considering the space comprised of the states
$f(\Phi_-)\ket{\lambda}$ ($f(\Phi_+)\ket{\lambda}$), where $f$ is any
polynomial of the negative (positive) roots modulo algebraic
relations. The character of this module is defined to be
\cite{fuchs2003symmetries}
\begin{align}\label{Verma module character}
\chi_{\mathcal{V}} = \frac{e^{\lambda}}{\prod_{\alpha \in \Phi_{-}}(1-e^{\alpha})}\;.
\end{align}
The character of the representation labelled by $\Lambda$ is
recovered by summing over the Weyl group action on the roots
\begin{align}\label{Weyl character from Verma}
\chi_{\lambda}=\sum_{w\in W}w(\chi_{\mathcal{V}})=\sum_{w\in W}\frac{e^{w(\lambda)}}{\prod_{\alpha \in \Phi_{-}}(1-e^{w(\alpha)})}.
\end{align}
We can utilise the identity
$w(e^{-\rho-\lambda}\chi_\mathcal{V})=\text{sgn}(w)e^{-\rho-\lambda}\chi_\mathcal{V}$
along with the fact that $w$ acts naturally---i.e. we may take
$w(e^{-\rho-\lambda}\chi_\mathcal{V})=w(e^{-\rho-\lambda})w(\chi_\mathcal{V})$
\cite{fuchs2003symmetries}---to show that
\begin{align}
 \chi_{\lambda}=\sum_{w\in W}w(\chi_{\mathcal{V}})=\frac{\sum_{w\in W}(-1)^{l(w)}e^{w(\lambda + \rho)}}{e^{\rho+\lambda}}\chi_\mathcal{V}=\frac{\sum_{w\in W}(-1)^{l(w)}e^{w(\lambda + \rho)}}{e^{\rho}\prod_{\alpha \in \Phi_{-}}(1-e^{\alpha})}.
\end{align}

The formulation of the Weyl character formula \eqref{Weyl character
  from Verma} is particularly useful in the context of UIRs of the
SCA \cite{Beem:2014kka}.

\subsection{Characters of 5D $\mathcal{N}=1$ Multiplets}

We are now in a position to compute the characters of
$F(4)$.\footnote{A summary of this discussion for the case of the
  (2,0) SCA can be found in App.~C of \cite{Beem:2014kka}.} Let us
consider a representation the highest weight of which has conformal
dimension $\Delta$ with $\mathfrak{so}(5)$ Lorentz quantum numbers
$(d_1, d_2)$ and $\mathfrak{su}(2)_R$ quantum numbers $K$. Note that
these are expressed in the Dynkin basis and hence are integer. The
highest weight can be decomposed as 
\begin{align} 
\lambda =
  \omega^\beta_1 d_1 + \omega^\beta_2 d_2+\omega^\alpha K\;, 
\end{align}
where $\omega^\beta_i$ ($i=1,2$) are the fundamental weights
associated with the $\mathfrak{so}(5)$ simple roots $\beta_i$, while
$\omega^\alpha$ is the fundamental weight associated with the
$\mathfrak{su}(2)_R$ simple root $\alpha$. We may in turn  express the fundamental
weights in terms of the simple roots (for reasons which will become
apparent) by using the Cartan matrix $A_{ij}$
\begin{align}\label{coroots 5d}\begin{split} \omega^\beta_i &=
    (A^{-1}_{B_2})_{ij}\beta_j=\left( \begin{array}{cc}
 1 & 1 \\
 \frac{1}{2} & 1 \\
\end{array}
\right)\begin{pmatrix}\beta_1 \\ \beta_2 \end{pmatrix}\;,\\
\omega^\alpha &= \frac{1}{2}\alpha\;.
\end{split}
\end{align}
Let us next consider $e^{\lambda}$ by defining the fugacities
\begin{align}\label{5d N=1 fugacities}\begin{split}
b_1 &= e^{\beta_1 + \beta_2}\;,\hspace{2mm}b_2 = e^{\frac{1}{2}(\beta_1 + 2\beta_2)}\;,\hspace{2mm}a=e^{\frac{1}{2}\alpha}\;, 
\end{split}
\end{align}
hence
\begin{align}
e^{\lambda} = b_1^{d_1}b_2^{d_2}a^{K}\;.
\end{align}
The character of a particular representation $\mathcal{R}$ is then defined to be 
\begin{align}
\chi_\mathcal{R} (a,\mathbf{b},q)=\Tr_\mathcal{R}(q^\Delta b_1^{d_1}b_2^{d_2}a^{K})\;,
\end{align}
where we have included the $\mathfrak{so}(2)$ Cartan, $\Delta$. For example we can read off the character for the supercharges
$\QQ_{{\bf A}a}$ as (recall that we are in the Dynkin basis)
\begin{align}
\sum_{ a = 1}^{4}\chi ( \QQ_{{\bOn}a}) =  a b_2 q^{\frac{1}{2}}+\frac{a b_1 q^{\frac{1}{2}}}{b_2}+\frac{a b_2 q^{\frac{1}{2}}}{b_1}+\frac{a q^{\frac{1}{2}}}{b_2}\;.
\end{align}
We can then apply this to specific representations of the SCA.

\subsubsection{Long Representations}\label{long reps 5d}

For the long representation $\mathcal{L}_{[\Delta; d_1,d_2;K]}$ we
construct the superconformal multiplet by acting on the highest weight
state $\ket{\Delta; d_1,d_2;K}^{hw}$ with momentum operators and
supercharges as in Eq.~\eqref{supconfbasis1}.  Thus for a generic long
representation we can decompose the character of the superconformal
Verma module using \eqref{Verma module character} as
\begin{align}\label{long}
\chi_\mathcal{L}(a,{ \bf b},q)=q^\Delta \chi_{[d_1,d_2]}({\bf b})
  \chi_{[K]}(a)f(a,{\bf b},q)\;.
\end{align}
The polynomial appearing above can be decomposed as
$f(a,\mathbf{b},q)=Q(a,\mathbf{b},q)P(\mathbf{b},q)$ since the
momentum operators and supercharges commute. Explicitly these
functions are
\begin{align}\label{q and p product 5D}
Q(a,\mathbf{b},q)&=\prod_{{\bf A},a}(1+\chi(\QQ_{{\bf A} a}))\;,\cr
P(\mathbf{b},q)&=\prod_{\mu=1}^{5}(1-\chi(\mathcal{P}_\mu))^{-1}\;.
\end{align}

The characters $ \chi_{[d_1, d_2]}({\bf b})$ and $ \chi_{[K]}({a})$
can be obtained through their Weyl orbits.  As a result one has
\begin{align}\label{5Dlong}
\chi_{[d_1,d_2]}({\bf b})&=\sum_{w\in W_{\SO(5)}}w(b_1)^{d_1}w(b_2)^{d_2}M(w({\bf b}))\;,\hspace{3mm}\chi_{[K]}(a)=\sum_{w\in W_{\SU(2)}} w(a)^K R(w(a))\;,
\end{align}
where the $M({\bf b})$ and $R(a)$ are the products of the characters
of negative roots, explicitly
\begin{align}
M({\bf b}) &=\frac{1}{\left(1-\frac{1}{b_1}\right) \left(1-\frac{1}{b_2^2}\right) \left(1-\frac{b_1}{b_2^2}\right) \left(1-\frac{b_2^2}{b_1^2}\right)}\;,\cr
R(a)&=\frac{a^2}{a^2-1}\;.
\end{align}

As an aside it will be worthwhile to explicitly demonstrate the Weyl
group actions appearing in \eqref{5Dlong}.  In the orthogonal basis
the generators of $W_{\SO(5)}= \mathcal{S}_2 \ltimes (\mathbb{Z}_2)^2$
and $W_{\SU(2)}=\mathcal{S}_2$ have the form
\begin{align}
w^B_1 = \begin{pmatrix}0&1\\1&0\end{pmatrix}\;,\hspace{2mm}w^B_2 =  \begin{pmatrix}1&0\\0&-1\end{pmatrix}\;,\hspace{2mm}w^A=\begin{pmatrix}0&1\\1&0\end{pmatrix}\;.
\end{align} 
There are eight elements in $\mathcal{S}_2 \ltimes (\mathbb{Z}_2)^2$
and two in $\mathcal{S}_2$. These act on the simple roots as
\begin{align}
w_i^B \beta_j = f(\beta_1,\beta_2)\;,\hspace{3mm}w^A \alpha = g(\alpha)\;,
\end{align}
where the RHS is a combination of simple roots depending on the
particular example. In fact, the simple reflections on the simple
roots will generate every other root minus the Cartans of the
algebra. For example, the simple roots of $\mathfrak{so}(5)$ in the
orthogonal basis are $\beta_1 = (1,-1)$ and $\beta_2 = (0,1)$. Acting
on them with $w^B_2$ produces
\begin{align}
w^B_2 \beta_1 &= (1,1)= \beta_1 + 2\beta_2\;,\cr
w^B_2 \beta_2 &= (0,-1) = -\beta_2\;.
\end{align}
Similarly, acting with $w_1^B$ will produce $-\beta_1$ and
$\beta_1+\beta_2$ respectively. Furthermore, since the Weyl group has
a natural action on $e^{\lambda}$
(i.e. $w(e^{\lambda})=e^{w(\lambda)}$), this action can be directly
translated to the fugacities.  Following the same example
\begin{align}
w^B_2(b_1) &= e^{w^B_2(\beta_1 + \beta_2)}=e^{(\beta_1 + \beta_2)}=b_1\;,\cr
w^B_2(b_2)&=e^{\frac{1}{2}w^B_2(\beta_1 + 2\beta_2)}=e^{\frac{1}{2}\beta_1}=\frac{b_1}{b_2}\;.
\end{align}
Combining the Weyl groups leads to $W_{\SO(5) \times \SU(2)}$, which
has sixteen elements acting on $a$ and $b_i$.  The $Q(a,\mathbf{b},q)$
and $P({\bf b},q)$ are both invariant under the action of any element
of this combined Weyl group. Using this fact, the character for long
representations can be rewritten in terms of
\begin{align}\label{longrepchar}
\chi_\mathcal{L} (a, \mathbf{b},q) =\big{\llbracket} q^\Delta b_1^{d_1}b_2^{d_2}a^K M({\bf b}) R(a) P({\bf b}, q) Q(a,\mathbf{b},q)\big{\rrbracket}_W,
\end{align}
where $[\! [ \cdots ]\!]_W$ is shorthand for the Weyl symmetriser. 

\subsubsection{Short Representations}\label{short reps 5d char}

Consider now the short multiplets of Table \ref{Tab:5DN1}. In order to
calculate their characters, one is instructed
\cite{Bhattacharya:2008zy,Dolan:2005wy,Bianchi:2006ti} to remove
certain combinations of $\QQ$s and $\PP$s from the expressions
$Q(a,\mathbf{b},q)$ and $P({\bf b},q)$ given in \eqref{q and p product
  5D}.\footnote{There is a subtlety with removing momentum operators,
  which will be addressed in the following section.}  We explicitly
consider a few examples to elucidate this point:

\begin{enumerate}

\item[a.]  Take the most basic short multiplet,
  $\mathcal{A}[d_1,d_2;K]$, with $d_1, d_2, K>0$. Its superconformal
  primary is annihilated by the supercharge $\QQ_{\bOn 4}$ and the
  associated character would be
\begin{align}
\chi_\mathcal{A} (a, \mathbf{b},q) =\left[\! \left[ q^\Delta b_1^{d_1}b_2^{d_2} a^K M({\bf b}) R(a) P({\bf b}, q) Q(a,\mathbf{b},q)\left( 1 + \frac{a q^{\frac{1}{2}}}{b_2}\right)^{-1}\right]\! \right]_W\;,
\end{align}
with $\Delta = 4 + 3K + d_1+d_2$. Notice that this includes the
character for the product over $\QQ_{\bA a}$ but now with the
$\QQ_{\bOn 4}$ contribution removed; hence the Weyl symmetrisation
removes descendant states associated with the action of
$\QQ_{\bOn 4}$.

\item[b.] Take the $\mathcal{B}[d_1,0;K]$ multiplet with $K\neq 0$.
  One is instructed to remove $\QQ_{\bOn 3}$ and $\QQ_{\bOn 4}$ and
  the supercharacter is
\begin{align}
\chi_\mathcal{B} (a,\mathbf{b},q) = &\left[\! \left[ q^\Delta
                                    b_1^{d_1}a^K
                                    M({\bf b}) R(a) P({\bf b},
                                    q) Q(a,\mathbf{b},q) \left( 1 + \frac{a b_2 q^{\frac{1}{2}}}{b_1}\right)^{-1}\left( 1 + \frac{aq^{\frac{1}{2}}}{b_2}\right)^{-1}\right]\! \right]_W\;,
\end{align}
with $\Delta = 3+3K+d_1$. 

\item[c.]  Suppose now we consider the multiplet
  $\mathcal{B}[d_1,0;0]$. In this case the $R$-symmetry lowering
  operator in the Dynkin basis $\mathcal R^-$ also annihilates the
  superconformal primary and two additional shortening conditions are
  generated
\begin{align}\label{generated shortening cons}
\mathcal R^- \QQ_{\bOn 3} \Psi_{\rm aux} = \QQ_{\bTw 3}\Psi_{\rm aux} &= 0\;,\cr
\mathcal R^-\mathcal{M}_2^- \QQ_{\bOn 3} \Psi_{\rm aux} = \QQ_{\bTw 4}\Psi_{\rm aux}&= 0\;,
\end{align}
where we remind the reader that $\mathcal M^-_2$ is a Lorentz lowering
operator in the Dynkin basis. Therefore, for the purposes of building the Verma module we can
remove both of these from the basis of Verma-module generators. As a result, the
modified product over supercharges, now indicated by
$\hat{Q}({\bf a,b},q)$, is
\begin{align}
\hat{Q}(a,\mathbf{b},q)=Q(a,\mathbf{b},q)\left( 1 + \frac{a b_2 q^{\frac{1}{2}}}{b_1}\right)^{-1}\left( 1 + \frac{aq^{\frac{1}{2}}}{b_2}\right)^{-1}\left( 1 + \frac{ b_2 q^{\frac{1}{2}}}{b_1 a}\right)^{-1}\left( 1 + \frac{q^{\frac{1}{2}}}{b_2 a}\right)^{-1}\;.
\end{align}
\end{enumerate}
The last thing to take into account is the possible removal of $\PP$s
from $P(\mathbf{b},q)$ when some components of the multiplet
correspond to operator constraints. This will be discussed at length
in App.~\ref{App:5DRS}.

\subsubsection{The 5D Superconformal Index}\label{superconformal index from superchar}

The supercharacter for a given multiplet can be readily converted into
the superconformal index. The five-dimensional superconformal index,
as we have previously defined it in the Dynkin basis in
Sec.~\ref{5DIndMain}, is given by\footnote{The fermion number in this
  case is
  $F = 2d_1+d_2 \simeq d_2$, since $d_1$ is always integer.}
\begin{align}
  \label{eq:18}
  \II(x,y)=\Tr_{\HH}(-1)^F e^{-\beta \delta}  x^{\frac{2}{3}\Delta +\frac{1}{3}(d_1+d_2)}y^{d_1}\;.
\end{align}
The states that are counted satisfy $\delta = 0$, where
\begin{align}
  \label{eq:17}
  \delta :=\{ \QS_{\bTw 1},\QQ_{\bOn 4}\}=\Delta - \frac{3}{2}K -d_1-d_2\;.
\end{align}
In order to make contact between the character of a 5D superconformal
representation and this index, one can simply make the following fugacity reparametrisations 
\begin{align}\label{char to index}
  q\to x^{2/3}\;,\hspace{3mm} b_1\to x^{1/3} y\;,\hspace{3mm}b_2\to x^{1/3}
\end{align}
and introduce a factor of $(-1)^F$. The resulting object is precisely
the index since every state without $\delta =0$ pairwise cancels.

\subsection{Characters of 6D $(\mathcal{N},0)$ Multiplets}\label{sec: supercharacters app 6D}

Consider a representation, the highest weight of which has conformal
dimension $\Delta$ with $\mathfrak{su}(4)$ quantum numbers
$(c_1, c_2, c_3)$ and $R$-symmetry quantum numbers $\mathbf{R}$. For
$\mathcal{N}=1$ we have that $\mathbf{R} = K$, the Dynkin label of
$\mathfrak{su}(2)_R$, while for $\mathcal{N}=2$ the Dynkin labels of
$\mathfrak{so}(5)_R$ are $\mathbf{ R} = (d_1,d_2)$. The highest weight
can be decomposed as
\begin{align}
\lambda = \omega^\alpha_1 c_1 +  \omega^\alpha_2 c_2 +  \omega^\alpha_3 c_3 + \mathbf{\omega}_i^R R^i\;,
\end{align}
where the index in $\omega^R_i R^i$ is summed over. The
$\omega^\alpha_i$ are the fundamental weights associated with the
$\mathfrak{su}(4)$ simple roots $\alpha_i$ and $\vec{\omega}^R$ are
the fundamental weights associated with the simple roots of the
$R$-symmetry algebra, $\vec{\beta}$.\footnote{Recall that in the previous
  section, we had defined the simple root of $\mathfrak{su}(2)_R$ as
  $\alpha$. In order to avoid confusion with the $\mathfrak{su}(4)$
  Lorentz-algebra simple roots, we label all simple roots associated with the
  6D $R$-symmetry algebras as $\vec{\beta}$.} For $\mathfrak{su}(2)_R$
$\vec{\beta}$ has only one component, while for $\mathfrak{so}(5)_R$
$\vec{\beta}$ has two components, $(\beta_1, \beta_2)$. Again, we
can express the fundamental weights in terms of the simple roots by
using the Cartan matrix $A_{ij}$
\begin{align}
\omega^\alpha_i &= (A^{-1}_{A_3})_{ij}\alpha_j=\begin{pmatrix}\frac{3}{4}&\frac{1}{2}&\frac{1}{4}\\ 
\frac{1}{2}&1 &\frac{1}{2}\\
\frac{1}{4}&\frac{1}{2}&\frac{3}{4}\end{pmatrix}\begin{pmatrix}\alpha_1 \\ \alpha_2 \\ \alpha_3 \end{pmatrix}\;,
\end{align}
alongside the expressions for $\mathfrak{su}(2)_R$ and $\mathfrak{so}(5)_R$ given in \eqref{coroots 5d}.

This allows for a rewriting of $e^{ \lambda}$ by defining
the $\mathfrak{su}(4)$ fugacities
\begin{align}
a_1 = e^{\frac{1}{4}(3\alpha_1 + 2 \alpha_2 + \alpha_3)}\;,&\hspace{2mm}a_2=e^{\frac{1}{2}(\alpha_1 +2 \alpha_2 + \alpha_3)}\;,\hspace{2mm}a_3=e^{\frac{1}{4}(\alpha_1 + 2\alpha_2 + 3\alpha_3)}\;.
\end{align}
Similarly, we define the $R$-symmetry fugacities $\mathbf{b}$ using \eqref{5d N=1 fugacities}. For $\mathcal{N}=1$ we have that 
\begin{align}
\mathbf{b}=b= e^{\frac{1}{2}\beta}\;,
\end{align}
while for $\mathcal{N}=2$ we have $\mathbf{b}=(b_1,b_2)$ with
\begin{align}
b_1 &= e^{\beta_1 + \beta_2}\;,\hspace{2mm}b_2 = e^{\frac{1}{2}(\beta_1 + 2\beta_2)}\;.
\end{align}
hence
\begin{align}
e^{ \lambda} = a_1^{c_1}a_2^{c_2}a_3^{c_3}\prod_{i}b_i^{R^i}.
\end{align}
The character of a particular representation $\mathcal{R}$ is then
given by
\begin{align}
\chi_\mathcal{R} (\mathbf{a},\mathbf{b},q)=\Tr_\mathcal{R}\left(a_1^{c_1}a_2 ^{c_2}a_3^{c_3}\prod_{i}b_i^{R^i}\right)
\end{align}
and for $\mathcal{N}=1,2$ respectively we have:
\begin{align}
\chi^{(1,0)}_\mathcal{R} (\mathbf{a},b,q)&=\Tr_\mathcal{R}\left(a_1^{c_1}a_2 ^{c_2}a_3^{c_3}b^K\right)\;,\cr
\chi^{(2,0)}_\mathcal{R} (\mathbf{a},\mathbf{b},q)&=\Tr_\mathcal{R}\left(a_1^{c_1}a_2 ^{c_2}a_3^{c_3}b_1^{d_1}b_2^{d_2}\right)\;.
\end{align}
We can then apply this to specific irreducible representations of the
6D $(\mathcal{N},0)$ SCA.

\subsubsection{Long Representations}\label{long reps 6d}

For the long representation
$\mathcal{L}_{[\Delta; c_1,c_2,c_3;\mathbf{R}]}$ we construct the
superconformal multiplet by acting on the highest weight state
$\ket{\Delta; c_1,c_2,c_3;\mathbf{R}}^{hw}$ with momentum operators and
supercharges
$f(\mathcal{Q},\mathcal{P})\ket{\Delta; c_1,c_2,c_3;\mathbf{R}}^{hw}$.
This polynomial can be factorised as
$f(\mathbf{a},\mathbf{b},q)=Q(\mathbf{a},\mathbf{b},q)P(\mathbf{a},q)$,
since the momentum operators and supercharges commute. Explicitly
these functions are
\begin{align}\label{q and p product 6D}
Q(\mathbf{a},\mathbf{b},q)&=\prod_{\mathbf{A},a}(1+\chi(\QQ_{{\bf A}a})),\cr
P({\bf a},q)&=\prod_{\mu=1}^{6}(1-\chi(\mathcal{P}_\mu))^{-1}\;,
\end{align}
where the  range of the sum over $\bf A$ depends on the amount of
supersymmetry.

Thus for a generic long representation we can decompose the character
of the superconformal Verma module using \eqref{Verma module character} as 
\begin{align}\label{long}
\chi_\mathcal{L}({\bf a},{ \bf b},q)=q^\Delta \chi_{[c_1, c_2, c_3]}({\bf a}) \chi_{[d_1, d_2]}({\bf b}) P({\bf a}, q) Q ({\bf a},{ \bf b}, q). 
\end{align}
The characters $ \chi_{[c_1, c_2, c_3]}({\bf a})$ and
$ \chi_{[d_1, d_2]}({\bf R})$ can in turn be obtained through their
Weyl orbits.  The relevant Weyl groups are
$W_{\SU(4)} = \mathcal{S}_4$,
$W_{\SO(5)}=\mathcal{S}_2 \ltimes (\mathbb{Z}_2)^2$ and
$W_{\SU(2)}=\mathbb{Z}_2$ so one has
\begin{align} 
\chi_{[c_1, c_2, c_3]}({\bf a})&=\sum_{w\in W_{\SU(4)}}w(a_1)^{c_1}w(a_2)^{c_2}w(a_3)^{c_3}M(w(\mathbf{a})),\\
\chi_{[\mathbf{R}]}({\bf b})&=\sum_{w\in W_R}\left(\prod_i w(b_i)^{R^i}\right) R^{(\mathcal{N},0)}(w({\bf b})).
\end{align}
We use $W_R$ to indicate the Weyl group appropriate for the $R$ symmetry of the
$(\mathcal{N},0)$ SCA. The $M({\bf a})$ and
$R^{(\mathcal{N},0)}({\bf b})$ are the products of characters of
negative roots as defined in \eqref{Verma module character};
explicitly
\begin{align}
M({\bf a}) &=\frac{1}{\left(1-\frac{a_2}{a_1^2}\right) \left(1-\frac{a_2}{a_3^2}\right) \left(1-\frac{1}{a_1 a_3}\right) \left(1-\frac{a_1}{a_2 a_3}\right) \left(1-\frac{a_1 a_3}{a_2^2}\right) \left(1-\frac{a_3}{a_1 a_2}\right)}\;,\cr
R^{(2,0)}({\bf b}) &=\frac{1}{\left(1-\frac{1}{b_1}\right) \left(1-\frac{1}{b_2^2}\right) \left(1-\frac{b_1}{b_2^2}\right) \left(1-\frac{b_2^2}{b_1^2}\right)}\;,\cr
R^{(1,0)}(b)&=\frac{b^2}{b^2 -1}\;.
\end{align}

Again, we note that both $Q(\mathbf{a},\mathbf{b},q)$ and
$P(\mathbf{a},q)$ are invariant under the appropriate Weyl
symmetrisations and one can write for $\mathcal{N}=2$
\begin{align}
\chi^{(2,0)}_\mathcal{L} ({\bf a, b},q) =\big{\llbracket} q^\Delta a_1^{c_1}a_2^{c_2}a_3^{c_3}b_1^{d_1}b_2^{d_2} M({\bf a}) R^{(2,0)}({\bf b}) P({\bf a}, q) Q({\bf a,b},q)\big{\rrbracket}_W\;,
\end{align}
 matching \cite{Beem:2014kka}, while for $\mathcal{N}=1$ 
 \begin{align}
\chi^{(1,0)}_\mathcal{L} ({\bf a}, b,q) =\big{\llbracket} q^\Delta a_1^{c_1}a_2^{c_2}a_3^{c_3}b^K M({\bf a}) R^{(1,0)}(b) P({\bf a}, q) Q({\bf a},b,q)\big{\rrbracket}_W\;,
\end{align}
where $[\! [ \cdots ]\!]_W$ denotes the Weyl symmetriser. 

\subsubsection{Short Representations}

Consider now the short multiplets of Table \ref{Tab:6D(1,0)} for
$\mathcal N = 1$ or Table \ref{Tab:6D(2,0)} for $\mathcal N =2$. To
calculate their characters we remove certain combinations of $\QQ$s
from the expressions $Q({\bf a},\mathbf{b},q)$ given in \eqref{q and p
  product 6D}. This discussion is completely analogous to
Sec.~\ref{short reps 5d char}. There are a few cases when one is also
prescribed to remove certain $\PP_\mu$ from $P(\mathbf{a},q)$. This is
discussed at length in App.~\ref{App:5DRS}.

\subsubsection{The 6D $(1,0)$ Superconformal Index}\label{superconformal index from superchar}

Once we have obtained the full supercharacter we can readily covert it
into the superconformal index. For $\mathcal N =1$ the index as
defined in \eqref{6D10IndMain} is given by\footnote{The fermion number
  in this case is $F = c_1 + c_3$.} 
\begin{align}
  \II(p,q,s)=\Tr_{\HH}(-1)^F e^{-\beta \delta}  q^{\Delta-\frac{1}{2}K}p^{c_2}s^{c_1}\;.
\end{align}
The states that are counted satisfy $\delta = 0$, where
\begin{align}
  \delta =\Delta- 2 K -\frac{1}{2}(c_1+2c_2+3c_3)\;.
\end{align}
We can therefore write the character of a representation as an index
via the following fugacity reparametrisations
\begin{align}
  a_1\to s\;,\hspace{3mm}a_2\to p\;,\hspace{3mm}a_3\to 1\;,\hspace{3mm}b\to \frac{1}{q^{\frac{1}{2}}}
\end{align}
and inserting $(-1)^F$. The resulting object is precisely the index
since every state without $\delta =0$ pairwise cancels.

\subsubsection{The 6D $(2,0)$ Superconformal Index}

The 6D $(2,0)$ superconformal index, as previously defined in
\eqref{6D20IndMain}, is given by 
\begin{align}
  \II(p,q,s,t)=\Tr_{\HH}(-1)^F e^{-\beta \delta}q^{\Delta-d_1 - \frac{1}{2}d_2}p^{c_2+ c_3 -
  d_2}t^{-d_1 -d_2}s^{c_1+c_2}\;.
\end{align}
The states that are counted satisfy $\delta = 0$, where
\begin{align}
  \delta = \Delta - 2d_1-\frac{1}{2}c_1 - c_2 - \frac{3}{2}c_3\;.
\end{align}
In order to make contact between the character and this index, we make
the following fugacity reparametrisations
\begin{align}
  a_1\to s\;,\hspace{3mm}a_2\to p s\;,\hspace{3mm}a_3\to p\;,\hspace{3mm}b_1\to \frac{1}{q t}\;,\hspace{3mm}b_2\to \frac{q^{\frac{1}{2}}}{ p t}
\end{align}
and insert $(-1)^F$. The resulting object is precisely the index since
every state previously counted by the character without $\delta =0$
pairwise cancels.

\section{The Racah--Speiser Algorithm and Operator Constraints}\label{App:5DRS}

In this appendix we provide the details needed to carry out the RS
algorithm for the 5D $\mathcal{N}=1$, 6D $(1,0)$ and 6D $(2,0)$ SCAs.
Fortunately, we need only discuss three different algebras,
$\mathfrak{so}(5)$, $\mathfrak{su}(4)$ and $\mathfrak{su}(2)$. We will
assume some familiarity with the description of the RS algorithm from
App.~B of \cite{Dolan:2002zh}, the notation of which we use.

First, let us consider $\so(5)$. This is the Lorentz Lie algebra for 5D
$\mathcal{N}=1$ and the $R$-symmetry Lie algebra for 6D $(2,0)$. The highest
weight ($ \lambda$) identifications resulting from Weyl reflections
($\sigma$) are given by
\begin{align}\label{racah B}
    [d_1, d_2]&=-[-d_1-2,2d_1+ d_2+2]\;,\cr
    &=-[d_1+d_2 + 1,- d_2-2]\;,\cr
    &=-[-d_1 -d_2 -3, d_2]\;,\cr
    &=-[d_1 ,-4-2d_1- d_2]\;.
\end{align} 
We see that
$\lambda^\sigma =\lambda$ with
sign$(\sigma)=-$ under the following conditions:
\begin{align}\label{so5cond} d_1 = -1\;, \hspace{3mm} d_2 =
  -1\;,\hspace{3mm} 2d_1 + d_2 = -3\;,\hspace{3mm}d_1 + d_2 = -2\;.
\end{align} Therefore representations which satisfy any one of these
conditions are labelled by a highest weight on the boundary of the
Weyl chamber, in which case they have zero multiplicity and need to be removed. Notice that the
conditions \eqref{so5cond} correspond to the zeros of the Weyl
dimension formula for irreducible representations of
$\mathfrak{so}(5)$: \begin{align}\label{Weyl Dim Form S05} \mathrm
                      d(d_1,d_2) =\frac{1}{6}(d_1+1)(d_2+1)(d_1 + d_2+2)(2d_1 + d_2+3)\;.
\end{align} Next we consider $\su(4)$, the Lorentz Lie algebra of 6D
theories. The highest weight identifications resulting from Weyl
reflections are 
\begin{align}\label{racah
    A}
    [c_1, c_2, c_3] &= -[-c_1 -2, c_1+c_2 + 1, c_3]\;,\cr
    & =-[c_1 +c_2+1,- c_2-2,c_2+ c_3+1]\;,\cr
    &= -[c_1, c_2+c_3 + 1, -c_3-2]\;,\cr
    &= -[-c_2-2, -c_1-2, c_1 + c_2 + c_3 +2]\;,\cr
    &= -[c_1 + c_2 + c_3 +2, -c_3-2,-c_2 - 2]\;,\cr
    &= -[-c_2 - c_3 -3,c_2,-c_1 - c_2 - 3]\;.
  \end{align} 
The representations we delete are the ones where the following conditions are met:
\begin{align} c_1 = -1,\hspace{2mm}c_2 = -1, \hspace{2mm}c_3 = -1,
  \hspace{2mm}c_1+c_2 = -2, \hspace{2mm}c_2+c_3 -
  2,\hspace{2mm}c_1+c_2 + c_3 =-3. \end{align} Again, these correspond
to the zeros of the Weyl dimension formula for $\su(4)$:
\begin{align}\label{Weyl Dim Form SU4} \mathrm d(c_1,c_2,
  c_3)=\frac{1}{12}(c_1+1)(c_2+1)(c_3+1)(c_1 +
  c_2+2)(c_2+c_3+2)(c_1+c_2+c_3+3). \end{align}

Lastly, the $R$-symmetry Lie algebra for the minimally supersymmetric theories
is $\su(2)$, which involves identifying highest weights via the reflection:
\begin{align}\label{racah A1}
[K]&= -[-K-2]\;.
\end{align}
We therefore see that the representations we delete are the ones where
the following condition is met:
\begin{align}
K=-1\;.
\end{align}
Clearly this corresponds to the zero for the Weyl dimension formula of
$\su(2)$ 
\begin{align}
\mathrm{d}(K)=K+1\;.
\end{align}

We may now simply combine the set of Weyl reflections appropriate for
our SCA in order to dictate which states survive when building
representations. Our method will involve generating all possible
highest weight states, even in cases where the Dynkin labels become
negative, and then performing the RS algorithm.

It is interesting to note that after implementing the RS algorithm one
often encounters pairs of superconformal descendants with exactly the
same Dynkin labels but opposite multiplicities. These cancel out to
leave behind a much simpler set of superconformal representations with
only positive multiplicities. If after performing this step there are
also negative representations that have not been cancelled, they are
interpreted as constraints for operators in the multiplet
\cite{Dolan:2002zh}. 

There are instances when this general procedure leads to ambiguities,
i.e. there is more than one choice for performing the cancellations;
see also \cite{Cordova:2016xhm}. However, these can be resolved
uniquely by the requirement that all physical states should be reached
by the successive action of allowed supercharges on the superconformal
primary. For the examples that we investigate in this article, this
phenomenon only appears in the (2,0) SCA
($\mathcal B[c_1, c_2,0;0,0]$, $\mathcal C[c_1, 0,0;0,0]$), in which
case we have constructed the multipet spectra by also using successive
$\QQ$-actions.

\subsection{Operator Constraints Through Racah--Speiser}

We now further explore this concept. Since $R$-symmetry quantum
numbers will not play a role in this analysis we will simplify our
discussion by denoting all of their quantum numbers by $R$. The only
distinction we need to make is between the $\so(5)$ and $\su(4)$
Lorentz Lie algebras. It will be instructive to proceed by first providing
an example and then the result in full generality.

Consider the Lorentz vector representation in 6D, $[\Delta;0,1,0;R]$,
corresponding to an operator $\mathcal{O}_\mu$. In terms of quantum numbers, its components are given by 
\begin{align}
\mathcal{O}_1 \sim [0,1,0]\;,\hspace{2mm} \mathcal{O}_2 \sim [1,-1,1]\;,\hspace{2mm}\mathcal{O}_3 \sim [-1,0,1]\;,\hspace{2mm}\cr
\mathcal{O}_4 \sim [1,0,-1]\;,\hspace{2mm} \mathcal{O}_5 \sim [-1,1,-1]\;,\hspace{2mm}\mathcal{O}_6 \sim [0,-1,0]\;.
\end{align}
One can envisage a conservation-equation state for this object being
$-[\Delta +1;0,0,0;R]$, that is we pick components of $\PP_\mu$ and
$\mathcal{O}_\mu$ such that their combination has Lorentz quantum
numbers $[0,0,0]$. This combination is
\begin{align} \PP_6 \mathcal{O}_{1}+\PP_5
  \mathcal{O}_{2}+\PP_4
  \mathcal{O}_{3}+\PP_3
  \mathcal{O}_{4}+\PP_2
  \mathcal{O}_{5}+\PP_1
  \mathcal{O}_{6}=0\;, 
\end{align} 
which can be concisely interpreted as the conservation equation\footnote{One has that $(\PP_\mu)^\dagger
  =\PP^\mu$. Then it can be
  straightforwardly checked that $\PP^\mu= \PP_{7-\mu}$.}
\begin{align} \partial^\mu
  \mathcal{O}_{\mu}=0\;. 
\end{align} 
This is precisely what we see e.g. for the $R$-symmetry currents of
the $(1,0)$ and $(2,0)$ stress-tensor multiplets,
$\mathcal{B}[0,0,0;0]$ and $\mathcal{D}[0,0,0;2,0]$.

More generally, there are only two ways in which operator constraints
manifest themselves in 6D using the approach employed in this
paper. These are
\begin{align}\label{eom 6d gen}
[\Delta; c_1,c_2,c_3; R]&\xrightarrow{} -[\Delta +1;c_1,c_2 -1,c_3;R]\;,\cr
[\Delta; c_1,0,0; R]&\xrightarrow{} -[\Delta +1;c_1-1,0,1;R]+[\Delta +2;c_1-2,1,0;R]\cr&\hspace{40mm}-[\Delta+3,c_1-2,0,0;R]\;.
\end{align}
The first equation is a contraction of a vector index; this is a
conservation equation. The other requires that
$c_2 = c_3 =0$.  These are the ``generalised Dirac equations'' or
``generalised equations of motion" mentioned in
\cite{Dolan:2005wy}. When $c_1=0$ the primary is a scalar
$[2;0,0,0;R]$, and the only state that survives the RS algorithm is
$-[4;0,0,0;R]$, namely a Klein--Gordon equation. When $c_1 =1$ the
primary is a fermion $[5/2;1,0,0;R]$, with the only state surviving RS
being a Dirac equation $-[7/2;0,0,1;R]$. For $c_1 \geq 2$, all states
survive RS, leaving a Bianchi identity for a higher $p$-form field;
this is usually endowed with additional constraints, i.e. self-duality
in the free-tensor case, $c_1=2$.

The analogous states in 5D are given by 
\begin{align}\label{eom 5d gen}
[\Delta;d_1,d_2; R]&\xrightarrow{} -[\Delta +1;d_1-1,d_2;R]\;,\cr
[\Delta; 0,d_2; R]&\xrightarrow{} -[\Delta +1;0,d_2;R]+[\Delta +2;1,d_2-2;R]-[\Delta+3,0,d_2-2;R]\;. 
\end{align}
Again, the first expression is the contraction of a vector index and
the second expression recovers the Klein--Gordon equation ($d_2=0$),
the Dirac equation ($d_2=1$) and the Bianchi identity ($d_2\geq 2$).

\subsection{The Dictionary Between Racah--Speiser and Momentum-Null States}\label{conserveq}

The expressions \eqref{eom 6d gen} and \eqref{eom 5d gen} can also be
understood from a different perspective. When the superconformal
primary saturates both a conformal and a superconformal bound, certain
components in the multiplet satisfy operator constraints. One is then
instructed to remove appropriate $\PP_\mu$ generators from the
auxiliary Verma-module basis \cite{Dolan:2005wy}. We call the set
contained in the resulting module ``reduced states''.

Following \cite{Dolan:2005wy}, in 5D $\mathcal{N}=1$ we are instructed
to remove $\PP_5$ for the multiplets $\mathcal{B}[0,0;0]$ and
$\mathcal{D}[0,0;2]$. For $\mathcal{D}[0,0;1]$ one removes $\PP_3$,
$\PP_4$ and $\PP_5$. In 6D, where $R=K$ for $\mathcal{N}=1$ and
$R=d_1 +d_2$ for $\mathcal{N}=2$, one removes $\PP_6$ from
$\mathcal{B}[c_1,c_2,0;R=0]$, $\mathcal{C}[c_1,0,0;R=1]$ and
$\mathcal{D}[0,0,0;R=2]$. Likewise we remove $\PP_3$, $\PP_5$ and
$\PP_6$ from $\mathcal{C}[c_1,0,0;R=0]$ and $\mathcal{D}[0,0,0;R=1]$.

The connection between the momentum-null states and the
negative-multiplicity representations is clear when considering the
supercharacter: Taking the character of all states---including the
negative-multiplicity representations---produces the same result as
doing so for all reduced states with the appropriate
$\PP$s removed from the basis of generators. Hence it is not
appropriate to do both. This is reflected in the fact that the following character identity holds in 6D: 
\begin{align}\label{racah to dolan 6d}
\hat{\chi}[\Delta;c_1,c_2,c_3;R] \equiv \chi[\Delta; c_1,c_2,c_3;R]-\chi[\Delta+1;c_1,c_2-1,c_3;R]\;,
\end{align}
where a hat denotes using the $P$-polynomial from \eqref{q and p
  product 6D} after having removed $\PP_6$; the operator constraints
that one recovers in this fashion are conservation
equations. Likewise when we remove $\PP_3$, $\PP_5$ and $\PP_6$ we
obtain:
\begin{align}\label{racah to dolan bianchi 6d}
\hat{\chi}[\Delta;c_1,0,0;R] \equiv \chi[\Delta; c_1,0,0;R]-\chi[\Delta+1;c_1-1,0,1;R]+\chi[\Delta+2;c_1-2,1,0;R]\cr
-\chi[\Delta+3;c_1-2,0,0;R]\;.
\end{align}
The operator constraints recovered in this case are equations of
motion.

In 5D one can make the analogous identifications:
\begin{align}
\hat{\chi}[\Delta;d_1,d_2;R] \equiv \chi[\Delta; d_1,d_2;R]-\chi[\Delta+1;d_1-1,d_2;R]\;,
\end{align}
where we have removed $\PP_5$ from $\hat{\chi}$. Likewise, when we
remove $\PP_3$, $\PP_4$ and $\PP_5$ we obtain:
\begin{align}
\hat{\chi}[\Delta;0,d_2;R] \equiv \chi[\Delta;0,d_2;R]-\chi[\Delta+1;0,d_2;R]+\chi[\Delta+2;1,d_2-2;R]\nonumber\\- \chi[\Delta+3;0,d_2-2;R]\;.
\end{align}

This effectively endows us with a choice for how to treat the
operator constraints: We can either take the character of all
\emph{reduced} states---not including states associated with
constraints---with the appropriate $\PP_\mu$ removed, or we
can take the character of all states (including
negative-multiplicity representations) without removing any $\PP_\mu$.

Note that if one has to remove momenta from the basis of generators
for a given multiplet then not every component contained within need
have an associated operator constraint. The stress-tensor multiplets
in 6D $\mathcal B[0,0,0;0]$ and $\mathcal{D}[0,0,0;2,0]$ are examples
of such superconformal representations: they contain superconformal
descendants which do not contain operators satisfying
constraints. However, according to the arguments of the previous
subsection, we are still instructed to remove $\PP_6$ in one approach
of evaluating the supercharacter. At first glance this is perplexing.

The resolution to this small puzzle is that the associated characters
are actually invariant under the removal of $\PP_6$. Hence such states
satisfy
\begin{align}
\hat{\chi}[\Delta;c_1,c_2,c_3;R]=\chi[\Delta;c_1,c_2,c_3;R]\;,
\end{align}
in 6D, or in 5D
\begin{align}
\hat{\chi}[\Delta;d_1,d_2;R]=\chi[\Delta;d_1,d_2;R]\;.
\end{align}
The method for evaluating characters of conformal UIRs presented in
\cite{Dolan:2005wy} makes no distinction between states that do/do not
obey constraints due to the above invariance.

This knowledge is useful when the absent generators of the Verma
module are such that the removed $\QQ$s anticommute into $\PP$s; c.f.
Sec.~\ref{HSB5D}, Sec.~\ref{higher spin 6d (1,0)} and
Sec.~\ref{HS6D20}. In that case, one is effectively projecting out the
states associated with operator constraints from the very
beginning. Therefore what is generated under these circumstances is
the spectrum of \emph{reduced} states. It is still possible to
reconstruct the full multiplet spectrum using character relations, as
e.g. in \eqref{racah to dolan 6d}. This lets us recover all the
negative-multiplicity states.

\section{Summary of Superconformal Indices}\label{App:Index}

In this appendix we list the superconformal indices for multiplets in
5D $\mathcal{N}=1$ and 6D $(1,0)$, alongside the Schur limit of the
indices for multiplets in 6D $(2,0)$. All the indices, including the
fully refined indices for the (2,0) multiplets, are given in the
accompanying Mathematica file. We use the $\mathfrak{su}(2)$ character for the
spin-$\frac{l}{2}$ representation in our expressions: 
\begin{align}
  \chi_{l}(y)=\frac{y^{l+1}-y^{-l-1}}{y-y^{-1}}\;. 
\end{align}

\subsection{5D $\mathcal{N}=1$}

The distinct indices are provided below:

\begin{itemize}

\item
$\mathcal{A}[d_1,d_2;K]$ (for $d_1,d_2 \geq 0$, $K\geq 0$):
\begin{align}
\mathcal I(x,y) = (-1)^{d_2+1}\frac{x^{d_1+d_2+K+4}}{\left(1-x y^{-1}\right) \left(1-x y\right)}\chi_{d_1}(y)\;.
\end{align}

\item
$\mathcal{B}[d_1,0;K]$ (for $d_1 \geq 0$, $K\geq 0$):
\begin{align}
\mathcal I(x,y) = \frac{x^{ d_1+K+3}}{\left(1-x y^{-1}\right) \left(1-x y\right)}\chi_{d_1+1}(y)\;.
\end{align}

\item
$\mathcal{D}[0,0;K]$ (for $K\geq 1$):
\begin{align}
\mathcal I(x,y) =\frac{x^{K}}{\left(1-xy^{-1}\right) \left(1- x y\right)}\;.
\end{align}

\end{itemize}

\subsection{6D $(1,0)$}

The distinct indices are provided below:

\begin{itemize}

\item
$\mathcal{A}[c_1,c_2,c_3;K]$ (for $c_1, c_2, c_3 \geq 0$, $K\geq 0$):
\begin{align}
\mathcal I(p,q,s) =  (-1)^{c_1+c_3}q^{6+\frac{3K}{2}+\frac{1}{2}(c_1+2c_2+3c_3)}\Bigg\{\frac{p^{-c_2+1} s^{c_2+4} \left(p^{-c_1}-p s^{c_1+1}\right)-p^{-c_1+2} s^{2-c_2}}{(p q-1) \left(p^2-s\right) (p s-1) \left(p-s^2\right) (q-s) (p-q s)}\nonumber\\
+ \frac{p^{c_2+4} s^{c_1+4}+p^{c_1+4} s^{-c_1+1} \left(s^{-c_2}-s p^{c_2+1}\right)}{(p q-1) \left(p^2-s\right) (p s-1) \left(p-s^2\right) (q-s) (p-q s)}\Bigg\}\;.
\end{align}

\item
$\mathcal{B}[c_1,c_2,0;K]$ (for $c_1, c_2 \geq 0$, $K\geq 0$):
\begin{align}
\mathcal I(p,q,s) =  (-1)^{c_1+1}q^{4+\frac{3K}{2}+\frac{1}{2}(c_1+2c_2)}\Bigg\{\frac{p^{-c_2} s^{c_2+5} \left(p^{-c_1}-p s^{c_1+1}\right)-p^{-c_1+2} s^{1-c_2}}{(p q-1) \left(p^2-s\right) (p s-1) \left(p-s^2\right) (q-s) (p-q s)}\nonumber\\
+ \frac{p^{c_2+5} s^{c_1+4}+p^{c_1+4} s^{-c_1} \left(s^{-c_2}-s^2 p^{c_2+2}\right)}{(p q-1) \left(p^2-s\right) (p s-1) \left(p-s^2\right) (q-s) (p-q s)}\Bigg\}\;.
\end{align}

\item
$\mathcal{C}[c_1,0,0;K]$ (for $c_1 \geq 0$, $K\geq 1$):
\begin{align}
\mathcal I(p,q,s) =  (-1)^{c_1}q^{2+\frac{3K}{2}+\frac{c_1}{2}} \frac{  p^{ c_1+5} s^{-c_1} (p s-1)+p^{2} s^{c_1+5} \left(s-p^2\right)+p^{-c_1} s^{2} \left(p-s^2\right)}{(p q-1) \left(p^2-s\right) (p s-1) \left(p-s^2\right) (q-s) (q s-p)}\;.
\end{align}

\item
$\mathcal{C}[c_1,0,0;0]$ (for $c_1 \geq 0$):
\begin{align}
\mathcal I(p,q,s) =  (-1)^{c_1}q^{2+\frac{c_1}{2}}\Bigg\{\frac{p^{c_1+6} s^{1-c_1}-p^{c_1+5} s^{-c_1}+s^2 p^{-c_1} \left(p-s^2\right)-p^4 s^{c_1+5}+p^2 s^{c_1+6}}{(p q-1) \left(p^2-s\right) (p s-1) \left(p-s^2\right) (q-s) (q s-p)}\nonumber\\
+ \frac{  p^{ c_1+5} s^{-c_1} (p s-1)+p^{2} s^{c_1+5} \left(s-p^2\right)+p^{-c_1} s^{2} \left(p-s^2\right)}{(p q-1) \left(p^2-s\right) (p s-1) \left(p-s^2\right) (q-s) (q s-p)}\Bigg\}\;.
\end{align}

\item
$\mathcal{D}[0,0,0;K]$ (for $K\geq 1$):
\begin{align}
\mathcal I(p,q,s) = \frac{q^{\frac{3 K}{2}}}{(1-p q) (1-q s^{-1}) (1-p^{-1} q s)}\;.
\end{align}

\end{itemize}

\subsection{6D $(2,0)$}

The non-vanishing Schur indices are valid for all $c_i$ and $d_1$ values. These are provided below:

\begin{itemize}

\item
$\mathcal{B}[c_1,c_2,0;d_1,0]$:
\begin{align}
  \mathcal I^{Schur}(q,s) =  (-1)^{c_1}\frac{q^{4+d_1+\frac{c_1}{2}+c_2} }{1-q}\chi_{c_1 }(s)\;.
\end{align}

\item
$\mathcal{C}[c_1,0,0;d_1,1]$:
 \begin{align}
  \mathcal I^{Schur}(q,s) = (-1)^{c_1+1}\frac{q^{\frac{7}{2}+d_1+\frac{c_1}{2}}}{1-q}\chi_{c_1 + 1}(s)\;.
\end{align}

\item
$\mathcal{C}[c_1,0,0;d_1,0]$:
\begin{align}
  \mathcal I^{Schur}(q,s) = (-1)^{c_1}\frac{q^{2+d_1+\frac{c_1}{2}}}{1-q}\chi_{c_1 + 2}(s)\;.
\end{align}

\item
$\mathcal{D}[0,0,0;d_1,2]$:
\begin{align}
  \mathcal I^{Schur}(q,s) =\frac{q^{3+d_1}}{1-q}\;.
\end{align}

\item
$\mathcal{D}[0,0,0;d_1,1]$:
\begin{align}
  \mathcal I^{Schur}(q,s) =-\frac{q^{\frac{3}{2}+d_1}}{1-q}\chi_{1}(s)\;.
\end{align}

\item
$\mathcal{D}[0,0,0;d_1,0]$:
\begin{align}
  \mathcal I^{Schur}(q,s) =\frac{q^{d_1}}{1-q}\;.
\end{align}

\end{itemize}

The refined versions of the above indices, $ \mathcal I(p,q,s,t) $,
are provided in the accompanying Mathematica notebook.

\section{6D $(2,0)$ Spectra}\label{(2,0) spectra}

We will finally provide the spectra for short multiplets in 6D (2,0)
with generic quantum numbers. Since long representations are
cumbersome and standard to obtain we will not list them here.
Moreover, we will only write down \emph{distinct} spectra, that cannot
be obtained by fixing $c_i$ or $d_i$ and performing the RS
algorithm. For book-keeping purposes we have grouped the supercharges
into $Q = (\QQ_{\bTw a},\QQ_{\bTh a})$ and
$\tilde{Q} = (\QQ_{\bOn a},\QQ_{\bFo a})$, with $a = 1,\cdots 4$.

\subsection{$\mathcal{A}$-type Multiplets}\label{A (2,0) spectra}

The superconformal primary null-state condition for a generic $\mathcal A$-type
multiplet is $A^4 \Psi =0$. This corresponds to removing
$\QQ_{\bOn 4}$ from the basis of auxiliary Verma-module generators \eqref{6d (2,0) conf
  basis}. Starting from a superconformal primary given by
$[ c_1,c_2,c_3;d_1,d_2]$ with conformal dimension
$\Delta = 6+2(d_1+d_2)+\frac{1}{2}(c_1+2c_2+3c_3)$, we will provide
the states at each level $l$; the conformal dimension of the
superconformal descendants will be equal to $\Delta + \frac{l}{2}$.
Thus the spectrum for $\mathcal{A}[c_1,c_2,c_3;d_1,d_2]$ is given by
acting with all $Q$s and $\tilde{Q}$s, resulting in the two chains,
Table~\ref{full a spec (2,0) Q} and Table~\ref{full a spec (2,0) QT}
respectively. Note that this is a complete description for generic
$d_1$ and $d_2$. The reason being that, as described in
Sec.~\ref{6D20A}, once we have $\mathcal{A}[c_1,c_2,c_3;d_1,d_2]$ we
can obtain the spectra for $\mathcal{A}[c_1,c_2,0;d_1,d_2]$,
$\mathcal{A}[c_1,0,0;d_1,d_2]$ and $\mathcal{A}[0,0,0;d_1,d_2]$ by
dialling $c_i$ appropriately and running the RS algorithm.

It turns out that the same also holds for dialling $d_1$ and $d_2$ to
zero. Thus the complete spectra for all $\mathcal{A}$-type multiplets
can be obtained exclusively by considering the set of $Q$ and $\tilde{Q}$
actions (Table~\ref{full a spec (2,0) Q} and Table~\ref{full a spec
  (2,0) QT}) and then substituting in the desired quantum numbers
followed by implementing the RS algorithm.

\subsection{$\mathcal{B}$-type Multiplets}\label{B (2,0) spectra}

For generic quantum numbers, the superconformal primary for this
multiplet type obeys the null-state condition $A^3\Psi =0$. This
corresponds to removing $\QQ_{\bOn 3}$ and $\QQ_{\bOn 4}$ from the
auxiliary Verma-module basis \eqref{6d (2,0) conf
  basis}. Starting from a generic primary $[c_1, c_2, c_3;d_1,d_2]$
with conformal dimension
$\Delta = 4+2(d_1+d_2)+\frac{1}{2}(c_1+2c_2)$, the action of the
$\tilde{Q}$ supercharges remains the same as in Table~\ref{full a spec
  (2,0) QT}, with the difference that one needs to apply RS after
setting $c_3=0$. The $Q$--chain, however, is significantly shorter and
given in Table~\ref{full b spec (2,0) Q}.

\subsubsection{$\mathcal{B}[c_1,c_2,0;d_1,1]$}

In this case the action of the $Q$ supercharges is the same as in
Table \ref{full b spec (2,0) Q}. However, recall from Sec.~\ref{(2,0)
  B mult} that we also have the null-state condition
$(\mathcal R_2^-)^2 A^3 A^4\Psi=0$. Thus we are also prescribed to
remove $\QQ_{\bTw 3}\QQ_{\bTw 4}$ from the $\tilde{Q}$-spectrum. The
action of the $\tilde{Q}$ supercharges is therefore adjusted and
described in Table~\ref{full b spec (2,0) QT d1=1}.

\subsubsection{$\mathcal{B}[c_1,c_2,0;d_1,0]$}

For this case the action of the $Q$ supercharges is the same as in
Table~\ref{full b spec (2,0) Q}, after replacing $d_2=0$ and running
the RS algorithm. We are also prescribed to remove $\QQ_{\bTw 3}$ and
$\QQ_{\bTw 4}$ from the $\tilde{Q}$-spectrum. The action of the
$\tilde{Q}$ supercharges is therefore adjusted and described by
Table~\ref{full b spec (2,0) QT d2=0}.

\subsection{$\mathcal{C}$-type Multiplets}\label{C (2,0) spectra}

For generic quantum numbers the superconformal primary for this
multiplet obeys the null-state condition $A^2\Psi =0$. This
corresponds to removing $\QQ_{\bOn 2}$, $\QQ_{\bOn 3}$ and
$\QQ_{\bOn 4}$ from the basis of auxiliary Verma-module generators. Starting from a generic
primary $[c_1, c_2, c_3;d_1,d_2]$ with conformal dimension
$\Delta = 2+2(d_1+d_2)+\frac{c_1}{2}$, the action of the $\tilde{Q}$
supercharges remains the same as in Table~\ref{full a spec (2,0) QT},
with the difference that one needs to apply RS after setting
$c_2=c_3=0$. The action of the $Q$ supercharges is given in
Table~\ref{full c spec (2,0) Q}.

\subsubsection{$\mathcal{C}[c_1,0,0;d_1,2]$}

Recall that in this case  we should also remove $\QQ_{\bTw 1}\QQ_{\bTw
  2}\QQ_{\bTw 3}$ from the basis of auxiliary Verma-module generators. The resulting set
of $\tilde{Q}$ actions are given in Table~\ref{full c spec (2,0) QT d2=2}.

\subsubsection{$\mathcal{C}[c_1,0,0;d_1,1]$}

For this multiplet we also need to remove $\QQ_{\bTw a}\QQ_{\bTw b}$ for $a\neq b\neq 1$. The resulting set of $\tilde{Q}$ actions are given in Table~\ref{full c spec (2,0) QT d2=1}.

\subsubsection{$\mathcal{C}[c_1,0,0;d_1,0]$}

For this multiplet we also need to remove $\QQ_{\bTw a}$ for $a\neq 1$. The resulting set of $\tilde{Q}$ actions are given in Table~\ref{full c spec (2,0) QT d2=0}.

\subsubsection{$\mathcal{C}[c_1,0,0;1,0]$}\label{C (2,0) spectra EoM}

This multiplet has the same $Q$ and $\tilde{Q}$ actions as
$\mathcal{C}[c_1,0,0;d_1,0]$ with the difference that one needs to
apply RS after setting $d_1 =1$. This multiplet contains generalised
conservation equations and is small enough for us to detail its entire
spectrum in Table~\ref{full c spec (2,0) (1,0) Q and QT}.

\subsubsection{$\mathcal{C}[c_1,0,0;0,1]$}

For this case we can simply substitute $d_1=0$ into the spectrum of
$\mathcal{C}[c_1,0,0;d_1,1]$ and run the RS algorithm to get the
spectrum of Table~\ref{full c spec (2,0) (0,1) Q and QT}. This
multiplet also contains generalised conservation equations.

\subsection{$\mathcal{D}$-type Multiplets}\label{D (2,0) Spectra}

Since the action of the $Q$ supercharges have been given in Sec.~\ref{D20}, we need only provide the $\tilde{Q}$ chains for specific $d_2$ values. 

\subsubsection{$\mathcal{D}[0,0,0;d_1,3]$}

Recall that in this case we remove $\QQ_{\bTw 1}\QQ_{\bTw 2}\QQ_{\bTw
  3}\QQ_{\bTw 4}$. We summarise the actions of the $\tilde{Q}$ supercharges in Table~\ref{D spec (2,0) d2=3 QT}.

\subsubsection{$\mathcal{D}[0,0,0;d_1,2]$}

For this multiplet we remove $\QQ_{\bTw a}\QQ_{\bTw b}\QQ_{\bTw c}$
from the basis of auxiliary Verma-module generators. We summarise the actions of the
$\tilde{Q}$ supercharges in Table~\ref{D spec (2,0) d2=2 QT}.

\begin{table}[h]
\tiny
\begin{center}
\begin{tabular}{|l|}
\hline
{\hspace{52mm} \tiny Q actions for $\mathcal{A}[c_1,c_2,c_3;d_1,d_2]$}\\
\hline
\hline
{\scriptsize Level: 0}\\
\hline 
{  $\left[c_1,c_2,c_3;d_1,d_2\right]$}\\
\hline
\hline
{\scriptsize   Level: 1}\\
\hline
{  $ \left[c_1+1,c_2,c_3;d_1,d_2+1\right] $},{  $ \left[c_1+1,c_2,c_3;d_1,d_2-1\right] $},{  $ \left[c_1-1,c_2+1,c_3;d_1,d_2+1\right] $},{  $ \left[c_1-1,c_2+1,c_3;d_1,d_2-1\right] $},\\{  $ \left[c_1,c_2-1,c_3+1,d_1,d_2+1\right] $},{  $ \left[c_1,c_2-1,c_3+1,d_1,d_2-1\right] $},{  $ \left[c_1,c_2,c_3-1;d_1,d_2-1\right] $}\\
\hline
\hline
{\scriptsize Level: 2}\\
\hline
{  $\left[c_1+2,c_2,c_3;d_1,d_2\right]$},{  $\left[c_1,c_2+1,c_3;d_1,d_2+2\right]$}, {  $2\left[c_1,c_2+1,c_3;d_1,d_2\right]$},{  $\left[c_1,c_2+1,c_3;d_1,d_2-2\right]$},\\
\hline
\end{tabular}
\end{center}
\caption{\label{full a spec (2,0) Q} The spectrum of $\mathcal{A}[c_1,c_2,c_3;d_1,d_2]$ associated with the action of $Q$s.}
\end{table}

\begin{table}
\tiny
\begin{center}
\begin{tabular}{|l|}
\hline
{\hspace{52mm} \tiny $\tilde{Q}$ actions for $\mathcal{A}[c_1,c_2,c_3;d_1,d_2]$}\\
\hline
\hline
{\scriptsize Level: 0}\\
\hline 
{  $\left[c_1,c_2,c_3;d_1,d_2\right]$}\\
\hline
\hline
{\scriptsize   Level: 1}\\
\hline
$\left[c_1+1,c_2,c_3;d_1+1,d_2-1\right]$,$\left[c_1+1,c_2,c_3;d_1-1,d_2+1\right]$,$\left[c_1-1,c_2+1,c_3;d_1+1,d_2-1\right]$,$\left[c_1-1,c_2+1,c_3;d_1-1,d_2+1\right]$,\\$\left[c_1,c_2-1,c_3+1;d_1+1,d_2-1\right]$,$\left[c_1,c_2-1,c_3+1;d_1-1,d_2+1\right]$,$\left[c_1,c_2,c_3-1;d_1+1,d_2-1\right]$,$\left[c_1,c_2,c_3-1;d_1-1,d_2+1\right]$\\
\hline
\hline
{\scriptsize   Level: 2}\\
\hline
$\left[c_1+2,c_2,c_3;d_1,d_2\right]$,$\left[c_1,c_2+1,c_3;d_1+2,d_2-2\right]$,2$\left[c_1,c_2+1,c_3;d_1,d_2\right]$,$\left[c_1,c_2+1,c_3;d_1-2,d_2+2\right]$,\\
\hline
\end{tabular}
\end{center}
\caption{\label{full a spec (2,0) QT} The spectrum of $\mathcal{A}[c_1,c_2,c_3;d_1,d_2]$ associated with the action of $\tilde{Q}$s.}
\end{table}

\begin{table}[t]
\tiny
\begin{center}
\begin{tabular}{|l|}
\hline
\multicolumn{1}{|c|}{\tiny $Q$ actions for $\mathcal{B}[c_1,c_2,0;d_1,d_2]$}\\
\hline
\hline
{\scriptsize Level: 0}\\
\hline 
{  $\left[c_1,c_2,c_3;d_1,d_2\right]$}\\
\hline
\hline
{\scriptsize   Level: 1}\\
\hline
$\left[c_1+1,c_2,c_3;d_1,d_2+1\right]$,$\left[c_1+1,c_2,c_3;d_1,d_2-1\right]$,$\left[c_1-1,c_2+1,c_3;d_1,d_2+1\right]$,$\left[c_1-1,c_2+1,c_3;d_1,d_2-1\right]$,\\$\left[c_1,c_2-1,c_3+1;d_1,d_2-1\right]$,$\left[c_1,c_2,c_3-1;d_1,d_2-1\right]$\\
\hline
\hline
{\scriptsize Level: 2}\\
\hline 
$[c_1+2,c_2,c_3;d_1,d_2]$,$[c_1,c_2+1,c_3;d_1,d_2+2]$,2$[c_1,c_2+1,c_3;d_1,d_2]$,$[c_1,c_2+1,c_3;d_1,d_2-2]$,$[c_1,c_2-1,c_3;d_1,d_2-2]$,\\
\hline
\end{tabular}
\end{center}
\caption{\label{full b spec (2,0) Q} The spectrum of
  $\mathcal{B}[c_1,c_2,0;d_1,d_2]$ associated with the action of $Q$s.}
\end{table}

\begin{table}
\tiny
\begin{center}
\begin{tabular}{|l|}
\hline
\multicolumn{1}{|c|}{\tiny $\tilde{Q}$ actions for $\mathcal{B}[c_1,c_2,0;d_1,1]$}\\
\hline
\hline
{\scriptsize Level: 0}\\
\hline 
{  $\left[c_1,c_2,0;d_1,1\right]$}\\
\hline
\hline
{\scriptsize   Level: 1}\\
\hline
$[c_1+1,c_2,c_3;d_1+1,d_2-1]$,$[c_1+1,c_2,c_3;d_1-1,d_2+1]$,$[c_1-1,c_2+1,c_3;d_1+1,d_2-1]$,$[c_1-1,c_2+1,c_3;d_1-1,d_2+1]$,\\$[c_1,c_2-1,c_3+1;d_1+1,d_2-1]$,$[c_1,c_2-1,c_3+1;d_1-1,d_2+1]$,$[c_1,c_2,c_3-1;d_1+1,d_2-1]$,$[c_1,c_2,c_3-1;d_1-1,d_2+1]$\\
\hline
\hline
{\scriptsize Level: 2}\\
\hline 
$[c_1+2,c_2,c_3;d_1,d_2]$,$[c_1,c_2+1,c_3;d_1+2,d_2-2]$,2$[c_1,c_2+1,c_3;d_1,d_2]$,$[c_1,c_2+1,c_3;d_1-2,d_2+2]$,\\
\hline
\end{tabular}
\end{center}
\caption{\label{full b spec (2,0) QT d1=1} The spectrum of
  $\mathcal{B}[c_1,c_2,0;d_1,1]$ associated with the action of $\tilde{Q}$s.}
\end{table}

\begin{table}
\tiny
\begin{center}
\begin{tabular}{|l|}
\hline
\multicolumn{1}{|c|}{\tiny $\tilde{Q}$ actions for $\mathcal{B}[c_1,c_2,0;d_1,0]$}\\
\hline
\hline
{\scriptsize Level: 0}\\
\hline 
{  $\left[c_1,c_2,c_3;d_1,d_2\right]$}\\
\hline
\hline
{\scriptsize   Level: 1}\\
\hline
$[c_1+1,c_2,c_3;d_1+1,d_2-1]$,$[c_1+1,c_2,c_3;d_1-1,d_2+1]$,$[c_1-1,c_2+1,c_3;d_1+1,d_2-1]$,$[c_1-1,c_2+1,c_3;d_1-1,d_2+1]$,\\$[c_1,c_2-1,c_3+1;d_1-1,d_2+1]$,$[c_1,c_2,c_3-1;d_1-1,d_2+1]$\\
\hline
\hline
{\scriptsize Level: 2}\\
\hline 
$[c_1+2,c_2,c_3;d_1,d_2]$,$[c_1,c_2+1,c_3;d_1+2,d_2-2]$,2$[c_1,c_2+1,c_3;d_1,d_2]$,$[c_1,c_2+1,c_3;d_1-2,d_2+2]$,\\
\hline
\end{tabular}
\end{center}
\caption{\label{full b spec (2,0) QT d2=0} The spectrum of
  $\mathcal{B}[c_1,c_2,0;d_1,0]$ associated with the action of $\tilde{Q}$s. }
\end{table}

\begin{table}[t]
\tiny
\begin{center}
\begin{tabular}{|l|}
\hline
\multicolumn{1}{|c|}{\tiny $Q$ actions for $\mathcal{C}[c_1,0,0;d_1,d_2]$}\\
\hline
\hline
{\scriptsize Level: 0}\\
\hline 
{  $\left[c_1,c_2,c_3;d_1,d_2\right]$}\\
\hline
\hline
{\scriptsize   Level: 1}\\
\hline
$[c_1+1,c_2,c_3;d_1,d_2+1]$,$[c_1+1,c_2,c_3;d_1,d_2-1]$,$[c_1-1,c_2+1,c_3;d_1,d_2-1]$,$[c_1,c_2-1,c_3+1;d_1,d_2-1]$,\\$[c_1,c_2,c_3-1;d_1,d_2-1]$\\
\hline
\hline
{\scriptsize Level: 2}\\
\hline 
$[c_1+2,c_2,c_3;d_1,d_2]$,$[c_1,c_2+1,c_3;d_1,d_2]$,$[c_1,c_2+1,c_3;d_1,d_2-2]$,$[c_1+1,c_2-1,c_3+1;d_1,d_2]$,$[c_1,c_2-1,c_3;d_1,d_2-2]$,\\$[c_1+1,c_2-1,c_3+1;d_1,d_2-2]$,$[c_1-1,c_2,c_3+1;d_1,d_2-2]$,$[c_1+1,c_2,c_3-1;d_1,d_2]$,$[c_1+1,c_2,c_3-1;d_1,d_2-2]$,\\$[c_1-1,c_2+1,c_3-1;d_1,d_2-2]$\\
\hline 
\hline 
{\scriptsize Level: 3}\\
\hline
$[c_1+1,c_2+1,c_3;d_1,d_2-1]$,$[c_1+2,c_2-1,c_3+1;d_1,d_2-1]$,$[c_1,c_2,c_3+1;d_1,d_2-1]$,$[c_1,c_2,c_3+1;d_1,d_2-3]$,\\$[c_1+2,c_2,c_3-1;d_1,d_2-1]$,$[c_1,c_2+1,c_3-1;d_1,d_2-1]$,$[c_1,c_2+1,c_3-1;d_1,d_2-3]$,$[c_1+1,c_2-1,c_3;d_1,d_2-1]$,\\$[c_1+1,c_2-1,c_3;d_1,d_2-3]$,$[c_1-1,c_2,c_3;d_1,d_2-3]$\\
\hline
\hline
{\scriptsize Level: 4}\\
\hline
$[c_1+1,c_2,c_3+1;d_1,d_2-2]$,$[c_1+1,c_2+1,c_3-1;d_1,d_2-2]$,$[c_1+2,c_2-1,c_3;d_1,d_2-2]$,$[c_1,c_2,c_3;d_1,d_2-2]$,$[c_1,c_2,c_3;d_1,d_2-4]$\\
\hline
\hline
{\scriptsize Level: 5}\\
\hline
$[c_1+1,c_2,c_3;d_1,d_2-3]$\\
\hline
\end{tabular}
\end{center}
\caption{\label{full c spec (2,0) Q} The spectrum of
  $\mathcal{C}[c_1,0,0;d_1,d_2]$ associated with the action of $Q$s.}
\end{table}

\begin{table}
\tiny
\begin{center}
\begin{tabular}{|l|}
\hline
{\hspace{52mm} \tiny $\tilde{Q}$ actions for $\mathcal{C}[c_1,0,0;d_1,2]$}\\
\hline
\hline
{\scriptsize Level: 0}\\
\hline 
{  $\left[c_1,c_2,c_3;d_1,d_2\right]$}\\
\hline
\hline
{\scriptsize   Level: 1}\\
\hline
$\left[c_1+1,c_2,c_3;d_1+1,d_2-1\right]$,$\left[c_1+1,c_2,c_3;d_1-1,d_2+1\right]$,$\left[c_1-1,c_2+1,c_3;d_1+1,d_2-1\right]$,$\left[c_1-1,c_2+1,c_3;d_1-1,d_2+1\right]$,\\$\left[c_1,c_2-1,c_3+1;d_1+1,d_2-1\right]$,$\left[c_1,c_2-1,c_3+1;d_1-1,d_2+1\right]$,$\left[c_1,c_2,c_3-1;d_1+1,d_2-1\right]$,$\left[c_1,c_2,c_3-1;d_1-1,d_2+1\right]$\\
\hline
\hline
{\scriptsize   Level: 2}\\
\hline
$\left[c_1+2,c_2,c_3;d_1,d_2\right]$,$\left[c_1,c_2+1,c_3;d_1+2,d_2-2\right]$,2$\left[c_1,c_2+1,c_3;d_1,d_2\right]$,$\left[c_1,c_2+1,c_3;d_1-2,d_2+2\right]$,\\
\hline
\end{tabular}
\end{center}
\caption{\label{full c spec (2,0) QT d2=2} The spectrum of
  $\mathcal{C}[c_1,0,0;d_1,2]$ associated with the action of $\tilde{Q}$s.}
\end{table}

\begin{table}
\tiny
\begin{center}
\begin{tabular}{|l|}
\hline
{\hspace{52mm} \tiny $\tilde{Q}$ actions for $\mathcal{C}[c_1,0,0;d_1,1]$}\\
\hline
\hline
{\scriptsize Level: 0}\\
\hline 
{  $\left[c_1,c_2,c_3;d_1,d_2\right]$}\\
\hline
\hline
{\scriptsize   Level: 1}\\
\hline
$\left[c_1+1,c_2,c_3;d_1+1,d_2-1\right]$,$\left[c_1+1,c_2,c_3;d_1-1,d_2+1\right]$,$\left[c_1-1,c_2+1,c_3;d_1+1,d_2-1\right]$,$\left[c_1-1,c_2+1,c_3;d_1-1,d_2+1\right]$,\\$\left[c_1,c_2-1,c_3+1;d_1+1,d_2-1\right]$,$\left[c_1,c_2-1,c_3+1;d_1-1,d_2+1\right]$,$\left[c_1,c_2,c_3-1;d_1+1,d_2-1\right]$,$\left[c_1,c_2,c_3-1;d_1-1,d_2+1\right]$\\
\hline
\hline
{\scriptsize   Level: 2}\\
\hline
$\left[c_1+2,c_2,c_3;d_1,d_2\right]$,$\left[c_1,c_2+1,c_3;d_1+2,d_2-2\right]$,2$\left[c_1,c_2+1,c_3;d_1,d_2\right]$,$\left[c_1,c_2+1,c_3;d_1-2,d_2+2\right]$,$\left[c_1,c_2,c_3-2;d_1,d_2\right]$,\\
\hline
\end{tabular}
\end{center}
\caption{\label{full c spec (2,0) QT d2=1} The spectrum of
  $\mathcal{C}[c_1,0,0;d_1,1]$ associated with the action of $\tilde{Q}$s.}
\end{table}

\begin{table}[h]
\tiny
\begin{center}
\begin{tabular}{|l|}
\hline
{\hspace{52mm} \tiny $\tilde{Q}$ actions for $\mathcal{C}[c_1,0,0;d_1,0]$}\\
\hline
\hline
{\scriptsize Level: 0}\\
\hline 
{  $\left[c_1,c_2,c_3;d_1,d_2\right]$}\\
\hline
\hline
{\scriptsize   Level: 1}\\
\hline
$[c_1+1,c_2,c_3;d_1+1,d_2-1]$,$[c_1+1,c_2,c_3;d_1-1,d_2+1]$,$[c_1-1,c_2+1,c_3;d_1-1,d_2+1]$,$[c_1,c_2-1,c_3+1;d_1-1,d_2+1]$,\\$[c_1,c_2,c_3-1;d_1-1,d_2+1]$\\
\hline
\hline
{\scriptsize   Level: 2}\\
\hline
$[c_1+2,c_2,c_3;d_1,d_2]$,$[c_1,c_2+1,c_3;d_1,d_2]$,$[c_1,c_2+1,c_3;d_1-2,d_2+2]$,$[c_1+1,c_2-1,c_3+1;d_1,d_2]$,\\$[c_1+1,c_2-1,c_3+1;d_1-2,d_2+2]$,$[c_1-1,c_2,c_3+1;d_1-2,d_2+2]$,$[c_1+1,c_2,c_3-1;d_1,d_2]$,$[c_1+1,c_2,c_3-1;d_1-2,d_2+2]$,\\$[c_1-1,c_2+1,c_3-1;d_1-2,d_2+2]$,$[c_1,c_2-1,c_3;d_1-2,d_2+2]$\\
\hline
\hline
{\scriptsize Level: 3}\\
\hline
$[c_1+1,c_2+1,c_3;d_1-1,d_2+1]$,$[c_1+2,c_2-1,c_3+1;d_1-1,d_2+1]$,$[c_1,c_2,c_3+1;d_1-1,d_2+1]$,$[c_1,c_2,c_3+1;d_1-3,d_2+3]$,\\$[c_1+2,c_2,c_3-1;d_1-1,d_2+1]$,$[c_1,c_2+1,c_3-1;d_1-1,d_2+1]$,$[c_1,c_2+1,c_3-1;d_1-3,d_2+3]$,$[c_1+1,c_2-1,c_3;d_1-1,d_2+1]$,\\$[c_1+1,c_2-1,c_3;d_1-3,d_2+3]$,$[c_1-1,c_2,c_3;d_1-3,d_2+3]$\\
\hline
\hline
{\scriptsize Level: 4}\\
\hline 
$[c_1+1,c_2,c_3+1;d_1-2,d_2+2]$,$[c_1+1,c_2+1,c_3-1;d_1-2,d_2+2]$,$[c_1+2,c_2-1,c_3;d_1-2,d_2+2]$,$[c_1,c_2,c_3;d_1-2,d_2+2]$,\\$[c_1,c_2,c_3;d_1-4,d_2+4]$\\
\hline
\hline
{\scriptsize Level: 5}\\
\hline
$[c_1+1,c_2,c_3;d_1-3,d_2+3]$\\
\hline
\end{tabular}
\end{center}
\caption{\label{full c spec (2,0) QT d2=0} The spectrum of
  $\mathcal{C}[c_1,0,0;d_1,0]$ associated with the action of $\tilde{Q}$s.}
\end{table}

\begin{table}
\tiny
\begin{center}
\begin{tabular}{|l|}
\hline
{\hspace{52mm} \tiny Full spectrum for $\mathcal{C}[c_1,0,0;1,0]$}\\
\hline
\hline
{\scriptsize Level: 0}\\
\hline 
{  $\left[c_1,0,0;1,0\right]$}\\
\hline
\hline
{\scriptsize   Level: 1}\\
\hline
$[c_1+1,0,0;0,1]$,$[c_1-1,1,0;0,1]$,$[c_1+1,0,0;1,1]$\\
\hline
\hline
{\scriptsize   Level: 2}\\
\hline
$[c_1+2,0,0;0,0]$,$[c_1,1,0;0,0]$,$[c_1-2,2,0;0,0]$,$[c_1+2,0,0;1,0]$,$[c_1,1,0;1,0]$,$[c_1+2,0,0;2,0]$,$[c_1+2,0,0;0,2]$,$[c_1,1,0;0,2]$\\
\hline
\hline
{\scriptsize Level: 3}\\
\hline
-$[c_1-1,0,0;0,1]$,$[c_1+3,0,0;0,1]$,2$[c_1+1,1,0;0,1]$,$[c_1-1,2,0;0,1]$,$[c_1+3,0,0;1,1]$,$[c_1+1,1,0;1,1]$\\
\hline
\hline
{\scriptsize Level: 4}\\
\hline 
-$[c_1,0,0;0,0]$,-$[c_1-2,1,0;0,0]$,$[c_1+2,1,0;0,0]$,$[c_1,2,0;0,0]$,-$[c_1,0,0;1,0]$,$[c_1+4,0,0;1,0]$,$[c_1+2,1,0;1,0]$,$[c_1,2,0;1,0]$,-$[c_1,0,0;0,2]$,\\$[c_1+2,1,0;0,2]$\\
\hline
\hline
{\scriptsize Level: 5}\\
\hline
-2$[c_1+1,0,0;0,1]$,-$[c_1-1,1,0;0,1]$,$[c_1+3,1,0;0,1]$,$[c_1+1,2,0;0,1]$,-$[c_1+1,0,0;1,1]$\\
\hline
\hline
{\scriptsize Level: 6}\\
\hline
-$[c_1+2,0,0,0,0]$,-$[c_1,1,0,0,0]$,$[c_1+2,2,0,0,0]$,-$[c_1+2,0,0,1,0]$,-$[c_1,1,0,1,0]$,-$[c_1+2,0,0,0,2]$\\
\hline
\hline
{\scriptsize Level: 7}\\
\hline
$-[c_1+3,0,0;0,1]$,-$[c_1+1,1,0;0,1]$\\
\hline
\hline
{\scriptsize Level: 8}\\
\hline
-$[c_1+2,1,0;0,0]$\\
\hline
\end{tabular}
\end{center}
\caption{\label{full c spec (2,0) (1,0) Q and QT} The spectrum of $\mathcal{C}[c_1,0,0;1,0]$.}
\end{table}

\begin{table}
\tiny
\begin{center}
\begin{tabular}{|l|}
\hline
{\hspace{52mm} \tiny Full spectrum for $\mathcal{C}[c_1,0,0;0,1]$}\\
\hline
\hline
{\scriptsize Level: 0}\\
\hline 
{  $\left[c_1,0,0;0,1\right]$}\\
\hline
\hline
{\scriptsize   Level: 1}\\
\hline
$[c_1+1,0,0;0,0]$,$[c_1-1,1,0;0,0]$,$[c_1+1,0,0;1,0]$,$[c_1-1,1,0;1,0]$,$[c_1+1,1,0;0,2]$\\
\hline
\hline
{\scriptsize   Level: 2}\\
\hline
2$[c_1+2,0,0;0,1]$,2$[c_1,1,0;0,1]$,$[c_1-2,2,0;0,1]$,$[c_1+2,0,0;1,1]$,$[c_1,1,0;1,1]$\\
\hline
\hline
{\scriptsize Level: 3}\\
\hline
-$[c_1-1,0,0;0,0]$,$[c_1+3,0,0;0,0]$,2$[c_1+1,1,0;0,0]$,$[c_1-1,2,0;0,0]$,$[c_1-3,3,0;0,0]$,-$[c_1-1,0,0;1,0]$,$[c_1+3,0,0;1,0]$,2$[c_1+1,1,0;1,0]$,\\$[c_1-1,2,0;1,0]$,$[c_1+1,1,0;2,0]$,$[c_1+3,0,0;0,2]$,$[c_1+1,1,0;0,2]$,$[c_1-1,2,0;0,2]$\\
\hline
\hline
{\scriptsize Level: 4}\\
\hline 
-2$[c_1,0,0;0,1]$,$[c_1+4,0,0;0,1]$,-$[c_1-2,1,0;0,1]$,2$[c_1+2,1,0;0,1]$,2$[c_1,2,0;0,1]$,$[c_1-2,3,0;0,1]$,-$[c_1,0,0;1,1]$,$[c_1+2,1,0;1,1]$,\\$[c_1,2,0;1,1]$\\
\hline
\hline
{\scriptsize Level: 5}\\
\hline
-2$[c_1+1,0,0;0,0]$,-$[c_1-1,1,0;0,0]$,$[c_1+3,1,0;0,0]$,-$[c_1-3,2,0;0,0]$,$[c_1+1,2,0;0,0]$,$[c_1-1,3,0;0,0]$,-2$[c_1+1,0,0;1,0]$,-$[c_1-1,1,0;1,0]$,\\$[c_1+3,1,0;1,0]$,$[c_1+1,2,0;1,0]$,$[c_1-1,3,0;1,0]$,-$[c_1+1,0,0;2,0]$,-$[c_1+1,0,0;0,2]$,-$[c_1-1,1,0;0,2]$,$[c_1+1,2,0;0,2]$\\
\hline
\hline
{\scriptsize Level: 6}\\
\hline
-2$[c_1+2,0,0;0,1]$,-2$[c_1,1,0;0,1]$,-$[c_1-2,2,0;0,1]$,$[c_1+2,2,0;0,1]$,$[c_1,3,0;0,1]$,-$[c_1+2,0,0;1,1]$,-$[c_1,1,0;1,1]$\\
\hline
\hline
{\scriptsize Level: 7}\\
\hline
-$[c_1+3,0,0;0,0]$,-$[c_1+1,1,0;0,0]$,-$[c_1-1,2,0;0,0]$,$[c_1+1,3,0;0,0]$,-$[c_1+3,0,0;1,0]$,-$[c_1+1,1,0;1,0]$,-$[c_1-1,2,0;1,0]$,-$[c_1+1,1,0;0,2]$\\
\hline
\hline
{\scriptsize Level: 8}\\
\hline
-$[c_1+2,1,0;0,1]$,-$[c_1,2,0;0,1]$\\
\hline
\hline
{\scriptsize Level: 9}\\
\hline
-$[c_1+1,2,0;0,0]$\\
\hline
\end{tabular}
\end{center}
\caption{\label{full c spec (2,0) (0,1) Q and QT} The spectrum of $\mathcal{C}[c_1,0,0;0,1]$.}
\end{table}

\begin{table}[t]
\tiny
\begin{center}
\begin{tabular}{|l|}
\hline
{\hspace{52mm} \tiny $\tilde{Q}$ actions for $\mathcal{D}[0,0,0;d_1,3]$}\\
\hline
\hline
{\scriptsize Level: 0}\\
\hline 
{  $\left[c_1,c_2,c_3;d_1,d_2\right]$}\\
\hline
\hline
{\scriptsize   Level: 1}\\
\hline
$\left[c_1+1,c_2,c_3;d_1+1,d_2-1\right]$,$\left[c_1+1,c_2,c_3;d_1-1,d_2+1\right]$,$\left[c_1-1,c_2+1,c_3;d_1+1,d_2-1\right]$,$\left[c_1-1,c_2+1,c_3;d_1-1,d_2+1\right]$,\\$\left[c_1,c_2-1,c_3+1;d_1+1,d_2-1\right]$,$\left[c_1,c_2-1,c_3+1;d_1-1,d_2+1\right]$,$\left[c_1,c_2,c_3-1;d_1+1,d_2-1\right]$,$\left[c_1,c_2,c_3-1;d_1-1,d_2+1\right]$\\
\hline
\hline
{\scriptsize   Level: 2}\\
\hline
$\left[c_1+2,c_2,c_3;d_1,d_2\right]$,$\left[c_1,c_2+1,c_3;d_1+2,d_2-2\right]$,2$\left[c_1,c_2+1,c_3;d_1,d_2\right]$,$\left[c_1,c_2+1,c_3;d_1-2,d_2+2\right]$,\\
\hline
\end{tabular}
\end{center}
\caption{\label{D spec (2,0) d2=3 QT} The spectrum of
  $\mathcal{D}[0,0,0;d_1,3]$ associated with the action of $\tilde{Q}$s.}
\end{table}

\begin{table}
\tiny
\begin{center}
\begin{tabular}{|l|}
\hline
{\hspace{52mm} \tiny $\tilde{Q}$ actions for $\mathcal{D}[0,0,0;d_1,2]$}\\
\hline
\hline
{\scriptsize Level: 0}\\
\hline 
{  $\left[c_1,c_2,c_3;d_1,d_2\right]$}\\
\hline
\hline
{\scriptsize   Level: 1}\\
\hline
$\left[c_1+1,c_2,c_3;d_1+1,d_2-1\right]$,$\left[c_1+1,c_2,c_3;d_1-1,d_2+1\right]$,$\left[c_1-1,c_2+1,c_3;d_1+1,d_2-1\right]$,$\left[c_1-1,c_2+1,c_3;d_1-1,d_2+1\right]$,\\$\left[c_1,c_2-1,c_3+1;d_1+1,d_2-1\right]$,$\left[c_1,c_2-1,c_3+1;d_1-1,d_2+1\right]$,$\left[c_1,c_2,c_3-1;d_1+1,d_2-1\right]$,$\left[c_1,c_2,c_3-1;d_1-1,d_2+1\right]$\\
\hline
\hline
{\scriptsize   Level: 2}\\
\hline
$\left[c_1+2,c_2,c_3;d_1,d_2\right]$,$\left[c_1,c_2+1,c_3;d_1+2,d_2-2\right]$,2$\left[c_1,c_2+1,c_3;d_1,d_2\right]$,$\left[c_1,c_2+1,c_3;d_1-2,d_2+2\right]$,\\
\hline
\end{tabular}
\end{center}
\caption{\label{D spec (2,0) d2=2 QT} The spectrum of
  $\mathcal{D}[0,0,0;d_1,2]$ associated with the action of $\tilde{Q}$s.}
\end{table}

\end{appendix}

\clearpage


\bibliography{6d4danomaly}

\begin{thebibliography}{10}
\ifx\href\asklfhas\newcommand{\href}[2]{#2}\fi
\ifx\arxivref\asklfhas\newcommand{\arxivref}[2]{\href{http://arxiv.org/abs/#1}{#2}}\fi
\ifx\doiref\asklfhas\newcommand{\doiref}[2]{\href{http://dx.doi.org/#1}{#2}}\fi
\parskip 0pt
\normalsize

\bibitem{Montonen:1977sn}
C.~Montonen \& D.~I. Olive,
\textit{``{Magnetic Monopoles as Gauge Particles?}''},
\doiref{10.1016/0370-2693(77)90076-4}{Phys.~Lett. \textbf{B72}, 117
  (1977)\ignorespaces}\ignorespaces
\bibitem{Goddard:1976qe}
P.~Goddard, J.~Nuyts \& D.~I. Olive,
\textit{``{Gauge Theories and Magnetic Charge}''},
\doiref{10.1016/0550-3213(77)90221-8}{Nucl.~Phys. \textbf{B125}, 1
  (1977)\ignorespaces}\ignorespaces
\bibitem{Witten:1978mh}
E.~Witten \& D.~I. Olive,
\textit{``{Supersymmetry Algebras That Include Topological Charges}''},
\doiref{10.1016/0370-2693(78)90357-X}{Phys.~Lett. \textbf{B78}, 97
  (1978)\ignorespaces}\ignorespaces
\bibitem{Seiberg:1994aj}
N.~Seiberg \& E.~Witten,
\textit{``{Monopoles, duality and chiral symmetry breaking in N=2
  supersymmetric QCD}''},
\doiref{10.1016/0550-3213(94)90214-3}{Nucl.~Phys. \textbf{B431}, 484
  (1994)\ignorespaces}\ignorespaces,
\normalsize{\texttt{\arxivref{hep-th/9408099}{hep-th/9408099}}}\ignorespaces
\bibitem{Argyres:2007cn}
P.~C. Argyres \& N.~Seiberg,
\textit{``{S-duality in N=2 supersymmetric gauge theories}''},
\doiref{10.1088/1126-6708/2007/12/088}{JHEP \textbf{0712}, 088
  (2007)\ignorespaces}\ignorespaces,
\normalsize{\texttt{\arxivref{0711.0054}{arXiv:0711.0054}}}\ignorespaces
\bibitem{Gaiotto:2009we}
D.~Gaiotto,
\textit{``{N=2 dualities}''},
\doiref{10.1007/JHEP08(2012)034}{JHEP \textbf{1208}, 034
  (2012)\ignorespaces}\ignorespaces,
\normalsize{\texttt{\arxivref{0904.2715}{arXiv:0904.2715}}}\ignorespaces
\bibitem{Green:2010da}
D.~Green, Z.~Komargodski, N.~Seiberg, Y.~Tachikawa \& B.~Wecht,
\textit{``{Exactly Marginal Deformations and Global Symmetries}''},
\doiref{10.1007/JHEP06(2010)106}{JHEP \textbf{1006}, 106
  (2010)\ignorespaces}\ignorespaces,
\normalsize{\texttt{\arxivref{1005.3546}{arXiv:1005.3546}}}\ignorespaces
\bibitem{Seiberg:1994pq}
N.~Seiberg,
\textit{``{Electric - magnetic duality in supersymmetric nonAbelian gauge
  theories}''},
\doiref{10.1016/0550-3213(94)00023-8}{Nucl.~Phys. \textbf{B435}, 129
  (1995)\ignorespaces}\ignorespaces,
\normalsize{\texttt{\arxivref{hep-th/9411149}{hep-th/9411149}}}\ignorespaces
\bibitem{Intriligator:1996ex}
K.~A. Intriligator \& N.~Seiberg,
\textit{``{Mirror symmetry in three-dimensional gauge theories}''},
\doiref{10.1016/0370-2693(96)01088-X}{Phys.~Lett. \textbf{B387}, 513
  (1996)\ignorespaces}\ignorespaces,
\normalsize{\texttt{\arxivref{hep-th/9607207}{hep-th/9607207}}}\ignorespaces
\bibitem{Nahm:1977tg}
W.~Nahm,
\textit{``{Supersymmetries and their Representations}''},
\doiref{10.1016/0550-3213(78)90218-3}{Nucl.~Phys. \textbf{B135}, 149
  (1978)\ignorespaces}\ignorespaces
\bibitem{Romelsberger:2005eg}
C.~Romelsberger,
\textit{``{Counting chiral primaries in N = 1, d=4 superconformal field
  theories}''},
\doiref{10.1016/j.nuclphysb.2006.03.037}{Nucl.~Phys. \textbf{B747}, 329
  (2006)\ignorespaces}\ignorespaces,
\normalsize{\texttt{\arxivref{hep-th/0510060}{hep-th/0510060}}}\ignorespaces
\bibitem{Kinney:2005ej}
J.~Kinney, J.~M. Maldacena, S.~Minwalla \& S.~Raju,
\textit{``{An Index for 4 dimensional super conformal theories}''},
\doiref{10.1007/s00220-007-0258-7}{Commun.Math.Phys. \textbf{275}, 209
  (2007)\ignorespaces}\ignorespaces,
\normalsize{\texttt{\arxivref{hep-th/0510251}{hep-th/0510251}}}\ignorespaces
\bibitem{Bhattacharya:2008zy}
J.~Bhattacharya, S.~Bhattacharyya, S.~Minwalla \& S.~Raju,
\textit{``{Indices for Superconformal Field Theories in 3,5 and 6
  Dimensions}''},
\doiref{10.1088/1126-6708/2008/02/064}{JHEP \textbf{0802}, 064
  (2008)\ignorespaces}\ignorespaces,
\normalsize{\texttt{\arxivref{0801.1435}{arXiv:0801.1435}}}\ignorespaces
\bibitem{Beem:2014kka}
C.~Beem, L.~Rastelli \& B.~C. van~Rees,
\textit{``{$ \mathcal{W} $ symmetry in six dimensions}''},
\doiref{10.1007/JHEP05(2015)017}{JHEP \textbf{1505}, 017
  (2015)\ignorespaces}\ignorespaces,
\normalsize{\texttt{\arxivref{1404.1079}{arXiv:1404.1079}}}\ignorespaces
\bibitem{Beem:2013sza}
C.~Beem, M.~Lemos, P.~Liendo, W.~Peelaers, L.~Rastelli \& B.~C. van~Rees,
\textit{``{Infinite Chiral Symmetry in Four Dimensions}''},
\doiref{10.1007/s00220-014-2272-x}{Commun.~Math.~Phys. \textbf{336}, 1359
  (2015)\ignorespaces}\ignorespaces,
\normalsize{\texttt{\arxivref{1312.5344}{arXiv:1312.5344}}}\ignorespaces
\bibitem{Cordova:2015fha}
C.~Cordova, T.~T. Dumitrescu \& K.~Intriligator,
\textit{``{Anomalies, Renormalization Group Flows, and the a-Theorem in
  Six-Dimensional (1,0) Theories}''},
\normalsize{\texttt{\arxivref{1506.03807}{arXiv:1506.03807}}}\ignorespaces
\bibitem{Louis:2015mka}
J.~Louis \& S.~Lüst,
\textit{``{Supersymmetric AdS$_{7}$ backgrounds in half-maximal supergravity
  and marginal operators of (1, 0) SCFTs}''},
\doiref{10.1007/JHEP10(2015)120}{JHEP \textbf{1510}, 120
  (2015)\ignorespaces}\ignorespaces,
\normalsize{\texttt{\arxivref{1506.08040}{arXiv:1506.08040}}}\ignorespaces
\bibitem{Cordova:2016xhm}
C.~Cordova, T.~T. Dumitrescu \& K.~Intriligator,
\textit{``{Deformations of Superconformal Theories}''},
\normalsize{\texttt{\arxivref{1602.01217}{arXiv:1602.01217}}}\ignorespaces
\bibitem{Beem:2015aoa}
C.~Beem, M.~Lemos, L.~Rastelli \& B.~C. van~Rees,
\textit{``{The (2, 0) superconformal bootstrap}''},
\doiref{10.1103/PhysRevD.93.025016}{Phys.~Rev. \textbf{D93}, 025016
  (2016)\ignorespaces}\ignorespaces,
\normalsize{\texttt{\arxivref{1507.05637}{arXiv:1507.05637}}}\ignorespaces
\bibitem{Minwalla:1997ka}
S.~Minwalla,
\textit{``{Restrictions imposed by superconformal invariance on quantum field
  theories}''},
Adv.Theor.Math.Phys. \textbf{2}, 781 (1998)\ignorespaces\ignorespaces,
\normalsize{\texttt{\arxivref{hep-th/9712074}{hep-th/9712074}}}\ignorespaces
\bibitem{Dobrev:1985qv}
V.~K. Dobrev \& V.~B. Petkova,
\textit{``{All Positive Energy Unitary Irreducible Representations of Extended
  Conformal Supersymmetry}''},
\doiref{10.1016/0370-2693(85)91073-1}{Phys.~Lett. \textbf{B162}, 127
  (1985)\ignorespaces}\ignorespaces
\bibitem{Dobrev:1985vh}
V.~K. Dobrev \& V.~B. Petkova,
\textit{``{On the Group Theoretical Approach to Extended Conformal
  Supersymmetry: Classification of Multiplets}''},
\doiref{10.1007/BF00397755}{Lett.~Math.~Phys. \textbf{9}, 287
  (1985)\ignorespaces}\ignorespaces
\bibitem{Dobrev:1985qz}
V.~K. Dobrev \& V.~B. Petkova,
\textit{``{Group Theoretical Approach to Extended Conformal Supersymmetry:
  Function Space Realizations and Invariant Differential Operators}''},
Fortsch.~Phys. \textbf{35}, 537 (1987)\ignorespaces\ignorespaces
\bibitem{Dobrev:2002dt}
V.~K. Dobrev,
\textit{``{Positive energy unitary irreducible representations of D = 6
  conformal supersymmetry}''},
\doiref{10.1088/0305-4470/35/33/308}{J.~Phys. \textbf{A35}, 7079
  (2002)\ignorespaces}\ignorespaces,
\normalsize{\texttt{\arxivref{hep-th/0201076}{hep-th/0201076}}}\ignorespaces
\bibitem{Witten:1995zh}
E.~Witten,
\textit{``{Some comments on string dynamics}''},
in \textit{``{Future perspectives in string theory. Proceedings, Conference,
  Strings'95, Los Angeles, USA, March 13-18, 1995}''}\bibitem{Seiberg:1996bd}
N.~Seiberg,
\textit{``{Five-dimensional SUSY field theories, nontrivial fixed points and
  string dynamics}''},
\doiref{10.1016/S0370-2693(96)01215-4}{Phys.~Lett. \textbf{B388}, 753
  (1996)\ignorespaces}\ignorespaces,
\normalsize{\texttt{\arxivref{hep-th/9608111}{hep-th/9608111}}}\ignorespaces
\bibitem{Morrison:1996xf}
D.~R. Morrison \& N.~Seiberg,
\textit{``{Extremal transitions and five-dimensional supersymmetric field
  theories}''},
\doiref{10.1016/S0550-3213(96)00592-5}{Nucl.~Phys. \textbf{B483}, 229
  (1997)\ignorespaces}\ignorespaces,
\normalsize{\texttt{\arxivref{hep-th/9609070}{hep-th/9609070}}}\ignorespaces
\bibitem{Douglas:1996xp}
M.~R. Douglas, S.~H. Katz \& C.~Vafa,
\textit{``{Small instantons, Del Pezzo surfaces and type I-prime theory}''},
\doiref{10.1016/S0550-3213(97)00281-2}{Nucl.~Phys. \textbf{B497}, 155
  (1997)\ignorespaces}\ignorespaces,
\normalsize{\texttt{\arxivref{hep-th/9609071}{hep-th/9609071}}}\ignorespaces
\bibitem{Intriligator:1997pq}
K.~A. Intriligator, D.~R. Morrison \& N.~Seiberg,
\textit{``{Five-dimensional supersymmetric gauge theories and degenerations of
  Calabi-Yau spaces}''},
\doiref{10.1016/S0550-3213(97)00279-4}{Nucl.~Phys. \textbf{B497}, 56
  (1997)\ignorespaces}\ignorespaces,
\normalsize{\texttt{\arxivref{hep-th/9702198}{hep-th/9702198}}}\ignorespaces
\bibitem{Heckman:2015bfa}
J.~J. Heckman, D.~R. Morrison, T.~Rudelius \& C.~Vafa,
\textit{``{Atomic Classification of 6D SCFTs}''},
\doiref{10.1002/prop.201500024}{Fortsch.~Phys. \textbf{63}, 468
  (2015)\ignorespaces}\ignorespaces,
\normalsize{\texttt{\arxivref{1502.05405}{arXiv:1502.05405}}}\ignorespaces
\bibitem{Bhardwaj:2015xxa}
L.~Bhardwaj,
\textit{``{Classification of 6d $ \mathcal{N}=\left(1,0\right) $ gauge
  theories}''},
\doiref{10.1007/JHEP11(2015)002}{JHEP \textbf{1511}, 002
  (2015)\ignorespaces}\ignorespaces,
\normalsize{\texttt{\arxivref{1502.06594}{arXiv:1502.06594}}}\ignorespaces
\bibitem{Maldacena:2011jn}
J.~Maldacena \& A.~Zhiboedov,
\textit{``{Constraining Conformal Field Theories with A Higher Spin
  Symmetry}''},
\doiref{10.1088/1751-8113/46/21/214011}{J.~Phys. \textbf{A46}, 214011
  (2013)\ignorespaces}\ignorespaces,
\normalsize{\texttt{\arxivref{1112.1016}{arXiv:1112.1016}}}\ignorespaces
\bibitem{Maldacena:2012sf}
J.~Maldacena \& A.~Zhiboedov,
\textit{``{Constraining conformal field theories with a slightly broken higher
  spin symmetry}''},
\doiref{10.1088/0264-9381/30/10/104003}{Class.~Quant.~Grav. \textbf{30}, 104003
  (2013)\ignorespaces}\ignorespaces,
\normalsize{\texttt{\arxivref{1204.3882}{arXiv:1204.3882}}}\ignorespaces
\bibitem{Alba:2015upa}
V.~Alba \& K.~Diab,
\textit{``{Constraining conformal field theories with a higher spin symmetry in
  d $>$ 3 dimensions}''},
\normalsize{\texttt{\arxivref{1510.02535}{arXiv:1510.02535}}}\ignorespaces
\bibitem{Buican:2014hfa}
M.~Buican, S.~Giacomelli, T.~Nishinaka \& C.~Papageorgakis,
\textit{``{Argyres-Douglas Theories and S-Duality}''},
\doiref{10.1007/JHEP02(2015)185}{JHEP \textbf{1502}, 185
  (2015)\ignorespaces}\ignorespaces,
\normalsize{\texttt{\arxivref{1411.6026}{arXiv:1411.6026}}}\ignorespaces
\bibitem{DelZotto:2015rca}
M.~Del~Zotto, C.~Vafa \& D.~Xie,
\textit{``{Geometric engineering, mirror symmetry and $
  6{\mathrm{d}}_{\left(1,0\right)}\to
  4{\mathrm{d}}_{\left(\mathcal{N}=2\right)} $}''},
\doiref{10.1007/JHEP11(2015)123}{JHEP \textbf{1511}, 123
  (2015)\ignorespaces}\ignorespaces,
\normalsize{\texttt{\arxivref{1504.08348}{arXiv:1504.08348}}}\ignorespaces
\bibitem{Buican:2016arp}
M.~Buican \& T.~Nishinaka,
\textit{``{Conformal Manifolds in Four Dimensions and Chiral Algebras}''},
\normalsize{\texttt{\arxivref{1603.00887}{arXiv:1603.00887}}}\ignorespaces
\bibitem{CordovaTalk1}
C.~C\'ordova,
\textit{``{Deformations of Superconformal Field Theories}''},
{Autumn Symposium on String/M Theory 2014; Princeton University Seminar
  2014}\ignorespaces,
\href{http://media.kias.re.kr/detailPage.do?pro\_seq=564\&type=p}{\texttt{http://media.kias.re.kr/detailPage.do?pro\_seq=564\&type=p}}
\bibitem{Dolan:2002zh}
F.~A. Dolan \& H.~Osborn,
\textit{``{On short and semi-short representations for four-dimensional
  superconformal symmetry}''},
\doiref{10.1016/S0003-4916(03)00074-5}{Annals~Phys. \textbf{307}, 41
  (2003)\ignorespaces}\ignorespaces,
\normalsize{\texttt{\arxivref{hep-th/0209056}{hep-th/0209056}}}\ignorespaces
\bibitem{Dolan:2005wy}
F.~A. Dolan,
\textit{``{Character formulae and partition functions in higher dimensional
  conformal field theory}''},
\doiref{10.1063/1.2196241}{J.~Math.~Phys. \textbf{47}, 062303
  (2006)\ignorespaces}\ignorespaces,
\normalsize{\texttt{\arxivref{hep-th/0508031}{hep-th/0508031}}}\ignorespaces
\bibitem{Penedones:2015aga}
J.~Penedones, E.~Trevisani \& M.~Yamazaki,
\textit{``{Recursion Relations for Conformal Blocks}''},
\normalsize{\texttt{\arxivref{1509.00428}{arXiv:1509.00428}}}\ignorespaces
\bibitem{Yamazaki:2016vqi}
M.~Yamazaki,
\textit{``{Comments on Determinant Formulas for General CFTs}''},
\normalsize{\texttt{\arxivref{1601.04072}{arXiv:1601.04072}}}\ignorespaces
\bibitem{Oshima:2016gqy}
Y.~Oshima \& M.~Yamazaki,
\textit{``{Determinant Formula for Parabolic Verma Modules of Lie
  Superalgebras}''},
\normalsize{\texttt{\arxivref{1603.06705}{arXiv:1603.06705}}}\ignorespaces
\bibitem{Cordova:2016}
C.~Cordova, T.~T. Dumitrescu \& K.~Intriligator,
\textit{``{Multiplets of superconformal symmetry in diverse dimensions.}''},
{to appear}\ignorespaces
\bibitem{IntriligatorTalk}
K.~Intriligator,
\textit{``{Anomalies, RG flows, and the a-theorem in six-dimensional (1,0)
  theories}''},
{Strings 2015}\ignorespaces,
\href{https://strings2015.icts.res.in}{\texttt{https://strings2015.icts.res.in}}
\bibitem{DumitrescuTalk}
T.~Dumitrescu,
\textit{``{Anomalies, RG Flows, and the a-theorem in 6d — Part I}''},
{2015 Simons Summer Workshop}\ignorespaces,
\href{{http://scgp.stonybrook.edu/archives/category/videos}}{\texttt{{http://scgp.stonybrook.edu/archives/category/videos}}}
\bibitem{CordovaTalk2}
C.~C\'ordova,
\textit{``{Anomalies, RG Flows, and the a-theorem in 6d — Part II}''},
{2015 Simons Summer Workshop}\ignorespaces,
\href{{http://scgp.stonybrook.edu/archives/category/videos}}{\texttt{{http://scgp.stonybrook.edu/archives/category/videos}}}
\bibitem{CordovaTalk3}
C.~C\'ordova,
\textit{``{Anomalies RG-Flows and the a-Theorem in Six-Dimensions}''},
{London Triangle Seminar, December 2015}\ignorespaces
\bibitem{Bianchi:2006ti}
M.~Bianchi, F.~A. Dolan, P.~J. Heslop \& H.~Osborn,
\textit{``{N=4 superconformal characters and partition functions}''},
\doiref{10.1016/j.nuclphysb.2006.12.005}{Nucl.~Phys. \textbf{B767}, 163
  (2007)\ignorespaces}\ignorespaces,
\normalsize{\texttt{\arxivref{hep-th/0609179}{hep-th/0609179}}}\ignorespaces
\bibitem{Kim:2012gu}
H.-C. Kim, S.-S. Kim \& K.~Lee,
\textit{``{5-dim Superconformal Index with Enhanced En Global Symmetry}''},
\doiref{10.1007/JHEP10(2012)142}{JHEP \textbf{1210}, 142
  (2012)\ignorespaces}\ignorespaces,
\normalsize{\texttt{\arxivref{1206.6781}{arXiv:1206.6781}}}\ignorespaces
\bibitem{Rodriguez-Gomez:2013dpa}
D.~Rodríguez-Gómez \& G.~Zafrir,
\textit{``{On the 5d instanton index as a Hilbert series}''},
\doiref{10.1016/j.nuclphysb.2013.11.006}{Nucl.~Phys. \textbf{B878}, 1
  (2014)\ignorespaces}\ignorespaces,
\normalsize{\texttt{\arxivref{1305.5684}{arXiv:1305.5684}}}\ignorespaces
\bibitem{Tachikawa:2015mha}
Y.~Tachikawa,
\textit{``{Instanton operators and symmetry enhancement in 5d supersymmetric
  gauge theories}''},
\doiref{10.1093/ptep/ptv040}{PTEP \textbf{2015}, 043B06
  (2015)\ignorespaces}\ignorespaces,
\normalsize{\texttt{\arxivref{1501.01031}{arXiv:1501.01031}}}\ignorespaces
\bibitem{Kac:1977qb}
V.~G. Kac,
\textit{``{A Sketch of Lie Superalgebra Theory}''},
\doiref{10.1007/BF01609166}{Commun.~Math.~Phys. \textbf{53}, 31
  (1977)\ignorespaces}\ignorespaces
\bibitem{Frappat:1996pb}
L.~Frappat, P.~Sorba \& A.~Sciarrino,
\textit{``{Dictionary on Lie superalgebras}''},
\normalsize{\texttt{\arxivref{hep-th/9607161}{hep-th/9607161}}}\ignorespaces
\bibitem{Hwang:2014uwa}
C.~Hwang, J.~Kim, S.~Kim \& J.~Park,
\textit{``{General instanton counting and 5d SCFT}''},
\doiref{10.1007/JHEP07(2015)063, 10.1007/JHEP04(2016)094}{JHEP \textbf{1507},
  063 (2015)\ignorespaces}\ignorespaces,
\normalsize{\texttt{\arxivref{1406.6793}{arXiv:1406.6793}}}\ignorespaces,
[Addendum: JHEP04,094(2016)]\ignorespaces
\bibitem{Passias:2016fkm}
A.~Passias \& A.~Tomasiello,
\textit{``{Spin-2 spectrum of six-dimensional field theories}''},
\normalsize{\texttt{\arxivref{1604.04286}{arXiv:1604.04286}}}\ignorespaces
\bibitem{Howe:1987ik}
P.~S. Howe \& A.~Umerski,
\textit{``{Anomaly Multiplets in Six-Dimensions and Ten-Dimensions}''},
\doiref{10.1016/0370-2693(87)90158-4}{Phys.~Lett. \textbf{B198}, 57
  (1987)\ignorespaces}\ignorespaces
\bibitem{Kuzenko:2015xiz}
S.~M. Kuzenko, J.~Novak \& I.~B. Samsonov,
\textit{``{The Anomalous Current Multiplet in 6D Minimal Supersymmetry}''},
\doiref{10.1007/JHEP02(2016)132}{JHEP \textbf{1602}, 132
  (2016)\ignorespaces}\ignorespaces,
\normalsize{\texttt{\arxivref{1511.06582}{arXiv:1511.06582}}}\ignorespaces
\bibitem{Buican:2013ica}
M.~Buican,
\textit{``{Minimal Distances Between SCFTs}''},
\doiref{10.1007/JHEP01(2014)155}{JHEP \textbf{1401}, 155
  (2014)\ignorespaces}\ignorespaces,
\normalsize{\texttt{\arxivref{1311.1276}{arXiv:1311.1276}}}\ignorespaces
\bibitem{Buican:2014qla}
M.~Buican, T.~Nishinaka \& C.~Papageorgakis,
\textit{``{Constraints on chiral operators in $ \mathcal{N}=2 $ SCFTs}''},
\doiref{10.1007/JHEP12(2014)095}{JHEP \textbf{1412}, 095
  (2014)\ignorespaces}\ignorespaces,
\normalsize{\texttt{\arxivref{1407.2835}{arXiv:1407.2835}}}\ignorespaces
\bibitem{Ohmori:2014kda}
K.~Ohmori, H.~Shimizu, Y.~Tachikawa \& K.~Yonekura,
\textit{``{Anomaly polynomial of general 6d SCFTs}''},
\doiref{10.1093/ptep/ptu140}{PTEP \textbf{2014}, 103B07
  (2014)\ignorespaces}\ignorespaces,
\normalsize{\texttt{\arxivref{1408.5572}{arXiv:1408.5572}}}\ignorespaces
\bibitem{Intriligator:2014eaa}
K.~Intriligator,
\textit{``{6d, $ \mathcal{N}=\left(1,\;0\right) $ Coulomb branch anomaly
  matching}''},
\doiref{10.1007/JHEP10(2014)162}{JHEP \textbf{1410}, 162
  (2014)\ignorespaces}\ignorespaces,
\normalsize{\texttt{\arxivref{1408.6745}{arXiv:1408.6745}}}\ignorespaces
\bibitem{fuchs2003symmetries}
J.~Fuchs \& C.~Schweigert,
\textit{``Symmetries, Lie Algebras and Representations: A Graduate Course for
  Physicists''},
Cambridge University Press (2003)\ignorespaces
\end{thebibliography}

\end{document}